\documentclass[twoside]{book}
\usepackage{amsmath}
\usepackage{graphicx}
\usepackage{amsfonts}
\usepackage{amssymb}
\usepackage[dvips]{epsfig}
\usepackage{palatino,euler}
\usepackage{fancyheadings}
\begin{document}

\frontmatter

\thispagestyle{empty}
\centerline{
\psfig{file=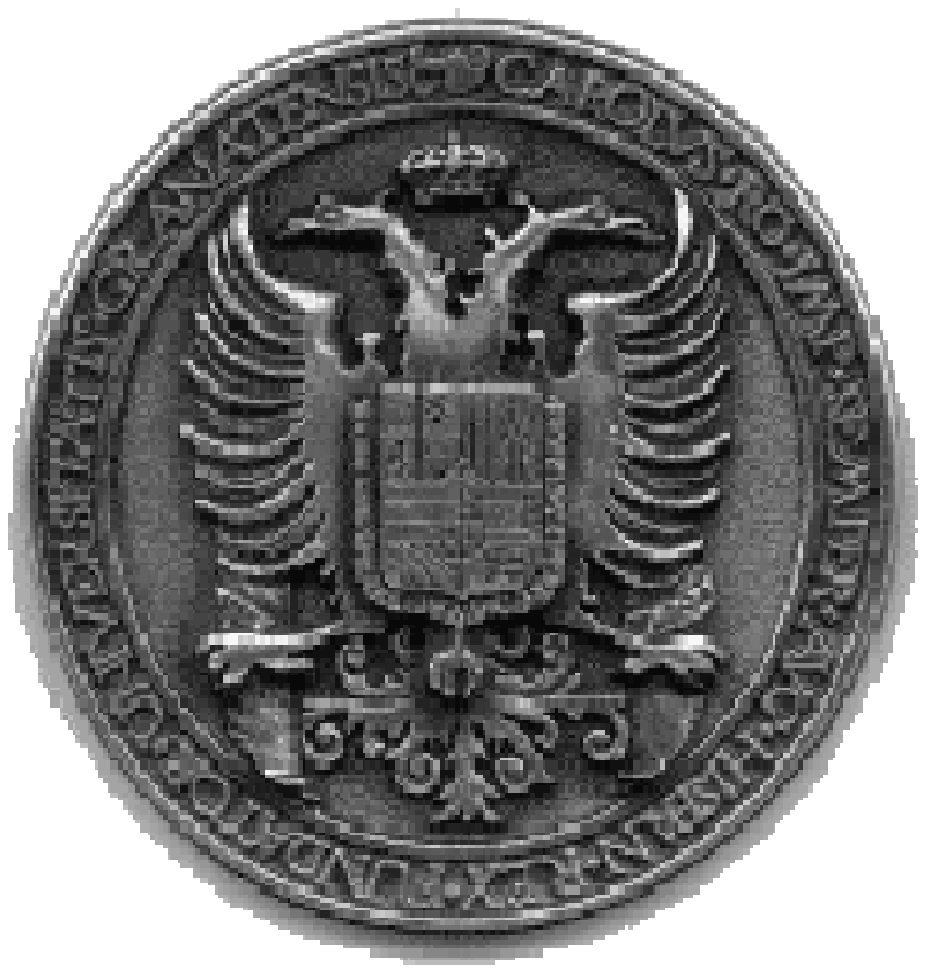,width=3.5cm}
}
\begin{center}
{\Large
Instituto Carlos I de F\'{\i}sica Te\'orica y Computacional \\
y Depto. de Electromagnetismo y F\'{\i}sica de la Materia \\
Universidad de Granada \\} 
\vspace*{2.0cm}
{\Huge \bf Some Aspects on Dynamics of Nonequilibrium Systems \\}
\vspace*{1.0cm}
{\Large \bf Metastability, Avalanches, Phase Separation, \\ 
Absorbing States and Heat Conduction}
\end{center}
\vspace*{1.5cm}
\begin{center}
\Large
PABLO HURTADO FERN\'ANDEZ \\
\end{center}
\vspace*{1.5cm}
\begin{center}
\Large
Ph.D. THESIS \\
Advisor: Prof. J. Marro Borau \\
Granada, November $6^{th}$, 2002
\end{center}

\newpage
\thispagestyle{empty}
\vspace*{10cm}

\newpage
\thispagestyle{empty}
\vspace*{6cm}
\begin{flushright}
\large
{\it To Ana}
\end{flushright}

\newpage
\thispagestyle{empty}
\vspace*{10cm}



\tableofcontents
\listoffigures
\listoftables
\mainmatter
\chapter{Introduction}
\label{capIntro}

The beauty of Physics underlies on the striking simplicity of its fundamental laws. 
Maxwell equations, Hamiltonian mechanics, Einstein and Schr\"odinger equations, can each be expressed 
in a few lines. Not only the laws of Physics, but also the philosophic ideas behind these laws are 
simple. Everything is simple, except the world these laws intend to describe.\cite{Kadanoff} Nature is 
complex at all levels. Think for instance about a living organism (say yourself). You are ultimately made of 
many elementary particles, whose behavior can be accurately described within the framework of the
Standard Model. However, we do not understand at all how the collective dynamics of such a large set of 
(quantum) particles can give rise to the amazing phenomenon of life. The whole does not behave as a simple 
superposition of its parts.

Complexity arises due to the non-trivial underlying structure of all natural systems, and the non-linear interactions
among their constituents. The obvious question now is: how the complex world emerges from
the simple laws of Physics?. The search for an answer to this question was the seed that gave rise to
the development of Statistical Mechanics during the second half of XIX century and the beginning of XX
century. In a general sense, Statistical Mechanics is a branch of Physics aimed to describe the
macroscopic (complex) properties of matter from the interactions between its microscopic constituents. The most 
successful achievement of Statistical Mechanics is Ensemble Theory\cite{Balescu}, which yields the 
connection between the microscopic fundamental physics and the macroscopic behavior of equilibrium systems.
An isolated system which shows no hysteresis and reaches a steady state is said to stay at 
equilibrium.\cite{Biel} Starting from a few basic postulates, Ensemble Theory provides us with a 
well-defined ``canonical'' formalism in order to obtain the stable equilibrium properties of 
macroscopic matter.\cite{Pnros} In particular, it allows us to understand how complex situations, as for 
instance phase transitions, arise in many-body interacting systems.

However, most of the systems we find in Nature are out of equilibrium: they are open, hysteretic  
systems, subject to thermal or energetic gradients, mass and/or energy fluxes, which suffer the action of 
external agents, or are subject to several sources of non-thermal noise. Think again about yourself:
can you feel your breath?. This is a pure non-equilibrium process, where a flux of air from the external 
medium towards your lungs appears due to a pressure gradient. Nonequilibrium processes are also essential
for cell functioning, brain processing, etc. What is more intriguing, all living organisms are
nonequilibrium structures: nonequilibrium conditions are essential for life.
On the other hand, nonequilibrium structures
appear at all scales. For instance, the Sun exhibits temperature gradients, mass and energy 
transport, convection, etc. On the opposite limit one finds for example 
magnetic nanoparticles, where quantum tunneling
acts as a non-thermal noise source inducing nonequilibrium conditions, or turbulent fluids. 
Nature abounds in such examples. In fact, non-equilibrium phenomena are the rule, being equilibrium systems 
a rather unlikely exception. It seems that nonequilibrium is a fundamental ingredient for the
observed structure in Nature. As in equilibrium 
systems, those systems out of equilibrium also show instabilities which give rise to spatio-temporal
patterns, dissipative structures, self-organization, time oscillations, spontaneous symmetry-breaking, etc.,
all of them commonly observed in Nature. All these instabilities are generally known as {\it nonequilibrium phase 
transitions}, and their properties are much richer than in equilibrium systems.
\begin{figure}
\centerline{
\psfig{file=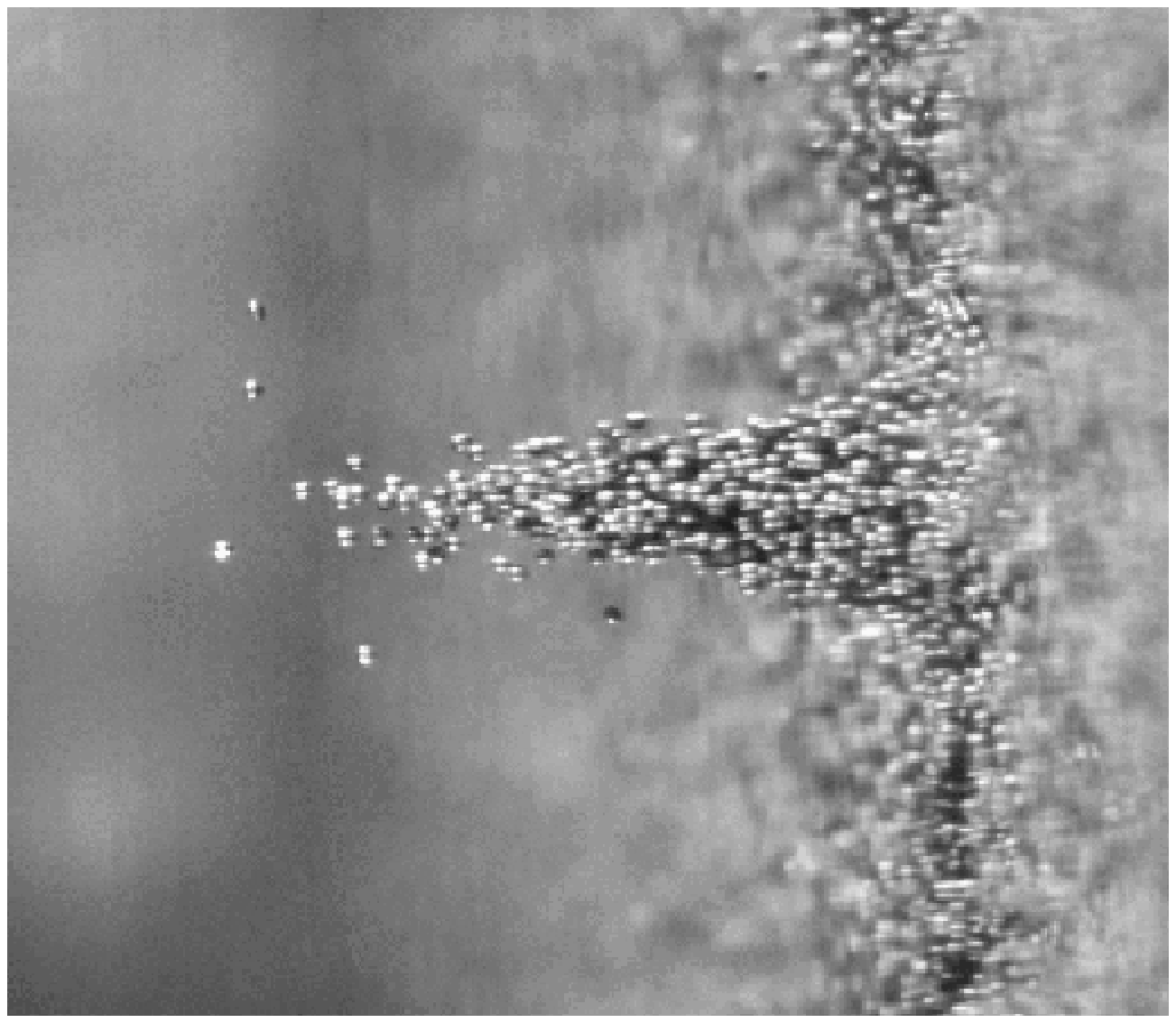,width=6cm,angle=-90}
\psfig{file=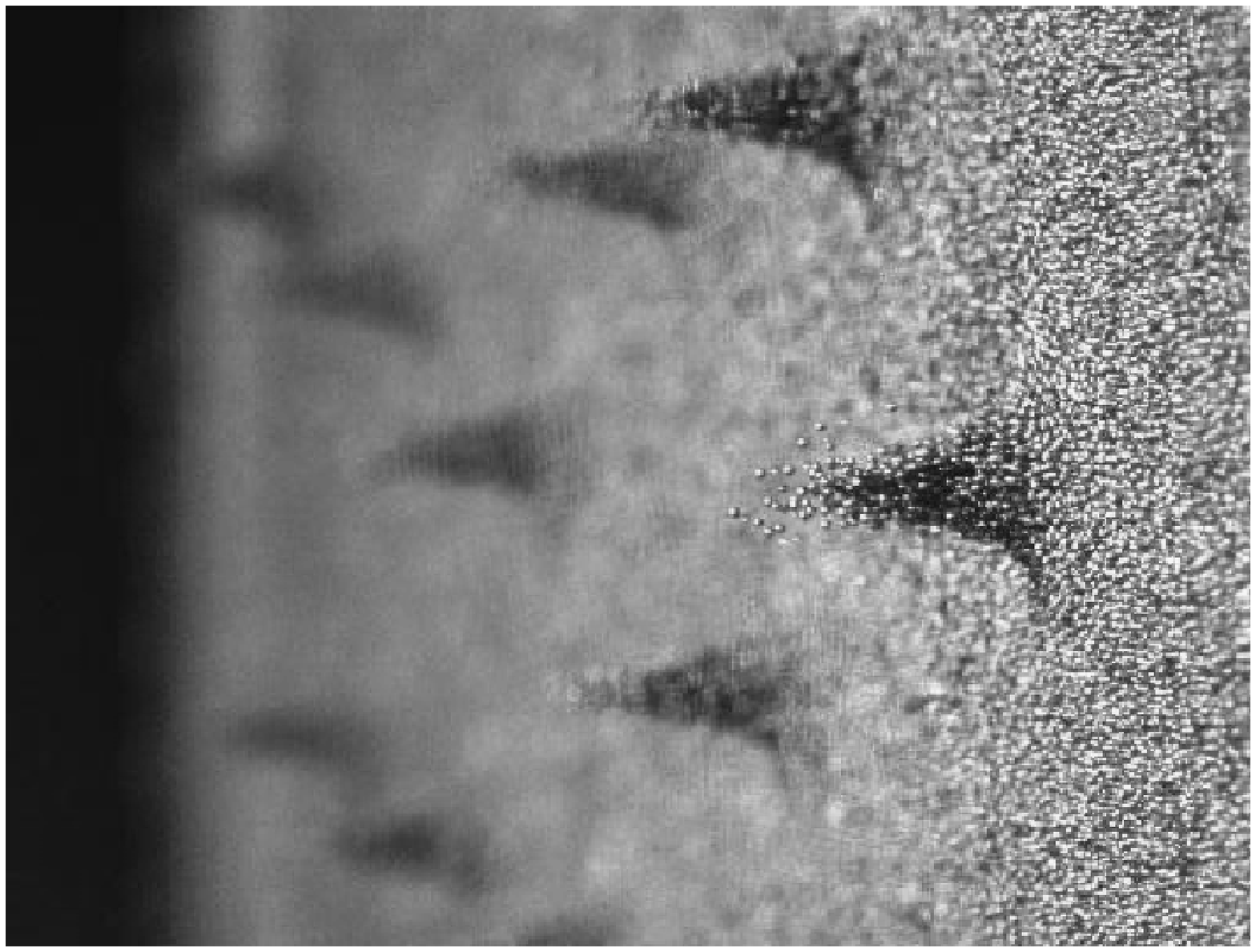,width=6cm,angle=-90}}
\caption[Pattern formation in vertically vibrated granular materials.]
{\small {\it Complexity in Nature:} pattern formation in vertically vibrated granular materials.
In particular, here we observe localized wave structures called {\it oscillons}.\cite{Umbanhowar}}
\label{oscilones}
\end{figure}

However, we are still lacking a general theory, equivalent to Statistical Mechanics for equilibrium systems, 
which allows us to classify and understand systems far from equilibrium, connecting their macroscopic 
phenomenology with their microscopic properties. In general, we only have at our disposal a set of 
{\it ad hoc} theoretical approximations which describe in a partial and incomplete manner the physics governing
these systems. Hence nonequilibrium systems constitute a challenge for theoretical physicists, besides
being very interesting from the practical point of view due to their ubiquity in Nature. The simplest situation
in a nonequilibrium system is that of a steady state. In this case, the properties of the system do not depend
on time, thus simplifying the analysis. Nonequilibrium steady states (NESS) have been studied in
depth during the last 20 years,\cite{Pedrotesis} finding that their properties are much richer than those for 
equilibrium steady states. In general NESS do not obey the Boltzmann distribution. Moreover, they are not 
unique for a given system and fixed parameters, depending on the specific dynamics and the previous history of 
the system. On the other hand, the dynamics of nonequilibrium systems has been poorly studied.

\begin{quotation}
{\it \noindent The study of the dynamical properties of some nonequilibrium systems is the aim of this thesis.}
\end{quotation}

Many of the most interesting dynamic phenomena in complex systems are
usually related to the transformation of one phase into another. Think for
instance on the evolution from a uniform mixture of chemical constituents to a phase-separated pattern of
precipitates. Other examples are the evolution from a disordered phase to an ordered one, or the exit
from a metastable phase towards the truly stable one, etc. These transformations are not abrupt, but they
involve a temporal (and usually inhomogeneous) evolution. The mechanisms behind 
these dynamical processes often involve the minimization of certain {\it privileged} fundamental observables 
-as the free energy in equilibrium systems- which contain the information about the system state at any time, 
and shed light on the relevant ingredients we must take into account to build up a theoretical description of 
the process. These dynamic transformations involve many features commonly observed in Nature: pattern formation
and morphogenesis\cite{Cross,Gollub}, avalanche-like dynamics and self-organization\cite{Jensen}, etc. 

The dynamic evolution between two different phases is a nonequilibrium process since the system is not in
a steady state. If the system under study is isolated and shows no hysteresis, we know that the steady state
it will finally reach is an equilibrium state.\cite{Biel} 
In this case the evolution between different phases, in spite
of being a nonequilibrium process, can be understood and described using (equilibrium) Statistical Mechanics.
This is the case of metastability and phase separation in equilibrium systems, which are described using 
appropriate extensions of the equilibrium free energy. We are interested in this thesis in dynamic processes in 
systems which asymptotically converge towards a NESS. These systems are essentially far from equilibrium (even 
in the steady state) and hence cannot be described by (equilibrium) Statistical Mechanics.

In addition to the phase-transformation dynamic processes described above, there are some dynamic phenomena 
in nonequilibrium systems which have no equilibrium counterparts. 
For instance, this is the case of systems suffering a 
dynamic phase transition between an active phase, characterized by the existence of non-trivial dynamics,
and an absorbing phase, where the system is frozen, without any dynamics and hence no chance of escape.
Another example is that of heat conduction, where a temperature gradient induces a steady energy flow from the
hot reservoir to the cold one. 

The study of dynamic phenomena in nonequilibrium systems and the underlying mechanisms driving these processes
yields much information about the relevant observables which define the system evolution. Such observation,
together with the comparison to similar results in equilibrium systems, help us in the search of a general
theory for nonequilibrium systems (which should describe equilibrium systems as a limiting case).

In general, the study of the dynamic and/or static properties of real systems with many degrees of 
freedom is a formidably complicated task. Hence, approaching the study of these systems from first principles
is not usually feasible in practice. For this reason we must study simplified models of reality that,
while capturing the fundamental ingredients of real systems, are much more easily tractable. Usually these
models are defined on a lattice, instead of being defined on the continuous space, and the interactions
among their constituents are modeled in a very simple way. The obvious question now is: what do these models
have to do with real systems?. In order to answer this question we must introduce the concept of 
{\it universality}, which is one of the most important philosophic ideas of modern physics. This concept
is based on the observation that disparate systems often display strikingly similar features and behavior.
It is observed in Nature that the large scale structure and 
behavior of a system do not depend on its microscopic details, 
but only on the fundamental features defining the system, as for instance dimensionality, symmetries, conservation 
laws, range of forces, kind of order parameter, etc. In this way all systems sharing the same essential
features exhibit the same kind of behavior, in spite of being apparently very different (say a lattice gas 
and a saturated real vapor). Therefore these systems exhibit universal 
behavior\footnote{In particular, universality appears in second order
phase transitions. There the correlation length diverges, so that all scales are equally relevant and the 
microscopic details are no longer important. All systems sharing the same values for the critical exponents
associated to a critical point belong to the same {\it universality class}. In a similar way, {\it dynamic
universality classes} can be defined which include systems sharing the same dynamic exponents.\cite{Cardy}
However, in this chapter we are presenting the concept of universality from a wider point of view, 
not only restricted to critical phenomena.}. This universal
behavior allows us to design {\it minimal} models of reality that capture all the relevant ingredients of
real systems, while they {\it maximally} simplify the microscopic irrelevant details. The universality 
property guarantees that the behavior of the system is not sensitive to the microscopic details, and hence
that our results for the oversimplified model will also describe the behavior of the real system, provided that 
both systems share the same fundamental features. As an example, think for a while on the Navier-Stokes
equation, which describes the macroscopic behavior of a incredibly large set of different fluids.\cite{Navier}
This equation is a simplified model of real fluids, based on several symmetries and conservation laws, which
describes the macroscopic behavior of fluids with different compositions, interatomic forces, 
molecular weights, etc., but which share some fundamental features as mass and energy conservation, 
dimensionality, etc.
\begin{figure}
\centerline{
\psfig{file=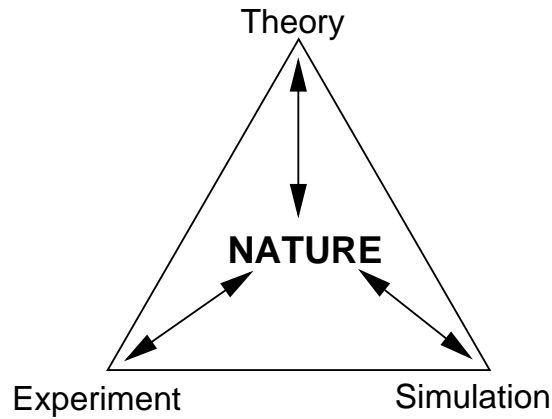,width=6cm,angle=-90}}
\caption[The three ways of doing physics research nowadays.]
{\small {\it Physics research:} The three ways of doing physics research nowadays. All three are
complementary.\cite{LandauBinder}}
\label{research}
\end{figure}

In spite of being oversimplified versions of real systems, the models we are going to study in this thesis 
exhibit a highly non-trivial and complicated behavior. For this reason we must use computer simulations in 
order to investigate such systems in detail, in addition to approximate theoretical tools. Computers allow
us to simulate systems which would be intractable in other way. In addition, simulations help us to obtain the 
intuition we need in order to solve and understand the behavior of complex many-body systems. Historically,
Physics has been called {\it natural philosophy}, since the research was only done via purely theoretical 
(i.e. philosophical) investigations. Eventually the experimental method was accepted as a second way of doing 
physics research, although it is limited by the scientists ability to design the appropriate experiments, 
prepare the system and accurately measure the desired magnitudes. Nowadays computer simulations have become a 
third way of doing physics research, yielding a new perspective. Sometimes computer simulations provide the
theoretical basis we need in order to understand some experimental results, and other times simulations
serve as {\it experiments} to which compare a theory. In any case, simulations complement the classical 
theoretical and experimental approaches to Nature.

The most used simulation method in this thesis will be the Monte Carlo method, although we will also
implement other methods, as the Molecular Dynamics one. Monte Carlo method is very useful
when studying the temporal evolution of models which have no deterministic dynamics (in the sense of Newton 
equations) but are subject to stochastic dynamics. In practice it is usually 
not possible to make a $100\%$ atomistic description of the system we want to model. Roughly speaking,
we do not need (and we are not able) to take into account the individual 
quark behavior when modeling a macroscopic material. 
On the contrary, one builds up a coarse-grained description of the system, taking into account only the
relevant variables for the problem under study (say spins in a magnetic material), and letting the fast
degrees of freedom that we forget about in our coarse-grained description to act as a heat bath, thus inducing 
stochastic transitions on the relevant degrees of freedom. This method makes use of a sequence of pseudo-random
numbers (this is the reason why it is called {\it Monte Carlo}).\cite{LandauBinder} On the other hand, the 
Molecular Dynamics method is based on the numerical integration of Newton equations of motion.

As we said before, the objective of this thesis is to study the dynamics of some nonequilibrium systems. In 
particular, we will pay attention to {\it metastability} and {\it avalanches} in a nonequilibrium ferromagnetic 
spin system, {\it phase segregation} in a driven (anisotropic) lattice gas, phase transitions in a system
with {\it (super)absorbing states}, and {\it heat conduction} in a one-dimensional particle system. In this way 
we want to cover a wide variety of dynamic phenomena appearing in nonequilibrium systems. Of course, the list is 
not complete, lacking some fundamental phenomena, as for instance hydrodynamics. However, and in spite of the 
heterogeneity of this thesis, we think that the studied systems and processes yield a comprehensive overview of 
the effects that nonequilibrium conditions induce on dynamic phenomena in complex systems. The thesis is divided 
into two parts. The first part, which comprises chapters 2, 3, 4, 5 and 6, 
is devoted to the study of metastability (chapters 3, 4 and 5) 
and avalanches (chapter 6) in a nonequilibrium ferromagnetic spin model. The second part of this work is devoted 
to the study of anisotropic phase separation (chapter 7), systems with superabsorbing states (chapter 8), and heat
conduction and Fourier's law (chapter 9).

Metastability is a crucial concept in many branches of Science. It has been observed in fluids, plasmas, quantum
field theory, superconductors and superfluids, magnetic systems, atmospheric dynamics, cosmology, etc. It usually
determines the system behavior. In particular, we are interested in metastability in nonequilibrium systems with 
short range interactions. In this way we study here metastability in a nonequilibrium ferromagnetic spin model,
which is relevant for the problem posed by magnetic storage of information. On the other hand, from the 
theoretical point of view, studying metastability in this impure ferromagnet will allow us to investigate the
existence of a nonequilibrium potential, equivalent to the equilibrium free energy, which controls the exit
from the metastable state. 

In this way, in {\bf chapter \ref{capMotiv}} we motivate the study of metastability in nonequilibrium systems,
presenting the ferromagnetic model we will investigate in the first part of this work. We also discuss some
of the properties that characterize this model, paying special attention to the way in which nonequilibrium
conditions enter the model definition.

In {\bf chapter \ref{capMedio}} we perform a mean field study of the metastability phenomenon. In particular we 
apply the Pair Approximation\cite{MarroDickman} to our model in order to obtain its static and dynamic 
properties. This study uncovers very interesting properties related to the 
non-linear interplay between the thermal noise and the non-thermal fluctuations induced by the nonequilibrium 
conditions. 

{\bf Chapter \ref{capSOS}} is devoted to the study of the properties of the interface in the nonequilibrium
model. The inhomogeneous character of the metastable-stable transition implies that the interface between
the metastable and stable phases plays a determining role in this process. In this chapter we generalize the
Solid-On-Solid approximation of Burton, Cabrera and Frank\cite{BCF} for an equilibrium discrete interface
in order to take into account the effect induced by nonequilibrium conditions. This generalization is based
on the concept of effective temperature. We find very interesting results at low temperatures showing, for 
instance, that the nonequilibrium surface tension converges to zero in this limit. We also study in this
chapter the shape of a spin droplet using the Wulff construction\cite{Wulff}.

In {\bf chapter \ref{capNuc}} we extend the equilibrium nucleation theory\cite{Rikvold} to the nonequilibrium system,
hypothesizing the existence of a nonequilibrium potential, similar in form to the equilibrium free energy, which 
controls the exit from the metastable state. Applying the results obtained in chapters \ref{capMedio} and
\ref{capSOS} for the bulk and interfacial properties, we find surprising results for the metastable-state
mean lifetime, the critical droplet size, the domain wall velocity and 
the metastable-stable transition morphology in the nonequilibrium
case, which are fully confirmed via Monte Carlo simulations. In addition to its theoretical importance, these 
results may be technologically relevant.

In {\bf chapter \ref{capAval}} we observe that under the combined action of both free boundaries and 
nonequilibrium conditions, the evolution from the metastable phase towards the stable one proceeds through
well-defined avalanches. These avalan-ches are shown to follow power-law, i.e. scale free distributions.
However, a detailed study reveals that this scale free behavior is a consequence of a finite superposition
of well-defined, gap-separated typical scales, instead of being a consequence of any underlying critical point.
The excellent comparison of our results with some Barkhausen experiments led us to suspect that Barkhausen Noise
in particular, and $1/f$ noise in general, might also come from a superposition of elementary events.

{\bf Chaper \ref{capDLG}} is devoted to the study of phase separation in nonequilibrium anisotropic lattice 
gases. Phase separation appears in system with conserved number of particles. It is a dynamic process which
has been largely studied in equilibrium systems. In addition to theoretically challenging, the details are
of great practical importance. However, as we previously discussed, most systems in Nature are out of 
equilibrium. Therefore, extending the concepts involved in the phase separation process to more realistic 
situations is very interesting. This is the case, for example, for mixtures under a shear flow, whose study
has attracted considerable attention.\cite{critic}-\cite{corberi} Hence in this chapter we study anisotropic 
phase separation in a driven lattice gas. We propose a cluster effective diffusion theory in order to explain
the late stage coarsening in this system. This theory describes correctly the grain growth process and the 
different growth regimes found during the evolution. In addition, we also demonstrate dynamical scaling of the 
structure factor, and generalize Porod's law to anisotropic systems. Finally we also study the dynamics of a 
continuous field equation, showing qualitatively its validity to describe the dynamics of the microscopic model.

In {\bf chapter \ref{capLipo}} we study a system showing a phase transition between an active phase, 
characterized by a nontrivial dynamics, and an absorbing phase, which is completely frozen. In this sense
this is a {\it dynamic phase transition}. There are many examples in Nature of systems with absorbing states
and absorbing phase transitions: chemical reactions in autocatalytic reaction-diffusion systems and heterogeneous 
catalysis models, problems related with directed percolation, fire and 
epidemic spreading, etc. Absorbing states appear in situations where certain observable can proliferate or
die, but never generate spontaneously. In this way the essential physics  comes from the competition between 
the growth and dead of the relevant observable. There are two main universality classes in systems with 
absorbing states: the directed percolation (DP) universality class, and the multiplicative noise universality 
class. In this chapter we study how a new, hidden symmetry in a system with absorbing states (known as Lipowski
model\cite{Lip1,Lip2}), namely the presence of the so-called {\it superabsorbing states}, is relevant at the 
critical point, thus defining a new scaling behavior. 

In {\bf chapter \ref{capFou}} we study heat conduction and Fourier's law in a one-dimensional particle model. 
Heat conduction is just one particular aspect of transport phenomena, which are dynamic processes that appear 
ubiquitously in Nature. Some classical examples are heat and mass transport in fluids, diffusion, electric
conduction, stellar convection, etc. In spite of being very important, their microscopic understanding is far 
from clear. In particular, in this chapter we want to investigate the microscopic basis of heat conduction.
Therefore we present computer simulation results for a chain of hard-point particles with alternating masses
subject to a temperature gradient. We find, performing different, complementary numerical analysis, 
that the system obeys Fourier's law at the Thermodynamic Limit.
This result is against the actual belief that one-dimensional systems with momentum conservative dynamics and 
non-zero pressure have an infinite thermal conductivity.\cite{Prosen} It seems that thermal resistivity occurs in 
our system due to a cooperative behavior in which light particles tend to absorb much more energy than heavier ones.

Finally, in {\bf chapter \ref{capConc}} we present our conclusions, summing up the results obtained along this
work and pointing out the possible research lines to follow in order to continue these investigations.

In what follows we summarize the original contributions contained in this thesis:
\begin{itemize}
\item In chapter \ref{capMedio} we calculate in mean field approximation and using computer simulations 
the {\it intrinsic coercive field}, $h^*$, associated to the nonequilibrium spin model. We show that the behavior 
of $h^*$ for strong nonequilibrium conditions signals the existence of a non-linear cooperative effect between the
thermal noise and the nonequilibrium fluctuations, which involve the disappearance of metastable states at low
temperatures, as opposed to what happens in equilibrium systems.

\item In chapter \ref{capMedio} we also present a natural way to introduce fluctuations in a dynamic mean field
theory. This method allows us to study the dynamics of the metastable-stable transition in mean field approximation.

\item In chapter \ref{capSOS} we generalize the Solid-On-Solid approximation\cite{BCF} in order 
to investigate the effects that nonequilibrium conditions induce on the system interface. This generalization,
based on the concept of effective temperature, accurately describes the properties of the nonequilibrium model 
interface.

\item In chapter \ref{capNuc} we introduce a hypothesis about the existence of a nonequilibrium potential, 
equivalent to the equilibrium free energy, which controls the exit from the metastable state. This hypothesis
allows us to properly describe the nonequilibrium metastable-stable transition.

\item In chapter \ref{capAval} we measure scale free avalanches, and identify their origin, which is based on a
finite superposition of avalanches with well-defined typical scales. Comparing our results with some Barkhausen 
experiments, we are able to propose a new explanation for Barkhausen Noise in particular and the ubiquitous $1/f$ 
Noise in general.

\item We propose in chapter \ref{capDLG} a cluster effective diffusion theory in order to explain coarsening
in the driven lattice gas, based on two different types of monomer events. In the same way, we demonstrate
dynamical scaling of the structure factor, and generalize Porod's law to anisotropic systems. We also study
the dynamics of a field theoretical equation whose dynamical properties have never been studied.

\item In chapter \ref{capLipo} we identify the presence of superabsorbing sites as a new relevant symmetry
in systems suffering absorbing phase transitions. In this way we define a novel scaling behavior.

\item Finally, in chapter \ref{capFou} we propose a new cooperative mechanism which gives rise to normal thermal
conductivity in one-dimensional particle systems.
\end{itemize}

\part{ Metastability and Avalanches in Ferromagnetic Systems under Nonequilibrium Conditions}


\chapter{Motivation and Model Definition}
\label{capMotiv}

\section{Introduction}
\label{capMotiv_apIntro}

The concept of metaestability is a cornerstone in many different branches of Science. In spite of its importance, it
is very difficult to obtain a precise and general definition for it. The British Encyclopedia defines
a metastable state in the following way: {\it in Physics and Chemestry, a metastable state is a particular excited state 
of an atom, nucleus or other system, such that its lifetime is larger than that of usual excited states, but generally lower than 
that of the ground state, which is often stable. A metastable state can be then considered as a temporal energetic trap or an 
intermediate stable state}. This definition, in spite of its ambiguity, captures the essence of what a metastable state is: it is 
a local, non-global stable state since the system finally evolves 
towards the ground state\footnote{This is not strictly true. There are  metastable states with infinite 
lifetime in systems subject to long range interactions. This problem points out, once more, the difficulties 
found in order to establish a precise definition of metastability.}, but it is also a state very similar to the stable one, due to 
the long time the system spends wandering around it. There are many other different definitions of metastability in literature, 
some of them much more precise from the mathematical point of view, although they are always restricted to particular systems. For 
instance, we should say that a metastable state in Equilibrium Thermodynamics corresponds to a local, not global free energy minimum.

Metastability is observed in fluids, solids, plasmas and many other systems, and it usually determines their behavior. 
The metastability phenomenon is often related to the existence of an underlying first order phase transition. A prototypical
example is that of supercooled water: if we slowly cool down a glass of water, in such a way that the final temperature is 
slightly below the solidification point, water will remain liquid. This supercooled liquid is in a metastable state: although the
true stable state for this temperature is the crystallized one, an energy barrier exists that prevents water crystallization
(the system is confined in an {\it energetic trap}). If we add to the system a sufficient amount of energy from the outside
in order to overcome the energy barrier (for instance, hitting softly the glass), the system will evolve from the metastable state
to the stable one through the {\it nucleation} of crystals of the stable phase inside the metastable bulk.\cite{Kob}
The processes of energetic {\it activation} and {\it nucleation} here illustrated are crucial in order to understand how a
generic system is able to exit the metastable state and to evolve to the true stable state. 

There are many other systems which show metastability. In fact, metastability is observed from the smallest scales in the 
Universe to the largest ones. For instance, if the Higgs boson mass is as small as suggested by the latest experiments,
this should point out that the Standar Model ground state, called {\it vacuum} in Quantum Field Theory, is metastable, instead of 
being a true stable state. This observation allows us to give a lower bound for the Higgs boson mass, since the metastable
vacuum must have a long lifetime compared to the age of the Universe.\cite{Isidori} Moreover, some works have recently appear that
speculate with the possibility that heavy ion collisions at the Relativistic Heavy Ion Collider in the Brookhaven National 
Laboratory (U.S.A.) should trigger a transition towards the true Standar Model vacuum.\cite{Jaffe} Such transition would imply
an apocalyptic disaster\footnote{Fortunately, the probability of such catastrophic scenario is estimated to be negligible, although 
non vanishing \ldots}. Metastability is also observed in quark/gluon plasma\cite{quarks}, systems showing superconductivity and 
superfluidity\cite{supercond}, electronic circuits\cite{circ}, globular proteins\cite{proteinas}, magnetic systems\cite{mag}, 
climatic models \cite{clima}, black holes and protoneutronic stars\cite{estrellas}, cosmology\cite{cosmo}, etc. A better microscopic
understanding of this ubiquitous phenomenon is then of great theoretical and technological interest, besides a formidable mathematical
challenge.

A problem of particular importance is that posed by magnetic storage of information, which is intimately related with metastability.
A magnetic material is usually divided into magnetic monodomains. In order to store information on this material, we magnetize 
each individual domain using a strong magnetic field. In this way, each domain exhibits a well defined magnetization in the 
direction of the local applied field, thus defining a bit of information. A main concern is to retain the individual domain 
orientations for as long as possible in the presence of weak arbitrarily-oriented external magnetic fields, in order to keep 
unaltered the stored information. The interaction with these weak external fields often produces metastable states in the
domains. The resistance of stored information depends on the properties of these metastable states, including the details 
of their decay.

In general, complex systems have many degrees of freedom which make very complicated any first-principles theoretical approach 
to their behavior. In particular, this is the case for the aforementioned magnetic systems, where there is a macroscopic number of 
magnetic moments or spins which interact among them and with an external magnetic field. Therefore, we are forced to study simplified
models of real systems that, while capturing their relevant ingredients, are much more easily tractable.
There has been in last decades a huge amount of works
studying the problem of metastability in lattice models of classical spins. The most studied model has been the Ising model in
one, two and three dimensions.\cite{Ising,Ramos} The general interest in this model is two-fold. On one hand, it captures many of the
fundamental features of a wide class of real systems. On the other hand, many of its equilibrium properties are analytically known in 
one and two dimensions\cite{Onsager}. This fact makes much more easy any theoretical approach to the properties of this model.
In this way, continuous theories based on nucleation mechanisms have been proposed which successfully describe the evolution
from the metastable state to the stable one.\cite{Langer} Also the problem of metastability in the low temperature limit has been
exactly resolved.\cite{Schonmann} These theoretical results have been checked many times via computer 
simulations.\cite{simulaciones,Rikvold} Very interesting analytical and computational results have been obtained, showing the existence
of different parameter space regions in finite systems, each one characterized by a typical metastable-stable transition 
morphology.\cite{Rikvold} Likewise, the effects that open borders\cite{Cirillo,contorno}, quenched impurities\cite{impurezas} and
demagnetizing fields\cite{demag} have on the properties of metastable states in these systems have also been investigated.

With some exceptions\cite{Vacas}, most works on metastability in magnetic systems have been limited to equilibrium models.
For these models, the equilibrium Statistical Mechanics of Boltzmann, Gibbs, Einstein, etc. yields a clear-cut connection between
microscopic and macroscopic Physics in terms of the partition function.\cite{Balescu} In this way, steady states in equilibrium systems 
are characterized by the Boltzmann distribution, $Z^{-1} \textrm{exp}(-\beta E)$, where $\beta$ is the inverse temperature, $E$ is the
state energy and the normalization constant $Z$ is the system's partition function. Although metastability is a dynamic phenomenon
not included in the Gibbs formalism,\cite{Cirillo} so successful on the other hand when describing equilibrium states, it is possible
to understand such phenomenon extending dynamically the Gibbs theory using the thermodynamic potentials defined in this equilibrium 
theory and its connection with the microscopic parameters that characterize the system. In this way, nucleation theory, which 
correctly describes metastability in systems near a first order phase transition, is based on the concept of free energy of a 
droplet of the stable phase. This magnitude is an heterogeneous extension of the thermodynamic potential associated to the 
canonical partition function in an equilibrium system.

However, most of the systems we find in Nature are out of equilibrium: they are open systems, subject to thermal or energetic 
gradients, mass and/or energy currents, which suffer the action of external agents, or are subject to several sources of non-thermal 
noise. As an example, it has been observed that some properties of metastable states in certain mesoscopic 
magnetic particles are highly affected by quantum tunneling of individual spins, which is a pure nonequilibrium process since it 
breaks detailed balance.\cite{Tejada} Furthermore, there are nonequilibrium lattice spin models which reproduce these 
results.\cite{Vacas} For nonequilibrium systems there is no theory equivalent to equilibrium Statistical Mechanics that connects
their microscopic properties with their macroscopic phenomenology. If we want to understand metastability in real 
(i.e., nonequilibrium) systems we must study simplified nonequilibrium models. 
On the other hand, and from a theoretical point of view, the study of metastability 
in nonequilibrium systems will allow us, comparing with the well-established equilibrium results, to understand how nonequilibrium
conditions affect a dynamic process like metastability. This comparison will also allow us to study the changes that the nucleation
process suffers under nonequilibrium conditions, as well as the possible existence of some functional, similar to the equilibrium free
energy, which controls the relaxation from the metastable state.

In the following chapters we are going to study metastability in magnetic thin films under nonequilibrium conditions. On the analogy of
equilibrium systems, it seems sensible to model these magnetic systems using a bidimensional kinetic Ising lattice with nearest
neighbor interactions and periodic boundary conditions. In addition, we will consider a (very) weak random dynamic perturbation
competing with the usual thermal spin flip process. It has been shown that the presence of this weak perturbation could explain some
intriguing properties, as for instance the non-vanishing value of magnetic viscosity in the low temperature limit, of some real
magnetic materials.\cite{Vacas,Tejada} The impurity makes the system to reach asymptotically a nonequilibrium steady state. That is, 
we assume that a principal role of the microscopic disorder which is generally present in actual specimens consists in modifying 
the dynamics -in a way similar to that of an external non-Hamiltonian agent.\cite{MarroDickman} 

It is observed that, under the action of the dynamic perturbation and a weak magnetic field oriented opposite to the initial 
magnetization, the system's demagnetization process from the initial metastable state to the true stable one proceeds through the 
nucleation of one or several critical droplets of the 
stable phase in the metastable bulk, as observed in equilibrium systems. Although, oppositely to what happens in equilibrium,  
we cannot properly define here any free energy functional that controls the demagnetization process, we can however hypothesize the 
existence of some nonequilibrium potential, similar to the equilibrium free energy, where two terms compete. On one hand, there is
a surface term, which hinders the growth of the stable phase droplet. On the other hand, there is a bulk term, which favours
its growth. If this hypothesis is correct (as we will see later on) we should expect a good description of the 
metastable-stable transition in terms of this nonequilibrium potential once we understand the bulk and interfacial 
properties of our model under the action of the nonequilibrium random perturbation. With this aim we will propose in the following 
chapters approximate theories to study the effect of the dynamic random perturbation on the system's bulk and interface.
As a result, we will conclude that the bulk exhibits a very interesting non-linear cooperative phenomenon between the thermal
noise and the non-thermal (nonequilibrium) fluctuations when subject to strong nonequilibrium conditions, although bulk properties
are qualitatively similar in both the equilibrium and weak nonequilibrium cases. On the other hand,
the interfacial properties in the nonequilibrium system change in a fundamental way. In particular, we will observe that
while the surface tension in the equilibrium system monotonically grows as temperature decreases, the surface tension in the 
nonequilibrium case exhibits a maximum for a given temperature, decreasing for smaller temperatures. Using this result, and the
hypothesis of existence of a nonequilibrium potential that controls the metastable-stable transition, we are able to develop
a nonequilibrium nucleation theory analogous to the equilibrium one. However, the results obtained from this nonequilibrium theory 
are surprising, since the nonequilibrium conditions imply a completely different behavior of the system at low temperatures, as compared
with the equilibrium one. We observe that metastability tends to disappear at low temperatures. Even existing, the metastable lifetime
reaches a maximum for a given temperature, decreasing if we further decrease the temperature. These results point out that the general
belief which states that in order to prolong the lifetime of a metastable state we must cool the system is not true if the system is 
subject to any random perturbation as the one we implement (perturbation which, on the other hand, is usually present in
real systems -quantum tunneling, external noises, etc.-). Moreover, our theory predicts the existence of a low temperature phase
where the system demagnetizes through the nucleation of multiple stable phase critical droplets, as opposed to equilibrium systems, 
which demagnetize through the nucleation of a single critical droplet at low temperatures.\cite{Rikvold} All our theoretical results
are compared with extensive Monte Carlo simulations, showing very good agreement.

This and the following chapters are organized as follows. In section \ref{capMotiv_apModel} of the present chapter we describe
the model in detail, summarizing some of its properties. We also briefly explain in this section the computational scheme used
in our simulations. Chapter \ref{capMedio} is devoted to a first order dynamic mean field approximation. This approximation,
also called Pair Approximation or Bethe-Peierls Dynamic Approximation, will allow us for a first theoretical approach to the problem of 
metastability. In Chapter \ref{capSOS} we study the properties of the interface in the nonequilibrium model. In order to do so we 
generalize the Solid-on-Solid Approximation for discrete interfaces to take into account the effects that nonequilibrium conditions
induce on the interfacial properties. Finally, in Chapter \ref{capNuc} we propose a nonequilibrium Nucleation Theory for the 
nonequilibrium model, formally similar to that of equilibrium systems, but where we introduce the results obtained for the bulk and
interfacial properties affected by nonequilibrium conditions. In this chapter we also present our conclusions about the
problem of metastability in nonequilibrium systems, paying some attention to practical applications of our results and possible 
experimental implications.

\section{The Model}
\label{capMotiv_apModel}

In the following chapters we are going to study a bidimensional kinetic Ising model with periodic boundary
conditions and subject to a dynamic random perturbation. The two-dimensional Ising model\cite{Ising_orig} is defined on a 
square lattice of side $L$. On each lattice node a spin variable is defined, $s_i$, with $i \in [1,N]$, $N=L^2$. Each spin can 
take two different values: $s_i=+1$ (up spin) o $s_i=-1$ (down spin). The system is characterized by the Hamiltonian,
\begin{equation}
{\cal H}({\bf s}) = -J \sum_{\langle i,j \rangle} s_i s_j - h \sum_{i=1}^N s_i
\label{hamilt}
\end{equation}
where $J>0$ is the (ferromagnetic) coupling constant, ${\bf s} \equiv \{s_i,i=1, \ldots ,N \}$ is the system's configuration, 
and $h$ is an external magnetic field. The first sum runs over all
nearest neighbor pairs, $\langle i,j \rangle$, while the second sum runs over all spins. We endow this kinetic model with a dynamics
determined by the stochastic master equation,\cite{Glauber}
\begin{equation}
\frac{\textrm{d} P({\bf s}; t)}{\textrm{d} t} = \sum_{\bf s'} \big[\omega({\bf s'} \rightarrow {\bf s})
P({\bf s'};t) - \omega({\bf s} \rightarrow {\bf s'}) P({\bf s};t)\big]
\label{mastereq}
\end{equation}
where ${\bf s}$ and ${\bf s'}$ are system's configurations, $P({\bf s}; t)$ is the probability of
finding the system in a configuration ${\bf s}$ at time $t$, and $\omega({\bf s} \rightarrow {\bf s'})$ is the probability per 
unit time (or transition rate) for a transition from configuration ${\bf s}$ to ${\bf s'}$. In order to complete the 
definition of the model, we must precise the transition rate $\omega({\bf s} \rightarrow {\bf s'})$.
In our case we assume that the system evolves due to the superposition of two ``canonical'' dynamics. That is, we choose the
transition rate to be,
\begin{equation}
\omega ({\bf s} \rightarrow {\bf s}^i) = p + (1-p) \frac{\textrm{e}^{-\beta \Delta {\cal H}_i}}
{1 + \textrm{e}^{-\beta \Delta {\cal H}_i}}
\label{rate}
\end{equation}
({\it Glauber dynamics}). Here ${\bf s}^i$ stands for the configuration ${\bf s}$ after flipping the spin at node $i$, $\beta =1/k_BT$
is the inverse temperature, $k_B$ is the Boltzmann constant, and $\Delta {\cal H}_i \equiv {\cal H}({\bf s}^i) - {\cal H}({\bf s})$. 
We only allow single spin flip transitions between configurations. In what follows we 
fix Boltzmann constant to unity, $k_B=1$.

One can interprete the above dynamical rule as describing a spin flip process under the action of two competing heat baths:
with probability $p$ the spin flip is performed completely at random, independently of any energetic consideration (we can interpret 
in this case that ${\bf s}$ is in contact with a heat bath at {\it infinite} temperature), while the spin flip is performed 
at temperature $T$ with probability $(1-p)$ .

The dynamics we have chosen is a particular case of the general group of competitive dynamics.\cite{Pedrotesis} Let's suppose we
have two different dynamics $\omega _1$ and $\omega _2$ which independently satisfy detailed balance,
\begin{equation}
\frac{\omega_j({\bf s} \rightarrow {\bf s'})}{\omega_j({\bf s'} \rightarrow {\bf s})} = 
\textrm{e}^{-\beta \Delta {\cal H}}
\label{balance}
\end{equation}
where $j=1,2$ and in this case $\Delta {\cal H} = {\cal H}({\bf s'}) - {\cal H}({\bf s})$. The detailed balance condition ensures
that the stationary state under the action of one of these dynamics will be described by a Boltzmann distribution. 
That is, in the stationary state the probability of finding the system in a configuration ${\bf s}$, $P_{st}({\bf s})$, will
be proportional to $\textrm{exp}[-{\cal H}({\bf s})/T]$ (remember we fixed $k_B=1$). Hence, the detailed balance condition 
for the transition rate is sufficient, although not necessary, in order to make the system converge asymptotically to an equilibrium
stationary state. This is the case for dynamics $\omega _1$ and $\omega _2$ independently. However, if the stationary state to which
dynamics $\omega _1$ drives the system is different from the stationary state associated to dynamics $\omega _2$ (for instance,
if both dynamics work at different temperature), any competitive dynamics of the form $p \omega _1 + (1-p) \omega _2$ with $0<p<1$
will produce in the system what is called in literature {\it dynamical frustration}, and the competition between both dynamics 
generically drives the system towards a nonequilibrium  steady state\footnote{This is true except for some (in principle) 
nonequilibrium systems which can be mapped, under specific circumstances, 
to equilibrium systems with effective parameters. For these {\it special}
systems we can write an effective Hamiltonian in such a way that the competing dynamics $p \omega _1 + (1-p) \omega _2$ verify
detailed balance with respect to this effective Hamiltonian.\cite{hamefectivo,PedroMigueleffect}}.

As can be deduced from (\ref{rate}), in our case we have chosen $\omega _1 = 2\omega(\beta _1 \Delta {\cal H})$ and 
$\omega _2 = \omega(\beta _2 \Delta {\cal H})$, where 
$\omega (\beta \Delta {\cal H}) = \textrm{exp}(-\beta \Delta {\cal H})/[1+\textrm{exp}(-\beta \Delta {\cal H})]$
is the Glauber transition rate, with $\beta _1 =0$ and $\beta _2 =\beta \equiv 1/T$, and where only transitions between configurations
which differ in a single spin are allowed. This is only one of infinite possibilities when constructing a competing dynamics from two
canonical dynamics driving independently the system towards equilibrium. In principle, any of these infinite possibilities should be
equally valid in order to investigate how nonequilibrium conditions affect the properties of metastable states in Ising-like
systems. However, if we want to predict properties of real magnetic systems, we then have to choose carefully the dynamics.
Thus, Glauber dynamics, used here in our definition of $\omega$, can be derived from first principles for a system of  
$\frac{1}{2}$-spin fermionic quantum particles, each one subject to its own thermal bath.\cite{fermiones} On the other hand, the weak
dynamic perturbation parameterized by $p$ emulates in some 
sense the effect of quantum tunneling of individual spins in real magnetic systems. 
The existence of this small $p\neq 0$ allows the spins to flip independently of any energetic constraint imposed by their 
surroundings with a (very) low probability. This is roughly what quantum tunneling produces in real spins: the spin is able to flip
by {\it tunneling} through the energy barrier which impedes its classical flipping, i.e., independently of this energy barrier.
We also can interpret in a more general way the dynamic random perturbation parameterized by $p$ as a generic
source of disorder and randomness, i.e. as a simplified representation of the impure dynamic behavior typical of real 
systems.\cite{Vacas}

For $p=0$, the dynamics (\ref{rate}) corresponds to the canonical Ising case which converges asymptotically towards a Gibbs equilibrium
state at temperature $T$ and energy ${\cal H}$. In this case the model for $h=0$ exhibits a second order phase transition at a
critical temperature $T=T_c \approx 2.2691 J \equiv T_{ons}$.\cite{Onsager} The critical exponents associated to this phase 
transition define the Ising universality class. This universality class is one of the most robust, and all phase
transitions in monocomponent systems with up-down symmetry and without any extra symmetry or conservation law 
(model A in the Hohenberg-Halperin classification\cite{HohenHalp}) belong to this universality class. For $p\neq 0$ the conflict 
in (\ref{rate}) impedes canonical equilibrium, and as we mentioned above the system then evolves towards a nonequilibrium steady 
state whose nature essentially differs from a Gibbs state at temperature $T$. The system now, and always for $h=0$, exhibits a
second order phase transition at a critical temperature $T_c(p) < T_{ons}$ for small enough values of $p$. This critical point
disappears for values of $p$ above certain critical value $p_c$ defined by the condition $T_c(p_c)=0$ (for more details, see
next chapter). In general, critical phenomena in models with competition of dynamics have been studied as a paradigm of
nonequilibrium phase transitions. In particular, it has been proved for Ising-like models with dynamics of the type 
$p \omega(T_1) + (1-p) \omega (T_2)$ where two different temperatures $T_1$ and $T_2$ compete, and where $\omega(T)$ is
an equilibrium dynamics at temperature $T$, that the critical point observed at $T_{1,c}(p,T_2)$ belongs to the Ising universality
class.\cite{Tamayo,FattahJJMA} Notice that our model is just a particular realization of a system under competition of 
temperatures.

In order to study metastability in our model, we initialize the system in a state with all spins up, i.e. 
$s_i = +1$ $\forall i \in [1,N]$, under a weak magnetic field which favours the opposite orientation, $h=-0.1$ (we will keep constant
this magnetic field during our study). In what follows we will study several different values of temperature $T$ and dynamic 
perturbation $p$, such that the system is always in the ordered phase (i.e. below the critical point). Under these conditions the 
initial state  is metastable, and it eventually will decay towards the truly stable state, which in this case is a negative 
magnetization state, $m \equiv N^{-1} \sum _i s_i < 0$. 
In principle we could use a classic Monte Carlo scheme\cite{Binderlibro} in order to 
simulate such system. However, as a consequence of the strong local stability that characterizes metastable states, the time the system
needs to exit the metastable state and to evolve towards the stable one can be as long as $\tau \sim 10^{40}$ Monte Carlo Steps
per spin at low temperatures. A Monte Carlo Step per spin corresponds to a spin flip trial of all the spins in the lattice on the 
average (from now on we will denote this temporal unit as MCSS). In general, for Ising spins interacting with a phonon bath,
a MCSS corresponds roughly with a physical time of the order of the inverse phonon frequency, which is approximately $10^{-13}$
seconds.\cite{reviewMCAMC} 
If we suppose for a while that the classic Monte Carlo algorithm is able to execute a spin flip trial after every tic 
of the CPU clock (which has an approximate period of $10^{-9}$ seconds), we should wait for low temperatures a time of order
$N \times 10^{31}$ seconds in order to observe the metastable-stable transition ($N$ is the number of spins in the lattice).
This time is fairly larger than the age of the Universe. Hence, we need to use advanced, faster-than-real-time algorithms in order
to simulate the behavior of these systems. This is the case for the {\it Monte Carlo with Absorbing Markov Chains algorithms},
generically known as MCAMC algorithms.\cite{reviewMCAMC} They are {\it rejection-free} algorithms, based on the {\it n-fold way} 
algorithm\cite{nfold}, that without changing the system dynamics (they only rewrite in an efficient way the classic Monte Carlo 
algorithm) are many order of magnitude faster than standard algorithms. However, for low temperatures and weak magnetic fields, even
MCAMC algorithms are not efficient for the metastability problem. Hence we have to implement, together with MCAMC algorithms, 
the so-called {\it slow forcing approximation}.\cite{forcing} In this approximation the system is forced to evolve towards the 
stable state using a moving magnetization wall. That is, we define an upper bound for magnetization, which depends on time, 
$m_{lim}(t) = 1 - \phi t$, and we force the system magnetization to stay at any time below this magnetization threshold.
Although this constraint clearly modifies the dynamics, it has been proved that a slow forcing limit exists for $\phi$\cite{forcing},
in such a way that in this limit all observables are independent of the forcing, while the simulation is still significatively 
accelerated as compared to the system without forcing. In Appendix \ref{apendMCAMC} we explain in detail the MCAMC algorithms and the
slow forcing approximation.

\chapter{Mean Field Approximation to the Problem of Metastability}
\label{capMedio}

\section{Introduction}
\label{capMedio_apIntro}

In this chapter we study the problem of metastability in the nonequilibrium magnetic system taking as
starting point a first order dynamic mean field approximation. This approach is a generalization for the study of dynamic problems
of Kikuchi's method\cite{Kikuchi} known as {\it Cluster Variation Method}. This method has been reformulated for the study 
of some nonequilibrium systems by Dickman and other authors\cite{Dickmanpair,MarroDickman}, 
with the name of Pair Approximation. In this chapter
we will formulate in a first step the pair approximation applied to our model. Later on we will use this approximation to obtain 
information about the static properties that characterize the system. Finally, we will study the dynamics of the system using the 
mean field theory.

\section{Formulation of the Pair Approximation}
\label{capMedio_apForm}

The approximation we describe here, following reference \cite{Pedrotesis}, is a mean field approximation as far as 
it neglects correlations actually present in the system, and it builds, using this assumption, a set of equations for
averaged observables which describe the dynamical and statical behavior of the system.

\begin{figure}
\centerline{
\psfig{file=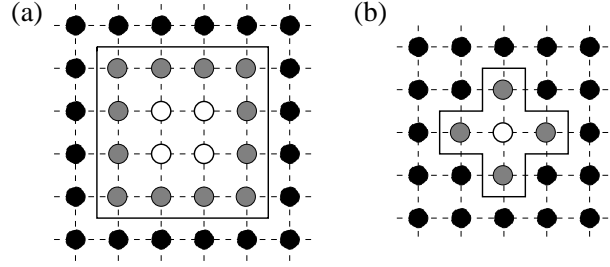,width=8cm}}
\caption[Different examples of spin domains.]
{\small Different examples of spin domains, each one characterized by a different kind of partition 
$\mathbb{P}(\Lambda)$. External spins are coloured in black, surface spins are gray, and internal spins are white.}
\label{dominio}
\end{figure}

As we saw in section \ref{capMotiv_apModel}, our system is defined on a square lattice 
$\Lambda=\{1, \ldots , L\}^2 \subset \mathbb{Z}^2$. A state in the system is described by a configuration vector 
$\mathbf{s}=\{s_i, i=1, \ldots , N\}$, where $s_i$ is the spin variable associated to lattice site $i$, and $N=L^2$ is
the number of spins in the system. The dynamics is given by the master equation,
\begin{equation}
\frac{\textrm{d}P(\mathbf{s};t)}{\textrm{d} t} = \sum_{i \in \Lambda} 
\Big[\omega({\bf s}^i \rightarrow {\bf s})P({\bf s}^i;t) - \omega({\bf s} \rightarrow {\bf s}^i) 
P({\bf s};t)\Big]
\label{mastereq2}
\end{equation}
where $P(\mathbf{s};t)$ is the probability of finding the system in a state $\mathbf{s}$ for time $t$, 
$\omega (\mathbf{s} \rightarrow \mathbf{s}^i)$ is the transition rate between states $\mathbf{s}$ and $\mathbf{s}^i$, and
$\mathbf{s}^i$ is a configuration exactly equal to $\mathbf{s}$ but with the spin at position $i$ overturned.
Our dynamics only allows transitions between configurations which differ in the state of a single spin. Let's
assume now that we perform a partition $\mathbb{P}$ of the lattice $\Lambda$, in such a way that domains $q_j$ resulting
from this partition will verify the following restrictions: $q_j \in \mathbb{P}(\Lambda)$ such that 
$q_j \cap q_{j'} = \varnothing$ if $j \neq j'$ and $\bigcup _{j} q_j = \Lambda$. Subindex $j$ indicates the domain lattice 
position. Given a domain $q_j$, its surface ${\cal S}_j$ is formed by all spins in the domain which have some nearest neighbor 
outside the domain. Equivalently, the domain's interior, ${\cal I}_j$, is formed by all spins in the domain whose nearest neighbor
spins are also inside the domain.\footnote{A more general definition of the domain's interior and surface can be written using
the transition rates. The domain spins whose flipping probability depends on spins outside the domain define the domain's surface.
The interior is defined via ${\cal I}_j = q_j - {\cal S}_j$. However, for our particular system, where the spin flipping probability 
depends on the value of the spin and its four nearest neighbors (see eq. (\ref{rate})), this general definition reduces to the one
expressed in the main text.} Thus $q_j={\cal I}_j \cup {\cal S}_j$. Fig. \ref{dominio} shows an example. Let's assume now that we have 
a local observable $A(\mathbf{s}_{q_j};j)$ which exclusively depends on spins belonging to domain $q_j$ (we denote these spins as
$\mathbf{s}_{q_j}$). The average of this observable at time $t$ is,
\begin{equation}
\langle A(j) \rangle _t = \sum_{\mathbf{s}} A(\mathbf{s}_{q_j};j) P(\mathbf{s};t)
\label{promedio}
\end{equation}
If we derive with respect to time this expression, and make use of eq. (\ref{mastereq2}), we obtain a temporal evolution
equation for the average,
\begin{equation}
\frac{\textrm{d} \langle A(j) \rangle _t}{\textrm{d}t} = \sum _{\mathbf{s}} \sum _{i \in \Lambda}
\Big[\omega (\mathbf{s}^i \rightarrow \mathbf{s}) P(\mathbf{s}^i;t) A(\mathbf{s}_{q_j};j) - 
\omega (\mathbf{s} \rightarrow \mathbf{s}^i) P(\mathbf{s};t) A(\mathbf{s}_{q_j};j)\Big]
\label{mastereq3}
\end{equation}
Now if we make a variable change in the first term of right hand side in this equation, $\mathbf{s}^i \rightarrow \mathbf{s}_c$
(which implies $\mathbf{s} \rightarrow \mathbf{s}_c^i$) and we notice that the index in the sum over configurations is a dumb index
(that is, it does not matter whether to sum over configurations $\mathbf{s}_c^i$ or sum over configurations $\mathbf{s}_c$, since
we go over {\it all} configurations), we can write in a compact way,
\begin{equation}
\frac{\textrm{d} \langle A(j) \rangle _t}{\textrm{d}t} = \sum _{\mathbf{s}} \sum _{i \in q_j}
\Delta A(\mathbf{s}_{q_j};j;i) \omega (\mathbf{s} \rightarrow \mathbf{s}^i) P(\mathbf{s};t)
\label{mastereqA}
\end{equation}
where we define,
\begin{equation}
\Delta A(\mathbf{s}_{q_j};j;i) = A(\mathbf{s}_{q_j}^i;j) - A(\mathbf{s}_{q_j};j)
\label{deltaA}
\end{equation}
and where the sum over $i \in \Lambda$ is now a sum over $i \in q_j$ since $\Delta A(\mathbf{s}_{q_j};j;i) = 0$ if the
spin at position $i$ is outside the domain $q_j$. We can rewrite eq. (\ref{mastereqA}) taking into account the definition
of surface and interior of domain $q_j$,
\begin{eqnarray}
\frac{\textrm{d} \langle A(j) \rangle _t}{\textrm{d}t} & = & \sum _{\mathbf{s}_{q_j}} \sum _{i \in 
{\cal I}_j} \Delta A(\mathbf{s}_{q_j};j;i) \omega (\mathbf{s}_{q_j} \rightarrow \mathbf{s}_{q_j}^i) 
Q(\mathbf{s}_{q_j};t)  \nonumber \\
 & + & \sum _{\mathbf{s}} \sum _{i \in {\cal S}_j}
\Delta A(\mathbf{s}_{q_j};j;i) \omega (\mathbf{s} \rightarrow \mathbf{s}^i) P(\mathbf{s};t)
\label{exacta}
\end{eqnarray}
where we have defined the projected probability,
\begin{equation}
Q(\mathbf{s}_{q_j};t)=\sum_{\mathbf{s}-\mathbf{s}_{q_j}} P(\mathbf{s};t)
\label{probQ}
\end{equation}
which is the probability of finding domain $q_j$ in a configuration $\mathbf{s}_{q_j}$ at time $t$.
When we write $\omega (\mathbf{s}_{q_j} \rightarrow \mathbf{s}_{q_j}^i)$ in the first term of right hand side in eq. 
(\ref{exacta}) we want to stress the fact that the probability of flipping a spin in the domain's interior depends exclusively
on the spins belonging to this domain.

The steps performed up to now do not involve approximations. As a first approximation, we assume from now on that our system is
{\it homogeneous}, i.e. its properties do not depend on the selected point in the system. Hence 
$\langle A(j) \rangle \equiv \langle A \rangle$, $q_j \equiv q$, ${\cal I}_j \equiv {\cal I}$ and ${\cal S}_j \equiv {\cal S}$.
Equivalently, we suppose that the partition is regular, so all domains are topologically identical. On the other hand, eq. 
(\ref{exacta}) shows two well-differentiated terms. The first one only depends on what happens in the domain interior, while the second
one involves the domain's surface, couples the domain dynamics with its surroundings, and makes the problem unapproachable in practice.
Our second approximation consists in neglecting the surface term in this equation. This approximation involves\cite{Pedrotesis}
that the domain is {\it kinetically isolated} from the exterior: the domain's exterior part does not induce any {\it net} variation 
on the local observables defined inside the domain.\footnote{Notice that this approximation is exact for equilibrium systems due to the
detailed balance condition.} Thus we are neglecting in practice correlations longer than the domain size. Under both {\it homogeneity} and
{\it kinetic isolation} approximations, the equation we must study reduces to,
\begin{equation}
\frac{\textrm{d} \langle A \rangle _t}{\textrm{d}t} = \sum _{\mathbf{s}_{q}} \sum _{i \in 
{\cal I}} \Delta A(\mathbf{s}_{q};i) \omega (\mathbf{s}_{q} \rightarrow \mathbf{s}_{q}^i) 
Q(\mathbf{s}_{q_j};t) 
\label{pairapprox}
\end{equation}
In order to go on, we must know the expression for the projected probability $Q(\mathbf{s}_q;t)$. It is known that this probability
can be written as\cite{Pedrotesis},
\begin{equation}
Q(\mathbf{s}_q;t) = 1 + \langle s \rangle _t \sum _{i \in q} s_i + 
\sum _{i,j \in q} \langle s_i s_j \rangle _t s_i s_j + \ldots + \langle \prod _{i \in q} s_i \rangle _t 
\prod _{i \in q} s_i
\label{probexpansion}
\end{equation}
This formula involves $n$-body correlation functions. In order to be coherent with the kinetic isolation 
approximation, which neglects long range correlations, we express the probability $Q(\mathbf{s}_q;t)$ as a function
of a reduced number of correlation functions. In particular, our third approximation consists in expressing all correlations
as functions of magnetization $\langle s \rangle$ and the nearest neighbors correlation function, $\langle s_i s_j \rangle$, 
with $i$ and $j$ nearest neighbors sites inside the domain. This is equivalent to writing $Q(\mathbf{s}_q;t)$ as a function of 
$\rho(+,+)$, $\rho(-,-)$ and $\rho(+,-)$, where $\rho(s,s')$ is the density of $(s,s')$ nearest neighbors pairs,
and as a function of the density of up spins, $\rho(+)$. We only have to define now the domain $q$ that we are going
to use in our study. Since we only take into account nearest neighbors correlations, we must choose a
domain with only one spin in its interior, and $2d$ spins (the nearest neighbors of the interior spin) on the surface, being
$d$ the system dimension (in our particular case, $d=2$). Fig. \ref{dominio}.b shows an example of this domain type.

The probability of finding this domain in a configuration defined by a central spin $s$ and $n$ up nearest neighbors can
be easily written,
\begin{equation}
Q(\mathbf{s}_q;t) \equiv Q(s,n) = Deg(n) \rho(s) \rho(+|s)^{n} \rho(-|s)^{2d-n} 
\label{binomial1}
\end{equation}
where $Deg(n)$, which is the number of domain configurations that are compatible with a central spin $s$ and $n$ up nearest neighbors,
is given by the combinatoric number $2d \choose n$. $\rho(s)$ is the probability of finding a central spin in state $s$, and 
$\rho(\pm |s)$ is the conditional probability of finding a neighboring spin in state $\pm$ given that the central spin is in state $s$.
Since,
\begin{equation}
\rho(\pm|s) = \frac{\rho(\pm,s)}{\rho(s)}
\label{prob}
\end{equation}
where $\rho(\pm,s)$ is the probability of finding a nearest neighbor pair in a state $(\pm,s)$, we can write,
\begin{equation}
Q(s,n)= {2d \choose n} \rho(s)^{1-2d} \rho(+,s)^{n} \rho(-,s)^{2d-n}
\label{binomial2}
\end{equation}
In order to simplify the equations, we use the following notation,
\begin{eqnarray}
\rho(+) & \equiv & x \nonumber \\
\rho(-) & \equiv & 1-x \nonumber \\
\rho(+,+) & \equiv & z \\
\rho(+,-) & \equiv & v = \rho(-,+) \nonumber \\
\rho(-,-) & \equiv & w \nonumber 
\label{notation}
\end{eqnarray}
Taking into account that $v=x-z$ and $w=1+z-2x$, we finally write,
\begin{eqnarray}
Q(+,n) & = & {2d \choose n} x^{1-2d} z^{n} (x-z)^{2d-n} \nonumber \\
Q(-,n) & = & {2d \choose n} (1-x)^{1-2d} (x-z)^{n} (1+z-2x)^{2d-n}
\end{eqnarray}
Inserting these formulas into eq. (\ref{pairapprox}) we arrive to the basic equation in Pair Approximation,
\begin{eqnarray}
\frac{\textrm{d} \langle A \rangle _t}{\textrm{d}t} & = & \sum_{n=0}^{2d} 
{2d \choose n} \Big[\Delta A(+,n) x^{1-2d} z^{n} (x-z)^{2d-n} \omega (+,n)  \nonumber \\
 & - & \Delta A(-,n) (1-x)^{1-2d} (x-z)^{n} (1+z-2x)^{2d-n} \omega (-,n)\Big] 
\label{pairapprox2}
\end{eqnarray}
We must notice that, since the domain state $\mathbf{s}_q$ is defined by the pair $(s,n)$, we have modified our notation
in such a way that $\Delta A (\mathbf{s}_q;i) \equiv \Delta A(s,n)$ and 
$\omega (\mathbf{s}_q \rightarrow \mathbf{s}_q^i) \equiv \omega (s,n)$. Here we also assume that the transition rate depends
only on the value of the central spin, $s$, and the number of up nearest neighbors of this spin, $n$. This is true in our
particular model, where transitions rates are local.

From eq. (\ref{pairapprox2}) we can study the dynamics of any local observable in the system at a first order mean field 
approximation level. However, in order to do so we must know the temporal dependence of both $x \equiv \rho(+)$ and 
$z \equiv \rho(+,+)$. With this aim in mind we apply eq. (\ref{pairapprox2}) to both $x$ and $z$. Thus we must write down two 
local microscopic observables, $A_1(s,n)$ and $A_2(s,n)$, such that their configurational averages correspond to $x$ and $z$, 
respectively. We can check that these observables are,
\begin{eqnarray}
A_1(s,n) & = & \frac{1+s}{2} \nonumber \\
A_2(s,n) & = & \frac{n}{2d} \times \frac{1+s}{2} 
\label{observables}
\end{eqnarray}
Hence, $\langle A_1(s,n) \rangle = x$ and $\langle A_2(s,n) \rangle = z$. From these expressions we can see that $\Delta A_1(s,n)=-s$ 
and $\Delta A_2(s,n)=-sn/2d$. Applying eq. (\ref{pairapprox2}) to both observables we find,
\begin{eqnarray}
\frac{\textrm{d} x}{\textrm{d}t} & = & - \sum_{n=0}^{2d} 
{2d \choose n} \Big[ x^{1-2d} z^{n} (x-z)^{2d-n} \omega (+,n) \nonumber \\
 & - & (1-x)^{1-2d} (x-z)^{n} (1+z-2x)^{2d-n} \omega (-,n)\Big] \label{xzpair1} \\
\frac{\textrm{d} z}{\textrm{d}t} & = & - \frac{1}{2d} \sum_{n=0}^{2d} 
{2d \choose n} n \Big[ x^{1-2d} z^{n} (x-z)^{2d-n} \omega (+,n) \nonumber \\
 & - & (1-x)^{1-2d} (x-z)^{n} (1+z-2x)^{2d-n} \omega (-,n)\Big]  
\label{xzpair2}
\end{eqnarray}
These two equations are the basic equations in Pair Approximation. Hence, once defined the transition rate $\omega (s,n)$ 
(see eq. (\ref{rate})), the general working method thus consists in calculating both $x(t)$ and $z(t)$ using the above equations, 
and using eq. (\ref{pairapprox2}) and the results for $x(t)$ and $z(t)$ calculate any other local magnitude.

\section{Static Properties}
\label{capMedio_apEstat}

In a first step we want to study the stationary solutions of eqs. (\ref{xzpair1}) and (\ref{xzpair2}) as well as their stability
for the nonequilibrium ferromagnetic system. We have two non-linear coupled differential equations,
\begin{eqnarray}
\frac{\textrm{d}x}{\textrm{d}t} = F_1(x,z) \nonumber \\
\frac{\textrm{d}z}{\textrm{d}t} = F_2(x,z)
\label{deff1f2}
\end{eqnarray}
where $F_1(x,z)$ and $F_2(x,z)$ are defined by eqs. (\ref{xzpair1}) and (\ref{xzpair2}), respectively, once we include in these 
equations the explicit form of the transition rate, eq. (\ref{rate}). This dynamics only depends on the energy increment 
$\Delta {\cal H}$ due to a spin change. Since the energy of the system is defined via the Ising Hamiltonian, eq. (\ref{hamilt}),
if we flip a spin $s$ with $n$ up nearest neighbors, then $\Delta {\cal H} (s,n) = 2s[2J(n-d)+h]$, so the transition rate can
be written as,
\begin{equation}
\omega (s,n) = p + (1-p)\frac{\textrm{e}^{- 2s[2J(n-d)+h]}}{1+\textrm{e}^{- 2s[2J(n-d)+h]}}
\label{rate2}
\end{equation}
Once we specify the dynamics, the stationary solutions of the previous coupled set of equations, $x_{st}$ and $z_{st}$, are
the solutions of the system,
\begin{equation}
F_1(x_{st},z_{st})=0 \qquad , \qquad F_2(x_{st},z_{st})=0
\label{estac}
\end{equation}
Both stable and metastable states in a generic system are {\it locally stable} under small perturbations. Hence we are interested 
in locally stable stationary solutions of this set of equations. We must give a local stability criterion. In order to
build such criterion, let's study what happens if we slightly perturb the steady solutions, that is, $x=x_{st} + \epsilon _x$ and
$z=z_{st} + \epsilon _z$, where $\epsilon _x$ and $\epsilon _z$ are very small. From eqs. (\ref{deff1f2}), taking into account
the stationarity condition, eq. (\ref{estac}), a expanding to first order as a function of $\epsilon _x$ and $\epsilon _z$,
we find,
\begin{eqnarray}
\frac{\textrm{d} \epsilon _x}{\textrm{d} t} & = & 
\Big(\frac{\partial F_1}{\partial x}\Big)_{st} \epsilon _x + \Big(\frac{\partial F_1}{\partial z}\Big)_{st} \epsilon _z
+ {\cal O} (\epsilon _x^2, \epsilon _z^2, \epsilon _x \epsilon _z) \nonumber \\
\frac{\textrm{d} \epsilon _z}{\textrm{d} t} & = &  
\Big(\frac{\partial F_2}{\partial x}\Big)_{st} \epsilon _x + \Big(\frac{\partial F_2}{\partial z}\Big)_{st} \epsilon _z
+ {\cal O} (\epsilon _x^2, \epsilon _z^2, \epsilon _x \epsilon _z)
\label{difepsilon}
\end{eqnarray}
We can write this set of equation in matrix form,
\begin{equation}
\frac{\textrm{d}}{\textrm{d} t}
\left( \begin{array}{c}
\epsilon _x \\
\epsilon _z
\end{array} \right)
\approx
\left( \begin{array}{cc} 
\displaystyle (\frac{\partial F_1}{\partial x})_{st} & \displaystyle (\frac{\partial F_1}{\partial z})_{st} \\
\displaystyle (\frac{\partial F_2}{\partial x})_{st} & \displaystyle (\frac{\partial F_2}{\partial z})_{st} \\
\end{array} \right)
\left( \begin{array}{c}
\epsilon _x \\
\epsilon _z
\end{array} \right)
\label{matrix}
\end{equation}
This system can be solved in a standard way, getting,
\begin{equation}
\epsilon _x \approx \epsilon _x^{0} \textrm{e}^{\lambda t} \qquad , \qquad 
\epsilon _z \approx \epsilon _z^{0} \textrm{e}^{\lambda t}
\label{soluc}
\end{equation}
where $\epsilon _x^{0}$ and $\epsilon _z^{0}$ are constants determined by initial conditions, and $\lambda$ is determined by
the solutions of the equation,
\begin{equation}
\left| \begin{array}{cc}
\displaystyle (\frac{\partial F_1}{\partial x})_{st}-\lambda & \displaystyle (\frac{\partial F_1}{\partial z})_{st} \\
\displaystyle (\frac{\partial F_2}{\partial x})_{st} & \displaystyle (\frac{\partial F_2}{\partial z})_{st}-\lambda \\
\end{array} \right| = 0
\label{determinante}
\end{equation}
This equation for the determinant is reduced to a second order polynomial equation for $\lambda$, with two solutions
$\lambda _+$ and $\lambda _-$. From eq. (\ref{soluc}) it is obvious that the steady solutions $x_{st}$ and $z_{st}$ will be 
locally stable if the real parts of both $\lambda _+$ and $\lambda _-$ are negative, since in this way we ensure that the perturbation
will disappear in the limit $t \rightarrow \infty$. The real parts of both $\lambda _+$ and $\lambda _-$ will be negative if 
the following condition is fulfilled,
\begin{eqnarray}
 & & \left( \frac{\partial F_1}{\partial x} \right)_{st} + 
\left( \frac{\partial F_2}{\partial z} \right)_{st} < 0 \nonumber \\
 & & \left( \frac{\partial F_1}{\partial x} \right)_{st} 
\left( \frac{\partial F_2}{\partial z} \right)_{st} - 
\left( \frac{\partial F_1}{\partial z} \right)_{st} 
\left( \frac{\partial F_2}{\partial x} \right)_{st} >  0
\label{Hurwitz}
\end{eqnarray}
This criterion, known as Hurwitz criterion\cite{Hurwitz}, states the necessary and sufficient conditions that a steady solution 
of our set of non-linear differential equations must fulfill in order to be locally stable under small perturbations.

\subsection{Phase Diagram}
\label{capMedio_apEstat_subPha}

We are also interested in simple necessary (although not sufficient) conditions that locally stable steady states must fulfill.
For instance, if we perturb the stationary state by only varying $x$ and keeping untouched $z$, that is, $x=x_{st} + \epsilon _x$ 
and $z=z_{st}$, we arrive to the following solution once we apply the previous stability analysis,
\begin{equation}
\epsilon _x \approx \epsilon_x^0 \textrm{e} ^{t (\frac{\partial F_1}{\partial x})_{st}}
\end{equation}
so the steady state defined by $(x_{st},z_{st})$ can be locally stable only if $(\frac{\partial F_1}{\partial x})_{st} < 0$. It
will be unstable if this derivative is larger than zero. The condition
\begin{equation}
\Big(\frac{\partial F_1(x,z)}{\partial x}\Big)_{st} =0
\label{condcrit}
\end{equation}
defines a point $(x_{st}^c,z_{st}^c)$ of incipient instability or marginal stability which signals the presence of an underlying
critical point or second order phase transition for $h=0$ between a disordered phase and an ordered phase.\cite{Pedrotesis} 
Just at this critical
point we have $x_{st}^c=\frac{1}{2}$, since it separates an ordered phase with non vanishing spontaneous magnetization from
a disordered phase with zero spontaneous magnetization. This observation trivially implies $z\equiv\rho(+,+)=\rho(-,-)\equiv(1+z-2x)$
at the critical point. We also have $z_{st}^c=\frac{1}{3}$ at the critical point.\cite{Juanjotesis} Using these values for 
$x_{st}^c$ y $z_{st}^c$ in eq. (\ref{condcrit}) once we substitute there the explicit form of $F_1(x,z)$, eq. (\ref{xzpair1}),
and solving for temperatures, we find,
\begin{figure}
\centerline{
\psfig{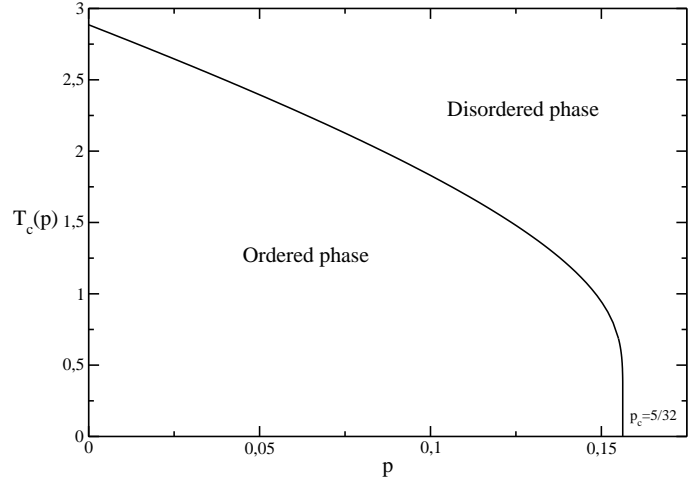}}
\caption[Phase diagram of the nonequilibrium magnetic system.]
{\small Critical temperature for the nonequilibrium ferromagnetic system as a function of $p$ in Pair Approximation.}
\label{diagfases}
\end{figure}
\begin{equation}
\frac{T_c(p)}{J}=\frac{-4}{\ln \big[-\frac{1}{2} + \frac{3}{4} \sqrt{\frac{1-4p}{1-p}}\big]}
\label{tempBethe}
\end{equation}
This equation yields the critical temperature for the nonequilibrium model in first order mean field approximation
as a function of parameter $p$, which characterizes the dynamic nonequilibrium perturbation present in the system.
We can also arrive to this expression from eqs. (\ref{Hurwitz}), which define the general stability criterion, applying 
the marginal stability condition. Fig. \ref{diagfases} shows $T_c(p)$ as a function of $p$. For $p=0$ the critical temperature 
$T_c(p)$ is just the Bethe temperature, $T_{Bethe}/J \approx 2.8854$, to be compared with the exact critical value for $p=0$,
which is the Onsager temperature, $T_{ons}/J \approx 2.2691$. For each value of $p$, temperature $T_c(p)$ signals the border,
always in mean field approximation, between the ordered phase at low temperatures ($T<T_c(p)$) and the disordered phase at
higher temperatures ($T>T_c(p)$). There is a critical value of $p$, $p_c$, such that for larger values of $p$ there is no ordered 
phase for any temperature.  This value $p_c$ can be obtained from the condition $T_c(p_c)=0$, yielding $p_c=\frac{5}{32}=0.15625$.
On the other hand, the phase transition we obtain in mean field approximation belongs to the mean field universality class, as 
expected, on the contrary to the real nonequilibrium system, which as we said in Chapter \ref{capMotiv}  belongs to the Ising 
universality class. Such discrepancy is due to the absence of fluctuations in our mean field approximation.

\subsection{Stable and Metastable States}
\label{capMedio_apEstat_subStat}

After this brief parenthesis about the model critical behavior, we turn back to study its locally stable steady states
in the ordered phase. These stationary states $(x_{st},z_{st})$ will be given by solutions of the set of non-linear 
differential equations (\ref{estac}), subject to the Hurwitz local stability condition, eqs. (\ref{Hurwitz}). Unfortunately,
the non-linearity of the set of eqs. (\ref{estac}) impedes any analytical solution, so we have to turn to numerical solutions.
\begin{figure}[t]
\centerline{
\psfig{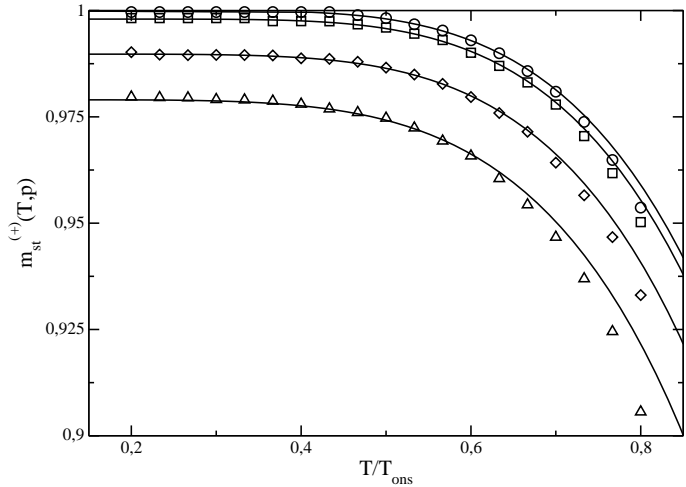}}
\caption[Stable state magnetization for $h=0$.]
{\small Locally stable steady state magnetization as a function of temperature (in units of Onsager temperature)
for different values of $p$ and for $h=0$. In particular, from top to bottom, $p=0$, $0.001$, $0.005$ and 
$0.01$. Points are results obtained from Monte Carlo simulations for a system with $L=53$. Continuous lines
are the solutions in Pair Approximation. Error bars in computational results are much smaller than the symbol sizes.}
\label{magh0}
\end{figure}

In a first step we center our attention on the study of stationarity for zero magnetic field, $h=0$. In this case, the system
exhibits up-down symmetry, so we will have two symmetrical branches of solutions in the ordered phase, one of positive magnetization
and another one with negative magnetization. Moreover, we can prove for $h=0$ that if the pair $(x_{st},z_{st})$ is a 
locally stable steady solution of the set of eqs. (\ref{estac}), then the pair $(1-x_{st},1+z_{st}-2x_{st})$ is also a locally
stable steady solution. If we solve the set of eqs. (\ref{estac}) using standard numerical 
techniques\cite{Recipes} and we keep only those solutions which fulfill Hurwitz criterion, we finally obtain the results shown in 
Fig. \ref{magh0}. There we compare the theoretical predictions for the positive magnetization branch with results obtained from
Monte Carlo simulations for different values of the dynamic random perturbation $p$. The agreement between theory
and computational results is excellent for low and intermediate temperatures for all studied values of $p$, failing gradually
as we approach the critical temperature. Furthermore, the differences between theory and simulations begin to be relevant
for temperatures higher than a $75 \%$ of the critical temperature for each case. Such inaccuracy of Pair Approximation for 
temperatures close enough to the critical one was expected a priori, since mean field theory neglects long range correlations, 
which on the other hand gradually arise as we approach the critical region. Fig. \ref{magh0} shows also that, as we increase
$p$ for a fixed temperature, the system's magnetization decrease in absolute value. Therefore, an increase of $p$ is equivalent to 
an increase of disorder in the system. On the other hand, the qualitative form of the curve $m_{st}^{(+)}(T,p)$ does not change for
$p \neq 0$ as compared to the equilibrium system ($p=0$).
\begin{figure}[t!]
\centerline{
\psfig{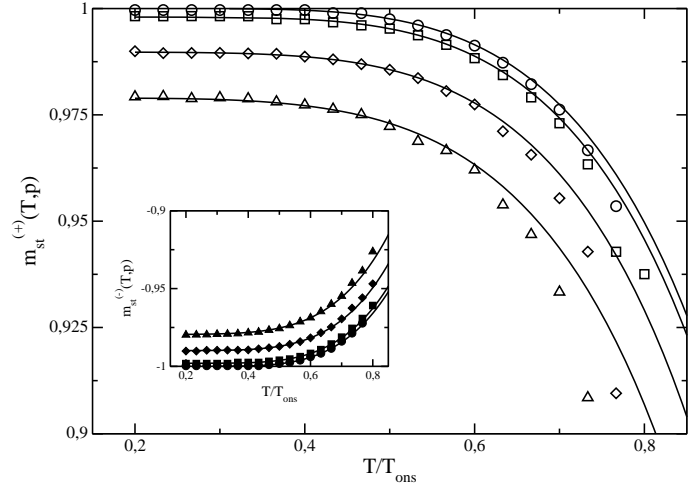}}
\caption[Stable and metastable state magnetization for $h=-0.1$.]
{\small Magnetization of the locally stable steady state of positive magnetization as a function of temperature (in units 
of Onsager temperature) for different values of $p$ and $h=-0.1$. From top to bottom, $p=0$, $p=0.001$, $p=0.005$ and 
$p=0.01$. Points are results obtained from Monte Carlo simulations for a system of size $L=53$. The continuous lines are 
Pair Approximation solutions. Error bars associated to computational results are much smaller than symbol sizes. In the inset we
show the results for the negative magnetization branch.}
\label{magh01}
\end{figure}

The Monte Carlo simulations whose results are shown in Fig. \ref{magh0} have been performed for a system with size $L=53$, subject 
to periodic boundary conditions (we use the same boundary conditions in all simulations within this chapter), with $h=0$ and different
values of $T$ and $p$. In order to measure the magnetization of the positive magnetization steady state, we put the system
in an initial state with all spins up. We let evolve this state with the dynamics (\ref{rate}) for certain values of $T$ and $p$.
After some relaxation time, the systems starts fluctuating around the steady state. We then measure magnetization at temporal
intervals $\Delta t$ larger than the correlation time, and we average over different measurements. The error associated to this 
average is the standard statistical error. A second method to measure the stationary state magnetization is based on the stable
phase growth and shrinkage rates, which we will define later on in this chapter. Both methods yield equivalent results.

We also can study the steady states for $h<0$. In particular, here we study the case $h=-0.1$. As opposed to the $h=0$ case, 
here there is no up-down symmetry since the magnetic field favours the negative orientation of spins. Therefore the negative
and positive magnetization branches are now different. Moreover, the locally stable steady state with positive magnetization is
now metastable. Numerically solving the set of eqs. (\ref{estac}) subject to the conditions (\ref{Hurwitz}) we obtain the results
shown in Fig. \ref{magh01}. In this figure we also show results from simulations analogous to the ones described above, but with 
$h=-0.1$, and where the initial state is defined with all spins up (down) if we want to measure the positive (negative) magnetization 
branch. Comparatively, these results are very similar to the results obtained for $h=0$.

\subsection{Hysteresis and the Intrinsic Coercive Field}
\label{capMedio_apEstat_subHys}

\begin{figure}[t]
\centerline{
\psfig{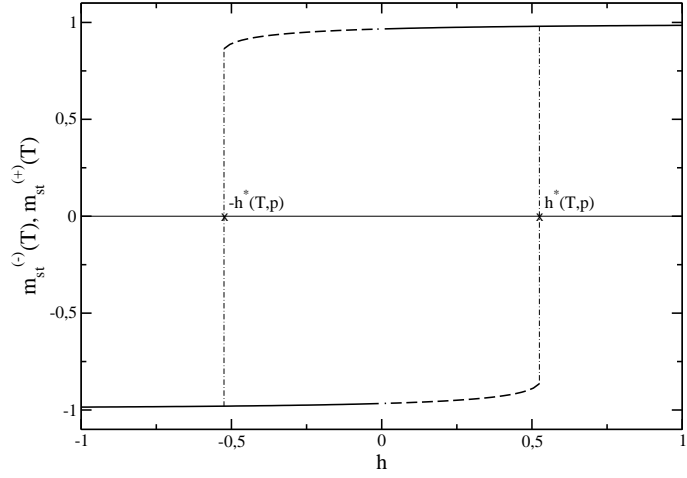}}
\caption[Hysteresis loop for the nonequilibrium magnetic system.]
{\small Locally stable steady state magnetization for both magnetization branches as a function of magnetic field $h$
for fixed $T=0.7T_{ons}$ and $p=0.005$. The continuous line represents stable states, the dashed line represents metastable states, 
and the dot-dashed line signals the discontinuous transition where metastable states disappear. This discontinuity appears for 
a magnetic field $h^*(T,p)$.}
\label{histeresis}
\end{figure}

An interesting question consists in knowing what happen to locally stable steady states as we change the magnetic field. In order to
answer this question we numerically solve again the set of eqs. (\ref{estac}) subject to Hurwitz conditions for fixed temperature 
and dynamic random perturbation $p$, varying the magnetic field between $h=-1$ and $h=+1$. In particular, Fig. \ref{histeresis} shows
the result for $T=0.7 T_{ons}$ and $p=0.005$. This curve forms what is generally known as a hysteresis loop. Hysteresis is a property
of many systems near a first order critical point, and it is intimately related to metastability. A system is said to exhibit 
{\it hysteresis} if its properties depend on its previous history. 
Thus systems showing hysteresis are systems with {\it memory}. In our case, 
as can be seen in Fig. \ref{histeresis} for fixed $T$ and $p$ and for a fixed magnetic field in the interval $[-h^*(T,p),h^*(T,p)]$,
the system properties (represented this time by the magnetization) depend on whether the system evolved along the positive 
or the negative magnetization branch, i.e. they depend on the previous system history. As we can see in Fig. \ref{histeresis},
this dependence on previous history is clearly due to metastability, i.e. locally stable steady states with magnetization opposed to
the magnetic field.

\begin{figure}[t!]
\centerline{
\psfig{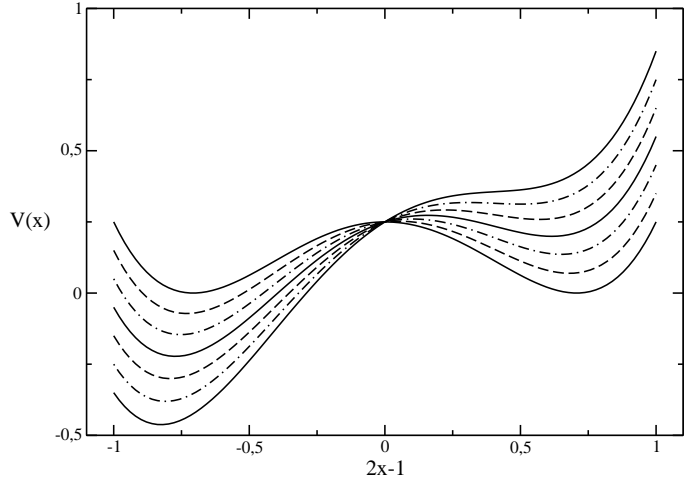}}
\caption[Schematic plot of the nonequilibrium potential.]
{\small Schematic plot of the potential $V(x)$ defined in the main text, for fixed temperature and $p$, and for several
different values of magnetic field $h<0$. Notice that the local minimum in the positive magnetization branch is attenuated as $|h|$
increases, up to its disappearance for large enough values of $|h|$.}
\label{pot}
\end{figure}

There is a magnetic field $h^*(T,p)$ such that for all $|h| > h^*(T,p)$ metastable states
disappear. This magnetic field $h^*(T,p)$ is known as {\it intrinsic coercive field}\footnote{The intrinsic coercive field $h^*$ 
is defined in a precise manner 
as the magnetic field for which magnetization is zero in a hysteresis loop. In our case, the hysteresis loop goes discontinuously
from positive to negative magnetization (and vice versa) when the metastable state disappears, so we can say that magnetization
{\it discontinuously} crosses $m=0$ at $h=h^*$.\cite{Intrinsic}}. 
As we increase the absolute value of the field, metastable states get weaker and weaker.
The reason underlies on the increase of the transition rate for spins in the metastable phase as we increase the magnetic field 
strength, see eq. (\ref{rate2}). Thus there is a value of the magnetic field for which the metastable state is no longer 
metastable and transforms into an unstable state. In order to study this item more carefully, let's suppose that we are able
to simplify eqs. (\ref{deff1f2}) in such a way that we know $z=z(x)$. Now we can rewrite eq. (\ref{xzpair1}) as,
\begin{equation}
\frac{\textrm{d}x}{\textrm{d}t} = - \frac{\delta V(x)}{\delta x}
\label{potencial}
\end{equation}
where $V(x)$ is a (nonequilibrium) potential which controls the system evolution. 
Fig. \ref{pot} shows a schematic plot of this potential
for the ordered phase at fixed temperature and $p$, and for several negative magnetic fields of increasing absolute value.
As we can see, the effect of the magnetic field is to attenuate the local minimum associated to the metastable state. 
For magnetic fields $|h| < h^*(T,p)$ this local minimum, although attenuated, exists. However, for magnetic fields $|h| > h^*(T,p)$
the metastable minimum disappears, and so the metastable state. Therefore, for $|h| > h^*(T,p)$ the set of eqs. (\ref{estac}) has
only one solution, with magnetization sign equal to that of the applied field.

In order to calculate $h^*(T,p)$ let's study how a metastable state changes under small perturbations of magnetic field. Let's 
assume then that $(x_{st}^{h_0},z_{st}^{h_0})$ is a locally stable stationary state for parameters $T$, $p$ and $h_0$, with
magnetization opposed to the external magnetic field. If we slightly perturb this magnetic field, $h=h_0 + \delta h$, also
the locally stable stationary solution will be modified, $x_{st}^{h} = x_{st}^{h_0} + \epsilon _x$ and 
$z_{st}^{h} = z_{st}^{h_0} + \epsilon _z$. We can write,
\begin{eqnarray}
\frac{\textrm{d}x_{st}^{h}}{\textrm{d}t} & = & F_1(x_{st}^{h},z_{st}^{h};h) \approx 
F_1(x_{st}^{h_0},z_{st}^{h_0};h_0) + \frac{\partial F_1}{\partial x} 
(x_{st}^{h_0},z_{st}^{h_0};h_0) \epsilon _x \nonumber \\
& + & \frac{\partial F_1}{\partial z} (x_{st}^{h_0},z_{st}^{h_0};h_0) \epsilon _z + 
\frac{\partial F_1}{\partial h} (x_{st}^{h_0},z_{st}^{h_0};h_0) \delta h
\label{xperth}
\end{eqnarray}
where we have specified the magnetic field dependence. Due to the steadiness of the initial state, 
$F_1(x_{st}^{h_0},z_{st}^{h_0};h_0)=0$, and since $(x_{st}^{h},z_{st}^{h})$ is also a steady state, we have that 
$\frac{\textrm{d}x_{st}^{h}}{\textrm{d}t} =0$. Hence,
\begin{equation}
\frac{\partial F_1}{\partial x} (x_{st}^{h_0},z_{st}^{h_0};h_0) \epsilon _x +
\frac{\partial F_1}{\partial z} (x_{st}^{h_0},z_{st}^{h_0};h_0) \epsilon _z = 
-\frac{\partial F_1}{\partial h} (x_{st}^{h_0},z_{st}^{h_0};h_0) \delta h
\label{sist1}
\end{equation}
In a similar way we can calculate the perturbation that $z_{st}^{h}$ suffers,
\begin{equation}
\frac{\partial F_2}{\partial x} (x_{st}^{h_0},z_{st}^{h_0};h_0) \epsilon _x +
\frac{\partial F_2}{\partial z} (x_{st}^{h_0},z_{st}^{h_0};h_0) \epsilon _z = 
-\frac{\partial F_2}{\partial h} (x_{st}^{h_0},z_{st}^{h_0};h_0) \delta h
\label{sist2}
\end{equation}
Solving the linear system formed by eqs. (\ref{sist1}) and (\ref{sist2}) for $\epsilon _x$ we obtain,
\begin{equation}
\epsilon _x = \Big[ \frac{\displaystyle \frac{\partial F_2}{\partial h} \frac{\partial F_1}{\partial z} -
 \frac{\partial F_1}{\partial h} \frac{\partial F_2}{\partial z}}{\displaystyle \frac{\partial F_1}{\partial x}
\frac{\partial F_2}{\partial z} - \frac{\partial F_2}{\partial x} \frac{\partial F_1}{\partial z}} 
\Big]_ {x_{st}^{h_0}, z_{st}^{h_0}, h_0,T, p} \delta h 
\label{respuesta}
\end{equation}
\begin{figure}
\centerline{
\psfig{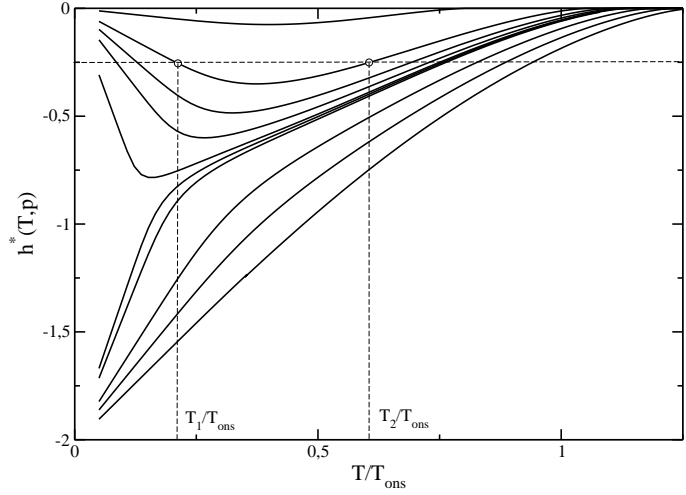}}
\caption[Intrinsic coercive field as obtained from mean field approximation.]
{\small Intrinsic coercive field, $h^*(T,p)$, as a function of temperature for different values of $p$. From bottom to top, 
$p=0$, $0.01$, $0.02$, $0.03$, $0.031$, $0.032$, $0.035$, $0.04$, $0.05$ and $0.1$. Notice that the qualitative change of behavior in
the low temperature limit appears for $p \in (0.031,0.032)$. Here we also show, for $h=-0.25$, the temperatures $T_1 < T_2$ such that
for $T_1 < T < T_2$ there are metastable states for $p=0.05$.}
\label{hcrit}
\end{figure}
This equation says that the metastable state magnetization response after a small variation of the magnetic field is proportional
to such perturbation in a first approximation. However, the magnetization response will be divergent when,
\begin{equation}
\Big[\frac{\partial F_1}{\partial x} \frac{\partial F_2}{\partial z} - 
\frac{\partial F_2}{\partial x} \frac{\partial F_1}{\partial z}\Big]_ 
{x_{st}^{h_0}, z_{st}^{h_0}, h_0,T, p} = 0
\label{condh}
\end{equation}
When this condition holds, there will be a discontinuity in the metastable magnetization as a function of $h$. For fixed $T$ and 
$p$ we thus identify the magnetic field $h_0$ for which condition (\ref{condh}) 
is fulfilled as the {\it intrinsic coercive field}, $h^*(T,p)$.
Unfortunately, we cannot analytically calculate $h^*(T,p)$, since we do not explicitely know the metastable solutions  $x_{st}^{h_0}$ 
and $z_{st}^{h_0}$. Solving again the problem with standard numerical methods, we obtain the results shown in Fig. \ref{hcrit}.
There we plot $h^*(T,p)$ as a function of temperature for different values of the nonequilibrium parameter $p$. The first conclusion
we draw from this family of curves is the existence of two different low temperature limits for $h^*(T,p)$, depending on the
value of $p$. For small enough values of $p$ (including the equilibrium case, $p=0$), the curve $h^*(T,p)$ extrapolates towards 
$-2$ in the limit $T \rightarrow 0$. In particular this is true for $p \in [0,0.031]$ (see Fig. \ref{hcrit}). On the contrary,
for large enough values of $p$, namely $p \in [0.032,\frac{5}{32})$, the curve $h^*(T,p)$ extrapolates towards $0$ in the
limit $T \rightarrow 0$. There is a critical value for $p$, that we estimate here to be $\pi _c \approx 0.0315$, which separates
both types of asymptotic behaviors. As we said before, the intrinsic coercive 
field $h^*(T,p)$ signals the magnetic field strength
above which there are no metastable states. As we see in Fig. \ref{hcrit}, for $p< \pi _c$ the behavior of $h^*(T,p)$ for the 
nonequilibrium system is qualitatively similar to that of the equilibrium one: $|h^*(T,p<\pi_c)|$ is a monotonously decreasing function
of $T$. Therefore, for $p<\pi_c$, if we cool the system we need a stronger magnetic field in order to {\it destroy} the metastable 
state. This result agree with intuition. In a metastable state there are two competing processes: a net tendency of the system to
line up in the direction of the field, and a net tendency in order to maintain the spin order, i.e. in order to keep all spins oriented
in the same direction (whatever this direction is). A metastable state survives a long time because the tendency towards 
maintaining the order in the system overcomes the tendency to line up along the field direction. Both the temperature $T$ and the 
nonequilibrium parameter $p$ are ingredients which introduce disorder in the system. Hence if we drop temperature, since in this
way order grows in the system, we would expect in this phenomenologic picture that the magnetic field needed in order to {\it destroy}
the metastable state should be stronger, as we effectively check for $p<\pi_c$. In the same way, as $p$ is increased, disorder
grows in the system, so $|h^*(T,p)|$ must decrease for a fixed temperature, as we again observe.

On the contrary, for $p>\pi_c$ the system exhibits an unexpected behavior, 
difficult to understand using the above phenomenologic picture.
Let's assume we fix the magnetic field to be $h=-0.25$ and the nonequilibrium parameter to be $p=0.05>\pi_c$. As we can see in 
Fig. \ref{hcrit}, we can define two different temperatures, $T_1 < T_2$, such that if $T<T_1$ or $T>T_2$ the system does not show
metastable states, while metastable states do exist if temperature lies in the interval $T \in (T_1,T_2)$. The fact that 
$h^*(T,p)$ extrapolates to zero in the low temperature limit for $p=0.05>\pi_c$ points out that the nonequilibrium parameter $p=0.05$,
which is the relevant source of disorder and randomness in the low temperature limit, takes a value in this case large enough in 
order to {\it destroy} on its own any metastable state. In principle, following the above phenomenologic picture, we would say that
increasing in this case temperature the metastable state should not ever exist, because we add disorder to the system. However,
we observe that a regime of intermediate temperatures exists, $T \in (T_1,T_2)$, where metastable states emerge. 
This observation involves the presence of a {\it non-linear cooperative phenomenon} between the thermal noise 
(parameterized by $T$) and the non-thermal fluctuation source (parameterized by $p$): 
although both noises add independently disorder to the system,
which involves the attenuation or even the destruction of the existing metastable states, the combination of both noise sources,
parameterized in the dynamics (\ref{rate}), not always implies a larger disorder, giving rise to regions in parameter space
$(T,p)$ where there are no metastable states for low and high temperatures, existing however metastability for intermediate 
temperatures. This counter-intuitive behavior resembles in some sense the behavior of some systems under the action of
multiplicative noise, as for instance the annealed Ising model\cite{Thorpe}, where a disordered phase exists for low and high 
temperatures, but there is an ordered phase for intermediate temperatures.\cite{reentrant}
\begin{table}[t!]
\centerline{
\begin{tabular}{|c||c|c|c|}
\hline \hline
Class & Central spin & Number of up neighbors & $\Delta {\cal H}$ \\
\hline \hline
1 & +1 & 4 & 8J+2h \\
\hline
2 & +1 & 3 & 4J+2h \\
\hline
3 & +1 & 2 & 2h \\
\hline
4 & +1 & 1 & -4J+2h \\
\hline
5 & +1 & 0 & -8J+2h \\
\hline \hline
6 & -1 & 4 & -8J-2h \\
\hline
7 & -1 & 3 & -4J-2h \\
\hline
8 & -1 & 2 & -2h \\
\hline
9 & -1 & 1 & 4J-2h \\
\hline
10 & -1 & 0 & 8J-2h \\
\hline \hline
\end{tabular}
}
\caption[Spin classes for the two-dimensional isotropic Ising model.]
{\small Spin classes for the two-dimensional isotropic Ising model with periodic boundary conditions. The last column shows
the energy increment involved by a spin flip for each class.} 
\label{tabclasesbis}
\end{table}

\subsection{Intrinsic Coercive Field from Monte Carlo Simulations: Stable Phase Growth and Shrinkage Rates}
\label{capMedio_apEstat_subCoer}

Now we want to check our theoretical predictions via computer simulations, so we need to discern when the system exhibits a metastable 
state. In order to establish a criterion, we must introduce the concept of spin class (see Appendix \ref{apendMCAMC}). 
For a spin $s$ in the lattice, the spin class to which this spin belongs to is defined once we know the spin orientation, $s=+1$
or $s=-1$, and its number of up nearest neighbors, $n\in [0,4]$. Therefore, for the two-dimensional isotropic Ising model subject 
to periodic boundary conditions there are $10$ different spin classes, schematized in Table \ref{tabclasesbis}. All spins belonging
to the same spin class involve the same energy increment $\Delta {\cal H} (s,n)$ when flipped (see Table \ref{tabclasesbis}),
so the transition rate for a spin to flip depends exclusively on the spin class $i \in [1,10]$ to which the spin belongs to,
$\omega _i \equiv \omega (s,n)$, see eq. (\ref{rate2}). If $n_k(m)$ is the number of spins in the system that belong to class $k$
when the system has magnetization $m$, then $n_k(m) \omega_k$ will be the number of spins in class $k$ which flip per unit time
when we have $n_{up}=N(1+m)/2$ up spins. Since in our convention (see Table \ref{tabclasesbis}) all classes $k \in [1,5]$ are 
characterized by a central spin with $s=+1$ and $n=4,3, \ldots ,0$ up nearest neighbors, the number of up spins which flip
per unit time when magnetization is $m$ will be,
\begin{equation}
g(m)=\sum_{k=1}^{5} n_k(m) \omega_k
\label{growing}
\end{equation}
This observable is the growth rate of the negative magnetization phase, and it depends on system's magnetization. In a similar way,
we define the shrinkage rate of the negative magnetization phase as,
\begin{equation}
s(m)=\sum_{k=6}^{10} n_k(m) \omega_k
\label{shrinking}
\end{equation}
Now $s(m)$ is the number of down spins which flip per unit time when system's magnetization is $m$. Since we are studying a system
subject to a negative magnetic field, we will name $g(m)$ and $s(m)$ {\it stable phase growth and shrinkage rates},
respectively.\cite{projective}
\begin{figure}
\centerline{
\psfig{file=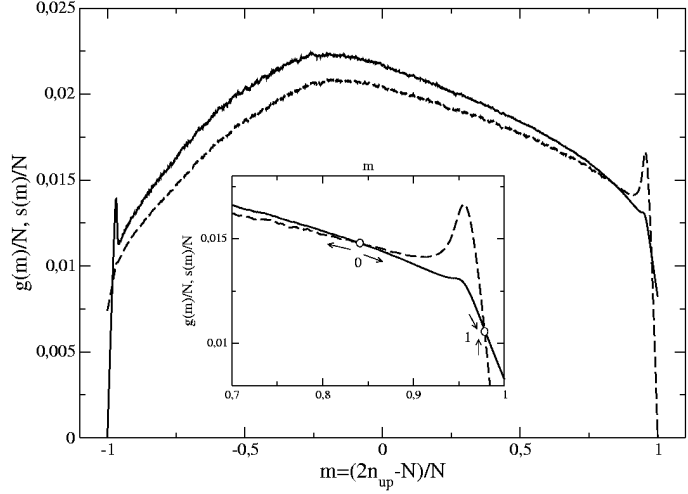,width=9cm}}
\caption[Measured stable phase growth and shrinkage rates.]
{\small Growth and shrinkage probabilities of the stable phase, $g(m)/N$ and $s(m)/N$ respectively, for a system of size 
$L=53$, with $T=0.6 T_{ons}$, $p=0.005$ and $h=-0.1$.  The continuous line represents $g(m)/N$, while the dashed line represents 
$s(m)/N$. The inset shows a detail of the positive magnetization region.}
\label{gs}
\end{figure}

If we have a state with magnetization $m$, the rate of change of magnetization will be,
\begin{equation}
\frac{\textrm{d}m}{\textrm{d}t} = \frac{2}{N}\big[s(m)-g(m)\big]
\label{magrates}
\end{equation}
Thus the system will show steady states for $g(m)=s(m)$. Fig. \ref{gs} shows $g(m)/N$ and $s(m)/N$ as measured in a system 
with size $L=53$, temperature $T=0.6 T_{ons}$, $p=0.005$ and $h=-0.1$, after averaging over $1000$ different demagnetization 
experiments. These demagnetization experiments begin with all spins up (such state is metastable for the studied parameters)
and finish once the negative magnetization stable state has been reached. As we can see in Fig. \ref{gs}, there are three points 
where the curves $g(m)$ and $s(m)$ intersect one each other. Two of this intersection points appear in the positive magnetization 
region, and the third one appears in the negative magnetization region. The points where $g(m)=s(m)$ indicate steady states of 
the real system, whose magnetization can be deduced from the intersection abscissa. Let's denote these magnetization values 
as $m_{-1}$, $m_0$ and $m_1$, being $m_{-1}$ the magnetization of the intersection point
in the negative magnetization region, $m_0$ the magnetization of the intermediate intersection point, and $m_1$ the largest
intersection point magnetization. In order to discern local stability, let's study what happen if we slightly perturb the
magnetization in these steady states. If we perturb for instance the steady state with the largest intersection point magnetization,
$m_1$, in such a way that the final state has magnetization $m=m_1 + \delta m$, we can see that if $\delta m > 0$ then 
$g(m_1 + \delta m) > s(m_1 + \delta m)$, while $g(m_1 + \delta m) < s(m_1 + \delta m)$ if $\delta m < 0$. In both cases,
as indicated by eq. (\ref{magrates}), the system tends to counteract the perturbation, coming back to the stationary state.
Hence the stationary state with the largest magnetization, $m_1$, is locally stable under small perturbations. The arrows in
the inset of Fig. \ref{gs} represent the tendency of the system immediately after the perturbation. We find something analogous
for the steady state with negative magnetization, $m_{-1}$, i.e. it is locally stable. Therefore
the stationary state represented by $m_1$ signals the metastable state, while the stationary state $m_{-1}$ signals the
stable state in this case (remember $h<0$). The steady state 
$m_0$ is unstable under small perturbations, as can be easily derived using the above arguments. This stationary state signals
the separation point between the region where the stable phase tends to disappear ($m>m_0$), and the region where the stable phase
tends to grow ($m<m_0$). As we will see in forthcoming chapters, this point defines the critical fluctuation needed in order to
exit the metastable state. This critical fluctuation is the magnitude that controls the demagnetization process.
Finally, we want to point out that measuring $g(m)$ y $s(m)$ in particular experiments, extracting the stable and metastable state
magnetizations, $m_{-1}$ and $m_1$ respectively, and averaging such measures over many different experiments, we can obtain 
a measure of the average stable and metastable state magnetizations. This measure compares perfectly  with the previously presented
results (see Fig. \ref{magh0} and \ref{magh01}, and complementary discussion).
\begin{figure}[t!]
\centerline{
\psfig{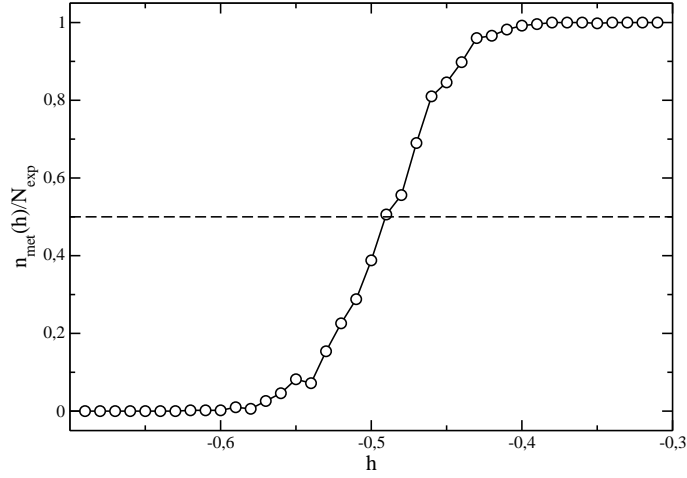}}
\caption[Probability of finding a metastable state as a function of $h$.]
{\small Probability of finding a metastable state, as defined in the main text, as a function of magnetic field $h<0$
for a system of size $L=53$, with temperature $T=0.7 T_{ons}$ and $p=0$, where we have performed $N_{exp}=500$ demagnetization 
experiments for each value of $h$. Error bars are smaller than symbol sizes.}
\label{probh}
\end{figure}

\begin{figure}[t!]
\centerline{
\psfig{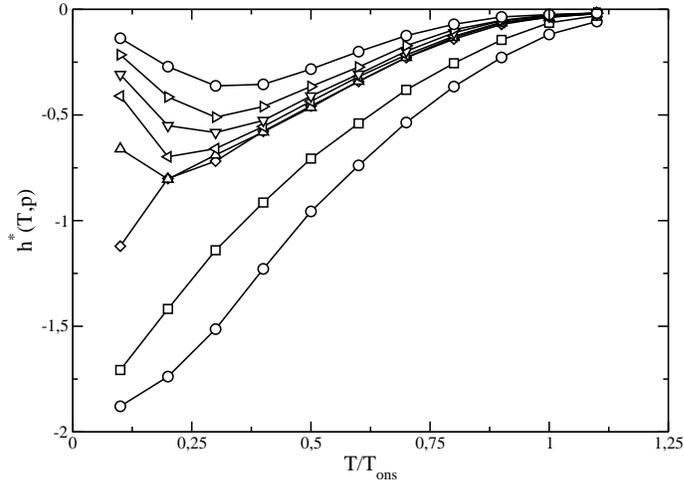}}
\caption[Intrinsic coercive field as obtained from Monte Carlo simulations.]
{\small Monte Carlo results for the intrinsic coercive field, $h^*(T,p)$, as a function of temperature for different values of $p$.
In particular, from bottom to top,  $p=0$, $0.01$, $0.03$, $0.0305$, $0.0320$, $0.0350$, $0.04$ and $0.05$. Notice the change of 
asymptotic behavior in the low temperature limit for $p \in (0.03,0.0305)$. This figure is to be compared with Fig. \ref{hcrit}.}
\label{hcritMC}
\end{figure}
It is clear from the previous discussion that if there is a magnetization interval inside the metastable region (in our case, the 
positive magnetization region) where $s(m)>g(m)$, that is, where the stable phase shrinkage rate is larger than its growth rate,
then a metastable state will exists. On the other hand, when $|h|>h^*(T,p)$ the metastable state will not exist. In this case it
is observed that the curve $s(m)$ does not intersect $g(m)$ in the positive magnetization region. Hence the existence or absence
of intersection between $s(m)$ and $g(m)$ in the positive magnetization region (for $h<0$) allows us to decide whether the system 
exhibits a metastable state or not. In the above discussion we have treated all states with the same magnetization as an single state.
However, there are many different microscopic states in the system which are compatible with a fixed magnetization. These states
may exhibit very different properties. In particular, the rates $g(m)$ y $s(m)$ depend not only on magnetization, but on the population
of all the spin classes. Thus, for a fixed set of parameters $T$, $p$ and $h<0$, we can  have experiments where $s(m)$ and $g(m)$
intersect one each other in the positive magnetization region, and for the same parameters we can observe other experiments
where they do not intersect. Therefore, instead of speaking about the existence or absence of a metastable state, we must speak about
the {\it probability} of existence of a metastable state. 
In this way we can define a method to measure the intrinsic coercive field  $h^*(T,p)$
in Monte Carlo simulations. For a fixed set of parameters $T$, $p$ and $h<0$, we perform $N_{exp}$ different demagnetization 
experiments, starting from a state with all spins up. We measure on each experiment the stable phase growth and shrinkage rates,
$g(m)$ and $s(m)$ respectively, as a function of magnetization. If $n_{met}$ of those $N_{exp}$ experiments are such that 
$g(m)$ and $s(m)$ intersect one each other in the positive magnetization region, we can define the probability of existence of 
a metastable state as $n_{met}(T,p,h)/N_{exp}$. If we repeat such process for
fixed values of temperature $T$ and nonequilibrium perturbation $p$, varying the magnetic field in a wide interval, we obtain
the results shown in Fig. \ref{probh}. Here we observe that the metastable state existence probability abruptly changes from
$+1$ to $0$ in a narrow magnetic field interval. Therefore we define in this case 
the intrinsic coercive field, $h^*(T,p)$, for fixed $T$ and $p$, as the magnetic field for which $n_{met}(T,p,h^*)/N_{exp}=0.5$.

Fig. \ref{hcritMC} shows $h^*(T,p)$, as measured from Monte Carlo simulations using the above explained method, as a function
of temperature for a system size $L=53$, and for varying values of $p$. Comparing this figure with
Fig. \ref{hcrit}, we observe that Monte Carlo results confirm both qualitatively and quantitatively\footnote{At the quantitative
level, Monte Carlo results match rather well with the theoretical predictions, although we observe, as expected, some systematic 
deviations for temperatures close enough to the critical one.} the theoretical predictions based on Pair Approximation.
In this way we observe that the low temperature asymptotic behavior of $h^*(T,p)$ depends on the nonequilibrium parameter $p$.
There is a critical value $\pi_c$ for $p$ which separates both asymptotic behaviors. We estimate from Monte Carlo simulations 
$\pi_c^{MC} \approx 0.03025$ (see Fig. \ref{hcritMC}). This critical value has to be compared with the result derived from Pair
 Approximation, $\pi_c^{pair} \approx 0.0315$. Hence we confirm that the system exhibits, 
as we discussed in the previous section, a non-linear cooperative phenomenon 
between the thermal noise, parameterized by $T$, and the non-thermal
noise, parameterized by $p$, for $p>\pi_c$, in such a way that there are no metastable states for low and high temperatures, but
there is an intermediate temperature region where metastable states emerge due to the non-linear coupling between both noises.

\section{Dynamics of Metastable States in Mean Field Approximation}
\label{capMedio_apDin}

In the previous section we have studied the {\it static} properties of our metastable system. These static properties
are related to the properties of the nonequilibrium potential, defined by eq. (\ref{potencial}), 
which controls the system behavior, see
Fig. \ref{pot}. In particular, we have calculated the local extrema positions, and their local stability against
small perturbations. Finally, we have investigated the intrinsic coercive magnetic field. 
These static investigations give us a good picture
of the nonequilibrium potential shape for different values of temperature $T$, nonequilibrium perturbation $p$ and magnetic field $h$.

However, the metastability phenomenon is, as we discussed in Chapter \ref{capMotiv}, a {\it dynamical} process
where the system, after wandering a long time around the metastable state, {\it evolves} towards the globally stable state.
Therefore we are also interested in understanding the properties of this evolution, the processes which give rise to the
metastable-stable transition, and the role played by nonequilibrium conditions in the system behavior. We also want to characterize
the system dynamics with observables that, as the metastable state lifetime, define the typical temporal scales of the process.

Due to the successful description that we have obtained from the dynamic mean field approximation for the static 
properties of the system, it seems sensible to derive also the system dynamic properties using this approximation.
However, this methodology faces up to a fundamental problem, implicit in the structure of the mean field approximation.
In the previous section we affirmed that, in mean field theory, all locally stable steady states with positive magnetization
are metastable for $h<0$. However, we can trivially prove that once the set of non-linear differential equations (\ref{deff1f2})
reaches a steady state, it {\it never} evolves to any other state. This fact does not fit with the definition we gave in Chapter
\ref{capMotiv} for a metastable state, where we identified as a key feature of such state its transient character, i.e. a
metastable state finally evolves towards the truly stable state. The reason why mean field theory fails when describing
the exit from a metastable state underlies in one of the basic hypothesis assumed in this approximation: the suppression of 
fluctuations. The real system, once situated in the metastable region, {\it fluctuates} around the metastable state,
due to the presence of noise in the system (in our case, noise has thermal -$T$- and non-thermal -$p$- origin). These fluctuations
temporarily separate the system from the metastable state, although it rapidly reacts coming back to the metastable state.
However, the {\it existence} of such fluctuations implies a non vanishing probability for one of these fluctuations to have
a large enough amplitude in order to allow the system to overcomes the energy barrier which separates the metastable state from 
the stable one. That is, fluctuations constitute the origin of the metastable-stable transition, so a theory which neglects
fluctuations cannot properly describe this dynamical process.

Hence, in order to theoretically investigate the dynamical aspects associated to metastability in our system  using mean field theory,
we must generalize the Pair Approximation in order to include the effect of fluctuations. In order to do so, 
let's remember one of the two basic equations in our approximation, namely eq. (\ref{xzpair1}),
\begin{eqnarray}
\frac{\textrm{d} x}{\textrm{d}t} & = & - \sum_{n=0}^{2d} 
{2d \choose n} \Big[ x^{1-2d} z^{n} (x-z)^{2d-n} \omega (+,n) \nonumber \\
 & - & (1-x)^{1-2d} (x-z)^{n} (1+z-2x)^{2d-n} \omega (-,n)\Big] 
\label{xzpair1bis}
\end{eqnarray}
This equation describes the time evolution of the probability of finding an up spin in the system, $x \equiv \rho(+)$, as a function
of both $x$ and $z$, where $z$ is the probability of finding a $(+,+)$ nearest neighbor pair. Let's remember now that the magnitudes,
\begin{eqnarray}
Q(+,n) & \equiv & {2d \choose n} x^{1-2d} z^{n} (x-z)^{2d-n} \nonumber \\
 & = & {2d \choose n} \rho(+) \rho(+|+)^n \rho(-|+)^{2d-n} \nonumber \\
Q(-,n) & \equiv & {2d \choose n} (1-x)^{1-2d} (x-z)^{n} (1+z-2x)^{2d-n} \nonumber \\
 & = & {2d \choose n} \rho(-) \rho(+|-)^n \rho(-|-)^{2d-n} 
\label{classprob}
\end{eqnarray}
defined in eq. (\ref{binomial2}) and entering eq. (\ref{xzpair1bis}), are the probabilities of finding a spin in state up or down,
respectively, and with $n$ up nearest neighbors. In the previous section we have specified that the spin class to which a spin 
belongs to is defined by the spin orientation and by its number of up nearest neighbors. Therefore, the magnitude $Q(s,n)$,
which we define in mean field theory for our particular choice of the domain topology, is the probability of finding a spin
in the system belonging to a spin class defined by the pair $(s,n)$. For instance, $Q(+,4)$ is the probability (in mean field
approximation) of finding a spin in the system belonging to the first spin class, as defined in our convention, see Table 
\ref{tabclasesbis}. Although it is not clearly stated in our notation, we must notice that $Q(s,n)$ depends on $x$ and $z$,
see eqs. (\ref{classprob}).

Once we have identified the magnitudes $Q(s,n)$ as the population densities for each spin class in Pair Approximation, we can 
trivially write down the stable phase growth and shrinkage probabilities, $\tilde{g}(x,z)$ and $\tilde{s}(x,z)$ respectively,
in this approximation,
\begin{eqnarray}
\tilde{g}(x,z) & \equiv & \sum_{n=0}^{2d} Q_{x,z}(+,n) \omega (+,n) \nonumber \\
 & = & \sum_{n=0}^{2d} {2d \choose n} x^{1-2d} z^{n} (x-z)^{2d-n} \omega (+,n) \nonumber \\
\tilde{s}(x,z) & \equiv & \sum_{n=0}^{2d} Q_{x,z}(-,n) \omega (-,n) \nonumber \\
 & = & \sum_{n=0}^{2d} {2d \choose n} (1-x)^{1-2d} (x-z)^{n} (1+z-2x)^{2d-n} \omega (-,n)
\label{gspares}
\end{eqnarray}
where we have stressed when using the notation $Q_{x,z}(s,n)$ the dependence of the spin class population densities on $x$ and $z$.
Now we can write eq. (\ref{xzpair1bis}) as,
\begin{equation}
\frac{\textrm{d}x}{\textrm{d}t} = \tilde{s}(x,z)-\tilde{g}(x,z)
\label{xratespares}
\end{equation}
which is formally equivalent to eq. (\ref{magrates}), taking into account that we spoke there
about stable phase growth and shrinkage {\it rates}, and here we speak about stable phase growth and shrinkage {\it probabilities}.

This reformulation of some of the Pair Approximation original equations in terms of spin class population densities will allow us to 
introduce fluctuations in the system in a very natural way. Other methods which introduce 
fluctuations in mean field theories have been proposed in the literature\cite{MAPedro}, although these methods need additional
assumptions. In order to implement fluctuations in mean field theory, let's have a look on the real system evolution. In our
model, each spin in the lattice belongs to one of $10$ possible spin classes. As the stochastic dynamics flips the spins, the 
population of the different spin classes changes. Moreover, as we explain in Appendix \ref{apendMCAMC}, the population of the different 
classes is the magnitude which defines, together with the microscopic dynamics of the system, the stochastic jumps in system's 
phase space. Our mean field theory yields an approximation for the spin class population densities for each value of $x$ and $z$.
Hence, in order to include fluctuations in our mean field theory we only have to build a stochastic process similar to the one
observed in the real system, but where instead of using the spin class population densities obtained from simulations, we will
use the spin class population densities derived from mean field theory. This process will give rise to {\it stochastic}
trajectories in the mean field phase space, defined by $(x,z)$, which will include fluctuations around the average behavior equivalent
to those observed in the real system.

\begin{figure}[t!]
\centerline{
\psfig{file=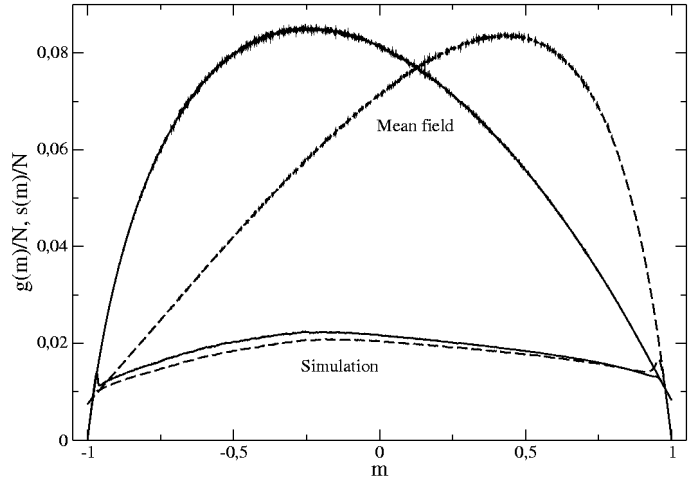,width=9cm}}
\caption[Comparison of growth and shrinkage rates.]
{\small Comparison of stable phase growth and shrinkage probabilities, $g(m)/N$ and $s(m)/N$ respectively, as obtained
from mean field stochastic dynamics (pair of curves with larger amplitude) and from Monte Carlo simulations of the real system.
The continuous line represents $g(m)/N$ in both cases, while the dashed line represents $s(m)/N$. The parameters are $T=0.6T_{ons}$, 
$p=0.005$ and $h=-0.1$, for a system size $L=53$, i.e. $N=2809$ spins. These parameters are the same than those used in Fig. \ref{gs}.
Notice that $m=2x-1$.}
\label{gscomp}
\end{figure}

Let's assume now that we want to study the system dynamics in mean field approximation for temperature $T$, nonequilibrium 
perturbation $p$, and magnetic field $h<0$. The algorithm we use in order to study the demagnetization process is the 
following\footnote{This algorithm is based on the s-1 MCAMC algorithm explained in Appendix \ref{apendMCAMC}. In general,
MCAMC algorithms, when applied to the real system, are exact reformulations of the standard Monte Carlo algorithm, although 
they are much more efficient since they are rejection-free. The philosophy used by these algorithms, based on the different spin 
class populations, allows us to write down a mean field stochastic dynamics from our theory, as we show in the main text.},
\begin{enumerate}
\item We put the mean field system in an initial state defined by $x=1=z$. This initial state is equivalent to a state with all 
spins up.
\item For those values of $x$ and $z$ we evaluate the spin class population densities for all the classes, $Q_{x,z}(s,n)$,
using eqs. (\ref{classprob}).
\item We calculate the time $\Delta t$ the system spends in the current state $(x,z)$  before going to some other state
(see Appendix \ref{apendMCAMC}). This time depends exclusively on the current state. We increment time, $t \rightarrow t + \Delta t$.
\item We randomly select one of the possible spin classes with probability proportional to the spin class flip probabilities, 
$Q_{x,z}(s,n) \times \omega(s,n)$. This spin class is defined by the pair $(s,n)$, where $s= \pm 1$ and $n=0,\ldots,4$.
\item We actualize the variables $x$ and $z$, in such a way that $x \rightarrow x+\delta x(s)$ and 
$z \rightarrow z + \delta z (s,n)$.
\item We come back to point 2 until we reach the stable state.
\end{enumerate}
The increments $\delta x(s)$ and $\delta z(s,n)$ obviously depend on the chosen spin class in order to actualize the system. Thus,
if we choose class 1, characterized by a central spin $s=+1$ and $n=4$ up nearest neighbors (see Table \ref{tabclasesbis}),
we know that $x$, which is the probability of finding an up spin in the system, will decrease in an amount $1/N$ when we flip a spin
in this spin class, where $N$ plays the role of the number of spins in the system. On the other hand, if we choose class 7,
characterized by a down central spin and $n=3$ up nearest neighbors, now $x$ will increase up to $x+1/N$. Therefore, the increment
induced on $x$ only depends on the values of the central spin which defines each class, and not on the number of up nearest 
neighbors. Thus we can write, for a fixed selected spin class characterized by a central spin $s$,
\begin{equation}
\delta x(s) = -\frac{s}{N}
\label{dx}
\end{equation}
In order to calculate the 
modification that the variable $z$ suffers after one step, $\delta z(s,n)$, let's assume again that we choose class 1 to actualize 
the system, where we already know that $s=+1$ and $n=4$. Variable $z$ is the probability of finding a $(+,+)$ pair in the system,
$\rho(+,+)$. Each spin in class 1 contributes to $z$ with $n=4$ different $(+,+)$ nearest neighbors pairs. The total number of pairs 
in a system with $N$ spins and dimension $d$ is $dN$. Hence when we flip a spin in class 1, the variable $z$ has to decrease in an 
amount $4/dN$. In general we can write,
\begin{equation}
\delta z(s,n)= - \frac{sn}{dN}
\label{dz}
\end{equation}
With eqs. (\ref{dx}) and (\ref{dz}) we completely define the mean field stochastic dynamics previously explained. We have checked 
that this stochastic dynamics yields the same results for the static properties of the system that we found using Pair 
Approximation in the previous section. On the other hand, this method allows us to obtain the dependence of $z$ on $x$ for any 
$x \in [0,1]$, $z=z(x)$, for any parameter space point $(T,p,h)$. This was not possible numerically integrating the 
set of eqs. (\ref{xzpair1}) and (\ref{xzpair2}), due to the presence of multiple stationary states. In particular, once we obtain
the dependence $z(x)$, we can compare the stable phase growth and shrinkage probabilities calculated using the mean field
stochastic dynamics, ecs. (\ref{gspares}), with the computational results for these observables. Fig. \ref{gscomp} shows this 
comparison. We use there the parameters $T=0.6 T_{ons}$, $p=0.005$ and $h=-0.1$ for a system size $L=53$, i.e. $N=2809$ spins.
These parameters were also used in Fig. \ref{gs}.

We observe in this figure that, while the comparison between the computational results and the mean field stochastic dynamics 
prediction is remarkably good in the relatively narrow magnetization intervals $m \in [-1,m_{-1}]$ and $m \in [m_1,1]$ (we defined 
$m_1$ and $m_{-1}$ in the previous section), this comparison turns over disastrous for magnetizations in the interval 
$m \in (m_{-1},m_1)$. In particular, the values  $m_{-1}$ and $m_1$ derived from the mean field stochastic dynamics nicely agree with 
the stable and metastable state magnetizations measured in the real system. 

The stable phase growth and shrinkage probabilities, $g(m)/N$ y $s(m)/N$ respectively, are the most relevant observables in the
metastable-stable transition. They determine most of the properties of this dynamic process between the metastable state ($m_1$)
and the stable one ($m_{-1}$). For instance, we can calculate the metastable state lifetime from these probabilities (see Appendix
\ref{apendMCAMC}). Therefore, the discrepancies between the theoretical prediction for both probabilities (based on the mean field
stochastic dynamics approximation) and the measured values for $g(m)/N$ y $s(m)/N$ in the interval $m \in (m_{-1},m_1)$ points out
that our theoretical approximation, in spite of taking into account fluctuations, is not able to describe the dynamics of the 
metastable state demagnetization process.

The key question now is: Why does the extended mean field approximation fail when describing the dynamics of the demagnetization 
process ?. The answer to this question underlies again in one of the basic hypothesis of mean field approximation: the hypothesis
of {\it homogeneity}. During the formulation of Pair Approximation (see section \ref{capMedio_apForm}) we assumed two basic 
hypothesis: homogeneity and kinetic isolation\footnote{We have to add a third hypothesis, namely suppression of fluctuations.}.
The kinetic isolation hypothesis involved neglecting long range correlations. Although these long range correlations are generally
relevant to understand the global behavior of the system, we can neglect them for temperatures far away the critical one,
as in our case. On the other hand, the homogeneity hypothesis involved the assumption that the properties of the system did not
depend on the position in the system. However, the metastable-stable dynamic transition is a highly inhomogeneous phenomenon:
fluctuations present in the system give rise to small {\it droplets} of the stable phase inside the metastable bulk. These compact 
droplets appear instead of non-compact fluctuations because compact fluctuations minimize in some sense the surface-volume ratio, 
and hence they are energetically favoured. These localized droplets constitute an inhomogeneity not taken into account in our
mean field theory. If one (or several) of these droplets reaches a size large enough such that the system is able to overcome the 
energy barrier which separates the metastable state from the stable one, then the system will rapidly evolve towards the stable state.
When we impose homogeneity in mean field theory, we force the system to behave coherently, that is, all spins in the system 
behave in the same way. Therefore, the fluctuations we have included in the theory via the mean field stochastic dynamics approach,
described above, are {\it coherent}, non-compact fluctuations, and hence energetically punished. This is the reason why we observe that
it is much more difficult for the system to exit the metastable well in mean field approximation than for the real 
system\footnote{In practice, in order to perform the simulation with mean field stochastic dynamics we must implement the
slow forcing approximation (see Appendix \ref{apendMCAMC}). In this way we are able to exit the metastable state.}.
In particular, it is observed in Fig. \ref{gscomp} that the area delimited by the curve $[s(m)-g(m)]/N$ between the points 
$m_0$ and $m_1$,  which is in some sense a measure of the strength of the metastable state, is much larger in mean field approximation 
than for the real system. On the other hand, our mean field theory describes in a proper way the static properties of both the stable 
and metastable states because these states are homogeneous, i.e. without any {\it preferred} point.

The above discussion implies that we need to build a inhomogeneous theory, based on the picture of nucleation of stable phase
compact droplets in the metastable bulk, in order to understand the dynamic properties of the metastable system. This theory was
formulated long ago for equilibrium systems \cite{Langer,Rikvold}, and it is based on the free energy of one of these droplets.
This droplet free energy can be written as the competition of two different terms. 
On one hand, there is a volume term, related to the properties 
of the pure homogeneous phase which constitutes the droplet, i.e. the stable phase. On the other hand, there is a surface term, 
related to the interface separating the stable phase inside the droplet from the metastable phase that surrounds it. This term is 
associated with the inhomogeneity which characterize the metastable-stable transition. In previous sections we have studied
the properties of both pure homogeneous phases (the metastable and the stable phases) for our nonequilibrium system. If we want to 
build a droplet theory valid for the nonequilibrium ferromagnetic system we must understand also the system interfacial properties, 
which will characterize the inhomogeneous, surface term which determines, in competition with the volume term, the
droplet properties. Therefore, our aim in the next chapter consists in studying the interface in our model\footnote{
This last section, devoted to the investigation of the system dynamics in mean field approximation, can be thought as a
waste of time, since it was obvious from the very beginning that the homogeneity hypothesis in mean field theory should impede any 
realistic description of the dynamics. However, apart from the pedagogical value of the discussion, in this section we have presented 
a novel method to include fluctuations in mean field theory in a natural way, which may be very useful in other kind of problems where
the system is homogeneous.}.

\section{Conclusion}
\label{capMedio_apConc}

In this chapter we have studied both the static and dynamic properties of metastable states in the nonequilibrium ferromagnetic model
using a first order dynamic mean field approximation.

In particular, we have applied the so-called Pair Approximation\cite{Dickmanpair,MarroDickman}, 
a dynamic analogous of the equilibrium Bethe-Peierls Approximation,
to the problem of metastability in our lattice spin system. This theory is based on a mean field approximation for the master 
equation governing the system dynamics, once this stochastic equation is reduced to local observables. The approximation is developed
using three fundamental hypothesis. In a first step, it neglects all fluctuations in the system, so in this approach we only study
the {\it average} behavior of local observables. On the other hand, this theory also neglects long range correlations.
In particular, we only have into account nearest neighbor correlations (this is why this approximation is termed
{\it first order}). The last hypothesis assumes that the system is homogeneous, which implies that all points in the lattice behave
in the same way, independently from their positions.

Taking into account these hypothesis, and taking as starting point the master equation, we obtain two coupled non-linear 
differential equations for the dynamics of $x$, the probability of finding an up spin in the system, and $z$, the probability
of finding a $(+,+)$ nearest neighbors pair in the system. We obtain numerically the locally stable steady solutions of
this set of differential equations, both for zero magnetic field and $h<0$. For $h=0$ we obtain theoretical predictions for
the stationary state magnetization as a function of temperature for different values of $p$. These predictions perfectly compare
with Monte Carlo results in the low and intermediate temperature regime, although some differences between
theory and simulation appear for temperatures near to the critical one, $T_c(p)$, since for these temperatures long range 
correlations become important. As the value of the nonequilibrium perturbation $p$ is increased, the stationary state magnetization
decreases in magnitude for $h=0$ for a fixed temperature, although the qualitative shape of curves $m_s^{(\pm)}(T,p)$ is similar to 
those of the equilibrium system. The critical temperature $T_c(p)$ signals a second order phase transition in the nonequilibrium 
systems between a disordered phase for high temperatures and an ordered phase for low temperatures. Applying the marginal 
stability condition to the dynamic equations, we are able to extract the phase diagram in first order mean field approximation
for the nonequilibrium model. The phase diagram yields the critical temperature $T_c(p)$ as a function of the nonequilibrium 
parameter $p$. The ordered phase disappears for all temperatures when $p>p_c$, where $p_c=\frac{5}{32}$ in this approximation.
Finally, for the locally stable steady magnetization for $h<0$ we obtain qualitatively similar results as compared with the 
$h=0$ case, although now the up-down symmetry which held for $h=0$ breaks up. The comparison of predicted curves with Monte 
Carlo results for both magnetization branches for $h<0$ is also excellent.

On the other hand, the system exhibits hysteresis due to the existence of metastable states. This implies that the system keeps memory
of the past evolution history. In particular, using mean field approximation we calculate the intrinsic coercive field $h^*(T,p)$, 
defined in this case as the magnetic field for which the metastable state becomes unstable. We observe that $h^*(T,p)$ shows
two different kinds of asymptotic behaviors in the low temperature limit, which depend on the value of 
$p$. There is a critical value for $p$, $\pi_c\approx 0.0315$, which separates both behaviors. For $p<\pi_c$ the intrinsic coercive 
field $h^*(T,p)$ increases in magnitude as temperature decreases, in the same way that in equilibrium systems.  However, for $p>\pi_c$
we predict that the intrinsic coercive field converges towards zero 
in the limit $T\rightarrow 0$, showing a maximum in magnitude for certain
intermediate temperature. This involves the existence of a {\it non-linear cooperative phenomenon} between the thermal noise 
(parameterized by $T$) and the non-thermal noise (parameterized by $p$): although both noise sources independently add disorder to the 
system, which implies the attenuation, or even destruction of existing metastable states, the combination of both noises parameterized
in the microscopic dynamics does not always involves a larger disorder, giving rise to parameter space regions where there
are no metastable states for low and high temperatures, but metastable states appear for intermediate temperatures. This theoretical 
prediction based on the mean field approximation is fully confirmed via Monte Carlo simulations.

Finally, apart from the mean field investigations on the static properties of both stable and metastable states in the system,
summarized in previous paragraphs, we have also attempted a description of the dynamics of the metastable-stable transition using Pair 
Approximation. However, one of the basic hypothesis in this approximation, namely the hypothesis of suppression of fluctuations, 
impedes any realistic description of this dynamic process using Pair Approximation. The reason underlies in that fluctuations 
constitute the basic mechanism which gives rise to the metastable-stable transition. Therefore, in order to describe the metastable
state demagnetization process, we relax the above hypothesis, including fluctuations in the dynamic mean field theory. This can be
done in a natural way using the concepts of stable phase growth and shrinkage rates, observables which are defined in a simple manner
in our approximation, and the philosophy underlying MCAMC algorithms (see Appendix \ref{apendMCAMC}). In this way we write a 
{\it mean field stochastic dynamics}, which includes fluctuations in a natural way. From this extended theory we
predict the dynamic (and static) properties of the system. However, while the static results obtained from the mean field stochastic 
dynamics are equivalent to those obtained in Pair Approximation and reproduce the measured properties in Monte Carlo simulations, 
the results on the dynamics of the metastable-stable transition are remarkably different from those obtained in simulations. 
This discrepancy is due to the failure of another basic hypothesis of mean field approximation: the homogeneity hypothesis.
This hypothesis implies that the exit from the metastable state in the mean field stochastic dynamics approximation is produced
by {\it coherent} fluctuations of all spins in the system, which are energetically punished. 
The metastable demagnetization process in the
real system is, on the other hand, a highly inhomogeneous process, where one or several stable phase droplets nucleate in the 
metastable bulk, since these compact structures minimize the system free energy for a fixed magnetization. Thus, in order to understand
the dynamics of the metastable-stable transition we must therefore write an inhomogeneous theory where the interface plays a very 
important role. This inhomogeneous theory, based on the droplet picture, will be developed in Chapter \ref{capNuc}. However,
in order to write such theory for the nonequilibrium system, we must first understand the interfacial properties in the model,
since they will play a fundamental role in the droplet nucleation process. We study this problem in the next chapter.

\chapter{Interfacial Properties}
\label{capSOS}

\section{Introduction}
\label{capSOS_apIntro}

In the previous chapter we have inferred the need of an inhomogeneous description, based on the picture of droplets of the stable 
phase nucleating in the metastable bulk, in order to describe properly the dynamics of the metastable-stable transition
in the nonequilibrium ferromagnetic system. One of the fundamental features characterizing the inhomogeneity 
in this process is the presence of interfaces separating the metastable phase from the stable one. Thereby, we must study the 
properties of an interface in the classic spin system under nonequilibrium conditions in order to develop an inhomogeneous theory that
explains the dynamics of the demagnetization process from the metastable state.

We devote this chapter to the study of the interfacial properties of the non-equilibrium spin model. In particular, we are going to
pay attention to both the microscopic properties of the interface and its surface tension. With this aim in mind we will first 
describe the Solid-On-Solid (SOS) approximation, first formulated by Burton, Cabrera and Frank\cite{BCF}. This approximation
studies the interfacial properties in the Ising model for zero magnetic field, and it is based on a microscopic description of 
the interface. In spite of the approximate character of this theory, the predictions derived from it nicely reproduce 
the properties of the equilibrium interface. Afterwards we will present a generalization of SOS approximation
which will allow us to study and predict the properties of the interface in our nonequilibrium spin model. Finally, using the results
obtained via the generalized SOS theory, we will study the shape of a spin droplet in the nonequilibrium system. This shape can be 
derived once we know the interface surface tension using Wulff construction\cite{Wulff}.

\section{The Solid-On-Solid Approximation}
\label{capSOS_apSOS}

\begin{figure}
\centerline{
\psfig{file=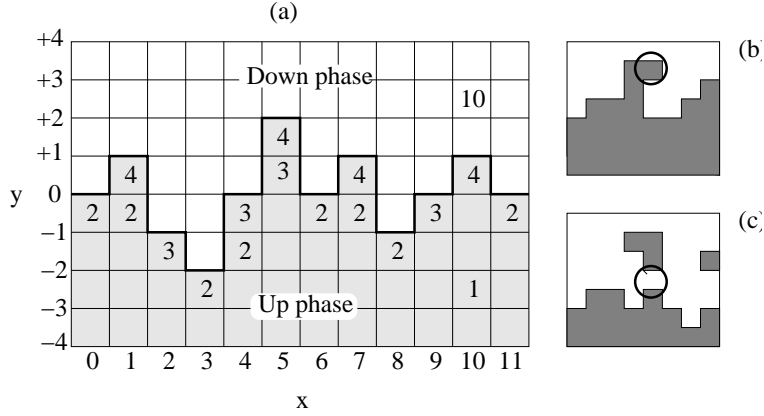,width=10cm}}
\caption[Schematic representation of a SOS interface.]
{\small (a) Schematic representation of a SOS interface between an up spin phase and a down spin phase. In this case $L=11$.
The numbers inside the spins indicate the spin class to which they belong to (see Table \ref{tabclasesbis}). Notice that interfacial 
spins belong to classes $2$, $3$ and $4$. (b) Example of configuration with overhangs, which is forbidden in SOS approximation. 
The marked spin involves a multivaluated interface for that $x$ coordinate. (c) Schematic representation of an interface close to 
interact with bulk fluctuations. These kind of configurations are also forbidden in SOS approximation.}
\label{interf}
\end{figure}

The Solid-On-Solid approximation (see ref. \cite{BCF}) describes the interface in a 
bidimensional equilibrium Ising model for zero magnetic field, $h=0$,
using a simple picture where the interface is an {\it univaluated} function. If we have, as in our case, a square lattice, the 
interface is completely defined by a set of integer height values, $y_x$, with $x=0,\ldots,L$, as Fig. \ref{interf}.a shows.
If $\mathbf{y} =\{y_i, i=0,\ldots,L\}$ is a configuration of the interface, the energy associated to this configuration is,
\begin{equation}
{\cal H}_I(\mathbf{y})=2J \sum_{i=1}^L(1+|y_i-y_{i-1}|)
\label{energyinterf}
\end{equation}
where $J$ is the spin coupling constant in Ising model (see eq. (\ref{hamilt})). If we compare this Hamiltonian ${\cal H}_I$ with
the original equilibrium Ising Hamiltonian, eq. (\ref{hamilt}), for zero magnetic field, $h=0$, we observe that 
${\cal H}_I(\mathbf{y})$ is just the energy contributed by all $(+,-)$ spin pairs forming the interface to the total Ising 
configuration energy. Since in the ferromagnetic Ising model spins tend to align with their nearest neighbor spins, we can say
that a $(+,-)$ spin pair constitutes a {\it broken bond}. The energy cost of a broken bond in the Ising model is just $2J$,
so the Hamiltonian ${\cal H}_I$ is just the energy associated to all the broken bonds which define the interface.

Let's introduce now the canonical ensemble for this model, which consists in all possible interfacial configurations in a system
with length $L$ and with boundary conditions specified by $y_L-y_0={\cal Y} \equiv L \tan \phi$, where $\phi$ is the angle formed
by the interface and the $\hat{x}$ axis. The associated partition function will be given by,
\begin{equation}
{\cal Z}(L,\phi)=\sum_{\mathbf{y}}\textrm{e}^{-\beta {\cal H}_I(\mathbf{y})} \delta\big[(y_N-y_0)-{\cal Y}\big]
\label{partic1}
\end{equation}
where $\beta$ is the inverse temperature, and $\delta(t)$ is the Kronecker Delta function ($\delta(t)=1$ if $t=0$ and 
$\delta(t)=0$ otherwise). The corresponding free energy per site is defined using the following limit,
\begin{equation}
\sigma_{pr}(\phi)= \lim_{L \rightarrow \infty} -\frac{1}{\beta L} \ln {\cal Z}(L,\phi)
\label{efree}
\end{equation}
This free energy is known as the projected surface tension\cite{Miracle}. The surface tension, which is the free energy per unit
length, can be written as,
\begin{equation}
\sigma(\phi)=|\cos \phi| \phantom{a} \sigma_{pr}(\phi)
\label{tension}
\end{equation}
Due to the boundary conditions, mathematically expressed via the Kronecker Delta function that appears in eq. (\ref{partic1}),
it is not possible to perform the sum needed to calculate the canonical partition function. In order to perform the 
calculation we introduce a new ensemble, conjugated to the previous one, where instead of fixing the height values at the borders
(that is, instead of fixing the angle $\phi$ formed between the interface and the $\hat{x}$ axis), we fix a new thermodynamic
parameter, $\gamma(\phi)$\footnote{This is similar to what happens when we introduce the macrocanonical ensemble in gases. While
the thermodynamic parameters in the canonical ensemble for gases are $(T,V,N)$, in the macrocanonical ensemble we introduce a new
thermodynamic parameter, namely the chemical potential $\mu$, which substitutes the number of particles $N$, which is not fixed now
in this new ensemble. The thermodynamic potentials derived from both ensembles are related via a Legendre transform based on
the thermodynamic parameters $\mu$ and $N$.}. If we define the {\it step variables} $\delta_i=y_i - y_{i-1}$, we can write the
Hamiltonian for an interface with a configuration $\mathbf{\delta}=\{\delta_i, i=1,\ldots,L\}$ as 
${\cal H}_I(\mathbf{\delta})=2J \sum_{i=1}^L(1+|\delta_i|)$. Thus boundary conditions, defined by $y_L-y_0={\cal Y}$, can be expressed
now as $\sum_{i=1}^L \delta_i={\cal Y}$. The partition function for the new ensemble is,
\begin{equation}
{\cal Z}_2(L,\gamma)= \sum_{\mathbf{\delta}}\textrm{e}^{-\beta {\cal H}_I(\mathbf{\delta})}
\textrm{e}^{\gamma(\phi)\sum_{i=1}^L \delta_i}
\label{partic2}
\end{equation}
where the new thermodynamic parameter $\gamma(\phi)$ is just the Lagrange multiplier which keeps the {\it average}
step variable in a $x$ independent value, that is, $\langle \delta_i \rangle = \tan \phi$ for all $i \in [1,L]$.\cite{RikvoldSOS}
If we use in eq. (\ref{partic2}) the previously derived expression for ${\cal H}_I(\mathbf{\delta})$, we obtain,
\begin{equation}
{\cal Z}_2(L,\gamma)=\textrm{e}^{-2\beta JL}[z(\gamma)]^L
\label{zzz}
\end{equation}
where $z(\gamma)$ is the partition function for a single step,
\begin{equation}
z(\gamma)=\sum_{\delta = -\infty}^{\infty} \textrm{e}^{-2\beta J |\delta|} \textrm{e}^{\gamma(\phi) \delta}
= \frac{1-X^2}{1+X^2-2X \cosh \gamma(\phi)}
\label{zdef}
\end{equation}
and where we have used the notation $X\equiv \textrm{e}^{-2\beta J}$. It is obvious from eq. (\ref{zzz}) that in SOS approximation
the different step variables $\delta_i$ are supposed to be statistically independent among them. We could also derive this idea from 
the additive character of the Hamiltonian ${\cal H}_I$ once written in terms of the step variables. In order to calculate the
$\phi$ dependence of the Lagrange multiplier $\gamma(\phi)$, we use the condition $\langle \delta \rangle = \tan \phi$. The probability
of finding a step with size $\delta$ in the new ensemble is (see the previous equation),
\begin{equation}
p(\delta)=\frac{1}{z(\gamma)} X^{|\delta|}\textrm{e}^{\gamma (\phi) \delta}
\label{probdelta}
\end{equation}
We thus can write,
\begin{equation}
\langle \delta \rangle = \frac{\partial \ln z(\gamma)}{\partial \gamma} = \tan \phi
\label{generatriz}
\end{equation}
where we use the single step partition function as a moment generating function. From this equality we find,
\begin{eqnarray}
\textrm{e}^{\gamma(\phi)}= \frac{(1+X^2)\tan \phi + R(\phi)}{2X(\tan \phi+1)} \nonumber \\
\textrm{e}^{-\gamma(\phi)}= \frac{(1+X^2)\tan \phi - R(\phi)}{2X(\tan \phi-1)}
\label{expgamma}
\end{eqnarray}
where $R(\phi)=[(1-X^2)^2 \tan ^2 \phi + 4X^2]^{1/2}$. Substituting these expressions in eq. (\ref{zdef}) we arrive to,
\begin{equation}
z(\phi)=\frac{(1-X^2)(1-\tan ^2 \phi)}{1+X^2 - R(\phi)}
\label{zfi}
\end{equation}
The thermodynamic potential associated to this second ensemble can be written as a function of $\phi$ in the following way,
\begin{eqnarray}
\varphi (\phi) & \equiv & \lim_{L \rightarrow \infty} - 
\frac{1}{\beta L} \ln {\cal Z}_2[L,\gamma (\phi)] \nonumber \\
 & = & - \frac{1}{\beta} \ln \big[X\frac{(1-X^2)(1-\tan ^2 \phi)}{1+X^2 - R(\phi)}\big]
\label{varfi}
\end{eqnarray}
This thermodynamic potential is related to the projected surface tension, eq. (\ref{efree}), via a Legendre transform which involves
the thermodynamic variables $\tan \phi$ and $\gamma$. If we perform such Legendre transform, and obtain from the so-calculated 
$\sigma_{pr}(\phi)$ the surface tension, we find,
\begin{eqnarray}
\sigma_{SOS}(\phi;T) & = & |\cos \phi | \Big\{ - T \ln \big[X\frac{(1-X^2)(1-\tan ^2 \phi)}{1+X^2 - R(\phi)}\big] 
 \nonumber \\ 
& & + T \tan \phi \ln \big[\frac{(1+X^2)\tan \phi + R(\phi)}{2X(\tan \phi+1)}\big] \Big\} 
\label{tensionsuperf}
\end{eqnarray}
where we have used that $\beta =1/T$, with $T$ the system temperature. The subindex {\it SOS} included in our notation for the
surface tension, $\sigma_{SOS}(\phi;T)$, points out that this is the solution in Solid-On-Solid approximation. This theory is approximate
in several ways. First, as we said before, the SOS approximation neglects correlations between neighboring step variables $\delta_i$.
On the other hand, when we assume that the interface is univaluated we are neglecting the possible presence of {\it overhangs}.
These overhangs give rise to non-analytic regions in the interface of the real Ising model, since for a fixed $x$ coordinate
there are several different values of the height $y(x)$. Fig. \ref{interf}.b shows an example of overhang. Furthermore,
the SOS approximation also neglects all possible interactions between the interface and the fluctuations or droplets appearing in the 
bulk. Fig. \ref{interf}.c shows an example of this possible interaction. Therefore it is surprising that, in spite of these
approximations, the surface tension derived in SOS approximation, eq. (\ref{tensionsuperf}), reproduces the known exact result
for $\phi=0$ \cite{Temperley}, and yields a very good approximation for $|\phi| \leq \pi/4$.\cite{Ziainterf} For values of $|\phi|$
larger that $\pi/4$ is more reasonable to use the SOS approximation taking as reference frame the $\hat{y}$ axis instead of the $\hat{x}$
axis.

\markboth{Interfacial Properties}{\ref{capSOS_apGen} Nonequilibrium Generalization of SOS Approximation}
\section{Generalization of the Solid-On-Solid Approximation for an Interface under Nonequilibrium Conditions}
\label{capSOS_apGen}
\markboth{Interfacial Properties}{\ref{capSOS_apGen} Nonequilibrium Generalization of SOS Approximation}

In this section we want to generalize the SOS approximation to our nonequilibrium model. In order to
perform such extension of the equilibrium interfacial theory we must understand deeply, in a first step, the effects that the 
nonequilibrium random perturbation $p$ induces on the system.

\subsection{Effective Temperature}
\label{capSOS_apGen_subTeff}

Our starting point to do that is the microscopic dynamics we have
imposed to the system. The transition rate in the model, eq. (\ref{rate}), is,
\begin{equation}
\omega (s,n) = p + (1-p) \frac{\textrm{e}^{-\beta \Delta {\cal H}(s,n)}}
{1 + \textrm{e}^{-\beta \Delta {\cal H}(s,n)}}
\label{ratebis}
\end{equation}
This dynamics, which we called Glauber dynamics, is local as far as it depends exclusively on the spin $s$ we are trying to flip, and 
on the number of its up nearest neighbor spins, $n$, through the energy increment involved in the transition, $\Delta {\cal H}(s,n)$.
As we discussed before, there are many other different competing dynamics. One of the most used dynamics in literature because of its 
appropriate features is Metropolis dynamics,
\begin{equation}
\omega (s,n) = p + (1-p) \textrm{min} [1,\textrm{e}^{-\beta \Delta {\cal H}(s,n)}]
\label{ratemetro}
\end{equation}
This dynamics is a bit faster than Glauber dynamics, although the qualitative 
results derived using Metropolis dynamics to study metastability in 
the nonequilibrium system are very similar to those obtained with Glauber dynamics. In order to analyze the effect that the 
nonequilibrium dynamic perturbation induces on the system behavior, it is appropriate to work initially using Metropolis dynamics 
instead of the original Glauber dynamics due to technical details which will be clear later on in this section. The results we obtain
are apparently independent of this detail.

Dynamics as those written in eqs. (\ref{ratebis}) and (\ref{ratemetro}), where
two canonical rates working at different temperatures compete, generically drive the system towards a nonequilibrium steady state.
We could think however that we can map this a priori nonequilibrium system to an equilibrium system with effective parameters.
The transition rate should verify now the detailed balance condition, eq. (\ref{balance}), for these effective parameters. In 
particular, for competing temperatures, we could think of mapping the nonequilibrium system to an equilibrium model with certain 
{\it effective temperature}. In order to calculate this effective temperature we use eq. (\ref{ratemetro}). Thus we want to 
write the transition rate (\ref{ratemetro}) as $\min[1,\textrm{e}^{-\beta_{eff}\Delta {\cal H}(s,n)}]$, which is a dynamics driving
the system towards an equilibrium steady state at inverse temperature $\beta_{eff}$. For $\Delta {\cal H}(s,n) > 0$,
\begin{equation}
p+(1-p)\textrm{e}^{-\beta \Delta {\cal H}(s,n)} \equiv \textrm{e}^{-\beta_{eff} \Delta {\cal H}(s,n)}
\label{teff0}
\end{equation}
where $\beta_{eff}=1/T_{eff}$, being $T_{eff}$ the effective temperature. From this equation we obtain an expression for the
effective temperature, as deduced from Metropolis rate, for $\Delta {\cal H}(s,n) > 0$,
\begin{equation}
T_{eff}^{(s,n)}(p)=\frac{-\Delta {\cal H}(s,n)}{\ln \big[p+(1-p)\textrm{e}^{-\beta \Delta {\cal H}(s,n)}\big]}
\label{teffmetro}
\end{equation}
For $\Delta {\cal H}(s,n) \leq 0$ we have that $T_{eff}^{(s,n)} = T$. The effective temperature defined in this way is not unique
for a system with $p \neq 0$, because it clearly depends on the spin class to which the flipping spin belongs to through the
energy increment $\Delta {\cal H}(s,n)$ involved in this transition, for $\Delta {\cal H}(s,n)>0$. All spins in the same spin class
share  the same energy increment $\Delta {\cal H}(s,n)$ when flipped. Hence, all spins in the same spin class suffer the same effective
temperature. However, for $0<p<1$ this effective temperature varies depending on the chosen spin class. In the limit $p=0$ the
effective temperature defined in (\ref{teffmetro}) reduces to the usual temperature, $T$.
\begin{figure}[th]
\centerline{
\psfig{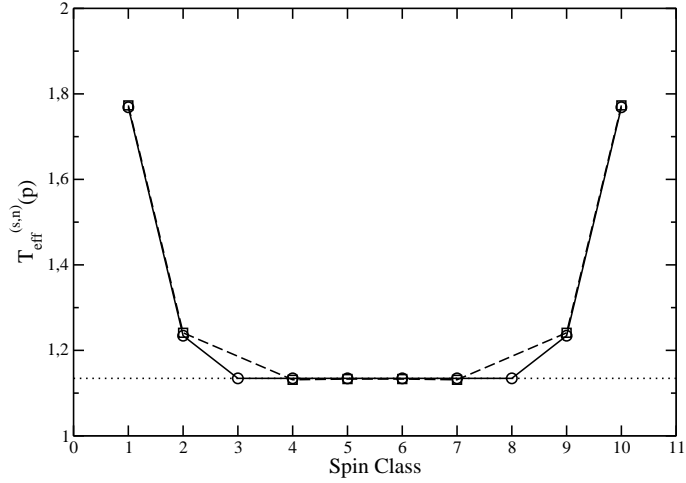}}
\caption[Effective temperature as a function of the spin class.]
{\small Effective temperature $T_{eff}^{(s,n)}$ as a function of the spin class $i \in [1,10]$ as obtained from Metropolis
dynamics ($\bigcirc$) and from Glauber dynamics ($\Box$), for a system with $T=0.5T_{ons}$ and $p=0.01$ ($h=0$ always in this section).
Notice that $T_{eff}^{(s,n)}$ is not defined for spin classes $3$ and $8$ (both corresponding with $\Delta {\cal H}(s,n)=0$)
for Glauber dynamics. The dotted line corresponds to the system temperature, $T$.}
\label{tefffig}
\end{figure}

First, the fact that the effective temperature depends on $\Delta {\cal H}(s,n)$ points out that we cannot exactly map the 
nonequilibrium system to an equilibrium model with well-defined effective temperature. On the other hand, the dependence of 
$T_{eff}^{(s,n)}$ on the energy increment (that is, on the spin class to which the chosen spin belongs to) will allow us to obtain
a clear physical picture of the effect that the nonequilibrium parameter $p$ induces on system's dynamics. Fig. \ref{tefffig} shows
the effective temperature for each spin class, as defined from Metropolis dynamics, see eq. (\ref{teffmetro}), for a system
with temperature $T=0.5T_{ons}$ and $p=0.01$ (the magnetic field $h$ is always zero in this section), taking into account that
the energy increment associated to each class is $\Delta {\cal H}(s,n)=2s[2J(n-d)+h]$. In this figure we observe 
that the effective temperature curve is fully symmetrical for up and down spins: that is, spin classes 1 and 10 suffer the same 
effective temperature, and the same happens for spin classes 2 and 9, 3 and 8, 4 and 7, 5 and 6. On the other hand, we see that
{\it more ordered} spin classes, i.e. spin classes characterized by a larger number of nearest neighbor spins pointing in the same
direction that the central spin, suffer a higher effective temperature. Thus, classes 1 and 10, where all four nearest neighbor spins
point in the same direction than the central spin, are subject to the highest effective temperature. This observation is compatible 
with intuition. As we have explained before, the presence of the nonequilibrium perturbation $p$ implies that, with a very small 
probability, spins in the system are able to flip independently from energetic restrictions imposed by the interaction with their 
nearest neighbors. The larger is the number of nearest neighbors pointing in the same direction than the central spin,
the larger is the energy barrier that is {\it violated} when flipping such central spin independently from any energetic constraint.
Hence, if we interpret the effect of the nonequilibrium parameter $p$ in terms of an effective temperature, it is obvious the need
of a larger effective temperature in order to overcome higher energy barriers. Therefore we would a priori expect that the larger is 
the local order that a spin feels, the larger is the effective temperature this spin suffers. This is in fact what we have obtained in 
the above calculation.

The effective temperature can be also defined using Glauber dynamics, eq. (\ref{ratebis}), in a way similar to the one followed when 
using Metropolis dynamics, eq. (\ref{teffmetro}). However, the definition of effective temperature based on Glauber dynamics shows a
singularity for $\Delta {\cal H}(s,n)=0$, that is, for spin classes 3 and 8, so using this definition in theoretical analysis is
not suitable. Fig. \ref{tefffig} also shows the effective temperature as obtained from Glauber dynamics. The effective temperature
obtained from Metropolis dynamics is almost indistinguishable from the one obtained via Glauber dynamics, and it does not show
any singularity for $\Delta {\cal H}(s,n)=0$. This is the reason why we have used from the very beginning Metropolis dynamics
to define the effective temperature, instead of using Glauber dynamics.

\subsection{Statistical Weight Associated to a Broken Bond in the Nonequilibrium Interface}
\label{capSOS_apGen_subWei}

Coming back to the interface problem, we observe in Fig. \ref{interf}.a that the spins that define the interface belong to 
different spin classes. If we fix our attention on the up phase interfacial spins\footnote{In order to define the interface
we must center our attention on all up interfacial spins, {\it or} on all down interfacial spins. In this case we choose the up 
interfacial spins.}, they belong to spin classes 2, 3 and 4. Spins in classes 1 and 5 cannot never belong to the interface.
Spins in class 1 are typical of the up phase bulk, and spins in class 5 are isolated up spins in the down phase bulk. If we call
$T_{eff}^{(i)}$ the effective temperature associated to class $i\in [1,10]$, then we have spins in the interface that suffer
effective temperatures $T_{eff}^{(2)} \geq T_{eff}^{(3)} \geq T_{eff}^{(4)}$ if they belong to classes 2, 3 and 4, respectively.

One of the central magnitudes in the Solid-On-Solid approximation for the equilibrium system ($p=0$) is 
$X \equiv \textrm{e}^{-2\beta J}$, which is the probabilistic weight associated to a broken bond in the interface. Thus the probability
of finding a step of height $\delta$ in the interface is proportional to $X^{|\delta|}$, see eq. (\ref{probdelta}). Using the
concept of effective temperature previously introduced, we can assume that the probabilistic weight associated to a broken bond
in the nonequilibrium system ($p \neq 0$) will depend on the spin class to which the interfacial spin whose bond is broken belongs to
through the effective temperature associated to this spin class. If this spin belongs to class $i$, we denote this statistical weight
as $X_i$. Hence we can write in this picture $X_i=\textrm{e}^{-2\beta _{eff}^{(i)} J}$, where $\beta_{eff}^{(i)}=1/T_{eff}^{(i)}$,
being $T_{eff}^{(i)}$ the effective temperature associated to the class $i$ to which the interfacial spin whose bond is broken 
belongs to. That is, we assume that each spin class $i$ behaves as an equilibrium system at effective temperature $T_{eff}^{(i)}$,
so the probabilistic weight associated to a broken bond in class $i$ is just the Boltzmann factor associated to this broken bond at
temperature $T_{eff}^{(i)}$, $X_i=\textrm{e}^{-2\beta _{eff}^{(i)} J}$. Moreover, we assume now that, as a first approximation,
the probabilistic weight associated to an interfacial broken bond in the system with $p \neq 0$ will be given by the following
average,
\begin{equation}
X_p = \Pi_2(T,p) X_2 + \Pi_3(T,p) X_3 + \Pi_4(T,p) X_4
\label{pesobond}
\end{equation}
where $\Pi_i(T,p)$ is the probability of finding an interfacial spin belonging to class $i$, and where $X_i$ is the statistical weight
for a broken bond associated to class $i$, as we said before. This last assumption constitutes a mean
field approximation, because we neglect the effect derived from presence of different spin classes in the interface, with different 
associated statistical weights $X_i$ for each broken bond. Instead we average such effect, building an average probabilistic weight 
$X_p$, identical for all broken bonds in the interface and independent of the spin class. Since all interfacial spins in the up phase
belong to classes 2, 3 and 4, it is obvious that the following normalization condition holds,
\begin{equation}
\Pi_2(T,p) + \Pi_3(T,p) + \Pi_4(T,p) = 1
\label{normalizpi}
\end{equation}

$X_p$ is our approximation for the statistical weight associated to a broken 
bond in the nonequilibrium interface. In order to generalize the SOS 
approximation to the system with $p\neq 0$, we only have to substitute in the results obtained in the previous section
the variable $X$ for the new generalized probabilistic weight $X_p$. Thus, for the nonequilibrium
system, the probability of finding a step in the interface with magnitude $\delta$ is,
\begin{equation}
p_p(\delta)=\frac{1}{z_p(\gamma_p)} X_p^{|\delta|}\textrm{e}^{\gamma_p (\phi) \delta}
\label{probdeltap}
\end{equation}
where now $z_p(\gamma_p)$ and $\gamma_p(\phi)$ are the corresponding generalized versions, substituting $X$ for $X_p$,
of the partition function associated to a single step and the thermodynamic parameter conjugated to the interface slope,
respectively. In the same way, the generalized surface tension is now,
\begin{eqnarray}
\sigma_{SOS}^{(p)}(\phi;T,p) & = & |\cos \phi | \Big\{ - T \ln \big[X_p\frac{(1-X_p^2)(1-\tan ^2 \phi)}{1+X_p^2 - R_p(\phi)}\big] 
 \nonumber \\ 
& & + T \tan \phi \ln \big[\frac{(1+X_p^2)\tan \phi + R_p(\phi)}{2X_p(\tan \phi+1)}\big] \Big\} 
\label{tensionsuperfp}
\end{eqnarray}
where $R_p(\phi)=[(1-X_p^2)^2 \tan ^2 \phi + 4X_p^2]^{1/2}$. In order to find the explicit form of all these magnitudes we only have 
to evaluate the probabilities $\Pi_i(T,p)$ of finding an interfacial spin in class $i$.

\begin{table}[htp!]
\centerline{
\begin{tabular}{|c|c|c|c|}
\hline
Notation & Configuration & Step variables & $p(\delta,\epsilon)$   \\
\hline \hline
A & \psfig{file=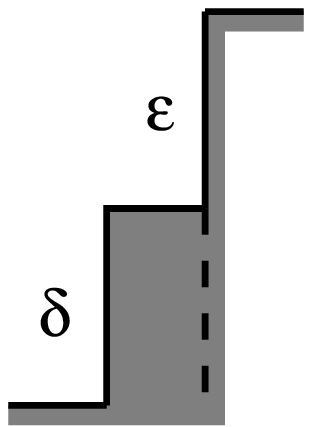,width=1cm}& $\delta > 0$, $\epsilon > 0$ & 
$[Y^{\delta+\epsilon}X_3^2 X_8 X_2^{\delta-1} X_9^{\epsilon-1}]/{\cal Q}$  \\
\hline
B & \psfig{file=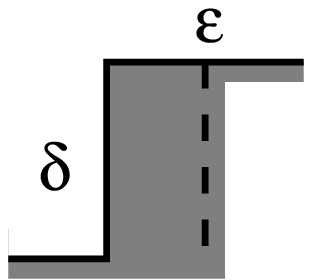,width=1cm}& $\delta > 0$, $\epsilon = 0$ &
$[Y^{\delta}X_2^{\delta-1}X_3^2]/{\cal Q}$  \\
\hline
C & \psfig{file=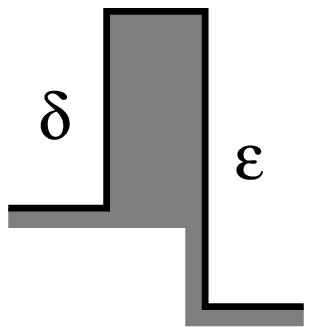,width=1cm}& $\delta > 0$, $\epsilon < 0$ &
$[Y^{\delta+\epsilon}X_4^3 X_3^{2(\alpha-1)} X_2^{\lambda - \alpha}]/{\cal Q}$  \\
\hline
D & \psfig{file=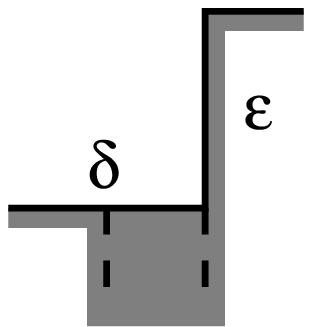,width=1cm}& $\delta = 0$, $\epsilon > 0$ &
$[Y^{\epsilon}X_8^2 X_9^{\epsilon-1}]/{\cal Q}$  \\
\hline
E & \psfig{file=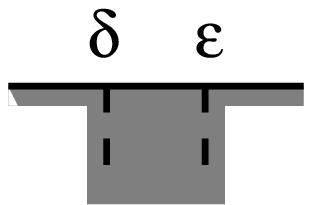,width=1cm}& $\delta = 0$, $\epsilon = 0$ &
$X_2/{\cal Q}$  \\
\hline
F & \psfig{file=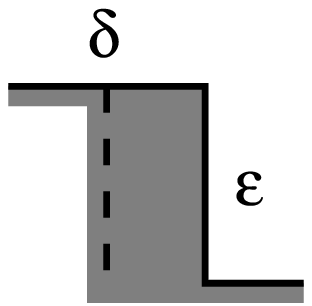,width=1cm}& $\delta = 0$, $\epsilon < 0$ &
$[Y^{\epsilon}X_3^2 X_2^{|\epsilon|-1}]/{\cal Q}$  \\
\hline
G & \psfig{file=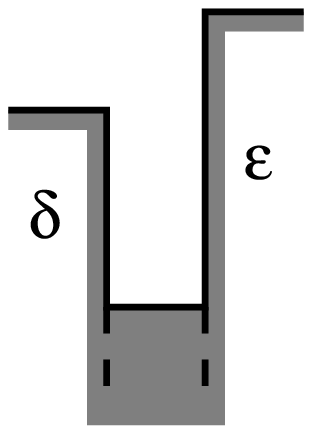,width=1cm}& $\delta < 0$, $\epsilon > 0$ &
$[Y^{\delta+\epsilon}X_7^3 X_8^{2(\alpha-1)}X_9^{\lambda - \alpha}]/{\cal Q}$  \\
\hline
H & \psfig{file=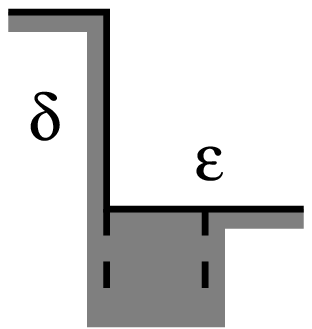,width=1cm}& $\delta < 0$, $\epsilon = 0$ &
$[Y^{\delta}X_8^2 X_9^{|\delta|-1}]/{\cal Q}$  \\
\hline
I & \psfig{file=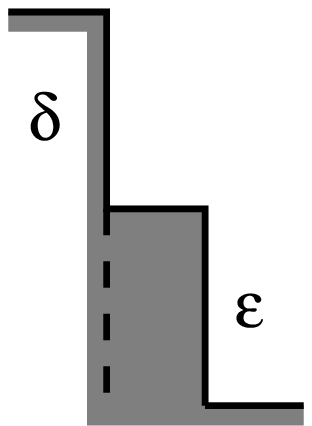,width=1cm}& $\delta < 0$, $\epsilon < 0$ &
$[Y^{\delta+\epsilon}X_3^2 X_8 X_2^{|\epsilon|-1}X_9^{|\delta|-1}]/{\cal Q}$  \\
\hline
\end{tabular}
}
\caption[Different typical configurations of an interfacial spin column.]
{\small In this table we present the 9 different typical configurations of an interfacial spin column in the SOS approximation.
These typical configurations are defined by the sign of the left step, $\delta$, and the sign of the right step, $\epsilon$.
The first column shows the notation we use for each typical configuration. The second column shows a schematic graphical 
representation for each typical configuration. The third column presents the characteristic values of $\delta$ and $\epsilon$ for
each case. Finally, the fourth column shows the two-body probability $p(\delta,\epsilon)$ of finding each typical configuration.
Notice that $Y=\textrm{e}^{\gamma_p(\phi)}$ and $X_i=\textrm{e}^{-2J\beta_{eff}^{(i)}}$. In some cases we have to define
$\alpha=\min(|\delta|,|\epsilon|)$ and $\lambda=\max(|\delta|,|\epsilon|)$.}
\label{configs}
\end{table}

\subsection{Population of Interfacial Spin Classes}
\label{capSOS_apGen_subPop}

In order to explicitely calculate the probabilities $\Pi_i(T,p)$ we must introduce the two-body probability function 
$p(\delta,\epsilon)$, which yields the probability that a step variable in the interface takes a value $\delta$ and its right
neighbor step variable takes a value $\epsilon$. Indeed, coming back to Fig. \ref{interf}.a, we observe there that, for a fixed
spin column, we know how many interfacial spins associated to this column belong to each spin class once we know the value of the 
left and right step variables associated to that interfacial spin column, $\delta$ and $\epsilon$ respectively. Thus, for instance, 
if we have a spin column where $\delta = +2$ and $\epsilon = -2$ (see column 5 in Fig. \ref{interf}.a) we know that, given the total
number $\delta = +2$ of interfacial spins associated to this column, there is one in class 4 and another one in class 3. On the other 
hand, if we have $\delta=-2$ and $\epsilon=+1$ (see column 6 in Fig. \ref{interf}.a) we find only one interfacial spin associated
to this column, and it belongs to class 2. In general, we have $9$ different typical configurations for an interfacial spin column
in the generalized SOS approximation, as shown in Table \ref{configs}. These $9$ typical configurations come from the possible 
sign combinations between $\delta$ and $\epsilon$, taking zero into account. Any interface in SOS approximation can be constructed
from these 9 typical column configurations.

Let's assume for a while that we have one of the previous typical configurations with $\delta=+2$ y $\epsilon=+2$, as the one found
in column 4 in Fig. \ref{interf}.a. This configuration corresponds to a configuration type A (see Table \ref{configs}). This spin
column formed by up and down spins in contact with the interface and flanked by a step $\delta=+2$ to the left and a step $\epsilon=+2$
to the right is composed by one spin in class 2, one spin in class 3, one spin in class 8 and one spin in class 9. Therefore there
are 5 broken bonds which define this interfacial column: one of these broken bonds belongs to the spin in class 2, two of them
belong to the spin in class 3, another one belongs to the spin in class 8 and the last one belongs to the spin in class 9.
The probabilistic weights associated to these broken bonds will be, respectively, $X_2$, $X_3^2$, $X_8$ y $X_9$. We assume now that
the probability of finding this configuration will be $p(\delta=+2,\epsilon=+2) \propto Y^4 X_2 X_3^2 X_8 X_9$, where 
$Y=\textrm{e}^{\gamma_p(\phi)}$ is the factor associated to the thermodynamic parameter $\gamma_p(\phi)$, which fixes the average
interface's slope. In general, for a configuration type A, we have,
\begin{equation}
p(\delta>0,\epsilon>0) = \frac{1}{\cal Q}[Y^{\delta+\epsilon}X_3^2 X_8 X_2^{\delta-1} X_9^{\epsilon-1}]
\label{tipoA}
\end{equation}
where ${\cal Q}$ is the associated normalization factor, which will be calculated later on.

Let's assume now that we have a configuration type C, with $\delta>0$ and $\epsilon<0$. If we define in this case 
$\alpha=\min(|\delta|,|\epsilon|)$ and $\lambda=\max(|\delta|,|\epsilon|)$, we have $\lambda$ interfacial spins associated to this 
spin column type C: $\lambda - \alpha$ of those interfacial spins are in class 2, each one with one associated broken bond,
$\alpha - 1$ of those interfacial spins are in class 3, each one with two associated broken bonds, and finally one of those $\lambda$ 
interfacial spins belongs to class 4, and it has three broken bonds. Hence the probability $p(\delta,\epsilon)$ of finding an
interfacial column with $\delta>0$ and $\epsilon<0$ is,
\begin{equation}
p(\delta>0,\epsilon<0) = \frac{1}{\cal Q}[Y^{\delta+\epsilon}X_4^3 X_3^{2(\alpha-1)} X_2^{\lambda - \alpha}]
\label{tipoC}
\end{equation}
In the same way we can build the probability $p(\delta,\epsilon)$ for the rest of typical configurations (see Table \ref{configs}).
In the limit where the nonequilibrium dynamic perturbation is zero, $p=0$, i.e. for the equilibrium system, where the effective
temperature $T_{eff}^{(i)}$ associated to class $i$ reduces to the usual temperature $T$, we have that $X_i (p=0)=X$, so
the two-body probability function reduces to $p(\delta,\epsilon) = p(\delta)p(\epsilon)$ for $p=0$, being $p(x)$ the probability
of finding a step variable of magnitude $x$ in the equilibrium SOS approximation, see eq. (\ref{probdelta}).
The two-body probability function $p(\delta,\epsilon)$ factorizes in the limit $p=0$ as a consequence of the statistical 
independence of neighboring steps in the equilibrium SOS approximation. However, for $p \neq 0$, although the two-body probability 
function $p(\delta,\epsilon)$ is written as the product of the probabilistic weights for the different broken bonds associated to
the different spin classes, $X_i$, such probability $p(\delta,\epsilon)$ includes nontrivial correlations, since the classes to which 
interfacial spins belong to depend strongly on the relative signs of $\delta$ and $\epsilon$. In this sense the use of 
$p(\delta,\epsilon)$ is beyond the SOS approximation for $p \neq 0$, where it is assumed no correlations between neighboring steps.

The normalization constant ${\cal Q}$ associated to the two-body probability function $p(\delta,\epsilon)$ can be calculated
from the normalization condition,
\begin{equation}
\sum_{\delta,\epsilon=-\infty}^{+\infty} p(\delta,\epsilon) = 1
\label{normapdos}
\end{equation}
The double sum in $\delta$ and $\epsilon$ must be divided depending on the sign of both step variables, $\delta<0$, $\delta=0$ and 
$\delta>0$, and in the same way for $\epsilon$, yielding 9 different sums where the probabilities of each column configuration type 
enter (see Table \ref{configs}). Taking into account the geometric sum, we obtain,
\begin{eqnarray}
{\cal Q} & = & \frac{X_3^3 Y^{2}}{(1-X_2Y)^2} + 2\frac{X_3^2Y}{1-X_2Y} + 
\frac{X_4^3}{(1-X_3^2X_2^{-1}Y^{-1})(1-X_2Y)} \nonumber \\ 
 & - & \frac{X_4^3X_3^2X_2^{-1}Y^{-1}}{(1-X_3^2X_2^{-1}Y^{-1})(1-X_3^2)} +
\frac{X_4^3X_2Y^{-1}}{(1-X_2Y^{-1})(1-X_3^2)} + X_2  \nonumber \\ 
 & + & 2\frac{X_3^2Y^{-1}}{1-X_2Y^{-1}} + \frac{X_4^3}{(1-X_3^2X_2^{-1}Y)(1-X_2Y^{-1})} 
+ \frac{X_3^3Y^{-2}}{(1-X_2Y^{-1})^2} \nonumber \\
 & - & \frac{X_4^3X_3^2X_2^{-1}Y}{(1-X_3^2X_2^{-1}Y)(1-X_3^2)} + 
\frac{X_4^3X_2Y}{(1-X_2Y)(1-X_3^2)} 
\label{Qnorma}
\end{eqnarray}
where we remind that $X_2=\textrm{e}^{-2J\beta_{eff}^{(2)}}$, $X_3=\textrm{e}^{-2J\beta_{eff}^{(3)}}$,
$X_4=\textrm{e}^{-2J\beta_{eff}^{(4)}}$ and $Y=\textrm{e}^{\gamma_p(\phi)}$. In order to write the above formula we have taken into
account that $X_9=X_2$, $X_8=X_3$ and $X_7=X_4$.

Once we have calculated the two-body probability functions $p(\delta,\epsilon)$, it is easy to evaluate the probability $\Pi_i(T,p)$
of finding an interfacial spin in class $i$. We assume for an interface in the SOS approximation that the probability of finding 
an interfacial spin in class $i$ is equal to the probability of finding a spin of an interfacial spin column in class $i$.
Assuming the last statement we neglect correlations between neighboring spin columns, following the Solid-On-Solid spirit.
Therefore, the probability of finding an interfacial spin belonging to class $i \in [2,4]$ is,
\begin{equation}
\Pi_i = \sum_{\delta,\epsilon=-\infty}^{+\infty} \pi_i(\delta,\epsilon) p(\delta,\epsilon)
\label{pii}
\end{equation}
where $\pi_i(\delta,\epsilon)$ is the probability of finding a spin belonging to class $i$ in an interfacial spin column characterized 
by the pair of steps $(\delta,\epsilon)$, and where we have not written explicitely the dependence on temperature $T$ and nonequilibrium 
perturbation $p$. In general, we can write,
\begin{equation}
\pi_i(\delta,\epsilon) = \frac{n_i(\delta,\epsilon)}{N(\delta,\epsilon)}
\label{probpii}
\end{equation}
$n_i(\delta,\epsilon)$ is the number of spins belonging to class $i$ in an interfacial spin column characterized by 
$(\delta,\epsilon)$, and $N(\delta,\epsilon)$ is the total number of interfacial spins associated to such spin column. Table 
\ref{clasesconfig} shows $n_i(\delta,\epsilon)$ and $N(\delta,\epsilon)$ for the different column configuration types.  Thus, for 
instance, for a column type A with $\delta,\epsilon > 0$, we have $N(\delta>0,\epsilon>0)=\delta$ (up) interfacial spins, from
which $\delta-1$ belong to class 2, and one belongs to class 3. Hence, in this case, $\pi_2(\delta>0,\epsilon>0)=(\delta-1)/\delta$, 
$\pi_3(\delta>0,\epsilon>0)=1/\delta$ and $\pi_4(\delta>0,\epsilon>0)=0$. For a column type C, where $\delta>0$ and $\epsilon<0$,
we must define the magnitudes $\alpha=\min(|\delta|,|\epsilon|)$ and $\lambda=\max(|\delta|,|\epsilon|)$. There are $\lambda$
(up) interfacial spins, from which $\lambda-\alpha$ belong to class 2, $\alpha-1$ belong to class 3 and only one belongs to class 4.
Therefore $\pi_2(\delta>0,\epsilon<0)=(\lambda-\alpha)/\lambda$, $\pi_3(\delta>0,\epsilon<0)=(\alpha-1)/\lambda$ and 
$\pi_4(\delta>0,\epsilon<0)=1/\lambda$. The rest of probabilities $\pi_i(\delta,\epsilon)$ are defined in the same way from the
different entries in Table \ref{clasesconfig}.
\begin{table}[t]
\centerline{
\begin{tabular}{|c||c|c|c|c|}
\hline
Notation & $n_2(\delta,\epsilon)$ & $n_3(\delta,\epsilon)$ & $n_4(\delta,\epsilon)$ & $N(\delta,\epsilon)$ \\
\hline \hline
A & $\delta-1$ & 1 & 0 & $\delta$ \\
\hline
B & $\delta-1$ & 1 & 0 & $\delta$ \\
\hline
C & $\lambda - \alpha$ & $\alpha - 1$ & 1 & $\lambda$ \\
\hline
D & 1 & 0 & 0 & 1 \\
\hline
E & 1 & 0 & 0 & 1 \\
\hline
F & $|\epsilon|-1$ & 1 & 0 & $|\epsilon|$ \\
\hline
G & 1 & 0 & 0 & 1 \\
\hline
H & 1 & 0 & 0 & 1 \\
\hline
I & $|\epsilon|-1$ & 1 & 0 & $|\epsilon|$ \\
\hline
\end{tabular}
}
\caption[Spin class populations for each interfacial column configuration.]
{\small In this table we show the populations $n_i(\delta,\epsilon)$ for each interfacial class, $i=2, 3, 4$, for each one
of the column configuration types summarized in Table \ref{configs}. Thus, in the second column we write the number of spins in 
class 2 for each column configuration type, in the third column we write the number of spins in class 3,
and in the fourth column, the number of spins in class 4. The last column shows the total number of interfacial (up) spins 
associated to each column configuration type, $N(\delta,\epsilon)$. The first column yields the notation for each type.
Remember that $\alpha=\min(|\delta|,|\epsilon|)$ and $\lambda=\max(|\delta|,|\epsilon|)$.}
\label{clasesconfig}
\end{table}

In order to calculate the probabilities $\Pi_i(T,p)$ we must perform the sums involved in eq. (\ref{pii}). We need the classic results
for the geometric sum and series to perform such sums, as well as some results derived from them taking into account the linear 
behavior of both the derivative and the Riemann integral operators,
\begin{eqnarray}
\sum_{k=0}^{\infty} x^k & = & \frac{1}{1-x} \quad , \quad |x|<1 \label{sum1} \\
\sum_{k=0}^n x^k & = & \frac{1-x^{n+1}}{1-x} \quad , \quad x \neq 1 \label{sum2} \\
\sum_{k=1}^{\infty} \frac{x^k}{k} & = & \ln (\frac{1}{1-x}) \quad , \quad |x|<1 \label{sum3} \\
\sum_{k=1}^n \frac{x^k}{k} & = & \ln (\frac{1}{1-x}) - \int_0^x \textrm{d}y \frac{y^n}{1-y} \quad , 
\quad |x|<1 \label{sum4} \\
\sum_{k=1}^{\infty} k x^k & = & \frac{x}{(1-x)^2} \quad , \quad |x|<1 \label{sum5} \\
\sum_{k=1}^n k x^k & = & \frac{x(1-x^n)}{(1-x)^2} - \frac{nx^{n+1}}{1-x} \quad , \quad x \neq 1 \label{sum6}
\label{sumatorias}
\end{eqnarray}
Using these equalities we find the desired results for the probabilities $\Pi_i(T,p)$. In particular, we obtain for $\Pi_2(T,p)$ the
following result,
\begin{eqnarray}
\Pi_2 & = & \frac{1}{\cal Q} \Big\{ \frac{X_3^3Y^2}{(1-X_2Y)^2} - \frac{X_3^3X_2^{-1}Y}{1-X_2Y}
\ln \big(\frac{1}{1-X_2Y}\big) + \frac{X_3^2Y}{1-X_2Y} \nonumber \\
 & - & \frac{X_3^2}{X_2}\ln \big(\frac{1}{1-X_2Y}\big) + \frac{X_4^3}{(1-X_3^2X_2^{-1}Y^{-1})(1-X_2Y)} \nonumber \\
 & + & \frac{X_4^3X_2^{-1}Y^{-1}}{(1-X_3^2X_2^{-1}Y^{-1})^2} \ln \big(\frac{1-X_2Y}{1-X_3^2}\big) 
- \frac{X_4^3X_2X_3^{-2}Y^{-1}}{1-X_2Y^{-1}} \nonumber \\
 & + & \frac{X_4^3X_2X_3^{-2}Y^{-1}}{(1-X_2Y^{-1})(1-X_3^2X_2^{-1}Y)} + 
\frac{X_3^2Y}{1-X_2Y} + X_2 \nonumber \\
 & + & \frac{X_4^3X_2^{-1}Y}{(1-X_3^2X_2^{-1}Y)^2}\ln \big(\frac{1-X_2Y^{-1}}{1-X_3^2}\big)
+ \frac{X_3^2Y^{-1}}{1-X_2Y^{-1}} \nonumber \\
 & - & \frac{X_3^2}{X_2}\ln \big(\frac{1}{1-X_2Y^{-1}}\big) +
\frac{X_4^3}{(1-X_3^2X_2^{-1}Y)(1-X_2Y^{-1})} \nonumber \\
 & - & \frac{X_4^3X_3^2X_2^{-1}Y}{(1-X_3^2X_2^{-1}Y)(1-X_3^2)} +
\frac{X_4^3X_2Y}{(1-X_2Y)(1-X_3^2)} \label{pi2} \\
 & + & \frac{X_3^2Y^{-1}}{1-X_2Y^{-1}} + \frac{X_3^3Y^{-2}}{(1-X_2Y^{-1})^2} -
\frac{X_3^3X_2^{-1}Y^{-1}}{1-X_2Y^{-1}}\ln \big(\frac{1}{1-X_2Y^{-1}}\big) \Big\} \nonumber
\end{eqnarray}
In the same way we can calculate $\Pi_4(T,p)$, obtaining,
\begin{eqnarray}
\Pi_4 & = & \frac{1}{\cal Q} \Big\{ \frac{X_4^3X_2^{-1}Y^{-1}}{1-X_3^2X_2^{-1}Y^{-1}}\ln \big(\frac{1}{1-X_2Y}\big) 
- \frac{X_4^3X_2^{-1}Y^{-1}}{1-X_3^2X_2^{-1}Y^{-1}} \ln \big(\frac{1}{1-X_3^2}\big) \nonumber \\
 & + & \frac{X_4^3X_2^{-1}Y}{1-X_3^2X_2^{-1}Y} \ln \big(\frac{1}{1-X_2Y^{-1}}\big) - 
\frac{X_4^3X_3^{-2}}{1-X_3^2X_2^{-1}Y} \ln \big(\frac{1}{1-X_3^2}\big) \Big\}
\label{pi4}
\end{eqnarray}
In order to obtain $\Pi_3$ we only have to apply the normalization condition, eq. (\ref{normalizpi}), which in this case implies
$\Pi_3=1-\Pi_2-\Pi_4$. The details about the calculation of the probabilities $\Pi_i$ are exposed in Appendix \ref{apendPi2},
where we write down in particular the detailed calculation for $\Pi_2$.

At this point we must notice that the probabilities $\Pi_i(T,p)$ depend in principle on the thermodynamic parameter $\gamma_p(\phi)$
(remember that $Y=\textrm{e}^{\gamma_p(\phi)}$), and via this parameter they implicitly depend on the average interface slope, 
$\tan \phi$. Hence we must use the notation $\Pi_i[T,p,\gamma_p(\phi)]$. On the other hand, this dependence was
expected, since the interfacial spin distribution among the different classes strongly depends on the typical values of the step
variables $\delta$ and $\epsilon$ for each column, and these values depend on the average interface slope. Hence $X_p$ will
depend on the parameter $\gamma_p(\phi)$ due to the explicit dependence of $\Pi_i[T,p,\gamma_p(\phi)]$ on such parameter, and it will 
be impossible to extract the explicit relation between the thermodynamic parameter $\gamma_p(\phi)$ and the interface slope, 
$\tan \phi$, on the contrary to what happened in the equilibrium case, where we calculated this explicit relation through eq. 
(\ref{generatriz}). However, we need to know the properties of the interface, i.e. its microscopic structure, codified in the
probability function $p_p(\delta)$, and its macroscopic structure, codified in the surface tension, $\sigma_{SOS}^{(p)}$, as a function
of the angle $\phi$ formed between the interface and the $\hat{x}$ axis. This knowledge will give us the possibility of studying the
properties and the shape of a spin droplet, which is one of the fundamental objects needed to write a nucleation theory for 
metastability (see next chapter). Therefore we must make an additional approximation at this stage. This approximation consists in 
assuming that the probabilities $\Pi_i$ of finding an interfacial spin belonging to class $i$ do not depend on the angle $\phi$
formed between the interface and the $\hat{x}$ axis. These probabilities $\Pi_i$ enter the definition of the statistical weight for
an interfacial broken bond in the generalized SOS approximation, $X_p$. Thus we assume that the probabilities $\Pi_i$ are the ones we
have obtained previously, see eqs. (\ref{pi2}) and (\ref{pi4}), particularized to an interface forming an angle $\phi=0$ with the 
$\hat{x}$ axis. In this case we will have $Y=\textrm{e}^{\gamma_p(\phi)}=1$ in the definition of the probabilities $\Pi_i$, so
all the dependence on $\gamma_p(\phi)$ disappears inside $X_p$. The dependence on the interface slope enters the theory through
the statistical weight $Y^{\gamma_p(\phi)\delta}$ that appears in the probability $p_p(\delta)$ of finding a step of size $\delta$
in the nonequilibrium interface, see eq. (\ref{probdeltap}). In this way we build in a simple way a generalization of the original 
SOS approximation through the definition of a new probabilistic weight associated to an 
interfacial broken bond, $X_p$, which exclusively depends on the temperature $T$ and the nonequilibrium perturbation $p$. This 
generalized theory keeps the elegant and simple structure of the original SOS approximation.

\section{Properties of the Nonequilibrium Interface}
\label{capSOS_apProp}

Once we have completed the generalization of SOS approximation to take into account the nonequilibrium character of our interface,
the next step consists in verifying the predictions of this theory.

\subsection{Interfacial Microscopic Structure}
\label{capSOS_apProp_subMicro}

In a first step, we wonder about the interfacial microscopic structure. 
Here the extended theory predicts that the probability of finding a step in the interface with size $\delta$ is given by eq. 
(\ref{probdeltap}), where $X_p$ is, as we know, the probabilistic weight associated to an interfacial broken bond. In order to check
this prediction, we perform Monte Carlo simulations of the model interface, paying special attention to its microscopic structure.
To do that we define the system on a rectangular lattice with shape $L_{\perp} \times L_{\parallel}$, where $L_{\perp}$ is the
lattice size in the direction perpendicular to the interface, and $L_{\parallel}$ is the lattice size in the direction of the 
interface. The initial condition for the simulation consists in a stripe formed by up spins which completely filles the first 
$\L_{\perp}/2$ rows of the system, and a second stripe formed by down spins filling the remainder of the system. In this way the initial
configuration exhibits a linear interface in the $\hat{x}$ direction between a phase with spins up (bottom) and a phase with spins down 
(top). Boundary conditions are periodic in the direction of the interface, and open in the direction perpendicular to the interface.
Thus spins in column 1, $(i,1)$ with $i \in [1,L_{\perp}]$, have as left neighbors the corresponding spins in column $L_{\parallel}$, 
$(i,L_{\parallel})$, and reversely. The open boundary conditions in the direction perpendicular to the interface involve that spins
in the first row,  $(1,j)$ with $j \in [1,L_{\parallel}]$, do not have down neighbors, and spins in the last row, $(L_{\perp},j)$,
do not have up neighbors.

In order to completely define the system to simulate, we must implement Glauber dynamics, previously defined in eq. 
(\ref{rate}). However we are only interested in computationally studying the interfacial structure. Thus we must eliminate the possible
interaction between the interface and the fluctuations present in the bulk. Moreover, the Solid-On-Solid approximation (both the 
equilibrium one and our generalization)  assumes that the interface does not interact with the bulk. Hence in order to properly 
compare the theoretical results with the computational ones we must suppress bulk dynamics, eliminating in this way the presence of 
bulk fluctuations. We can do that in practice making zero the transition rates for spins belonging to both spin classes 1 and 10, 
which are the spins that initially form the bulk. In this way the bulk remains frozen, fluctuation-free during the whole simulation
process, thus preventing the interference of bulk fluctuations on the interface properties. However, we can wonder about the effect
that these bulk fluctuations involve on interfacial properties. In order to investigate this effect, we have also simulated  the 
interface using the {\it full} Glauber dynamics, i.e. including bulk dynamics. We observe that bulk fluctuations are not
relevant for the interfacial properties as far as we are well below the critical temperature, where fluctuations of all scales appear
in the system. Since we are interested in the properties of the interface at temperatures far away the critical one, $T_c(p)$,
the suppression of bulk dynamics will not influence our results for the interface.
\begin{figure}
\centerline{
\psfig{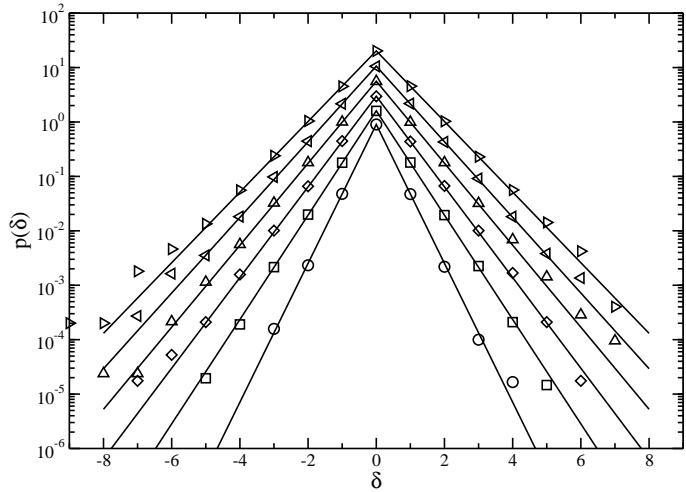}}
\caption[Probability of finding in the interface a step with size $\delta$.]
{\small Probability of finding in the interface a step with size $\delta$, $p_p(\delta)$, for a system size 
$L_{\perp} \times L_{\parallel} = 128 \times 256$ at temperature $T=0.3T_{ons}$ and several different values of the nonequilibrium
perturbation $p$. Namely, from bottom to top, $p=0$, $0.01$, $0.02$, $0.03$, $0.04$ and $0.05$. For the shake of clarity the
curves has been shifted a factor $2^i$, $i\in [0,5]$ in the vertical direction, where $i=100 \times p$.}
\label{prob-escalon}
\end{figure}

For the fixed initial condition, and using the {\it trunked} Glauber dynamics, the system rapidly evolves from the initial state 
with a completely flat interface towards a stationary state characterized by a rough non-trivial interface configuration. Such
configuration is defined by the step variables vector $\mathbf{\delta}=\{\delta_i, i=1,L_{\parallel}\}$\footnote{When we
characterize an interfacial configuration in the real system via the step variables vector $\mathbf{\delta}$ we are neglecting
the presence of overhangs (see Fig. \ref{interf}.b) in this interface. In fact it is observed that the presence of overhangs
in the interface affects very weakly its properties, at least for temperatures well below the critical one.}, where 
$\delta_i=y_i-y_{i-1}$, being $y_i$ the height of the interface for the spin column $i$. In order to measure the probability 
$p_p(\delta)$ we wait until the interface reaches the steady state. For a fixed interfacial configuration we measure the step sizes
$\delta_i$ using an algorithm for microscopic interfacial recognition\cite{Pablointerf}. From those values $\delta_i$ we accumulate
the histogram $p_p(\delta)$. However, as opposed to the SOS approximation, the real model shows correlations between close steps,
parameterized by certain typical correlation length $\xi$. In order to build the histogram $p_p(\delta)$ we must use statistically
independent measures of the step variable $\delta$. Therefore for a fixed interfacial configuration we measure step variables
separated by a distance larger that $\xi$, so in this way we ensure the statistical independence of measures. In our case we
sample the interfacial configuration extracting measures of $\delta$ separated by a distance $2\xi$. In order to accumulate enough
statistics we let the system evolve in time, repeating the histogram measurement process for $p_p(\delta)$ at time intervals 
$\Delta t$ larger than the typical correlation time in the interface (again, in order to ensure statistical independence of data).
The results of such measurements are shown in Fig. \ref{prob-escalon}, where we observe the probability of finding a step with size
$\delta$ in the interface, $p_p(\delta)$, for a system size $L_{\perp} \times L_{\parallel} = 128 \times 256$ and  a temperature 
$T=0.3T_{ons}$ with different values of $p\in [0,0.05]$. The interface we have studied was parallel to the $\hat{x}$ axis, so 
$\phi=0$. As we can observe in this figure, the distribution $p_p(\delta)$ is a decreasing exponential of $|\delta|$.
Furthermore, as we increase the nonequilibrium parameter $p$, the probability of finding larger step sizes increases, as expected. 
Fig. \ref{prob-escalon} also shows, for comparison, the theoretical prediction based on the generalized SOS approximation for
$p_p(\delta)$, eq. (\ref{probdeltap}) (notice that here we have $Y=1$ because $\phi=0$). As we see, the agreement between 
theory and simulations is nearly perfect for all studied values of $p$. This agreement points out that the generalized SOS approximation,
based on the concept of effective temperature, correctly reproduces the microscopic features of the interface subject to 
nonequilibrium conditions. We have also studied the microscopic properties of the interface for other temperature values below
the critical temperature, $T_c(p)$, always finding a very good agreement between theory and simulations. On the other hand, we have 
investigated the differences that the presence of overhangs and the interaction between the interface and the bulk induce in
our computational results. We conclude that, as far as we are well below the critical region, the extended SOS approximation
describes very well the observed real interface structure (including overhangs and bulk dynamics).

\subsection{Interfacial Macroscopic Properties}
\label{capSOS_apProp_subMacro}

\begin{figure}
\centerline{
\psfig{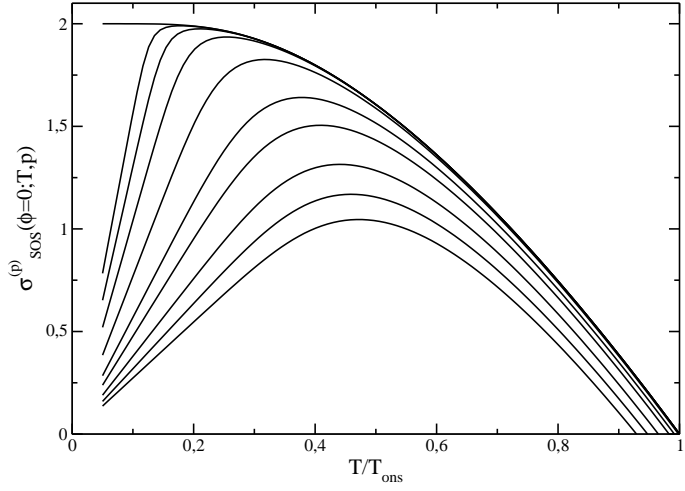}}
\caption[Surface tension for the nonequilibrium magnetic system.]
{\small Surface tension $\sigma_{SOS}^{(p)}(\phi=0;T,p)$ as a function of temperature for different values of $p$,
as derived from the generalized SOS approximation for an interface parallel to the $\hat{x}$ axis. In particular, from top to bottom,
$p=0$, $10^{-6}$, $10^{-5}$, $10^{-4}$, $10^{-3}$, $5\times 10^{-3}$, $10^{-2}$, $2\times 10^{-2}$, $3\times 10^{-2}$ and 
$4\times 10^{-2}$. Notice that for any $p \neq 0$ the surface tension shows a maximum for certain temperature, and it converges 
towards zero for smaller temperatures.}
\label{tensionsup}
\end{figure}

We next pay attention to the macroscopic properties of the interface, codified in the surface tension, $\sigma_{SOS}^{(p)}(\phi;T,p)$.
Fig. \ref{tensionsup} shows $\sigma_{SOS}^{(p)}(\phi=0;T,p)$, as defined in eq. (\ref{tensionsuperfp}), as a function of temperature
for different values of the nonequilibrium parameter $p$. These curves show the theoretical prediction for the surface tension in the
real system. The first conclusion derived from this figure is the fundamental difference between the surface tension for the
equilibrium system ($p=0$) and that for the nonequilibrium system ($p \neq 0$) in the low temperature limit.
We can observe in Fig. \ref{tensionsup} that for $p=0$ the surface 
tension grows monotonously as temperature decreases, converging towards
2 in the low temperature limit. However, for any $p \neq 0$ we observe 
that the surface tension exhibits a maximum for certain temperature, $T_{max}$. For $T<T_{max}$ the surface tension 
decreases as temperature decreases, while for $T>T_{max}$ the surface tension increases as temperature decreases. 
Moreover, $\sigma_{SOS}^{(p)}(\phi=0;T,p)$ linearly converges towards 0 in the limit $T \rightarrow 0$ 
for $p\neq 0$. As we will see in the next chapter, this fundamental difference in the macroscopic properties of the interface 
will involve very important differences in the properties of metastable states in the nonequilibrium system as compared to the 
equilibrium one in the low temperature region.

Let's rewrite $X_p$ in order to understand why this unexpected behavior emerges for the nonequilibrium system. As we know, $X_p$ is
the statistical weight associated to a interfacial broken bond in the generalized SOS approximation. In the limit $p=0$, the weight
$X_p$ converges to $X=\textrm{e}^{-2J\beta}$, where $\beta$ is the system's inverse temperature. Comparatively, for $p\neq 0$ we can
write $X_p$ as,
\begin{figure}
\centerline{
\psfig{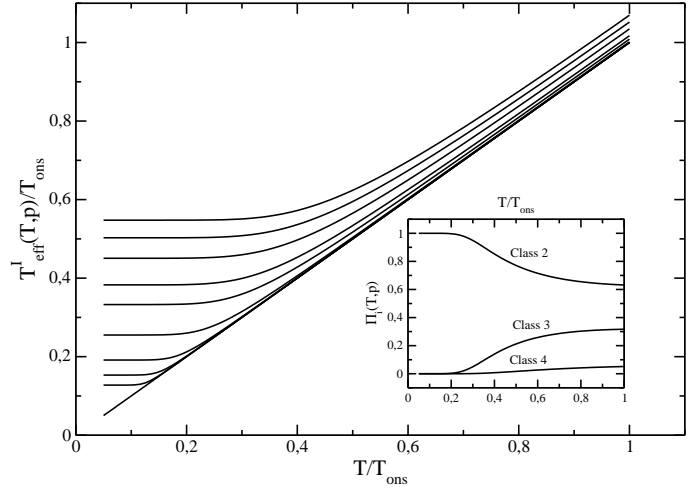}}
\caption[Interfacial effective temperature and spin class population.]
{\small The main plot shows the interface effective temperature, $T_{eff}^I$, obtained from our generalized SOS approximation,
as a function of system's temperature, $T$, for varying $p$. From bottom to top $p=0$, $10^{-6}$, $10^{-5}$, $10^{-4}$, 
$10^{-3}$, $5\times 10^{-3}$, 
$10^{-2}$, $2\times 10^{-2}$, $3\times 10^{-2}$ and $4\times 10^{-2}$. The inset shows the probabilities $\Pi_i(T,p)$ of finding an
interfacial spin belonging to class $i$ as a function of temperature for $p=0.01$. Notice that for low temperatures almost all spins
belong to class 2.}
\label{teffI}
\end{figure}
\begin{equation}
X_p \equiv \textrm{e}^{-2J\beta_{eff}^I}
\label{betaeffI}
\end{equation}
where $\beta_{eff}^I \equiv 1/T_{eff}^I$. The previous equation defines the interface effective temperature, $T_{eff}^I$. In some sense
this interface effective temperature yields the average of the effective temperatures associated to the different spin classes in the 
interface, taking into account the classes relative populations. In the limit  $p=0$ it is evident that the interface effective 
temperature, $T_{eff}^I$, reduces to the usual temperature, $T$. However, for any $p\neq 0$ both temperatures defer. Fig. \ref{teffI}
shows the interface effective temperature, as defined in eq. (\ref{betaeffI}), as a function of $T$ for different values of $p$.
The behavior of $T_{eff}^I(T)$ shown in Fig. \ref{teffI} helps us to understand the novel behavior of surface tension
for low temperatures in the nonequilibrium model. First we observe that for a fixed value of $p$ the relation between $T_{eff}^I$ and 
$T$ is linear for high temperatures. That is, in the high temperature limit, where the thermal noise ($T$) dominates over the
non-thermal noise ($p$), the interface effective temperature is completely coupled to the system's temperature.  In this limit the effect
induced by $p$ reduces to a slight increase of the interface effective temperature as compared to the system's temperature. However,
in the low temperature limit the noise with non-thermal origin dominates, so the interface effective temperature $T_{eff}^I$ 
completely decouples from the system's thermodynamic temperature, $T$, converging to a constant nonzero value, see Fig. \ref{teffI}. 
This low temperature limit for the interface effective temperature can be easily calculated, taking into account that, as shown 
in the inset of Fig. \ref{teffI}, almost all interfacial spins belong to class 2 in this limit and for moderate values
of $p$. Therefore $T_{eff}^I$ converges towards $T_{eff}^{(2)}(T=0)$ in the limit $T\rightarrow 0$, 
where $T_{eff}^{(2)}(T=0)$ is the effective temperature associated to class 2 for $T=0$. Hence,
\begin{equation}
\lim_{T\rightarrow 0} T_{eff}^I(T,p) \approx T_{eff}^{(2)}(T=0,p) = \frac{-4}{\ln (p)} > 0
\label{limteffI}
\end{equation}
Thus the interface suffers a nonzero effective temperature, independent of $T$, at low temperatures due to the 
non-thermal noise. Hence the statistical weight $X_p$ converges to a nonzero and constant value for $T\rightarrow 0$,
so the surface tension, which is directly proportional to $T$, converges linearly towards zero in this limit.

The next logical step should be the measurement of surface tension in Monte Carlo simulations for the interface of the discrete model,
in order to compare these computational results with the theoretical predictions. However, we do not know how to define in the system
a thermodynamic potential as the surface tension from the microscopic point of view, because the system is out of equilibrium.
Moreover, if we could microscopically define such potential, we still would have to face the additional problem of its explicit 
measurement. This problem comes from the need of a dense sampling of the complete system's phase space in order to obtain a 
reliable measurement of surface tension. The system presents many degrees of freedom, so its phase space is incredibly huge. Therefore
any feasible Monte Carlo simulation only samples a small region of this phase space, which implies an incorrect measurement 
of surface tension\footnote{Nowadays there are some methods which allow to calculate thermodynamic potentials as the free energy
of a complex system.\cite{Frenkel}}. As an alternative option, we can evaluate physical observables easy to define and measure from
the microscopic point of view, which have a direct and simple relation with surface tension. This is the case for the size of the 
critical droplet appearing when studying the problem of metastability in the ferromagnetic nonequilibrium system. As we explained
in previous sections, when the system is in a metastable state (for instance, a state with positive magnetization in the ordered phase
subject to a weak negative magnetic field) it eventually evolves from the metastable phase towards the truly stable state. 
This transition proceeds through the nucleation of a droplet of the stable phase in the metastable bulk. Small stable phase droplets
tend to disappear, while large enough droplets tend to grow. There is a critical size separating both tendencies, which is the critical
droplet size. The critical droplet controls the demagnetization process. Generalizing the results obtained for equilibrium systems,
we observe (see next chapter) that the critical droplet size is proportional to the surface tension in the model for zero magnetic
field. Thus, measuring the critical droplet size and comparing these measurements with the predictions based on the generalized SOS
approximation result for the surface tension, $\sigma_{SOS}^{(p)}(\phi=0;T,p)$, 
we will be able to quantify in an indirect way the agreement
between the prediction for the nonequilibrium surface tension and this observable in the real system.
Fig. \ref{gotacrit} in next chapter shows the computational results for the critical droplet size as a function of temperature for
different values of $p$, and the corresponding theoretical predictions based on the surface tension $\sigma_{SOS}^{(p)}(\phi=0;T,p)$. 
As we can see in this figure, the agreement between 
theory and simulations is rather good. Apart from the quantitative agreement, which is very good taking into account the approximate
character of the theory, it is remarkable that the predictions perfectly reproduce the existence of a maximum in the critical droplet
size as a function of temperature for any $p\neq 0$. This maximum is directly related to the maximum observed for the surface tension
in the nonequilibrium model.

Therefore we conclude that, from both the microscopic -$p_p(\delta)$- and the macroscopic -$\sigma_{SOS}^{(p)}(\phi;T,p)$- point of 
view, the generalized SOS approximation that we have derived in this chapter for the interface in the nonequilibrium magnetic system
correctly explains the properties of such interface, both qualitatively and quantitatively. This agreement between theory and 
simulations will allow us to deduce in the next section some properties for a spin droplet. These properties will be
very useful when trying to study and understand the dynamic properties of metastable states in the ferromagnetic nonequilibrium system.

\markboth{Interfacial Properties}{\ref{capSOS_apForma} Shape and Form Factor of a Spin Droplet}
\section{Shape and Form Factor of a Spin Droplet using Wulff Construction}
\label{capSOS_apForma}
\markboth{Interfacial Properties}{\ref{capSOS_apForma} Shape and Form Factor of a Spin Droplet}

In the previous section we have developed a theoretical approximation, based on the Solid-On-Solid approximation by Burton,
Cabrera and Frank\cite{BCF}. It has allowed us to calculate the surface tension associated to the interface in the ferromagnetic
system under nonequilibrium conditions. In this section we want to use the results obtained in the previous section
in order to calculate the shape of a spin droplet and the associated form factor.
\begin{figure}
\centerline{
\psfig{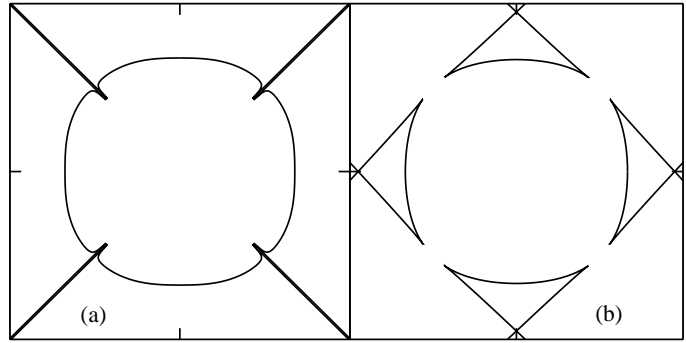}}
\caption[Polar plot of the surface tension and droplet radius.]
{\small (a) Polar plot of the surface tension as obtained in the generalized SOS approximation, $\sigma_{SOS}^{(p)}(\phi)$,
for temperature $T=0.3T_{ons}$ and $p=0.01$. Notice the underlying symmetry of surface tension, and the presence of singularities
for angles $\phi=(2n+1)\pi/4$, with $n=0,\ldots,3$; (b) Polar plot of the droplet radius $R(\theta)$ derived from the surface tension
plotted in (a) using the Wulff construction. The singularities in $\sigma_{SOS}^{(p)}(\phi)$ involve angular regions where $R(\theta)$
is not defined.}
\label{formasingular}
\end{figure}

\subsection{The Wulff Construction}
\label{capSOS_apForma_subWulff}

As we have stated above several times, the process of demagnetization from a metastable state in our system takes place through the 
nucleation of one or several droplets of the stable phase in the metastable bulk. In order to understand and analyze the nucleation
and consequent growth of these droplets we need to know in detail their shape. In general, the equilibrium shape of these droplets
is determined by minimization of the associated total surface tension for a fixed volume. For isotropic systems this process yields 
a droplet with spherical shape (circular in two dimensions). However, when the surface tension depends, as in our case, on the 
orientation of the interface with respect to certain privileged axis, the shape of the droplet will adjust to take advantage of the 
low free energy cost of certain interface orientations, and this shape will minimize interface orientations with high free energy cost.
This mechanism gives rise to droplets with typical crystal-like shape, which will depend on temperature and other system 
parameters.\cite{Rottman}

In our case we know the surface tension $\sigma_{SOS}^{(p)}(\phi)$ for all possible interface orientations, parameterized by the angle 
$\phi$. Hence the Wulff construction\cite{Wulff} will allow us to obtain the equilibrium droplet shape. This construction can be 
summarized as follows: through each point of the polar curve $\sigma_{SOS}^{(p)}(\phi)$, with $\phi \in [0,2\pi]$, we draw a line
perpendicular to the radius linking such point with the center. The interior envelope of all these perpendicular lines determines
$R(\theta)$, which is the droplet radius in polar coordinates. Mathematically speaking, the equilibrium shape $R(\theta)$ can be
calculated parametrically\cite{BCF2},
\begin{eqnarray}
R(\theta) & = & R_0\sqrt{x^2(\phi)+y^2(\phi)} \nonumber \\
x(\phi) & = & (\cos \phi)\sigma_{SOS}^{(p)}(\phi) - (\sin \phi) 
\frac{\textrm{d}\sigma_{SOS}^{(p)}(\phi)}{\textrm{d}\phi} \label{Rparam} \\
y(\phi) & = & (\sin \phi)\sigma_{SOS}^{(p)}(\phi) + (\cos \phi) 
\frac{\textrm{d}\sigma_{SOS}^{(p)}(\phi)}{\textrm{d}\phi} \nonumber \\
\tan \theta & = & \frac{y(\phi)}{x(\phi)} \nonumber
\end{eqnarray}
where $R_0$ is just a constant that fixes the droplet radius, and $\sigma_{SOS}^{(p)}(\phi)$ is the surface tension derived from
the SOS approximation generalized to the nonequilibrium case, eq. (\ref{tensionsuperfp}). In fact, in order to define the surface 
tension $\sigma_{SOS}^{(p)}(\phi)$ for angles outside the interval $\phi \in [-\pi/4,\pi/4]$ we use the symmetry that 
$\sigma_{SOS}^{(p)}(\phi)$ exhibits. As we previously discussed, in order to study the surface tension of an interface forming an angle 
$|\phi|>\pi/4$ with the $\hat{x}$ axis, it is convenient to change the reference axis to be the $\hat{y}$ axis. Thus if we suppose for 
instance that $\pi/4 < \phi < \pi/2$, we have that 
$\sigma_{SOS}^{(p)}(\phi) = \sigma_{SOS}^{(p)}(\pi/2-\phi)$, where now $0 < \pi/2-\phi < \pi/4$.
In the same way we can extend using equivalent symmetry arguments the definition of $\sigma_{SOS}^{(p)}(\phi)$ to the whole
circumference, $\phi \in [0,2\pi]$.

The surface tension $\sigma_{SOS}^{(p)}(\phi)$, as defined in the generalized SOS approximation, eq. (\ref{tensionsuperfp}), and
once extended by symmetry over the whole circumference, has a fundamental problem, because it is singular for angles 
$\phi=(2n+1)\pi/4$, with $n=0,\ldots,3$. Fig. \ref{formasingular}.a shows a polar plot of $\sigma_{SOS}^{(p)}(\phi)$
for temperature $T=0.3T_{ons}$ and $p=0.01$. In this figure we observe the aforementioned singularities.
These singularities found in $\sigma_{SOS}^{(p)}(\phi)$ induce the appearance of angular regions where the droplet radius, $R(\theta)$, 
obtained from the Wulff construction, eqs. (\ref{Rparam}), is not defined, see Fig. \ref{formasingular}.b. These undefined regions in
$R(\theta)$ constitute an important objection in our investigation, because one of the principal aims in this section consists in 
calculating the so-called form factor, $\Omega_d(T,p)$, which is the constant that relates the droplet radius ${\cal R}$
(to be defined later on) and the droplet volume. Thus $V=\Omega_d(T,p) {\cal R}^d$, where $d$ is the system dimension. The so-defined
form factor is an important magnitude in order to write a nucleation theory which correctly describes the dynamics of the 
nonequilibrium metastable-stable transition. To calculate $\Omega_d(T,p)$ we need to evaluate
the droplet shape, parameterized by $R(\theta)$, for all values of $\theta$. Hence we must regularize in some sense the function
$R(\theta)$ obtained from the generalized SOS approximation, in order to define this function over the whole circumference.

\subsection{Analytic Continuation of the Radial Function}
\label{capSOS_apForma_subCont}

In order to regularize $R(\theta)$ in the undefined angular regions, we propose an analytical continuation of the
radial function $R(\theta)$ inside such regions. We denote this analytical continuation as $R^{(c)}(\theta)$. In particular,
we analytically continue the radial function using a second order polynomial, $R^{(c)}(\theta)=a\theta^2+b\theta+c$. In order to
fix the three free coefficients in this polynomial we need three different conditions for it. In a first step, we notice that the
angular symmetry that surface tension exhibits is inherited by the radial function $R(\theta)$, as can be observed in 
Fig. \ref{formasingular}. This symmetry implies that the analytical continuation of the radial function must fulfill,
\begin{equation}
\frac{\textrm{d}R^{(c)}(\theta)}{\textrm{d}\theta}|_{\theta=\pi/4} = 0
\label{cond1}
\end{equation}
since we expect that $R^{(c)}(\pi/4+\epsilon)=R^{(c)}(\pi/4-\epsilon)$. This equation provides the first condition on $R^{(c)}(\theta)$.
The other two conditions are obtained requiring continuity and analyticity to the analytic continuation. Hence, if $\theta^*$ is the
angle where the connection between the radial function $R(\theta)$ and its analytical continuation $R^{(c)}(\theta)$ takes place,
the continuity and analyticity conditions reduce respectively to,
\begin{eqnarray}
R^{(c)}(\theta) |_{\theta=\theta^*} & = & R(\theta) |_{\theta=\theta^*} \\
\frac{\textrm{d}R^{(c)}(\theta)}{\textrm{d}\theta} |_{\theta=\theta^*} & = & 
\frac{\textrm{d}R(\theta)}{\textrm{d}\theta} |_{\theta=\theta^*}
\label{conds23}
\end{eqnarray}

Using these three conditions we are able to obtain the coefficients $a$, $b$ and $c$ which appear in the definition of the analytical
continuation $R^{(c)}(\theta)$ as a function of the connection angle $\theta^*$ and the values that $R(\theta)$ and its derivative take
for such connection angle. The only remaining problem is the determination of $\theta^*$. However, it is easy to solve such 
problem, because the nature of the singularity in the surface tension is such that it gives rise to a discontinuous change in the 
derivative of $R(\theta)$ for certain angle $\theta^*$, as observed in Fig. \ref{formasingular}.b. This angle $\theta^*$ where
the derivative of $R(\theta)$ is discontinuous signals the angle where the undefined region for $R(\theta)$ begins. Therefore
we define this angle (see Fig. \ref{formasingular}.b)
as the connection angle $\theta^*$. This connection angle can be calculated numerically in a simple way\footnote{For high temperatures,
the influence of the singularity appearing for the surface tension extends to certain interval around it. Hence the radial function
feels the singularity before it reaches the angular undefined region. In these cases we can get rid of the region where $R(\theta)$
{\it feels} the singularity by performing the analytical continuation for a connection angle previous to the angle defined by the 
discontinuous change in the derivative of $R(\theta)$. This new connection angle is detected looking for the angle where the angular 
derivative of $R(\theta)$ becomes negative, since such condition points out that the effect of the singularity is becoming relevant.}.
In this way we are able to continue analytically the radial function $R(\theta)$ in a safe manner. We only have to apply the analytical 
continuation once, inside the angular interval $\theta \in [0,\pi/4]$, due to the angular symmetry exhibited by the radial function 
(inherited from the angular symmetry shown by the surface tension). The remainder of the analytically extended angular function is
defined by symmetry taking as starting point this small angular interval.
\begin{figure}
\centerline{
\psfig{file=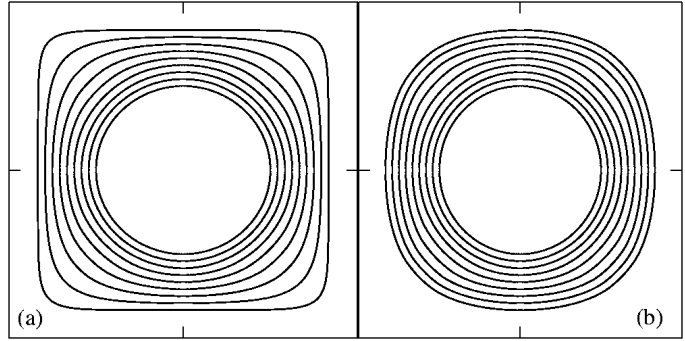,width=9cm}}
\caption[Shape of equilibrium and  nonequilibrium spin droplets.]
{\small (a) Shape of a spin droplet for the equilibrium system ($p=0$), and several different temperatures. In particular, 
from outside to inside, $T/T_{ons}=0.1$, $0.2$, $0.3$, $0.4$, $0.5$, $0.6$, $0.7$, $0.8$ and $0.9$. For the shake of clarity we
have rescaled the droplet radius depending on temperature; (b) The same than in (a), but for the nonequilibrium model,
with $p=0.01$.}
\label{formas}
\end{figure}

Using this analytical continuation technique we obtain the shape of a spin droplet for the nonequilibrium ferromagnetic model.
Fig. \ref{formas} shows the shape of such droplet for different temperatures, for an equilibrium system (with $p=0$) and for a 
nonequilibrium one, with $p=0.01$. We observe for $p=0$ that the shape of the droplet tends to be a square at low temperatures,
while for high temperatures we recover the circular shape associated to a bidimensional isotropic system. Thus
for high temperatures the observed differences in the surface tension for different orientations are very small as compared to 
thermal energy, so in practice the interface does not feel the existence of privileged interface orientations. This process gives rise 
to the observed isotropy. On the contrary, for low temperatures, the differences observed in the surface tension for different 
interface orientations are very important as compared to thermal energy, so interface orientations not parallel to the privileged
axes are highly punished. An important detail is that for temperatures of the order of $T=0.5T_{ons}$ the spin droplet shape
is already almost circular for $p=0$. This property points out that effective isotropy appears at not very high temperatures.
Fig. \ref{formas}.b shows the shape of a spin droplet for different temperatures, but now for the nonequilibrium system with $p=0.01$.
Equivalently to what we observed in the equilibrium system ($p=0$), the droplet for high enough temperatures takes circular 
(isotropic) shape. However, for low temperatures we observe that its shape converges towards an intermediate structure
between a circle and a square, where the underlying lattice anisotropy is reflected only partially. In order to understand
this difference between the equilibrium system ($p=0$) and the nonequilibrium one ($p\neq 0$) we must recall the concept of
interface effective temperature, $T_{eff}^I(T,p)$, see eq. (\ref{betaeffI}). In the previous section we observed that for $p\neq 0$
the interface effective temperature converged towards a nonzero value in the limit $T\rightarrow 0$, see Fig. \ref{teffI}.
Therefore, taking into account this interface effective temperature picture, we would expect that in this low temperature limit
for $p\neq 0$ the shape of the droplet should not depend on system's temperature. Instead we would expect this shape to converge towards 
a structure similar to that observed in the equilibrium system for a temperature approximately equal to the interface effective 
temperature. This is so because the shape of the droplet is exclusively defined by the interfacial properties. In fact,
such independence of the shape of the droplet on system's temperature is what we exactly observe for low temperatures, see 
Fig. \ref{formas}.b.
\begin{figure}
\centerline{
\psfig{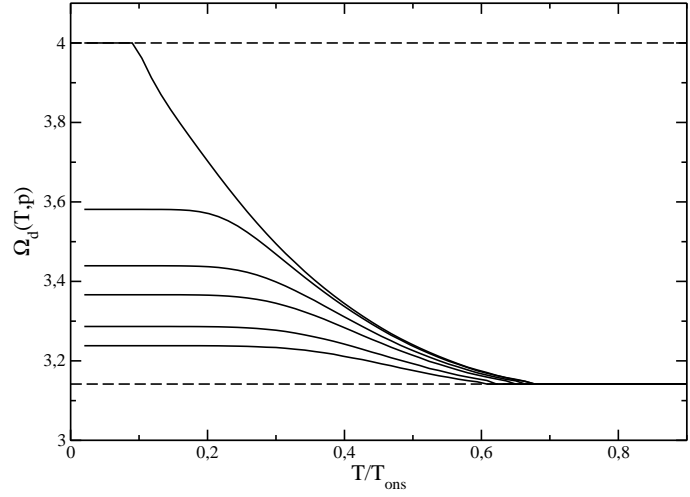}}
\caption[Form factor of a nonequilibrium spin droplet.]
{\small Form factor $\Omega_d(T,p)$ as a function of temperature for different values of $p$. In particular, from top
to bottom, $p=0$, $0.001$, $0.005$, $0.01$, $0.02$ and $0.03$. The upper dotted line signals $\Omega_d=4$, while the lower
dotted line signals $\Omega_d=\pi$.}
\label{fforma}
\end{figure}

\subsection{Droplet Form Factor}
\label{capSOS_apForma_subOmega}

Our last objective in this section consists in calculating the form factor, $\Omega_d(T,p)$, which relates the droplet radius,
${\cal R}$, with its volume, $V=\Omega_d(T,p) {\cal R}^d$. In order to calculate $\Omega_d(T,p)$ we need to precisely define the
droplet radius, ${\cal R}$, which is a typical length scale which characterizes the droplet size. In this case we choose the droplet 
radius ${\cal R}$ to be the radial function $R(\theta)$ for $\theta=0$, i.e. ${\cal R}\equiv R(\theta=0)$. In this way in the high
temperature limit, where the droplet is circular, the droplet radius ${\cal R}$ will coincide with the circle radius, while in
the low temperature limit, where the droplet is a square (for $p=0$), the radius ${\cal R}$ will be $\ell/2$, where $\ell$ is 
the square side. The droplet volume is defined by the following radial integral for a bidimensional system,
\begin{equation}
V=\Omega_d(T,p){\cal R}^d = 8\int_0^{\pi/4}\textrm{d}\theta \int_0^{R(\theta)} r \textrm{d}r
\label{vol1}
\end{equation}
so the form factor is calculated from,
\begin{equation}
\Omega_d(T,p) = 4\int_0^{\pi/4} \textrm{d}\theta \left[\frac{R(\theta)}{R(0)}\right]^2
\label{factorforma}
\end{equation}

Numerically solving the previous integral using the analytically extended radial function we obtain the results plotted in Fig. 
\ref{fforma}. In this figure we show the form factor $\Omega_d(T,p)$ as a function of temperature for different values of
$p$. For $p=0$ the form factor smoothly evolves from $\Omega_d=4$ for low temperatures to $\Omega_d=\pi$ for
high temperatures. The value $\Omega_d=4$ is typical of square droplets, where $V=4\times (\ell/2)^2=\ell^2$. On the other hand, 
the value $\Omega_d=\pi$ is typical of circular droplets, where $V=\pi{\cal R}^2$. As we see in Fig. \ref{fforma}, for $p\neq 0$ 
the form factor for low temperature converges towards a value smaller than $4$.
This fact confirms the above phenomenologic observation, deduced from the droplet shape for $p\neq 0$. The results derived in this 
section, and in particular those related to the form factor $\Omega_d(T,p)$, will be be very useful in the next chapter, where we will
formulate a nucleation theory to explain the dynamics of the metastable-stable transition in the ferromagnetic system subject to
nonequilibrium conditions.

\section{Conclusion}
\label{capSOS_apConc}

In this chapter we have studied both theoretically and computationally the microscopic and macroscopic properties of an interface
in the nonequilibrium model. In order to do so we have generalized the SOS approximation first introduced by Burton,
Cabrera and Frank\cite{BCF} for a discrete interface in an equilibrium spin system.

This theoretical approximation for the equilibrium system describes the structure of the interface from a microscopic 
point of view  using three basic hypothesis. On one hand, it neglects the presence of overhangs in the interface. 
Moreover, the approximation does not take into account the possible interactions between the interface and
bulk fluctuations. Finally, it assumes that the different step variables defining the interface are uncorrelated.
In spite of these three hypothesis, the SOS approximation for the equilibrium system correctly predicts the microscopic and macroscopic 
interfacial structure. In particular, it predicts a surface tension for the equilibrium system which coincides with the known exact
result obtained for the two dimensional Ising model for an angle $\phi=0$, and it is a very good approximation for nonzero angles.

The generalized SOS approximation for $p\neq 0$ is based on the concept of {\it effective 
temperature}. That is, the effect induced by the nonequilibrium perturbation $p$ on the system can be re-interpreted in terms of an 
effective temperature. We calculate that the higher is the order to which a spin is subject (i.e. the larger is the number
of neighboring spins pointing in the same direction that the given spin), the higher is the effective temperature this spin suffers for 
$p\neq 0$. The interface is composed by spins belonging to different classes (i.e. spins subject to different degrees of order),
which, consistently, suffer different effective temperatures. 
The central magnitude in the equilibrium SOS approximation is the probabilistic weight or Boltzmann factor associated to 
an interfacial broken bond, $X=\textrm{e}^{-2J\beta}$. 
In order to build the SOS approximation for $p\neq 0$ we generalize this statistical weight, denoted now as $X_p$, 
taking into account the effective temperature of the different spin classes. To do that we perform an average over the statistical 
weights associated to the different spin classes (each class having its own well defined effective temperature) using in this average 
the population densities for each interfacial spin class. We have calculated
these population densities using two-body probability functions.

We obtain from this generalization predictions for the microscopic structure of the 
interface . These predictions are codified in the probability function $p_p(\delta)$, which is the probability of finding a step of
size $\delta$ in the nonequilibrium interface. The comparison between the theoretical prediction for $p_p(\delta)$ 
and Monte Carlo results for the real system is excellent. This agreement supports the validity of our approach for
$p\neq 0$.

On the other hand, we obtain the surface tension of the interface for $p\neq 0$ using the extended theory. This surface tension
exhibits an unexpected behavior for the nonequilibrium system as compared 
to the equilibrium one. In particular, we observe that, while the
equilibrium surface tension monotonously increases as we decrease temperature, for the nonequilibrium model the surface tension
$\sigma^{(p)}_{SOS}$ shows a maximum for certain temperature, 
converging linearly towards zero for lower temperature. In order to understand
this behavior it is necessary to turn again to the concept of effective temperature. Thus we can define a global effective
temperature for the interface. We observe for $p\neq 0$ that, while 
for high temperatures the interface effective temperature, $T_{eff}^I$, is 
totally coupled to the system's temperature, in the $T\rightarrow 0$ limit $T_{eff}^I$ completely decouples 
from $T$, converging towards a constant, nonzero value, which involves that surface tension converges towards
zero for at low temperature for $p\neq 0$. Unfortunately, we cannot check this result through direct simulations for the
surface tension. However, as we will describe in the next chapter, there are certain observables, as the critical droplet size
(to be defined later on), which are directly related to the surface tension. As we will see, the theoretical predictions 
derived for this observable, based on the results for the surface tension of the nonequilibrium interface, $p\neq 0$, nicely
reproduce the measurements obtained via Monte Carlo simulations. The unexpected behavior of the surface tension for $p\neq 0$ will be
crucial for the properties of the metastable-stable transition, to be discussed in next chapter.

Finally, we have obtained the shape of a spin droplet in the nonequilibrium system from the surface tension we have derived in the
generalized SOS approximation, using the Wulff construction\cite{Wulff}. On one hand, we observe that for the equilibrium model
the droplet shape varies continuously from a square at low temperatures to a circle at high temperatures. On the contrary,
for the nonequilibrium system we obtain a droplet shape that, although it is circular for high temperature, in the low temperature
limit it converges towards an intermediate shape which is not a square, but partially exhibits the anisotropy inherited from the
underlying lattice. We have numerically calculated the form factor $\Omega_d$ associated to the droplet, confirming that while
for $p=0$ the form factor varies from $\Omega_d=4$ to $\Omega_d=\pi$, for $p\neq 0$ the form factor converges towards
a constant value $\Omega_d\in(\pi,4)$ for low temperatures.

The results obtained in this chapter on the effect of the nonequilibrium conditions on the interfacial properties are not only
relevant for the study of metastability, but they are also important for the study of many systems where one of the fundamental
ingredients consists in a nonequilibrium interface.

\chapter{Nucleation Theory for the Study of Metastability}
\label{capNuc}

\section{Introduction}
\label{capNuc_apIntro}

In this chapter we want to extend the nucleation theory, already developed for equilibrium spin systems\cite{Rikvold,Gunton}, to our 
nonequilibrium system subject to a magnetic field. Nucleation theory explains in a detailed way the (highly inhomogeneous) 
processes which make a ferromagnetic system to evolve from a metastable state to the corresponding stable state. 
This theory is based on the concepts of stable phase spin droplet and free energy cost of such droplet. 
In this chapter we will present in a first step the nucleation theory, as formulated for equilibrium magnetic 
systems. Once we understand the foundations of this theoretical approximation, we will extend such theory to the nonequilibrium system.
This extension will provide us with a good approximation for the dynamics of the metastable-stable transition in this case. Finally we
will analyze the effects that nonequilibrium conditions involve on the properties of such transition.

\section{Nucleation Theory for Equilibrium Magnetic Systems}
\label{capNuc_apEq}

In what follows we describe, following in part reference \cite{Rikvold}, 
the foundations of nucleation theory as applied to our system when $p=0$, i.e. for
the equilibrium case.\cite{Rikvold,Gunton} The Hamiltonian of this system is that of eq. (\ref{hamilt}), and the implemented
dynamics is given by the Glauber rate, eq. (\ref{rate}), once we fix $p=0$. An initial state with all spins up subject to a 
weak negative magnetic field for a temperature below the critical temperature, $T_{ons}$, is a metastable state. As we have explained
in previous chapters, a system in a metastable state eventually evolves towards the truly stable one. This evolution proceeds through
the nucleation of one or several droplets of the stable phase (down spins) in the metastable bulk (up spins). We could think of
other different ways in order to evolve from the metastable phase to the stable one. For instance, we could hypothesize that the
metastable-stable transition happens due to the {\it coherent} rotation of all spins in the system\footnote{In fact, the Ne\'el-Brown
theory\cite{Neel}, which is aimed to explain the process of demagnetization in metastable magnetic particles, is based on 
the concept of coherent rotation of spins.}. However, except for some marginal, sharply-defined cases, the free energy cost of such
coherent rotation is prohibitive (see section \ref{capMedio_apDin}), so this mechanism is not observed in practice.
The same argument helps us to understand why the system nucleates droplets: these compact configurations minimize in some way the free
energy of the system for a fixed magnetization, so they are observed with high probability during the metastable state 
demagnetization process.

The process of nucleation and growth of a droplet is controlled by two different, competing terms.\cite{Rikvold} On one hand we have 
a bulk term, related to the droplet volume, which favours the droplet growth, because the droplet belongs to the stable phase,
favoured by the magnetic field. On the other hand there is a surface term, which impedes the droplet growth due to the free energy cost
of the interface between the stable and metastable phases (the droplet bulk and the rest of the system, respectively). Due to the
competition between these two terms, it is observed that small droplets, where the surface term dominates over the bulk term, 
tend to shrink, while droplets with size larger than certain threshold size tend to grow (in this case, the bulk term dominates).
The droplet size that separates both typical behaviors is called critical droplet size, ${\cal R}_c$. This observable, ${\cal R}_c$,
yields a typical length scale for the metastable-stable transition. However, there are other different, well-defined typical length 
scales in the system. For instance, we have also the lattice spacing $a$, the typical 
correlation lengths of both the stable and metastable phases, $\xi_s$ and $\xi_{ms}$ respectively, the radius ${\cal R}_0$ up to which
a droplet can grow before interacting with another droplet (we name this radius ${\cal R}_0$ the {\it mean droplet
separation}), and the system size, $L$. However, only three of these six typical length scales will be relevant to our problem.
The lattice spacing is fixed to unity in this study, $a=1$. On the other hand, the typical correlation lengths $\xi_s$ and 
$\xi_{ms}$ will always be much smaller than the remaining length scales, since we are interested in temperatures 
well below the critical one, so both correlation lengths will be irrelevant for the investigation. Hence all characteristic
processes related to metastability in the equilibrium system will be a consequence of the interplay among the three relevant length
scales for this problem, namely ${\cal R}_c$, ${\cal R}_0$ and $L$.

Let's assume that we have a stable phase droplet with radius $R$ in a system at temperature $T$ in equilibrium, $p=0$. 
The droplet volume is $V(R)=\Omega_d(T) R^d$, where $\Omega_d(T)$ is the form factor defined in section 
\ref{capSOS_apForma}. The free energy associated to this droplet can be written as\cite{Rikvold},
\begin{equation}
F(R)=d\Omega_dR^{d-1}\sigma_0 - \Omega_dR^d\Delta
\label{elibre}
\end{equation}
where $\Delta$ is the free energy density difference between the metastable and stable phases, and $d\Omega_dR^{d-1}$ is the 
surface associated to a droplet with volume $\Omega_dR^d$. The factor 
$\sigma_0$ is the surface tension (or interfacial free energy per unit
length) along a primitive lattice vector (i.e. for an angle $\phi=0$ between the interface and one of the reference axis). We exactly 
know the surface tension $\sigma(\phi)$ for zero magnetic field in equilibrium. The solution obtained from the 
Solid-On-Solid approximation for the surface tension reproduces the exactly known result for $\phi=0$.\cite{Ziainterf} In principle,
we could expect that $\sigma_0$ depends on the magnetic field. However, many different studies based on 
analytic series expansions\cite{serie}, transfer matrix calculations\cite{transfer} and Monte Carlo simulations\cite{MC} point out
that the surface tension $\sigma_0$ entering the droplet free energy cost is the equilibrium surface tension for {\it zero} magnetic 
field. Hence we will use from now on the notation $\sigma_0(T)$ to emphasize the independence of this term on the magnetic field.

In eq. (\ref{elibre}) we observe the competence between the surface term, which impedes the droplet growth, and the bulk term, which
favours it. Since the system attempts to minimize the free energy, in order to determine the critical droplet radius ${\cal R}_c$,
which separates the growth tendency from the shrinkage tendency we only have to find the maximum
of the droplet free energy with respect to droplet radius. In this way we obtain,\cite{Gunton}
\begin{equation}
{\cal R}_c(T) = \frac{(d-1)\sigma_0(T)}{\Delta} \approx \frac{(d-1)\sigma_0(T)}{2m_s(T)|h|}
\label{rcrit}
\end{equation}
where he have substituted $\Delta \approx 2m_s(T)|h|$, with $h$ the magnetic field and $m_s(T)$ the equilibrium spontaneous 
magnetization (defined positive).

In order to understand the previous approximation for the free energy density difference between the metastable and stable phases,
$\Delta$, we start from the definition of free energy, $F=-k_BT\ln {\cal Z}(T)$, where $k_B$ is the Boltzmann constant, which we fix
to unity already in Chapter \ref{capMotiv}, and where ${\cal Z}(T)$ is the canonical partition function,
\begin{equation}
{\cal Z}(T) = \sum_{\mathbf{s}} \textrm{e}^{-\beta {\cal H}(\mathbf{s})}
\label{particion}
\end{equation}
Here  ${\cal H}(\mathbf{s})$ is the energy associated to a spin configuration $\mathbf{s}$ in the equilibrium system, eq. 
(\ref{hamilt}). In the low temperature limit, the only states with relevant statistical weight (Boltzmann factor) are the stable
and metastable states, $\textrm{e}^{-\beta {\cal H}_s}$ and $\textrm{e}^{-\beta {\cal H}_{ms}}$. Hence, in this very low temperatures 
limit the free energy associated to a metastable state will be approximately the energy associated to such state, 
$F_{ms} (T<<) \approx {\cal H}_{ms}$, and equivalently for the stable state, $F_s (T<<) \approx {\cal H}_s$. In this limit the 
metastable state will be given by a configuration where practically all spins are up, while in the stable state almost all spins will 
be down. The system Hamiltonian was ${\cal H}(\mathbf{s}) = -J\sum_{\langle i,j\rangle}s_is_j - h\sum_is_i$. For $T\rightarrow 0$
the exchange term will be approximately $\sum_{\langle i,j\rangle}s_is_j \approx 2N$, where $2N$ is the number of neighbor spin
pairs  in the system. Hence we can write $N\Delta = F_{ms}-F_s \approx {\cal H}_{ms}-{\cal H}_s \approx Nh[m_s(T,h)-m_{ms}(T,h)]$, 
where $m_s(T,h)$ and $m_{ms}(T,h)$ are the stable and metastable magnetizations, respectively, $m=N^{-1}\sum_is_i$. For weak
magnetic fields we have $|m_s(T,h)| \approx |m_{ms}(T,h)| \approx m_s(T,h=0)$, so, if we define $m_s(T) \equiv m_s(T,h=0)$, we
finally arrive to $\Delta \approx 2m_s(T)|h|$. This approximation is expected to remain valid even for temperatures near the 
critical one, although we have derived it in the low temperature limit.\cite{serie}

The free energy cost associated to the critical droplet is,
\begin{equation}
F_c({\cal R}_c) = \Omega_d(T)\sigma_0(T) {\cal R}_c^{d-1} = \sigma_0(T) \frac{V_c}{{\cal R}_c}
\label{elibrecrit}
\end{equation}
where $V_c\equiv \Omega_d{\cal R}_c^d$ is the critical droplet volume. The nucleation rate $I(T,h)$ per unit time and volume,
which is the probability per unit time and volume of nucleating a critical droplet in the system, can be determined from 
$F_c({\cal R}_c)$ using the Arrhenius law\cite{Langer,Langer1},
\begin{equation}
I(T,h) = A(T) |h|^{b+c} \textrm{e}^{-\frac{F_c}{T}} \approx 
A(T) |h|^{b+c} \textrm{e}^{-\frac{\sigma_0V_c}{{\cal R}_cT}}
\label{nucrate}
\end{equation}
The function $A(T)$ is non-universal, the exponent $b$ is an universal exponent related to the Goldstone modes present on the droplet 
surface\cite{Langer2}, and the exponent $c$ yields the dependence of the {\it kinetic prefactor} on the magnetic field\cite{Langer1},
being the only part of $I(T,h)$ which can explicitely depend on the specific spin dynamics. In particular, for a bidimensional system 
there are many numerical evidences pointing out that $b=1$, as predicted via field theory\cite{Langer2}, while the value of $c$
changes between $c\approx 1$ for sequential dynamics and $c\approx 2$ for random dynamics.\cite{Rikvold}

The nucleation rate $I(T,h)$ yields the rate at which the critical droplet nucleates in the system. Using the information codified in 
eq. (\ref{nucrate}), and once we determine the mean droplet separation, ${\cal R}_0$, and its associated time scale, $t_0$, which is 
the time the droplet needs in order to radially grow a distance ${\cal R}_0$, we will be able to calculate the metastable state mean
lifetime using Avrami law\cite{Langer,Avrami}. In order to calculate ${\cal R}_0$ and $t_0$ we need to know the radial growth velocity 
for a stable phase droplet, $v_{\perp}$. This growth velocity can be determined using the Allen-Cahn approximation\cite{Langer}.
This approximation is based on a phenomenologic motion equation written using thermodynamic arguments, which linearly relates
the rate of change of the local order parameter, which in our case is the magnetization $m(\vec{r})$, to the local thermodynamic force,
\begin{equation}
\frac{\textrm{d}m(\vec{r})}{\textrm{d}t} = -\Gamma' \frac{\delta F}{\delta m(\vec{r})}
\label{ecmov}
\end{equation}
From this equation, assuming that the free energy functional $F$ has a shape similar to that of the asymmetrical Ginzburg-Landau
functional\cite{Newman}, and assuming that the droplet has a spherical shape, we find the following result for the radial growth 
velocity of a stable phase droplet in Allen-Cahn approximation\cite{Langer},
\begin{equation}
v_{\perp} = (d-1)\Gamma (\frac{1}{{\cal R}_c} - \frac{1}{R}) \underset{R\rightarrow \infty}{\longrightarrow}
\frac{(d-1)\Gamma}{{\cal R}_c} \equiv v_0
\label{velradial}
\end{equation}
where the constant $\Gamma$ depends on the particular implemented dynamics. We can approximate $v_{\perp} \approx v_0$ always that,
as in our case, ${\cal R}_0 \gg {\cal R}_c$, so from now on we take $v_0$ as the radial growth velocity.

It is obvious that ${\cal R}_0 = v_0t_0$, since $t_0$ is the time the droplet needs to radially grow a distance ${\cal R}_0$. On the
other hand, we also know that ${\cal R}_0^dt_0I=1$, since, by definition of both magnitudes ${\cal R}_0$ and $t_0$, the probability 
of finding a droplet nucleating a volume ${\cal R}_0^d$ in a time $t_0$ is equal to unity. From these two relations we arrive to,
\begin{eqnarray}
t_0(T,h) & = & (v_0^d I)^{-\frac{1}{d+1}} = B(T) |h|^{-\frac{b+c+d}{d+1}} \textrm{e}^{\frac{\Xi(T)}{|h|^{d-1}}} 
\label{t0} \\
{\cal R}_0(T,h) & = & v_0t_0 = C(T) |h|^{-\frac{b+c-1}{d+1}} \textrm{e}^{\frac{\Xi(T)}{|h|^{d-1}}} \label{R0}
\end{eqnarray}
where we have defined the function $\Xi(T)$ as,
\begin{equation}
\Xi(T) = \frac{1}{d+1}\Omega_d(T) \Big(\frac{d-1}{2m_s(T)}\Big)^{d-1} \frac{\sigma_0^d(T)}{T} 
\label{Xi}
\end{equation}
and where both amplitudes are written in the following way,
\begin{eqnarray}
B(T) & = & \Big(\frac{2\Gamma m_s(T)}{(d-1)\sigma_0(T)}\Big)^{\frac{-d}{d+1}} A(T)^{-\frac{1}{d+1}} \\
C(T) & = & \frac{2\Gamma m_s(T)}{(d-1)\sigma_0(T)} B(T)
\label{amplitudes}
\end{eqnarray}
In the above equations we have carefully specified that $\Xi(T)$, and the amplitudes $B(T)$ and $C(T)$, do depend exclusively on
temperature. We must notice that the amplitudes $B(T)$ and $C(T)$ inherit the non-universal character of $A(T)$.

We can calculate the metastable state lifetime using the results obtained up to now. In general, this lifetime is defined as the 
average first passage time (in Monte Carlo Steps per spin, MCSS) to certain stable phase volume fraction. The stable phase volume
fraction at time $t$, $\Phi_s(t)$, is defined as the number of spins in the stable phase divided by the total number of spins, $N$.
We hence can interpret $\Phi_s(t)$ as the probability, at time $t$, that a spin already belongs to the stable phase. In order to
calculate $\Phi_s(t)$ it is convenient, on the other hand, to ask the opposite question: for a fixed point $O$ in the system,
which is the probability $P$ that the point $O$ does not belong to the stable phase at time $t$ ?. In order to calculate such 
probability, we know that a droplet of the stable phase will reach the point $O$ before a time $t$ if the following conditions
are fulfilled:
\begin{itemize}
\item The droplet nucleates a distance $r'$ away the point $O$ such that $0 \leq r' \leq v_0t$.
\item The time $t'$ which signals the beginning of the nucleation event must fulfill $0 \leq t' \leq t-\frac{r'}{v_0}$.
\end{itemize}
Since the nucleation rate $I(T,h)$ is the probability of nucleating a droplet per unit time and volume, the probability $P$
that point $O$ has not been reached by the stable phase at time $t$ can be written as,
\begin{equation}
P = \prod_{0 \leq r' \leq v_0t \atop 0 \leq t' \leq t-\frac{r'}{v_0}} \big[1-I(T,h)\textrm{d}\vec{r}'\textrm{d}t'\big]
\label{Pini}
\end{equation}
We have that $\textrm{e}^{-I\textrm{d}\vec{r}'\textrm{d}t'}=1-I\textrm{d}\vec{r}'\textrm{d}t'$, and since the product is 
continuous, we can write in general, for a $d$-dimensional system,
\begin{equation}
P = \textrm{exp}\Big[-I(T,h)\int_0^{v_0t}\textrm{d}r'd\Omega_d r'^{d-1} 
\int_0^{t-\frac{r'}{v_0}}\textrm{d}t' \Big] = \textrm{e}^{-\frac{\Omega_d}{d+1}\big(\frac{t}{t_0}\big)^{d+1}}
\label{P}
\end{equation}
Hence, the probability that the point $O$ belongs to the stable phase at time $t$, i.e. the stable phase volume fraction at time $t$,
will be given by $\Phi_s(t)=1-P$. The last equality constitutes the Avrami's law.\cite{Avrami} From this equation we can obtain 
the time the system needs to nucleate for first time certain stable phase volume fraction $\Phi_s$,
\begin{equation}
\tau(\Phi_s) = t_0(T,h) \Big[\frac{d+1}{\Omega_d(T)}\ln\big(\frac{1}{1-\Phi_s}\big)\Big]^{\frac{1}{d+1}}
\label{vidamedia}
\end{equation}
This equation yields the mean lifetime of the metastable state as a function of the nucleated volume fraction of the stable phase
that we fix as threshold for its measure, for an equilibrium ferromagnetic system of {\it infinite} size. We must take into account 
that we have assumed that the positions of the (possibly overlapping) growing droplets are uncorrelated when constructing the 
probability $P$. This {\it ideal gas} hypothesis will be valid when the total volume fraction occupied by the droplets is small enough,
so we can neglect correlations among them. This is the case in our study, due to the weakness of the magnetic field. For strong
magnetic fields the picture of localized nucleating droplets is no longer valid in order to explain the exit from the metastable state.
The metastable-stable transition is then observed to proceed via long-wavelength unstable modes reminiscent of spinodal 
decomposition.\cite{Rikvold}

\markboth{Nucleation Theory for the Study of Metastability}{\ref{capNuc_apExt} Nonequilibrium Extension of Nucleation Theory}
\section{Extension of Nucleation Theory to Nonequilibrium Ferromagnetic Systems}
\label{capNuc_apExt}
\markboth{Nucleation Theory for the Study of Metastability}{\ref{capNuc_apExt} Nonequilibrium Extension of Nucleation Theory}

In the previous section we have presented nucleation theory, which explains the dynamics of the metastable-stable transition for
an equilibrium ferromagnetic system, based on the droplet picture. The central magnitude in this theory is the droplet free
energy, eq. (\ref{elibre}). From such observable we have calculated the critical droplet size, and using Arrhenius law, the
nucleation rate, from which, applying Avrami's law, we obtained the mean lifetime of the metastable state.

In order to extend this theory to the nonequilibrium system ($p\neq 0$),
we first have to write the {\it free energy cost} of a droplet in this case. However, here we face some fundamental problems,
which impede any first-principles definition for the free energy in the nonequilibrium case. 
First, from a microscopic point of view, we do not know
for $p\neq 0$ the stationary probability distribution $P(\mathbf{s})$ of finding the magnetic 
system in a steady configuration $\mathbf{s}$. This steady probability function is solution of the master equation,
eq. (\ref{mastereq2}), once applied the stationarity condition. Since detailed balance does not hold for $p\neq 0$,
see eq. (\ref{balance}), the distribution $P(\mathbf{s})$ will be different from the Boltzmann distribution. Moreover, the system is 
open, subject to a continuous energy flux due to the nonequilibrium perturbation. We do not know how to precisely define
the energy of this system. We also do not know how to construct any statistical ensemble in the nonequilibrium case, and
we also do not know how to connect the microscopic properties of the system, which should be captured by those nonequilibrium 
statistical ensembles, with the system macroscopic behavior. Finally, there is not even an unique macroscopic theory, 
equivalent to equilibrium Thermodynamics, to connect with taking as a starting point the microscopic equations. 
This inexistent macroscopic theory ought to
satisfactorily describe the macroscopic phenomena that appear in nonequilibrium systems\footnote{Irreversible 
Thermodynamics\cite{deGroot} is an extension of usual Thermodynamics, based on conservation and balance equations, the maximum
entropy production postulate, and a series of phenomenological laws which postulate the proportionality among the fluxes and the
thermodynamic forces (for more references, see Chapter \ref{capFou} in this thesis). This macroscopic theory describes in a partial 
and approximate way some nonequilibrium situations, although it cannot be considered a complete macroscopic theory for nonequilibrium
systems.}. All these facts force us to propose a phenomenologic approximation, not based on first principles, to the problem 
of the nonequilibrium {\it free energy} associated to a stable phase droplet in the system with $p\neq 0$.

\subsection{ Nonequilibrium Potential and Critical Droplet Size}
\label{capNuc_apExt_subPot}
\begin{figure}
\centerline{
\psfig{file=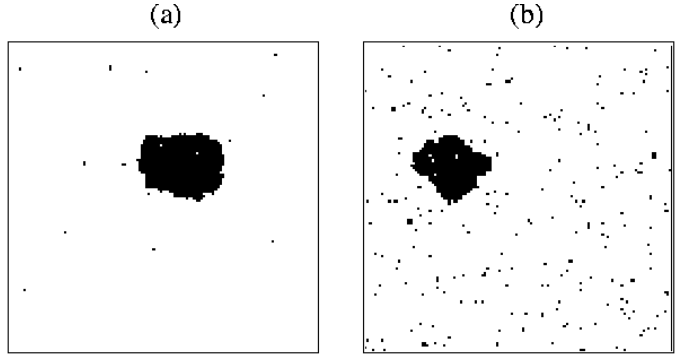,width=9cm}}
\caption[Snapshots of the metastable-stable transition.]
{\small (a) Snapshot of the metastable-stable transition for a system with $T=0.5T_{ons}$, $h=-0.1$, $p=0$ (equilibrium system)
and $L=128$. In black, stable phase spins (down spins), while metastable phase spins (up spins) are in white. (b) The same than in (a),
for the same magnetization, but for the nonequilibrium system, with $p=0.01$.}
\label{snapsgotas}
\end{figure}

Fig. \ref{snapsgotas}.a shows a typical escape configuration from the metastable state in an equilibrium system, with $T=0.5T_{ons}$, 
$h=-0.1$, $p=0$ and $L=128$, for certain fixed magnetization, while Fig. \ref{snapsgotas}.b shows an escape configuration for an
identical system, for the same magnetization, but with $p=0.01$, i.e. under nonequilibrium conditions. Comparing both figures we
realize that, in spite of the obvious differences due to the presence of the nonequilibrium perturbation, which introduces small
fluctuations in the bulk, the picture based on a stable phase droplet which nucleates in the metastable bulk is still valid.
Furthermore, we think that the shape and the growth of such droplet in the 
nonequilibrium system are again controlled by the competition between the
droplet bulk, which favours the droplet growth, and the droplet surface, which hinders such growth. Therefore we observe that, although
there are many fundamental problems in order to define from first principles a nonequilibrium potential which controls the exit
from the metastable state for $p\neq 0$, it is possible to establish from a phenomenological level a formal equivalence between the
equilibrium and nonequilibrium cases.

Taking into account this observation, we assume the existence of a nonequilibrium potential, ${\cal F}$, which 
controls the metastable-stable transition for the nonequilibrium system. We also assume that there are two different terms 
competing in ${\cal F}$: a surface term, which hinders the droplet growth process, and a bulk (or volume) term, which favours such 
growth. Moreover, we assume that this nonequilibrium potential for a stable phase droplet is {\it formally identical} to the free
energy of a droplet in the equilibrium case, eq. (\ref{elibre}), so for a droplet of size $R$,
\begin{equation}
{\cal F}(R) = d\Omega_d(T,p)R^{d-1}\sigma_0(T,p) - 2m_s(T,p)|h|\Omega_d(T,p)R^d
\label{elibrep}
\end{equation}
where now $\Omega_d(T,p)$ is the form factor associated to a droplet of the stable phase at temperature $T$ and nonequilibrium 
perturbation $p$, see Fig. \ref{fforma}, $\sigma_0(T,p)$ is the surface tension along one of the primitive lattice vectors for zero
magnetic field and parameters $T$ and $p$, and $m_s(T,p)$ is the spontaneous magnetization for the nonequilibrium system (defined 
positive) at temperature $T$ and perturbation $p$. This phenomenologic hypothesis, which is not justified from a formal point of
view, will be checked {\it a posteriori}, comparing the theoretical predictions derived from it with the results obtained from Monte 
Carlo simulations of the original model.

Although we do not have exact solutions for the observables $\Omega_d(T,p)$, $\sigma_0(T,p)$ and $m_s(T,p)$, we have developed in
previous chapters good approximations for all of them. Thus, we take the spontaneous magnetization
$m_s(T,p)$ as the steady solution for zero magnetic field of the set of eqs. (\ref{xzpair1})-(\ref{xzpair2}) which we obtained
in Pair Approximation, see section \ref{capMedio_apForm} and Fig. \ref{magh0}. We approximate
$\sigma_0(T,p)$ using the expression we derived for the surface tension in the generalized SOS approximation 
(section \ref{capSOS_apGen}), $\sigma_{SOS}^{(p)}(\phi;T,p)$ for an angle $\phi=0$, which we denote as 
$\sigma_{SOS}^{(p)}(T)\equiv\sigma_{SOS}^{(p)}(\phi=0;T,p)$, see eq. (\ref{tensionsuperfp}). The form factor $\Omega_d(T,p)$ has been
calculated in section \ref{capSOS_apForma} using the generalized SOS result for the surface tension, see Fig. \ref{fforma}.
\begin{figure}[t!]
\centerline{
\psfig{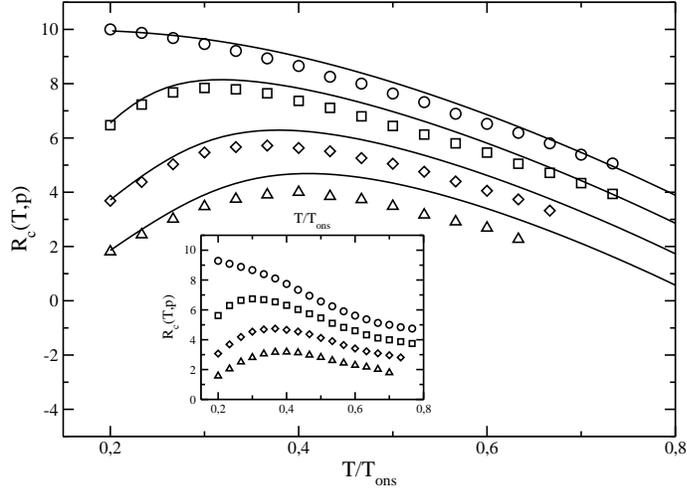}}
\caption[Critical droplet size as a function of temperature.]
{\small Critical droplet size, ${\cal R}_c$, as a function of temperature for different values of $p$, for a system size 
$L=53$, with periodic boundary conditions and subject to a magnetic field $h=-0.1$. In particular, from top to bottom, 
$p=0$($\bigcirc$), $0.001$($\Box$), $0.005$($\Diamond$) and $0.01$($\triangle$). The points are Monte Carlo results, measuring
${\cal R}_c(T,p)$ using the method based on the growth probability of an initial square droplet, after averaging over 
$N_{exp}=1000$ experiments. Lines correspond to the theoretical prediction, eq. (\ref{rcritp}). 
For the shake of clarity, results for the $n$-th value of $p$, $n=1,\ldots,4$ (using the above order) have been shifted 
$(1-n)$ units in the $\hat{y}$ axis. The inset shows Monte Carlo results for ${\cal R}_c$ as
those plotted in the main graph, but here we measure ${\cal R}_c$ using the stable phase growth and shrinkage rates. Data here have 
been also shifted. In all cases error bars are smaller than the symbol sizes.}
\label{gotacrit}
\end{figure}

Using this information, and taking into account the form of the nonequilibrium potential for a stable phase droplet, the
critical droplet size for $p\neq 0$ will be,
\begin{equation}
{\cal R}_c(T,p)= \frac{(d-1)\sigma_{SOS}^{(p)}(T)}{2m_s(T,p)|h|}
\label{rcritp}
\end{equation}
where the different magnitudes have been defined in the previous paragraph. Fig. \ref{gotacrit} shows the theoretical prediction for the
critical droplet size ${\cal R}_c(T,p)$ as a function of temperature for different values of $p$ and $h=-0.1$. 
This figure also shows the results of different Monte Carlo simulations for a system with size $L=53$,
with periodic boundary conditions and the same magnetic field. As we see there, the agreement between theory
and Monte Carlo results is rather good. From the qualitative point of view, we observe that in the equilibrium system 
($p=0$) the critical droplet size grows monotonously as temperature decreases, converging towards a
value ${\cal R}_c(T\rightarrow 0) \approx 1/|h|=10$. On the other hand, for any $p\neq 0$ the critical droplet size depicts a maximum
as a function of temperature, decreasing for lower temperatures. This non-monotonous behavior of ${\cal R}_c(T,p\neq 0)$ is clearly
inherited from the non-monotonous behavior of the surface tension which we derived in the generalized SOS approximation for the
nonequilibrium interface. Moreover, as we said previously, the good agreement shown in Fig. \ref{gotacrit}
verifies in an indirect way that the surface tension we derived in the generalized SOS 
approximation constitutes a good approximation for the nonequilibrium surface tension.

On the other hand, from the quantitative point of view, we observe that the theoretical predictions are very good for low enough 
temperatures, although they slightly overestimate the value of the measured ${\cal R}_c$ for larger temperatures, for any value
of $p$. It is remarkable that this overestimation for ${\cal R}_c$ at high temperatures also appears for the equilibrium system,
where $\sigma_{SOS}^{(p=0)}(T)$ perfectly reproduces the exact known result for the surface tension, and where $m_s(T,p=0)$
yields values almost indistinguishable from the real ones. In this case the observed differences between theoretical predictions
and computational results for the critical droplet size at high temperatures can be traced back to the lack of precision of the
approximation $\Delta \approx 2m_s(T)|h|$ in this temperature regime. For $p\neq 0$ it is observed that the differences appearing 
between the measured critical droplet size and the one predicted by the nonequilibrium nucleation theory are similar to those
observed in the equilibrium system, so again the approximation for the free energy density difference between the metastable and 
stable phases, $\Delta$, seems to be at the origin of the discrepancy. In spite of these slight differences, we can affirm that
our theoretical approximation nicely reproduces the simulation results.

\begin{figure}
\centerline{
\psfig{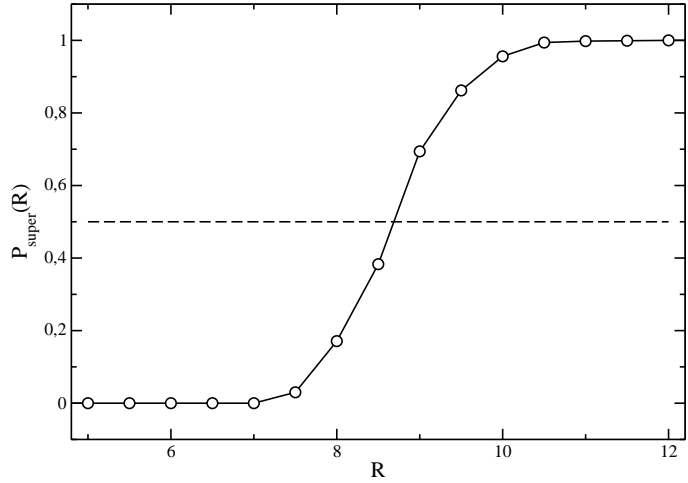}}
\caption[Probability of finding a supercritical droplet with radius $R$.]
{\small Probability that a initial (square) droplet with radius $R$ is supercritical, $P_{super}(R)$, as a function of $R$
for a system size $L=53$, at temperature $T=0.4T_{ons}$, $p=0$, $h=-0.1$, once we collect $N_{exp}=1000$ different experiments
for each value of $R$.}
\label{probgota}
\end{figure}

In order to obtain the critical droplet size from Monte Carlo simulations of the real system we have used two different methods which,
as we can observe in Fig. \ref{gotacrit}, yield equivalent results. In a first method, we simulate a system with size $L$,
with a total amount of $N=L^2$ spins, and subject to periodic boundary conditions. We initialize the system in a state with all spins
up (i.e. in the metastable phase), except for a {\it square} droplet of down spins (stable phase spins) with side $2R$ which we situate
on the lattice center. For this initial condition we let evolve the system under the usual Glauber dynamics, eq. (\ref{rate}), for fixed
temperature $T$, nonequilibrium perturbation $p$ and magnetic field $h<0$. The imposed initial condition is highly unstable.
If the initial state converges as time goes on to a state with magnetization $m\approx +1$ (metastable state), then the initial droplet,
with side $2R$, was subcritical (that is, for this droplet the surface term dominates over the bulk term, so it tends to shrink).
This means that the radius $R$ of the initial droplet, defined as half of the square droplet side, is smaller than the critical 
droplet radius ${\cal R}_c(T,p,h)$ for these parameters. On the other hand, if the initial droplet grows until the system reaches 
a state with magnetization near $m\approx -1$ (stable state), the radius $R$ of the initial square droplet was larger than the 
critical droplet radius, so the initial droplet was supercritical (now the bulk term dominates over the interfacial one).
Since the system is stochastic, the growth or shrinkage of a droplet depending on its size is not a deterministic process. Therefore
we can define a function $P_{super}(R)$ which yields the probability that a (square) stable phase droplet with radius $R$ is
supercritical. In our case, in order to evaluate such probability we perform $N_{exp}$ experiments as the above described, and we 
accumulate the number of times $n_{super}(R)$ that a droplet of size $R$ grows up to the stable state.  In this case we have that 
$P_{super}(R) = n_{super}(R)/N_{exp}$. Fig. \ref{probgota} shows the probability $P_{super}(R)$ for a system size $L=53$, with
periodic boundary conditions, at temperature $T=0.4T_{ons}$, nonequilibrium parameter $p=0$ and magnetic field $h=-0.1$, once we
collect a total amount of $N_{exp}=1000$ different experiments for each value of $R$. As expected, the function $P_{super}(R)$
abruptly changes from 0 to 1 in a narrow interval for the radius of the initial square droplet. Here we define the critical droplet 
radius ${\cal R}_c$ as the radius $R^*$ for which $P_{super}(R^*)=0.5$.\footnote{Due to the discontinuous character of variable $R$,
most of the times there is no sampled value $R^*$ such that $P_{super}(R^*)=0.5$. In these cases it is necessary to perform a
linear interpolation between the two values of the variable $R$ that flank the intersection point between the function $P_{super}(R)$
and the constant line $y=0.5$ (see Fig. \ref{probgota}).} Using this method we have obtained the results shown in the main plot 
in Fig. \ref{gotacrit}, after averaging over $N_{exp}=1000$ different experiments.

The second method we previously referred to in order to measure ${\cal R}_c(T,p,h)$ is based on the stable phase growth and shrinkage
rates, $g(m)$ and $s(m)$ respectively, defined in section \ref{capMedio_apEstat}, see eqs. (\ref{growing}) and (\ref{shrinking})
and complementary discussion. We said there that $g(m)$ was the number of spins in the system that change from the metastable phase 
to the stable one per unit time, when the system was in a state with magnetization $m$. Equivalently, $s(m)$ was the number of
spins in the system that change from the stable phase to the metastable one per unit time, when the system magnetization is $m$.
The rate of change associated to the order parameter was given by the difference $s(m)-g(m)$, see eq. (\ref{magrates}), so
stationary states was defined by the solutions of the equation $s(m)=g(m)$. We deduced that this equation has three different
solutions for a system showing metastability, $m_1$, $m_0$ y $m_{-1}$, where $m_1$ and $m_{-1}$ were magnetizations near $+1$ and $-1$,
respectively, and $m_0$ signaled a intermediate positive magnetization for $h<0$ (see Fig. \ref{gs}). While $m_1$ and $m_{-1}$ were
locally stable solutions which identified the metastable and stable states respectively, the solution $m_0$ was unstable against
small perturbations. Moreover, we deduced that if $m>m_0$ the system rapidly evolves towards the metastable state, while if $m<m_0$
the system evolves towards the stable state. In this sense the magnetization $m_0$ signals a critical magnetization separating 
the region where the stable phase tends to shrink ($m>m_0$) from the region where the stable phase tends to grow ($m<m_0$). This concept
is formally analogous to that of critical droplet. Furthermore, 
the magnetization $m_0$ signals the magnetization that the system exhibits 
when the critical droplet nucleates. As we will explain in the following, we are able to obtain the critical droplet size using this 
magnetization value, $m_0$, and the metastable state magnetization, $m_1$. We define $n_0^{-}=N(1-m_0)/2$ as the number of down spins
in the {\it critical} state characterized by $m_0$, where $N$ is the total number of spins in the system. Equivalently, we define 
$n_1^{-}=N(1-m_1)/2$ as the number of down spins in the metastable state. In the {\it critical} state we expect that 
$n_0^{-}=n_c^{-} + n_{bulk}^{-}$, where $n_c^{-}$ is the number of down spins belonging to the critical droplet, and $n_{bulk}^{-}$
is the number of down spins in the metastable phase bulk which has not been occupied by the critical droplet. We can assume that,
as a first approximation, the density of down spins in the metastable bulk when we are in the {\it critical} state is equal to
the density of down spins in the metastable state, which we define as $d_1^{-} = n_1^{-}/N$. In this case we have that 
$n_{bulk}^{-}=d_1^{-}(N-n_c^{-})$, so,
\begin{equation}
n_c^{-} = \frac{n_0^{-} - n_1^{-}}{1-\frac{n_1^{-}}{N}}
\label{nc}
\end{equation}
The variable $n_c^{-}$ yields the volume of the critical droplet. As a definition of the critical droplet size we choose,
\begin{equation}
{\cal R}_c \equiv \frac{\sqrt{n_c^{-}}}{2}
\label{Rcnc}
\end{equation}
Thus measuring the magnetizations $m_0$ and $m_1$ for which the curves $g(m)$ and $s(m)$ intersect one each other, we can obtain
a measure of the critical droplet size. In order to do so we perform a series of $N_{exp}$ experiments studying the demagnetization 
process from the metastable to the stable state. We measure in each experiment the rates $g(m)$ and $s(m)$. As we have already done
in other simulations, we initialize the system in a state with all spins up under the action of a negative magnetic field. Such state
is metastable, and the system eventually evolves up to the stable state. We measure the magnetizations $m_0$ and $m_1$ on each experiment,
and we calculate from them the critical droplet size ${\cal R}_c$ defined in eq. (\ref{Rcnc}). Finally, we average over the different
experiments in order to obtain good statistics. In this way we have obtained the results shown in the inset of Fig. \ref{gotacrit}
for the critical droplet radius as a function of temperature for different values of $p$. These data have been obtained for a system 
size $L=53$ with a magnetic field $h=-0.1$, and averaging over $N_{exp}=1000$ different experiments.

If we compare the computational results obtained using both methods for the critical droplet size, we observe that these results
are completely equivalent, up to a small amplitude factor, which varies slightly with the temperature and the nonequilibrium 
parameter $p$. In particular, the results for ${\cal R}_c$ obtained from the probability $P_{super}(R)$ of finding a supercritical 
square droplet with side $2R$ are slightly larger than the results obtained from the stable phase growth and shrinkage rates.
In order to collapse both measures we must multiply the results obtained via the first method by a scaling factor of order $0.9$. 
On the other hand, these tiny global differences observed between both measures were expected due to the different influence of the
form factor on each computational scheme. However, in spite of these small discrepancies, the results obtained from both measures 
depict the same behavior: ${\cal R}_c$ is monotonous in equilibrium, while ${\cal R}_c$ shows a maximum as a function of temperature 
for the nonequilibrium system. Finally, we must notice before going on that the two computational methods presented here in order to
measure the critical droplet size will be valid always that the metastable-stable transition proceeds through the nucleation of
a {\it single} critical droplet, and not for several critical droplets, as is observed in certain parameter space regions. However,
for the parameters we study, the system usually decays through the nucleation of a single critical droplet (see next sections).

\subsection{Radial Growth Velocity for a Stable Phase Droplet}
\label{capNuc_apExt_subVel}

In the following we study the radial growth velocity for a stable phase droplet in the nonequilibrium system. In equilibrium we 
deduced this velocity via the Allen-Cahn approximation. This approximation was based on a phenomenologic equation 
postulating the proportionality between the rate of change of the order parameter and the local thermodynamic force, eq. (\ref{ecmov}).
This thermodynamic force is just the variation induced on the system free energy by a perturbation of the local magnetization,
$\delta F/\delta m(\vec{r})$. For the nonequilibrium case we cannot properly define a free energy functional. However, if our
postulate about the existence of a nonequilibrium potential for $p\neq 0$ similar to the free energy in the equilibrium system is
correct, we can define the {\it thermodynamic} force the order parameter suffers in the same way, but now using the nonequilibrium 
potential ${\cal F}$ instead of the equilibrium free energy $F$. Furthermore, due to the similarities observed in the 
metastable-stable transition for both the equilibrium and nonequilibrium systems, we assume that the shape of the nonequilibrium 
potential matches in some sense that of the Ginzburg-Landau functional. Hence the same formal result we derived in equilibrium 
is assumed to be valid for $p\neq 0$,
\begin{equation}
v_0(T,p,h) = \frac{(d-1)\Gamma}{{\cal R}_c(T,p,h)}
\label{velradialp}
\end{equation}
where now ${\cal R}_c(T,p,h)$ is the critical droplet radius for the nonequilibrium system, see eq. (\ref{rcritp}).
\begin{figure}[t]
\centerline{
\psfig{file=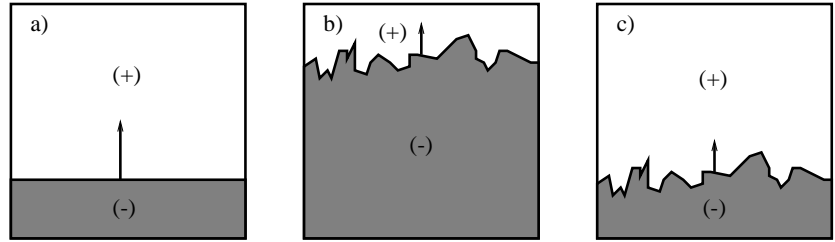,width=11cm}}
\caption[Schematic plot of the semi-infinite system evolution and shift.]
{\small In $a)$ we show the initial configuration for the semi--infinite system defined in the text. 
In $b)$ we show a possible 
schematic configuration before we shift the lattice. The fraction of up spins in this configuration has 
decreased to a value close to $\alpha$. Finally, in $c)$ it is shown the final system configuration after 
the shift. We have added some new up spins to the top of the system, and some down spins from the bottom 
of the system have been removed.}
\label{sketch-seminf}
\end{figure}

In order to check this prediction about the interface velocity we perform Monte Carlo simulations. We then build a system with
effective size $L\times \infty$. To do that in practice, we define the system in a square $L\times L$ lattice. We impose periodic 
boundary conditions in the horizontal direction ($\hat{x}$ axis), and open boundary conditions in the vertical one ($\hat{y}$ axis),
forming in this way a cylinder with height $L$. All spins in the upper row are fixed in the up state, while all spins in the lower row
are fixed in the down state. The initial configuration consists of a spin stripe of height $\alpha L$ where all spins are down, which 
fills the first $\alpha L$ rows, and a complementary stripe where all spins are up, which fills the remaining $(1-\alpha)L$ rows.
We choose in this case $\alpha=0.15$. The interface moves upwards for a negative magnetic field (i.e. the stable phase -down spins-
grows at the expense of the metastable one -up spins). In order to emulate an infinite system in the interface movement direction,
we shift the {\it observation window} as the interface advances, always keeping it inside the system. In practice we do that 
generating a new region with up spins in the upper part of the system as the interface advances, and eliminating an equivalent 
down spins region in the lower part of the system. In fact the shift is performed each time the fraction of up spins in the system 
is smaller than $\alpha$. The magnitude of the shift is such that once performed we approximately recover the fraction of up and down 
spins we had in the initial configuration, i.e. $15\%$ of up spins and $85\%$ of down spins. The choice $\alpha=0.15$ allows 
the added up spins to relax towards the typical state of the metastable bulk for the studied parameters $T$, $p$ and $h$ before the 
interface reaches them (see figure \ref{sketch-seminf}). 
In order to measure the interface velocity we calculate the system magnetization $m(t)$ as a function of time,
evaluating $m(t)$ without taking into account the variations in magnetization due to the added and removed spins in the upper and
lower parts of the system respectively. The slope of $m(t)$ yields the interface velocity, $v_0(T,p,h)$.
\begin{figure}[t]
\centerline{
\psfig{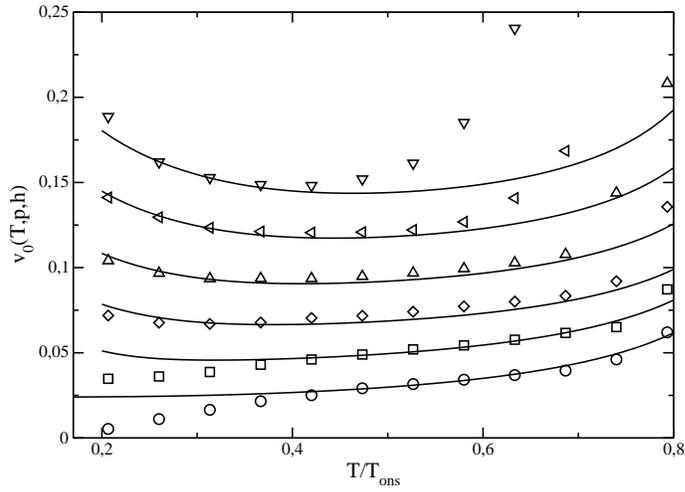}}
\caption[Monte Carlo results for the interface velocity.]
{\small Monte Carlo results for the interface velocity as a function of temperature for different values of $p$, and $h=-0.1$.
In particular, from bottom to top, $p=0$, $0.001$, $0.005$, $0.01$, $0.02$ and $0.03$. Continuous lines are our theoretical
prediction. The free parameter $\Gamma$ (see main text) is fixed in each case to yield the best fit, always obtaining 
$\Gamma \in [\frac{1}{5},\frac{1}{4}]$. For the shake of clarity, if the integer $n=1,\ldots,6$ characterizes the curve position (its
order from bottom to top), we have shifted the results for each $p$ an amount $(n-1)\times 0.02$ in the vertical axis. Error bars 
are smaller than symbol sizes.}
\label{veloc}
\end{figure}

We measure the interface velocity using this method for a system size $L=128$ as a function of temperature for different values of 
$p$ with fixed $h=-0.1$. Fig. \ref{veloc} shows the results of these measures. This figure also includes the predictions based on
eq. (\ref{velradialp}), once we fix the only free parameter, $\Gamma$, to yield the best fit  in each case (we obtain always 
$\Gamma\in[\frac{1}{5},\frac{1}{4}]$). The theory reproduces rather well the Monte Carlo results for most of the
values for $T$ and $p$. It is remarkable that for $p=0$ (and very small values of $p$, say $p=0.001$) the interface velocity 
prediction based on eq. (\ref{velradialp}) fails in the low temperature limit. This discrepancy is due to the underlying lattice 
anisotropy which induces effects that dominate the behavior of the system in the low temperature limit (remember the droplet shape in 
the low temperature limit for $p=0$, Fig. \ref{formas}.a). A continuous and macroscopic theory as the Allen-Cahn approximation,
which is isotropic and does not take into account the lattice details, fails when describing the interface velocity at low temperatures.
However this discrepancy is {\it healed} as the value of $p$ is increased, since the interface for $p\neq 0$ {\it feels} an effective
temperature $T_{eff}^I>T$, eq. (\ref{betaeffI}), due to the nonequilibrium perturbation, which is nonzero even in the low temperature 
limit. Hence the interface almost does not realize the presence of the underlying lattice for large enough $p\neq 0$, and so
the continuous approximation remains valid.

On the other hand we also observe in Fig. \ref{veloc} that the differences between the theoretical prediction for the interface velocity 
and the computational results grow as we approach the critical temperature $T_c(p)$. These differences are due to the interaction 
between the interface and bulk fluctuations, whose relative importance grows as we approach the critical point. Our theory does not 
take into account such possible interaction. We can remove these differences in the high temperature limit eliminating the bulk
dynamics, as we did when measuring the interface microscopic structure in the nonequilibrium system, see section \ref{capSOS_apGen} 
and Fig. \ref{prob-escalon}. In this way we recover the agreement between the theory and the measured velocities for high
temperatures.

It is remarkable that the interface velocity shows a minimum as a function of temperature for an intermediate temperature in the
nonequilibrium system ($p\neq 0$), growing the interface velocity if we further decrease temperature, see Fig. \ref{veloc}.
This surprising feature of the nonequilibrium system, measured in Monte Carlo simulations and compatible with our theory, 
points out that when there is non-thermal noise ($p$), a decrease of the thermal noise ($T$) when the system is at intermediate
temperatures favours the interface advance. The origin of this property underlies again on the non-monotonous behavior of surface
tension for $p\neq 0$, see Fig. \ref{tensionsup}.

\subsection{Mean Lifetime for the Metastable State}
\label{capNuc_apExt_subVida}

The hypothesis of existence of a nonequilibrium potential ${\cal F}$, formally identical
to the equilibrium free energy, which controls the dynamics of the metastable-stable transition for $p\neq 0$, has provided us
with correct theoretical predictions for both the critical droplet size and the droplet radial growth velocity. Once we have 
derived these magnitudes we are able to study the metastable state mean lifetime for $p\neq 0$. In this case, the nucleation
rate $I(T,p,h)$ can be written using Arrhenius law\cite{Langer,Langer1} as,
\begin{equation}
I(T,p,h) = A(T,p) |h|^{b+c} \textrm{e}^{-\frac{{\cal F}_c(T,p,h)}{T}} 
\label{nucratep}
\end{equation}
where ${\cal F}_c(T,p,h)={\cal F}(R={\cal R}_c)$, eq. (\ref{elibrep}). For the nonequilibrium system the non-universal amplitude factor
$A(T,p)$ will depend not only on temperature, but generically also on $p$. We can calculate for $p\neq 0$, in an equivalent way as we 
did in equilibrium, the mean droplet separation ${\cal R}_0(T,p,h)$ and the time $t_0(T,p,h)$ a droplet needs to grow radially a 
distance ${\cal R}_0$ once we know the nucleation rate, obtaining,
\begin{eqnarray}
t_0(T,p,h) & = & B(T,p) |h|^{-\frac{b+c+d}{d+1}} \textrm{e}^{\frac{\Xi(T,p)}{|h|^{d-1}}} 
\label{t0p} \\
{\cal R}_0(T,p,h) & = & C(T,p) |h|^{-\frac{b+c-1}{d+1}} \textrm{e}^{\frac{\Xi(T,p)}{|h|^{d-1}}} \label{R0p}
\end{eqnarray}
where now we define the function $\Xi(T,p)$ as,
\begin{equation}
\Xi(T,p) = \frac{1}{d+1}\Omega_d(T,p) \Big(\frac{d-1}{2m_s(T,p)}\Big)^{d-1} \frac{\big[\sigma_{SOS}^{(p)}(T)\big]^d}{T} 
\label{Xip}
\end{equation}
and where both amplitudes for the nonequilibrium system are written as,
\begin{eqnarray}
B(T,p) & = & \Big(\frac{2\Gamma m_s(T,p)}{(d-1)\sigma_{SOS}^{(p)}(T)}\Big)^{\frac{-d}{d+1}} A(T,p)^{-\frac{1}{d+1}} \\
C(T,p) & = & \frac{2\Gamma m_s(T,p)}{(d-1)\sigma_{SOS}^{(p)}(T)} B(T,p)
\label{amplitudesp}
\end{eqnarray}
These equations are structurally identical to those formulated for the equilibrium system, eqs. (\ref{t0})-(\ref{amplitudes}). The
difference between these equations for $p\neq 0$ and their counterparts in equilibrium rests on the nonequilibrium generalized 
observables $\Omega_d(T,p)$, $m_s(T,p)$ and $\sigma_{SOS}^{(p)}(T)$. Using Avrami's law\cite{Langer,Avrami} we can write the mean
lifetime of the metastable state as a function of the stable phase volume fraction for $p\neq 0$,
\begin{equation}
\tau(\Phi_s;T,p,h) = t_0(T,p,h) \Big[\frac{d+1}{\Omega_d(T,p)}\ln\big(\frac{1}{1-\Phi_s}\big)\Big]^{\frac{1}{d+1}}
\label{vidamediap}
\end{equation}
This equation yields the mean lifetime of a metastable state for a system with $p\neq 0$ and {\it infinite} size. However, the systems
we study are always finite, with a typical size $L$. In this case the system will evolve from the metastable to the stable state
nucleating one or several stable phase droplets (or via some other marginal mechanisms to be described later on), depending on
the value of $L$ and its relation to the other relevant length scales in the system, namely ${\cal R}_c(T,p,h)$ and 
${\cal R}_0(T,p,h)$.

In general, the mean droplet separation ${\cal R}_0$ for intermediate values of the parameters (as the ones we are interested in) 
is much larger than the critical droplet size, ${\cal R}_c \ll {\cal R}_0$. In this case the picture based on nucleation of droplets
is valid, because the droplet volume fraction is small enough so we can neglect correlations between droplets. For very strong magnetic 
fields (not studied in this thesis) the picture based on localized droplets is no longer valid, and now the system decays from the 
metastable state via long-wavelength Goldstone modes reminiscent of spinodal decomposition. The region of parameter space where
the localized droplets picture is no longer valid is known as the Strong Field Region (SF)\cite{Rikvold}, and the magnetic field
signaling the transition between the intermediate field region, where the metastable-stable transition proceed through the nucleation
of one or several critical droplets, and the Strong Field Region is known as {\it Mean Field Spinodal Point}, 
$|h_{MFSP}|$.\cite{Rikvold,Tomita}

We now assume a large system size, such that,
\begin{equation} 
L \gg {\cal R}_0 \gg {\cal R}_c
\label{condMD}
\end{equation}
In this case the system evolves from the metastable state to the stable one through the nucleation of many stable phase critical 
droplets. This region is known as {\it Multidroplet Region} (MD)\cite{Rikvold}. In order to calculate the mean lifetime in this case,
we perform a partition of the system in $(L/{\cal R}_0)^d$ cells of volume ${\cal R}_0^d$. Each cell decays from the
metastable state to the stable one in an independent Poisson process of rate ${\cal R}_0 I = t_0^{-1}$. This rate is the probability
per unit time that a cell with volume ${\cal R}_0^d$ will decay to the stable phase. The stable phase volume fraction is in this
case self-averaging, and hence $\Phi_s^{(i)}$ inside any cell $i$ coincides approximately with the total stable phase
volume fraction of the system at any time. Therefore the mean lifetime of the metastable state will be in this case approximately
equal to the result obtained for the infinite system,
\begin{equation}
\tau_{MD}(\Phi_s;T,p,h) = t_0(T,p,h) \Big[\frac{d+1}{\Omega_d(T,p)}\ln\big(\frac{1}{1-\Phi_s}\big)\Big]^{\frac{1}{d+1}}
\label{taupMD}
\end{equation}
This lifetime does no depend on system size $L$. Additionally, it can be shown that the relative standard deviation associated
to the lifetime in region MD, $r_{\tau}$, is very small (this is also observed in SF region), so both regions (MD and SF)
are termed {\it Deterministic Region} in literature\cite{Rikvold}.

For smaller system sizes, such that
\begin{equation}
{\cal R}_0 \gg L \gg {\cal R}_c
\label{condSD}
\end{equation}
the random nucleation of a {\it single} stable phase critical droplet in a Poisson process of rate $L^dI$ is the rate-determining
step in the metastable-stable transition. Therefore in this case a single stable phase droplet will nucleate and it will 
rapidly grow to cover the whole system much before any other critical droplet appears. The time the system needs to nucleate a critical
droplet is much larger than the time this critical droplet needs to grow and cover the whole system. Hence we can approximate in
this case the metastable state mean lifetime with $\tau_{SD} = (L^dI)^{-1}$, i.e.,
\begin{equation}
\tau_{SD}(T,p,h) = \frac{|h|^{-(b+c)}}{A(T,p)L^d} \textrm{e}^{\frac{\Xi(T,p)}{|h|^{d-1}}}
\label{taupSD}
\end{equation}
The parameter space region where the exit from the metastable state proceeds through the nucleation of a single droplet is called 
{\it Single Droplet Region} (SD)\cite{Rikvold}. The mean lifetime in this region only exhibits a slight dependence on the threshold
$\Phi_s$, since the time the system needs to nucleate a critical droplet is much larger than the time the droplet needs to grow
and cover the whole system. On the other hand, in this case it can be shown that the relative standard deviation associated to
the lifetime, $r_{\tau}$, is of order unity, so we call this region {\it Stochastic Region}\cite{Rikvold}.

The magnetic field which signals the transition between the Single Droplet Region (SD) and the Multidroplet Region (MD) is known
as {\it Dynamic Spinodal Point}, $|h_{DSP}|$.\cite{Rikvold,Tomita} This Dynamic Spinodal Point can be estimated from the
condition ${\cal R}_0(T,p,h_{DSP}) \propto L$. Taking into account the expression derived for ${\cal R}_0$ in the nonequilibrium system,
eq. (\ref{R0p}), we have in the limit $h \rightarrow 0$ that the Dynamic Spinodal Point is given by,
\begin{equation}
|h_{DSP}| = \big[\frac{\Xi(T,p)}{\ln(L)}\big]^{\frac{1}{d-1}}
\label{hdsp}
\end{equation}
where the function $\Xi(T,p)$ is defined in eq. (\ref{Xip}). $|h_{DSP}|$ converges towards 
zero in the limit $L\rightarrow \infty$, which involves that the only relevant process in this large system size limit is that of
nucleation of multiple droplets. However, this convergence is logarithmic, i.e. very slow, so the exit process from the metastable 
state via the nucleation of a single droplet of the stable phase will be measurably even for macroscopic systems. On the other hand,
the Mean Field Spinodal Point, $|h_{MFSP}|$, which separates the MD and SF regions, can be roughly estimated from the condition 
$2{\cal R}_c \approx \xi_{ms}$, where $\xi_{ms}$ is the correlation length in the metastable phase.\cite{Rikvold} When this condition
holds, the correlation between droplets is relevant, so the description based on the droplet nucleation process is no longer valid.
For temperatures well below the critical one, the correlation length $\xi_{ms}$ is small, of the same order of magnitude that
the lattice spacing, so $\xi_{ms}\approx 1$. Using this observation in the above condition, we obtain the Mean Field Spinodal Point,
\begin{equation}
|h_{MFSP}|= \frac{(d-1)\sigma_{SOS}^{(p)}(T)}{m_s(T,p)}
\label{hmfsp}
\end{equation}
Finally, there is a last possibility, captured by the condition,
\begin{equation}
{\cal R}_0 \gg {\cal R}_c \gg L
\label{condCE}
\end{equation}
In this case the critical droplet size is larger than the system size, and the demagnetization process from the metastable state
proceeds through the {\it coherent} rotation of all spins in the system. This mechanism is relevant only for very small system sizes
and for very weak magnetic fields. The parameter space region where this mechanism is observed is known as the {\it Coexistence
Region} (CE), and the magnetic field separating CE and SD regions is known as {\it Thermodynamic Spinodal Point}, 
$|h_{THSP}|$.\cite{Rikvold}
\begin{figure}[t!]
\centerline{
\psfig{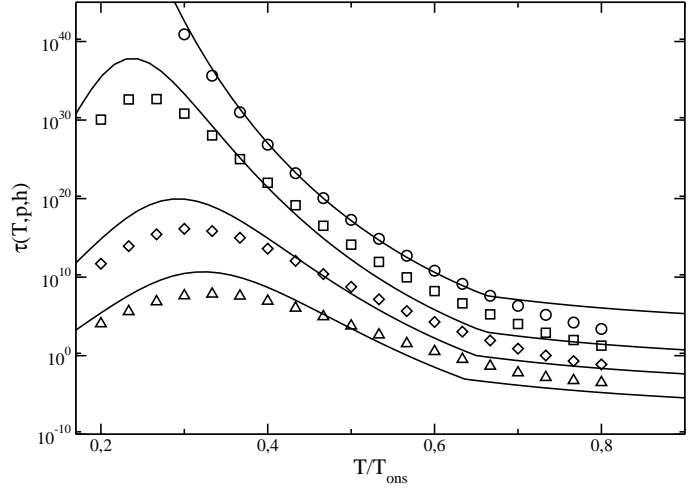}}
\caption[Equilibrium and nonequilibrium metastable state mean lifetime.]
{\small Semilogarithmic plot for the mean lifetime of the metastable state as a function of temperature for different values 
of $p$, as obtained in Monte Carlo simulations. 
We study here a system with size $L=53$, subject to periodic boundary conditions and with $h=-0.1$, and we average
over $N_{exp}=1000$ independent runs. In particular, from top to bottom, $p=0$($\bigcirc$), $0.001$($\Box$), $0.005$($\Diamond$) and 
$0.01$($\triangle$). The continuous lines are the theoretical predictions for each case. For the shake of clarity, we rescale the
$n$-th curve (in the aforementioned order) by a factor $10^{-2(n-1)}$. Error bars are smaller than symbol sizes.}
\label{taumedia}
\end{figure}

The above analysis allows us to investigate the finite size effects which affect the properties of the mean lifetime of the metastable 
state in the nonequilibrium system. In the following we want to check these theoretical predictions using Monte Carlo simulations
of the metastable-stable transition. We thus build a square system with size $L$, subject to periodic boundary conditions. We impose
as initial condition a state with all spins up, in such a way that under the action of a negative magnetic field this initial state
is metastable. We let evolve the system under these conditions using Glauber dynamics, eq. (\ref{rate}). The system rapidly evolves from
the initial state to a state in the metastable region, with magnetization close to $+1$. 
After this fast initial relaxation, the system spends 
a long time wandering around the metastable state. Eventually it will nucleate one or several critical droplets of the stable 
phase (assuming our parameters are such that the system is in the SD or MD regions), which will grow rapidly making the system to
evolve from the metastable state to the stable one, where it stays forever. We define the mean lifetime of the metastable state,
$\tau(T,p,h)$, as the average first passage time (in Monte Carlo Steps per spin, MCSS) to $m=0$. Hence in order to calculate the mean 
lifetime from our equations we must use $\Phi_s=0.5$ due to this convention. In practice the metastable state lifetime can be a long
as $10^{40}$ MCSS, so we need to use Monte Carlo with Absorbing Markov Chains (MCAMC) algorithms in order to perform the simulations.
In particular, we have used the $s-1$ MCAMC algorithm, together with the slow forcing approximation\footnote{Whenever we apply the 
slow forcing approximation, we ensure that the forcing rate $\phi$ (see Appendix \ref{apendMCAMC}) is slow enough so we have reached
the slow forcing limit, where the measured observables do not depend on the applied forcing. Notice on the other hand that when we apply
the slow forcing approximation, the metastable state mean lifetime is derived from the stable phase growth and shrinkage rates, see 
Appendix \ref{apendMCAMC}.}. These advanced algorithms are presented in Appendix \ref{apendMCAMC}. In order to improve our statistics
for the lifetime, we simulate $N_{exp}$ different metastable state demagnetization experiment, averaging the lifetime over all them.

Fig. \ref{taumedia} shows the mean lifetime of the metastable state, as defined in the above paragraph, in semilog scale, as a function
of temperature for different values of $p$, as measured in Monte Carlo simulations for a system with size $L=53$, with  $h=-0.1$
once we average over $N_{exp}=1000$ different runs. This figure also shows the theoretical predictions, based on eqs. (\ref{taupMD}) 
and (\ref{taupSD}). For each temperature $T$ and nonequilibrium parameter $p$ ($h=-0.1$ is fixed) we evaluate whether the system is 
in the MD or SD regions calculating ${\cal R}_0(T,p,h)$ and comparing it with the system size $L$. Consequently we use expressions
(\ref{taupMD}) or (\ref{taupSD}) to predict the lifetime. In practice we observe that most of the studied temperature interval 
in Fig. \ref{taumedia} lies on the SD region. The only free parameters in our theory are the constant $\Gamma$ which related the radial
growth velocity to the inverse of the critical droplet size, see eq. (\ref{velradialp}), and the non-universal amplitude factor
${\cal A}(T,p) \equiv A(T,p)|h|^{b+c}$  that appears in the nucleation rate $I(T,p,h)$, eq. (\ref{nucratep}). We fixed the constant
$\Gamma$ in each case when we studied the radial growth velocity in the previous section, where we obtained 
$\Gamma \in [\frac{1}{5},\frac{1}{4}]$ depending on the value of $p$. For the amplitude factor ${\cal A}(T,p)$ we assume that
its temperature dependence is weak as compared to the exponential dependence on temperature which dominates the behavior of $I(T,p,h)$.
Therefore we fix the value of ${\cal A}(T,p)$ to a constant which does not depend on temperature, ${\cal A}(p)$. For each $p$, 
${\cal A}(p)$ is derived from the best fit between the theoretical prediction and computational results (notice that 
${\cal A}(T,p)\approx {\cal A}(p)$ only produces a vertical shift, in semilogarithmic scale, in the mean lifetime curve).

As we can observe in Fig. \ref{taumedia}, in the equilibrium system the mean lifetime grows monotonously as temperature decreases
(moreover, $\tau(T,p=0,h)$ grows exponentially with  $1/T$, as predicted by nucleation theory). However, the mean lifetime for
$p\neq 0$ grows as temperature decreases, up to certain nonzero temperature $T_{max}^{(\tau)}(p,h)$ where the lifetime reaches a 
maximum, after which the lifetime change its tendency and decreases as temperature decreases. Hence we see that for any $p\neq 0$
the metastable state survives a maximum time for a nontrivial, nonzero temperature $T_{max}^{(\tau)}(p,h)$. 
Therefore if we need to maximally prolong the lifetime of a metastable state in a real magnetic thin film, 
which shows the kind of impure behavior parameterized by the nonequilibrium perturbation $p$
(think for instance of the problem posed by magnetic storage of information in technological applications), the most effective method
will not consists in a {\it blind} decrease of temperature, but in the search of the temperature $T_{max}^{(\tau)}(p,h)$ for which
the metastable states has the longest lifetime. On the other hand, the theoretical results for the lifetime reproduce rather well,
at least from the qualitative and semi-quantitative point of view, the Monte Carlo results. The theory predicts a maximum in
$\tau(T,p\neq 0,h)$ for certain nonzero temperature, which coincides with the one observed in simulations.
This non-monotonous behavior of the mean lifetime is inherited again from the non-monotonous behavior of surface tension in the
nonequilibrium system. Hence the interfacial properties of the system determine in a fundamental way the dynamical and statical 
processes related to metastability. On the other hand, from the quantitative point of view, our approximation for the lifetime of 
the metastable state for $p\neq 0$ differs somewhat from Monte Carlo results. 
In particular, we observe that although the general shape
of the theoretical curves is similar to Monte Carlo curves, the theoretical curves are more steep than the computational results.

\newpage

\subsection{Morphology of the Metastable-Stable Transition}
\label{capNuc_apExt_subMorf}

\begin{figure}[t!]
\centerline{
\psfig{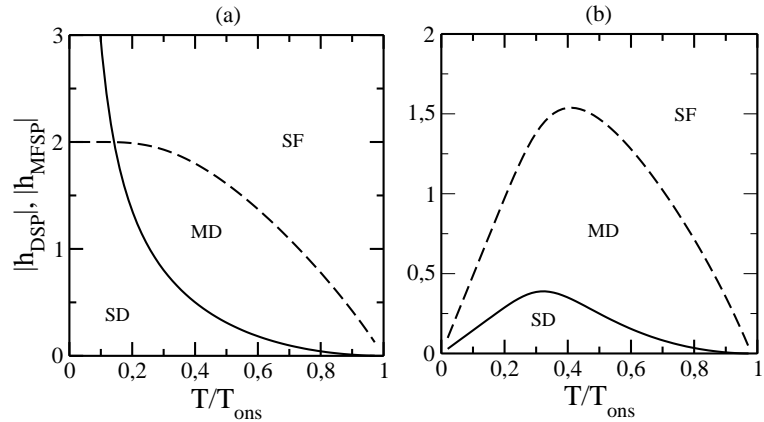}}
\caption[Dynamic spinodal point and mean field spinodal point.]
{\small Dynamic Spinodal Point, $|h_{DSP}|(T,p)$, represented by a  continuous line, and Mean Field Spinodal Point, 
$|h_{MFSP}|(T,p)$, represented by a dashed line, as a function of temperature for: (a) $p=0$, and (b) $p=0.01$. Notice the
fundamental change in behavior for low temperatures when we compare the equilibrium case ($p=0$) with the nonequilibrium one 
($p\neq 0$).}
\label{DSP}
\end{figure}

An interesting question is to study the morphology of the metastable-stable transition as a function of the system parameters. 
This study will allow us to divide the parameter space in different regions, each one characterized by a well-defined typical 
morphology for the process of demagnetization of the metastable state. In order to characterize the different morphologies we
study the Dynamic Spinodal Point, $|h_{DSP}|$, and the Mean Field Spinodal Point, $|h_{MFSP}|$, defined respectively in
eqs. (\ref{hdsp}) and (\ref{hmfsp}), as functions of temperature for different values of $p$. The field $|h_{DSP}|(T,p)$ separates
the Single Droplet Region from the Multidroplet Region. Hence if $|h|<|h_{DSP}|$ the metastable-stable transition proceeds through
the nucleation of a single droplet of the stable phase, while if $|h_{MFSP}|>|h|>|h_{DSP}|$ it proceeds through the nucleation
of multiple droplets. On the other hand, the Mean Field Spinodal Point , $|h_{MFSP}|$, separates the Multidroplet Region from the
Strong Field Region, where the nucleating droplet picture is no longer valid. The SF region is observed for $|h|>|h_{MFSP}|$.
\begin{figure}[t!]
\centerline{
\psfig{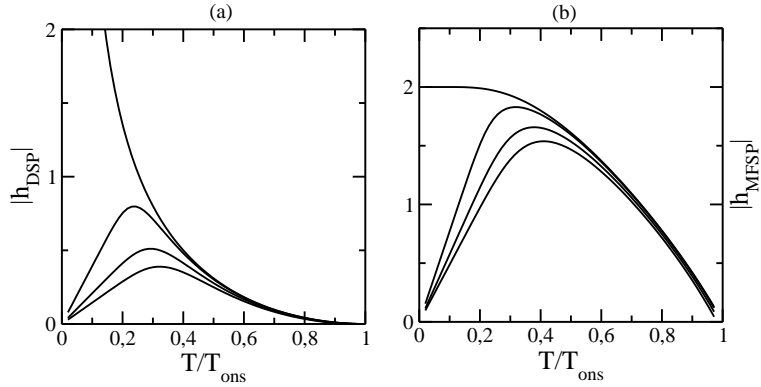}}
\caption[Nonequilibrium parameter dependence of $|h_{DSP}|$ and $|h_{MFSP}|$.]
{\small (a) $|h_{DSP}|(T,p)$ as a function of temperature for $L=53$ and several different values of $p$. In particular, from 
top to bottom, $p=0$, $0.001$, $0.005$ and $0.01$; (b) $|h_{MFSP}|(T,p)$ as a function of temperature for the same values of $p$.
Notice the fundamental change of behavior in both $|h_{DSP}|(T,p)$ and $|h_{MFSP}|(T,p)$ in the low temperature limit for any $p\neq 0$
as compared to the equilibrium case.}
\label{DSPMFSP}
\end{figure}

Fig.  \ref{DSP}.a shows the theoretical prediction for the fields $|h_{DSP}|(T,p)$ and $|h_{MFSP}|(T,p)$ as a function of temperature
for a equilibrium system ($p=0$) with size $L=53$. The parameter space is divided in three different regions. The SD region, 
characterized by the nucleation of a single critical droplet, dominates the morphology of the metastable-stable transition for low 
temperatures in this case ($p=0$). For intermediate temperatures, if the magnetic field is high enough, the metastable state 
demagnetization process proceeds through the nucleation of multiple critical droplets, which is the typical morphology of MD region.
Finally, for very strong magnetic fields the metastable-stable transitions exhibits a morphology typical of SF region, where
the concept of mutually independent nucleating droplets is not valid. These theoretical results about the morphology of the 
metastable-stable transition, first derived in refs. \cite{Rikvold,Ramos}, are perfectly verified in Monte Carlo simulations of the 
system with $p=0$.\cite{Rikvold,Ramos} Therefore, for a fixed (intermediate) magnetic field, the typical sequence of morphologies 
as temperature decreases from temperatures close to the critical one is SF$\rightarrow$MD$\rightarrow$SD.
\begin{figure}[p!]
\centerline{
\psfig{file=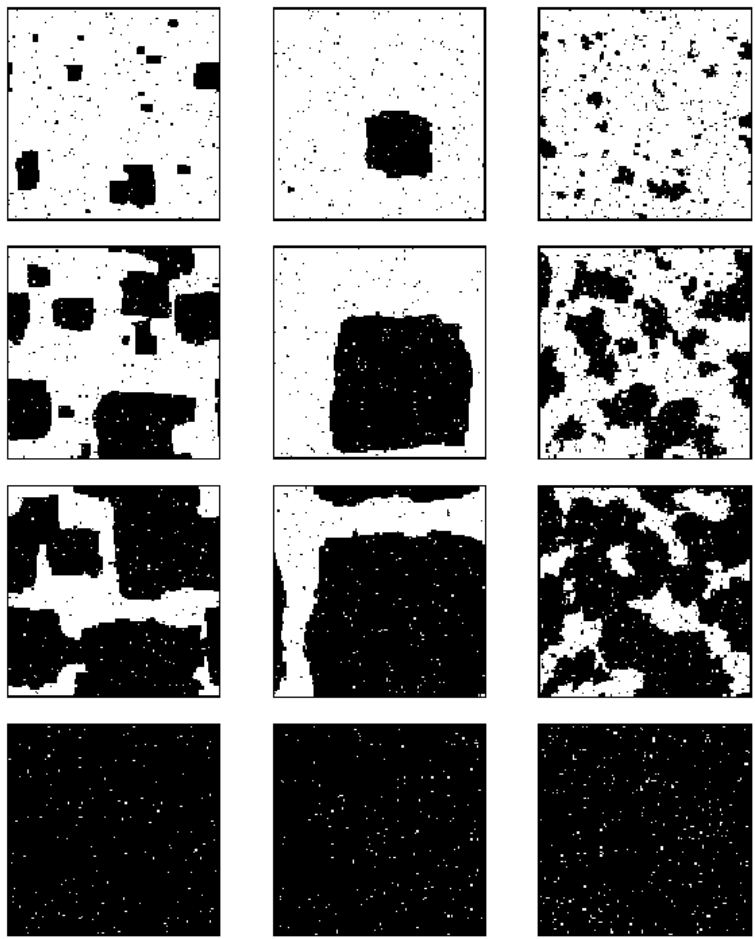,width=10cm}}
\caption[Different morphologies of the metastable-stable transition.]
{\small Snapshots of the metastable-stable transition for a system with size $L=128$, with $p=0.01$, $h=-0.25$ and
temperatures $T=0.1T_{ons}$ (left column), $T=0.3T_{ons}$ (central column) and $T=0.7T_{ons}$ (right column). The first row 
corresponds to a stable phase volume fraction $\Phi_s=0.1$, the second one to $\Phi_s=0.4$, the third one to $\Phi_s=0.7$ and
the last one tho the stable state for each temperature. Notice that, while for $T=0.1T_{ons}$ and $T=0.7T_{ons}$ the decay
from the metastable state proceeds through the nucleation of multiple droplets, for the intermediate temperature $T=0.3T_{ons}$
this transition proceeds through the nucleation of a single droplet, in total agreement with the theoretical predictions (see
the main text).}
\label{multisingle}
\end{figure}

Fig. \ref{DSP}.b shows the same results than Fig. \ref{DSP}.a but for a system with $p=0.01$, i.e. under nonequilibrium conditions.
As in equilibrium , the parameter space is divided in three different regions. However, for $p\neq 0$ there is a fundamental difference
as compared to the equilibrium case: both $|h_{DSP}|(T,p)$ and $|h_{MFSP}|(T,p)$ converge towards zero in the low temperature limit
for $p\neq 0$. This implies that the SD region does not dominates now at low temperatures. Instead, we observe that there is a region
for intermediate temperatures and weak magnetic fields where the SD region determines the morphology of the metastable-stable 
transition. This region disappears if we excessively increase or {\it decrease} the temperature, appearing instead the MD region. Hence,
for intermediate magnetic fields the characteristic morphology sequence in the metastable-stable transition will be 
SF$\rightarrow$MD$\rightarrow$SF as we decrease temperature from values near the critical one. For weak magnetic fields this sequence
will be SF$\rightarrow$MD$\rightarrow$SD$\rightarrow$MD$\rightarrow$SF.

Fig. \ref{DSPMFSP}.a shows the theoretical prediction for the Dynamic Spinodal Point, $|h_{DSP}|(T,p)$, as a function of temperature 
for varying $p$ in a system with size $L=53$. Apart from the fundamental change of behavior observed in $|h_{DSP}|(T,p)$ at low 
temperatures for $p\neq 0$, we observe that, as $p$ increases the SD region decreases, increasing the MD region. This behavior is 
easy to understand from a phenomenologic point of view, because the nonequilibrium perturbation $p$ acts as a noise with non-thermal 
origin, affecting mainly the most ordered regions in the system (remember the definition of effective temperature in 
Chapter \ref{capSOS}), so the presence of $p\neq 0$ favours the nucleation of new stable phase droplets in the metastable bulk, 
favouring in this way the MD morphology at the expense of the SD region. Analogously, Fig. \ref{DSPMFSP}.b shows the theoretical
prediction for $|h_{MFSP}|(T,p)$ as a function of $T$ for varying $p$. We observe here that as we increase $p$, the SF region
grows at the expense of the MD region.

Finally, if we take a constant magnetic field $h=-0.25$ in Fig. \ref{DSP}.b, where we plot $|h_{DSP}|(T,p)$ and $|h_{MFSP}|(T,p)$
as a function of $T$ for a system with $p=0.01$ and $L=53$, we observe that the theory points out that 
for a temperature $T=0.1T_{ons}$ the morphology associated to the metastable-stable transition will be that of the MD region. 
For $T=0.3T_{ons}$ this morphology will be the typical of SD region, while for $T=0.7T_{ons}$ we again will recover the
MD morphology. Fig. \ref{multisingle} shows a series of snapshots for the temporal evolution of a system with size 
$L=128$,\footnote{The theoretical predictions about the morphology of the metastable-stable transition that we perform using the 
curves associated to a system with size $L=53$ are almost identical to the predictions we would perform for a system size $L=128$.
This is so because the logarithmic dependence of $|h_{DSP}|(T,p)$ on the system size $L$ and the independence of $|h_{MFSP}|(T,p)$ 
on $L$. In Fig. \ref{multisingle} we use a system size $L=128$ for the shake of clarity.} with periodic boundary conditions and
with a magnetic field $h=-0.25$, for $p=0.01$ and temperatures $T=0.1T_{ons}$, $0.3T_{ons}$ and $0.7T_{ons}$. It is remarkable that,
in agreement with the theoretical predictions, the metastable-stable transition proceeds through the nucleation of multiple
droplets for both $T=0.1T_{ons}$ and $T=0.7T_{ons}$, while it proceeds through the nucleation of a single droplet for the intermediate
temperature $T=0.3T_{ons}$. Therefore, the fundamental differences that the approximation predicts in the low temperature
limit for the decay morphology between the equilibrium and the nonequilibrium cases are verified in Monte Carlo 
simulations.

\section{Conclusion}
\label{capNuc_apConc}

In this chapter we have developed a nucleation theory for the nonequilibrium ferromagnetic system, in order to understand
the dynamics of the metastable-stable transition in such system. 

We have generalized the equilibrium nucleation theory\cite{Rikvold,Gunton}, based on a picture where stable phase droplets nucleate
in the metastable bulk, in order to build the nonequilibrium dynamic theory. The central magnitude of nucleation theory is the free
energy cost of a droplet of the stable phase with radius $R$. In our generalization we have hypothesized
the existence of certain nonequilibrium potential ${\cal F}(R)$, which controls the exit from the metastable state, and plays in the
system the same role than the droplet free energy in equilibrium. Moreover, we have proposed, based
on phenomenological grounds, a particular expression for this nonequilibrium potential, which is {\it formally} identical
to the droplet equilibrium free energy.

There are two competing terms in the nonequilibrium potential ${\cal F}(R)$: a volume term, which favours the droplet growth, and
a surface term, which hinders such growth. The physical observables which determine the relative importance of both terms are 
$m_s(T,p)$, $\sigma_0(T,p)$ and $\Omega_d(T,p)$, where $m_s(T,p)$ is the spontaneous magnetization of the nonequilibrium system
for zero magnetic field, $\sigma_0(T,p)$ is the surface tension along one of the primitive lattice vectors for $h=0$, and 
$\Omega_d(T,p)$ is the form factor which relates the droplet volume to its radius, all these observables defined for temperature
$T$ and nonequilibrium perturbation $p$. Although we do not know the exact expression for any of these observables when $p\neq 0$,
we have obtained in previous chapters very good approximations to all of them.

We write a nucleation theory for $p\neq 0$ using all this information. From this theory we are able to correctly predict the behavior
and the properties of fundamental observables associated to the dynamic problem of metastability. For instance, the extended 
nucleation theory precisely predicts the dependence on $T$ and $p$ of the critical droplet radius, ${\cal R}_c(T,p,h)$. In particular,
we find that while the equilibrium system shows a critical droplet size which monotonously increases as temperature decreases, the
critical droplet size for the nonequilibrium system ($p\neq 0$) exhibits a maximum as a function of temperature, which depends on $p$.
In the same way, the generalized nucleation theory approximately predicts the growth velocity of a stable phase droplet under nonzero
magnetic field,  $v_0(T,p,h)$. We observe that for $p\neq 0$ this velocity shows a minimum for certain temperature, $T_{min}^{(v)}$,
growing for $T<T_{min}^{(v)}$. The mean lifetime of the metastable state, 
defined as the average first passage time to a zero magnetization
state, $\tau(T,p,h)$, is also correctly predicted by the theory. In the same way than for the previously discussed observables, the
fundamental feature distinguishing the equilibrium and nonequilibrium cases  is the fact that, for $p\neq 0$, the mean lifetime
$\tau(T,p,h)$ is non-monotonous with temperature, showing a maximum for a given nonzero temperature $T_{max}^{(\tau)}$, while in
the equilibrium case the lifetime grows monotonously as temperature decreases. Finally, the generalized theory also describes 
correctly the different typical morphologies which characterize the metastable-stable transition. The dominant typical morphology 
in the equilibrium system at low temperatures is that characterized by the nucleation of a single critical droplet, dominating
at higher temperatures the nucleation of multiple droplets. For the nonequilibrium system this behavior enriches considerably.
We observe for $p\neq 0$ that the morphology characterized by the nucleation of the single droplet dominates for intermediate
temperatures, being the characteristic morphology for low and high temperatures that associated to the nucleation of multiple 
droplets. All the theoretical predictions have been checked in Monte Carlo simulations of the nonequilibrium system.

The generalized nucleation theory allows us to describe in an approximate manner the dynamics of the metastable-stable transition
in the nonequilibrium ferromagnetic system. An important conclusion derived from this study is that the properties of the
interface determine in a fundamental way the exit dynamics from the metastable state. All the fundamental differences that we
observe (and predict) in this dynamic process between the equilibrium and nonequilibrium cases can be easily understood
once we know the properties of the interface for $p\neq 0$, which were studied in Chapter \ref{capSOS}, and how they compare
with the equilibrium interfacial properties. In particular, the non-monotonous behavior of surface tension as a function of temperature 
for $p\neq 0$ is inherited by all the observables associated to the metastable decay process, namely ${\cal R}_c$, $v_0$, $\tau$, 
$|h_{DSP}|$ and $|h_{MFSP}|$.

The extended nucleation theory is based on a phenomenologic, non-justi-fied hypothesis which assumes the existence and the
particular form of a non-equilibrium potential which controls the exit from the metastable state, in a similar way to that of the
free energy in equilibrium. This hypothesis allows us to obtain many results about the dynamics of the metastable-stable
transition, which correctly reproduce the behavior of the real system (as obtained from Monte Carlo simulations). The fundamental
ingredient in this hypothesis consist in assuming that, similarly to what happens in equilibrium systems, the droplet dynamics
is determined by the competition between the droplet bulk and its surface, in such a way that if we correctly capture the bulk and
interfacial behavior we will be able to obtain much information about the metastable state demagnetization process. This
observation, which yields very good results for the nonequilibrium ferromagnetic system here studied, may be generalizable to 
many other nonequilibrium systems showing metastability.

Finally, our results may also be relevant from the technological point of view. Think for a while on magnetic systems used for
magnetic storage of information (previously discussed in Chapter  \ref{capMotiv}). Such systems are magnetic thin films
composed by many small monodomain ferromagnetic particles. These particles, for which Ising-like models are in some 
cases a good description, generally show an impure behavior related to the presence of lattice, bond and/or spin disorder, 
quantum tunneling, etc. Therefore we expect that our simplified nonequilibrium system will model 
adequately (at least in a first approximation) the behavior of these
magnetic materials. On the other hand, a main concern in these magnetic systems is to retain for as long as possible the stored
information. The information is stored in these systems magnetizing with a strong magnetic field the particular domains,
defining in this way a bit of information for each magnetized particle. Due to the interaction with the external medium,
the different particles suffer small random magnetic fields, which involve the eventual appearance of metastable states in such
magnetic particles. The resistance of the stored information to these external perturbations strongly depends on the properties
of the underlying metastable states, including the details of their decay. As we have derived in this chapter, the presence of 
impurities, parameterized in our model by the nonequilibrium perturbation $p$, affects in a fundamental way the properties of 
metastable states. In particular, if we want to prolong as much as possible the metastable state lifetime, we must not
{\it blindly} decrease the system temperature, but we must look for the temperature $T_{max}^{(\tau)}$ for which the metastable state
lifetime is maximum.

\chapter{Scale Free Avalanches during Decay from Metastable States 
in Impure Ferromagnetic Nanoparticles}
\label{capAval}

\markboth{Scale Free Avalanches during Decay from Metastable States}{\ref{capAval_apIntro} Introduction}
\section{Introduction}
\label{capAval_apIntro}
\markboth{Scale Free Avalanches during Decay from Metastable States}{\ref{capAval_apIntro} Introduction}

In previous chapters we have studied the problem of metastability in a nonequilibrium 
ferromagnetic system which we think correctly models the behavior of some real impure magnetic materials.
We have developed a nonequilibrium nucleation theory which explains the dynamics of the metastable-stable 
transition in this system. This theory is based on a phenomenologic hypothesis about the existence of a nonequilibrium
potential that controls the exit from the metastable state in a way similar to that of the free energy
in equilibrium systems. Such nonequilibrium potential depends on the properties of both the bulk and the interface 
separating the stable phase from the metastable one. We have derived these properties using
mean field-like approximations.

This theoretical and computational study has shown that the presence of nonequilibrium conditions considerably
enriches the behavior of the system, mainly at low temperatures. In particular, we observe that the properties 
of metastable states strongly depend on the properties of the interface separating the metastable phase from the 
stable one. Thus, the non-monotonous behavior of surface tension as a function of temperature in the nonequilibrium 
model is inherited by most of the physical observables that characterize the dynamics of the metastable-stable transition, 
as for instance the metastable state mean lifetime $\tau$, the critical droplet size ${\cal R}_c$, etc.

As we have previously discussed, the results of this analysis, apart from their theoretical value, are relevant from the 
technological point of view. Consider again the problem posed by magnetic storage of information. We already know that
the resistance of stored information in magnetic recording media strongly depends on the properties of the underlying
metastable states of the magnetic particles that compose such materials. Retaining the orientation of
such domains for as long as possible is a main technological concern. Another principal aim during recording
is to maximize the amount of information stored. This requires manufacturing
very dense media which is also important for many other areas of present and
emergent technologies.\cite{Simo}
One needs in practice to create and control fine grains, i.e., magnetic
particles with {\it borders} whose size ranges from mesoscopic to atomic levels, namely, clusters
of $10^{4}$ to $10^{2}$ spins, and even smaller ones. Though 
experimental techniques are already accurate for the purpose,\cite{Shi,Wern} the underlying
physics is much less understood than for bulk properties. In particular, 
one cannot assume that such particles are neither \textit{infinite} nor
\textit{pure.} That is, they have free borders, which results in a large
surface/volume ratio inducing strong border effects, and impurities, which 
might dominate the behavior of near-microscopic particles; in fact, they are known to 
influence even macroscopic systems.

The effects that free borders induce on the properties of the metastable-stable transition have been already
studied in equilibrium systems.\cite{Cirillo,contorno} In this case it is observed that the system evolves from
the metastable state to the stable one through the {\it heterogeneous} nucleation of one or several critical
droplets which always appear at the system's border.\cite{Cirillo} That is, the free border acts as a droplet condenser. 
This is so because it is energetically favorable for the droplet to nucleate at the border. Apart from the observed 
heterogeneous nucleation, the properties of the metastable-stable transition in equilibrium ferromagnetic nanoparticles 
do not change qualitatively as compared to the periodic boundary conditions case.\cite{contorno} In our nonequilibrium
system we obtain similar results, namely heterogeneous nucleation and the same qualitative nucleation properties.
However, it is very interesting to study the fluctuations or noise that the nonequilibrium system exhibits as it evolves 
towards the stable state subject to the combined action of free borders and the nonequilibrium perturbation. 
As we will describe below, the metastable-stable transition in this case
proceeds through {\it avalanches}. These burst-like events, present in our
model case, characterize the dynamics of an enormous amount of nonequilibrium complex systems.\cite{Jensen}



In general, noise in magnetic systems has been shown to be of major importance in 
many technological applications, \cite{Safo,Sipahi1} as well as from the theoretical point 
of view\cite{Spaso}. The celebrated Barkhausen Noise, e.g. the magnetic noise by which a 
impure ferromagnet responds to a slowly varying magnetic field, has been profusely used as a 
non$-$invasive material characterization technique \cite{Sipahi1,Sipahi2}. The statistical 
properties of Barkhausen Noise are extremely sensitive to microstructural changes in the 
material, thus providing a sharp tool in order to characterize such system. Its applications 
include microstructure analysis, fatigue testing, measurement of fundamental properties of 
magnetic materials,  stress analysis, etc. \cite{Sipahi1} Theoretically, systems exhibiting 
Barkhausen Noise have been studied as a paradigm of complex spatio-temporal extended 
systems showing generically scale invariance.\cite{Spaso,Zapperi} Barkhausen systems show 
avalanche$-$like dynamics, where avalanches are scale-free. Furthermore, this property is 
found {\it naturally} in these magnetic materials, i.e. no fine tuning 
is needed in order to reach such scale invariant state. 

Scale-free noise is observed 
ubiquitously in Nature,\cite{Sethna} and its origin still remains unknown. It receives the generic name 
of {\it $1/f$ noise},\cite{Antal} and Barkhausen Noise is just a particular realization. This 
name reflects the power law behavior of the fluctuation's power spectrum, which also involves a 
power law behavior for avalanche distributions. There are many natural systems (besides 
Barkhausen materials) which show $1/f$ noise. Some examples are: biological (e.g. human cognition 
\cite{Gilden}), economical (e.g. the number of stocks traded daily \cite{Lillo}) and 
social systems, earthquakes \cite{Davidsen}, superconductors \cite{Field}, combustion, piles of rice 
\cite{Frette} and sand \cite{Schick}, crumpling paper \cite{Houle}, music \cite{Voss}, etc.
The \textit{universality} of this phenomenon points out the existence of a common underlying 
mechanism present in all the above apparently unrelated situations. In recent 
years it has been claimed that a new idea, namely \textit{Self Organized Criticality} 
(SOC) \cite{Bak,Jensen}, could be behind this universal, scale invariant phenomenon.
Self-Organized Criticality is based on the idea that complex, scale-free behavior can 
develop \textit{spontaneously} in many body systems whose dynamics is dominated by at least two very 
different time scales, and where metastability is observed.\cite{Jensen} In this picture, the system 
would evolve towards a \textit{critical} state where avalanches of all scales appear. In 
particular, Barkhausen Noise has been thought as an experimental realization of SOC, thus 
provoking an active and excited research in this field. However, experimental works are not able 
to conclude on the underlying mechanism responsible of this $1/f$ noise. Furthermore, 
experimentalists cannot even affirm the existence of any underlying critical point\cite{Spaso}, neither 
a plain old one nor a SOC one.

Summing up, the scenario to be investigated in this chapter involves the formation and evolution of a 
stable phase droplet inside the metastable parent phase in our nonequilibrium (i.e. impure) ferromagnetic 
system subject to free boundaries. Under the combined action 
of both the nonequilibrium perturbation and free boundaries, the formation of a nucleus of the stable phase 
turns out to proceed by avalanches. In addition to small events, which show up as a completely
random, thermal effect (\textit{extrinsic noise} \cite{Spaso}) having an exponential distribution, 
we describe well-defined \textit{critical avalanches.} These are typically much larger 
than the extrinsic noise, while they \textit{apparently} show no characteristic temporal and 
spatial scale. In particular, we find size and lifetime distributions that follow power laws, 
$P(  \Delta_m)  \sim \Delta_m^{- \tau}$ and $P(\Delta_t) \sim \Delta_t^{- \alpha}$ with 
$\tau \rightarrow 1.71(4)$ and $\alpha \rightarrow 2.25(3)$ for large enough systems. This holds up 
to an exponential cutoff which grows as a power law of system size. A detailed analysis of these
scale free avalanches reveals that they are in fact the combined result of many avalanches of different 
well-defined \textit{typical} size and duration. That is, the simplicity and versatility of our 
model system allows us to identify many different types of avalanches, each type characterized by a 
probability distribution with well defined typical size and duration, associated with a particular 
\textit{curvature} of the domain wall. Due to free borders and the microscopic impurity the system 
visits a broad range of domain wall configurations, and thus the combination of these avalanches 
generally results in a distribution which exhibits several decades of power law behavior and an exponential 
cutoff. However, this apparent scale-free behavior does not mean that avalanches are critical, in the sense of a 
second order phase transition where diverging correlation lengths appear. Instead, we find that avalanches in the
magnetic nanoparticle have a large (but finite) number of different, gap-separated typical sizes 
and durations.

These observations led us to suspect that Barkhausen Noise, as defined above, might also come from 
the superposition of more elementary events. In fact, the $1/f$ noise behavior in this case is assumed 
to reflect topological rearrangements of domain walls,\cite{Che_yotros} which result in practice in a 
series of jumps between different metastable states, which is the basic process in our model. We strongly 
support this conjecture because our avalanche model reproduces many features previously observed in 
different experimental situations. For instance, the avalanche exponents we obtain are almost identical 
to those measured in some Barkhausen experiments \cite{Spaso}, and our model system shows some properties, as for 
example \textit{reproducibility} \cite{Urbach}, observed in real magnetic materials. Up to now, 
theoretical explanations of Barkhausen Noise were based on the hypothesis of the existence of an 
underlying critical point, thus yielding the observed scale invariance. However, this assumption faces 
some fundamental problems, since experiments on Barkhausen Noise show no universality: experimental 
critical exponents are different for different conditions and materials, and expected universal 
observables in Barkhausen Noise are in practice very sensitive to microscopic details, which is against 
the concept of universality derived from a critical point. 
The conceptual framework we propose here for Barkhausen Noise solves this problem, 
since it does not imply the existence of an underlying critical point, and thus no universality is 
expected. The chances are that our observation that scale invariance originates in a combination of 
simple events, which we can prove in our model cases, is a general feature of similar phenomena in 
many complex systems \cite{Sethna}.

The remainder of the chapter is organized as follows. In section \ref{capAval_apModel} we briefly 
remind the model, introducing the free boundary conditions we use in this chapter. 
In section \ref{capAval_apMC} we present our simulations, and the analysis of the 
avalanche distributions obtained. Section \ref{capAval_apOrig} is devoted to the search of an explanation 
for the observed behavior. Finally, in Section \ref{capAval_apConc} we present the conclusions, paying special attention 
to the consequences derived from our results for the explanation of Barkhausen Noise in particular, and 
$1/f$ noise in general.

\newpage


\markboth{Scale Free Avalanches during Decay from Metastable States}{\ref{capAval_apModel} The Magnetic Particle Model}
\section{The Magnetic Particle Model}
\label{capAval_apModel}
\markboth{Scale Free Avalanches during Decay from Metastable States}{\ref{capAval_apModel} The Magnetic Particle Model}

In this chapter we consider a modification of the system studied in Chapters \ref{capMotiv}-\ref{capNuc}. 
This modification only involves a change on the implemented boundary conditions.
The system is defined again on a two-dimensional square lattice, with binary spins $s_i=\pm 1$ on each 
node, $i\in [1,N]$. These spins interact among them and with an external magnetic field via the Ising Hamiltonian, 
eq. (\ref{hamilt}). We choose the transition rate to be the Glauber rate, defined in eq. (\ref{rate}). This rule
describes a spin-flip mechanism under the action of two competing heat baths whose relative strength is characterized 
by the nonequilibrium parameter $p$, see eq. (\ref{rate}). For $p\neq 0$ the transition rate (\ref{rate}) 
asymptotically drives the system towards a nonequilibrium state, which essentially defers from the Gibbs (equilibrium) one.

\begin{figure}[t]
\centerline{
\psfig{file=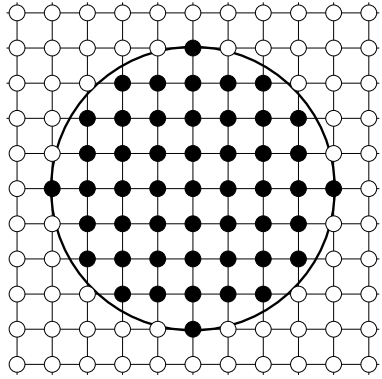,width=5cm}}
\caption[Schematic plot of a system subject to open circular borders.]
{\small Schematic plot of the system subject to open circular boundary conditions. The spins are 
represented by black points. White points are empty nodes. Spins at the border nodes do not have nearest
neighbor spins outside the circle.}
\label{opencircle}
\end{figure}

Motivated by the experimental situation, we choose to study a finite,
relatively small system subject to {\it open circular} boundary conditions. 
In order to implement these boundary conditions, we inscribe a circle of radius $R$ in the square 
lattice where the system is defined.
Sites outside this circle do not belong to the system. In this way we define the free boundary: 
spins at the border sites inside the circle do not have nearest-neighbor spins outside the 
circle. Fig. \ref{opencircle} shows an example.
The lattice is set initially with all spins up, $s_{i}=+1$ for $i=1,...,N$. 
Under a weak negative magnetic field, this ordered state is metastable, and it eventually decays 
to the stable state which, for low temperatures, corresponds to $m \equiv N^{-1} \sum_{i}s_{i} \simeq -1$.
Investigating the influence of different boundary conditions on the
relaxation, and how a small system compares with a macroscopic one, is a
crucial issue when trying to understand better the behavior of particles of
$\sim10^{3}$ spins. For this reason we will also use other different boundary conditions (to be defined later), 
in order to understand the role played by the free boundary.

We mainly report here on a set of fixed values for the model parameters,
namely, $J=1$, $h=-0.1$, $T=0.11T_{ons}$, and
$p=10^{-6}$, where $J$ is the (ferromagnetic) coupling constant, $h$ is the magnetic field,
$T$ is the system temperature and $p$ is the nonequilibrium perturbation.
This choice is dictated by simplicity and also because (after
exploring the behavior for other cases) we came to the conclusion that this
corresponds to an interesting region of the system parameter space. In fact,
the field, which only aims to produce a convenient metastable situation, needs
to be small $-$to avoid becoming a relevant parameter$-$ but not too small
that metastability lasts for a very long (unobservable) time. The restriction
to this small value for $h$ also guarantees that the system is in a regime in
which the decay is dominated by a single droplet (see section \ref{capNuc_apExt_subMorf}); 
otherwise a finite density
of droplets, or even a more complex situation, may occur, as we have concluded in the previous
chapter. The low value of $T$ allows for compact
configurations, which are more convenient for analysis of clusters. Finally,
our choice for $p$ corresponds to a \textit{small enough} perturbation which
induces interesting significant effects that are in fact comparable to the
ones by other stochastic sources. In particular, a measure of the relative
importance of the thermal ($T$) and non-thermal ($p$) noise sources can 
be obtained using the concept of interface effective temperature, $T_{eff}^I$, first introduced
in section \ref{capSOS_apProp_subMacro}. For high temperatures, where the thermal noise dominates,
$T_{eff}^I$ depends linearly on $T$, while for low temperatures, where the non-thermal noise source 
dominates, $T_{eff}^I$ is independent of $T$, converging to a nonzero value, see Fig. \ref{teffI}. 
For a fixed $p$ there is narrow range of temperatures where $T_{eff}^I$ changes its asymptotic 
tendency. In this temperature interval the relative importance of both noise sources is comparable.
Hence the temperature $T=0.11T_{ons}$ ensures for $p=10^{-6}$ that both $T$ and $p$ have a comparable effect,
see Fig. \ref{teffI}. For these values of the model parameters the 
system shows many metastable states between $m \approx +1$ and $m \approx -1$, 
which introduces interesting phenomena, as we will see below. Summing up, 
we believe that we are describing here typical behavior of our model, and the chances 
are that it can be observed in actual materials.

In spite of the free borders, which considerably accelerate the exit from the metastable state,
the simulations reported here required in practice using the $s-1$ MCAMC algorithm\cite{Novot,nfold},
together with the slow forcing approximation (see Appendix \ref{apendMCAMC}).\cite{forcing} 
We have checked that the results reported in this chapter do not depend on the implemented forcing.

\begin{figure}[t]
\centerline{
\psfig{file=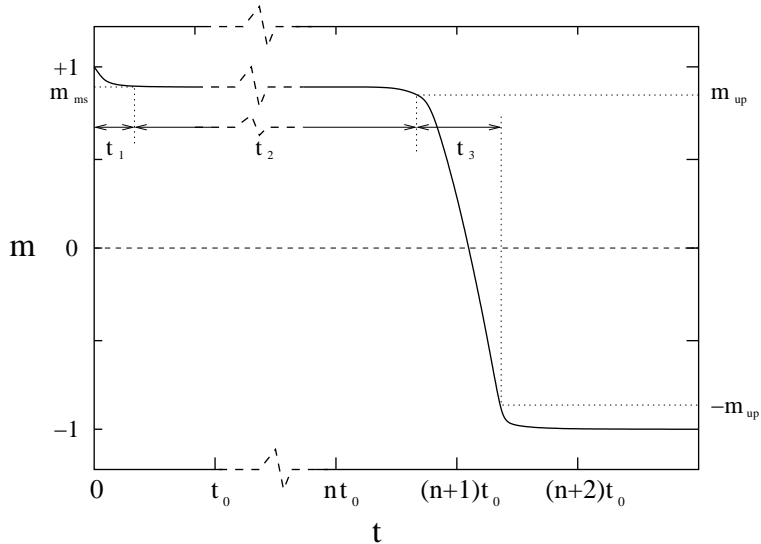,width=10cm,angle=0}}
\caption[Sketch of a typical decay from a metastable state.]
{\small Sketch of a typical decay from a metastable state showing the different time regimes.
Time units ($t_0$) are arbitrary. The threshold magnetizations, $m_{up}$ and $-m_{up}$, and the metastable 
magnetization, $m_{ms}$ are also shown. For the shake of clarity, time scales $t_1$, $t_2$ and $t_3$ have 
been plotted within the same order of magnitude, which is not the case in simulations and real systems.
}
\label{sketch}
\end{figure}

\markboth{Scale Free Avalanches during Decay from Metastable States}{\ref{capAval_apMC} Monte Carlo Results 
for Avalanche Statistics}
\section{Monte Carlo Results for Avalanche Statistics}
\label{capAval_apMC}
\markboth{Scale Free Avalanches during Decay from Metastable States}{\ref{capAval_apMC} Monte Carlo Results 
for Avalanche Statistics}

We are concerned with demagnetization from an initial fully$-$ordered state
with $m=+1$. This is subject to a negative (small) field, so that the system
eventually relaxes towards the stable state with $m \approx -1$. The temporal relaxation
typically shows three principal regimes (see Fig. \ref{sketch}):

\begin{enumerate}
\item  After a very short transient time, $t_{1}$, the system reaches the
metastable state. The magnetization $m(t)$ is then observed to
fluctuate, around its characteristic metastable value, $m_{\text{ms}}$, during
a long time interval, $t_{2}$.

\item  Suddenly at some time during such wandering, $m(t)$
decays in a time interval $t_{3}$ to a value near the stable magnetization,
$m_{\text{st}}$. As we know this evolution proceeds through the nucleation of
a critical droplet of the stable phase and its subsequent growth.
Time scales are $t_{1}<<t_{3}<<t_{2}$. It turns out
convenient to define the time $\tau_{0}$ at which this regime begins as the
last time for which $m(t) =m_{\text{up}},$ where $m_{\text{up}} < m_{\text{ms}}$ 
is a given magnetization threshold set $m_{\text{up}}=0.9$ here. 
We then define $t_{3}$ as the time interval between $\tau_{0}$ and the
time at which $m(t)  = -m_{\text{up}}$.

\item  After this fast decay, $m(t)  $ stays fluctuating around $m_{\text{st}}$.
\end{enumerate}

\begin{figure}[t]
\centerline{
\psfig{file=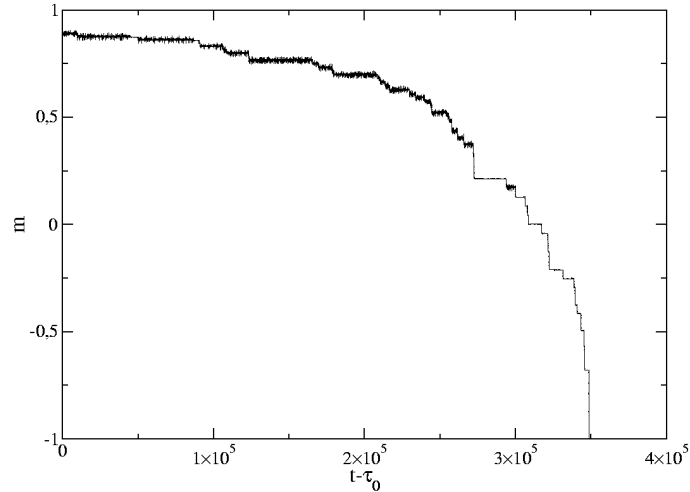,width=9cm,angle=0}}
\caption[Decay from a metastable state for the circular system.]
{\small Decay from a metastable state for the circular system with radius $R=30$, and the 
model parameters defined in the text. This figure corresponds to the $t_3$ region sketched in Fig. \ref{sketch}. 
The discrete, abrupt jumps in the magnetization (\textit{avalanches}) are seen by direct inspection. Time is in 
units of MCSS (Monte Carlo Steps per Spin). Here $\tau _0$ is the time for which magnetization crosses the 
upper magnetization threshold $m_{up} = 0.9$. In this case $\tau_0 =t_1 + t_2 \sim 10^{30}$ 
MCSS, and notice that $t_3 \sim 3.5 \times 10^5$ MCSS.
}
\label{evol}
\end{figure}

As it is clear in Fig. \ref{evol}, the relaxation of $m(t)$ occurs via
a sequence of well$-$defined abrupt jumps. That is, when the system relaxation
is observed after each MCSS, which corresponds to a `macroscopic' time scale,
strictly monotonic changes of $m(t)$ can be identified that we shall
call \textit{avalanches} in the following. One may think of other definitions 
of these avalanches, of course, so that we are somewhat arbitrary at this point, 
but we believe this does not significantly affect our results in this paper. Moreover, 
since our model system is subject to dynamical fluctuations, a clear$-$cut definition of 
avalanche is not possible, and thus this lack of precision will always exists. 
The important fact is that fluctuations on the microscopic, single$-$spin$-$flip 
scale are despised.

To be precise, consider the avalanche beginning at time $t_{a}$, when the system
magnetization is $m(t_{a})$, and finishing at $t_{b}$. We
define its \textit{size} and \textit{lifetime} or \textit{duration},
respectively, as $\Delta_{m}= | m(t_{b})-m(t_{a}) |$ and
$\Delta_{t}= | t_{b}-t_{a} |$. (We also studied $m(t_{b})-m(t_{a})$, 
i.e., positive and negative events. Such detail turns out to be
irrelevant for the purposes here, however; in fact, large events cannot be
positive in practice.). Alternatively, let us consider now the time evolution of the 
\textit{activity}, defined as $A(t)= m(t+1)-m(t)$. It is just the magnetization 
change in an unit time (measured in MCSS). This observable is the analogous to voltage pulses 
observed in experiments on magnetic noise. \cite{Kuntz, Durin} An avalanche is thus comprised between 
two successive crossings of $A(t)$ with the origin. In this way, the avalanche size is proportional to 
the area below this curve, and its duration or lifetime is defined as the time interval 
between such crossings. Our interest is on the histograms $P(\Delta_m)$, $P(\Delta_t)$ and 
$P(\Delta_t|\Delta_m)$.

\begin{figure}[t]
\centerline{
\psfig{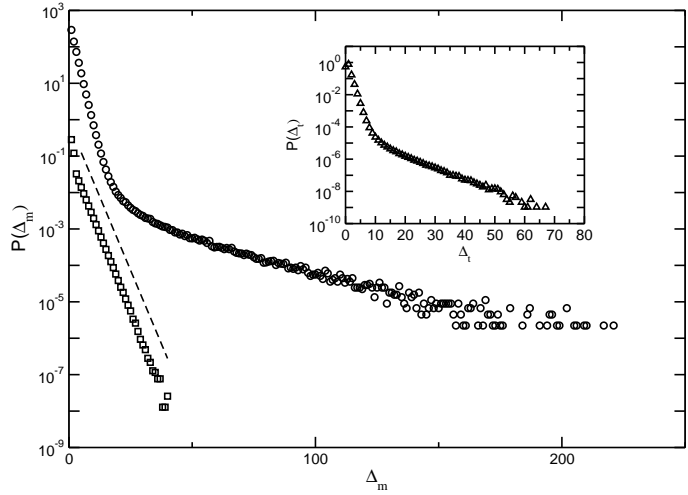}}
\caption[Semilog plot of the whole avalanche size distribution.]
{\small Here we show the avalanche size distribution for a circular magnetic nanoparticle of radius $R=30$ 
($\bigcirc$) and the same histogram for the semi--infinite system defined in the text ($\square$). 
Results have been shifted in the vertical direction for the shake of clarity.
The slope of the dashed line is a theoretical prediction for small avalanches. Notice the good accordance 
between theory and simulations. The inset shows the avalanche lifetime distribution for this circular particle. 
}
\label{comparacion}
\end{figure}

\subsection{Global Avalanche Distributions}
\label{capAval_apMC_subGlob}

Fig. \ref{comparacion} shows, among other things, a semilog plot of $P(\Delta_m)$ for a circular 
nanoparticle with radius $R=30$ (i.e. $N=2828$ spins, as compared 
to $N \in [10^3,10^4]$ spins in real magnetic nanoparticles), as obtained after performing $10^4$ different 
runs. Direct inspection of $P(\Delta_m)$ reveals the existence of two well$-$defined different regimes: 
$(i)$ An exponential regime for small avalanches ($\Delta_m < 20$), 
$P(\Delta_m)\sim \text{exp}(-\Delta_m / \bar{\Delta}_{small})$, where 
events have a typical size, namely $\bar{\Delta}_{small} \approx 1.4$, and $(ii)$ a second regime for
larger avalanches with no easily identifiable distribution. In order to elucidate the nature of both 
regimes, and the physical origin of this crossover, we now introduce a modification of our model, where 
boundary effects are effectively absent.

Let us study again our system on the $L \times \infty$ square lattice we introduced in the previous chapter 
(see section \ref{capNuc_apExt_subVel} and Fig. \ref{sketch-seminf}) when studying the growth velocity of the interface. 
There we defined the system in a $L \times L$,
with periodic boundary conditions along the $\mathbf{\hat{x}}$ (horizontal) direction, while the lattice is open along 
the $\mathbf{\hat{y}}$ direction. The initial configuration in this case consists in two horizontal stripes of height,
respectively, $(1-\alpha)L$ (upper part of the cylinder), in which all spins are set $s_{i}=+1$, and $\alpha L$, 
in which all spins are set $s_{i}=-1$, with $\alpha=0.15$. The interface moves upwards for the parameters
in our simulations (remember, $T=0.11T_C$, $h=-0.1$ and $p=10^{-6}$). In order to simulate an infinite system in
the interface movement direction, we perform a shift of the lattice in such a way that the interface does not {\it feel}
the presence of the boundary and it advances indefinitely (see section \ref{capNuc_apExt_subVel} for more details). We 
choose $L=53$, and thus $N=2809$, very similar to the number of spins in the circular system. This system 
simulates the temporal evolution of a model's domain wall in an semi-infinite bulk.

Domain wall motion in this system proceeds also by avalanches (as defined above). 
Fig. \ref{comparacion} also shows the 
histogram $P(\Delta_m)$ for this semi-infinite system, where boundary effects are negligible. Comparing 
this histogram with the same curve for the circular particle, we observe that although the initial 
exponential regime is almost identical in both cases, the tail of $P(\Delta_m)$ corresponding to the 
second regime does not exist for the semi-infinite system. A similar result holds if we define our model 
in a square lattice with periodic boundary conditions, as we did in previous chapters, or if we set $p=0$
in the circular nanoparticle. Thereby, the combined action of both free boundaries and impurities induces a new 
mechanism which is behind the large avalanches and essentially differs from the standard bulk noise driving the 
system and causing small, exponentially distributed avalanches only. It can be demonstrated analytically
(see Appendix \ref{apendAval} for the details) that small events are {\it local} random fluctuations of a growing 
{\it flat} domain wall. It can be also shown that this small avalanches follow an exponential distribution of the 
form $P_{small}(\Delta_m) \sim \text{exp}(-\Delta_m / \bar{\Delta}_{small})$, with a typical size,
\begin{equation}
\bar{\Delta}_{small} = \frac{1}{\displaystyle \ln \big[\frac{(1+p)(1+\text{e}^{2\beta|h|})}{p+
\text{e}^{2\beta|h|}}\big]}
\end{equation}
where $\beta = 1/k_{B}T$, and we set the Boltzmann constant $k_{B} = 1$. Fig. \ref{comparacion} also 
shows the analytical $P_{small}(\Delta_m)$ for comparison; the agreement with simulation is excellent.

\begin{figure}[t]
\centerline{
\psfig{file=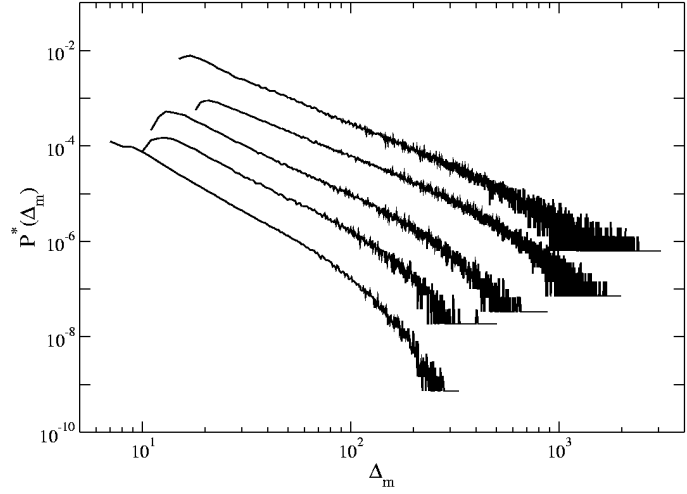,width=9cm,angle=0}}
\caption[Large avalanche size distribution showing power law behavior.]
{\small Large avalanche size distribution $P^*(\Delta_m)$ for the circular magnetic nanoparticle and for $5$ 
different sizes. From top to bottom, $R=120$, $84$, $60$, $42$ and $30$. Curves have been shifted in the vertical direction 
for visual convenience. The accumulated statistics goes from 15000 runs for the smallest size to 7000 experiments for the 
largest one.
}
\label{power-tam}
\end{figure}

\begin{figure}[t]
\centerline{
\psfig{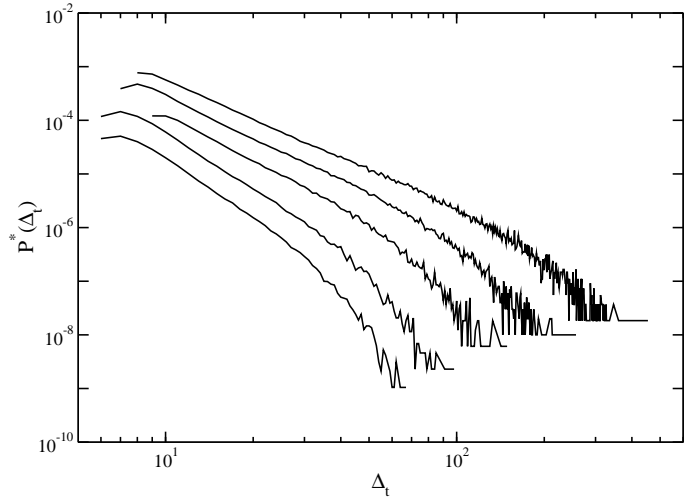}}
\caption[Large avalanche lifetime distribution with power law decay.]
{\small Large avalanche lifetime distribution $P^*(\Delta_t)$ for several different sizes of the circular 
magnetic nanoparticle. The sizes $R$ decrease from top to bottom, and take the same values as in the previous figure.
Again, curves have been shifted in the vertical direction for visual convenience.
}
\label{power-dur}
\end{figure}

\subsection{Large Avalanche Distributions}
\label{capAval_apMC_subLar}

Our aim in this chapter is to analyze large avalanches in the circular magnetic nanoparticle, its distribution and 
its origin. In order to do so, 
we must filter in some way the trivial small avalanches (\textit{extrinsic noise} \cite{Spaso}) reported 
above, since they soil the main signal. There are many different methods all over the literature to 
perform such filtering.\cite{Spaso,Frette} In our case, we just subtract the fitted exponential 
behavior for small avalanches from the global histogram%
\footnote{In practice, we must subtract two slightly different exponential distributions for small 
avalanches. As stated previously, small (trivial)  avalanches are random local fluctuations of a growing 
flat domain wall. However, the critical droplet which appears as the circular nanoparticle demagnetizes 
shows different flat fronts (see Fig. \ref{snap-circle}). Trivial avalanches near the corner formed 
between two different flat domain walls have a slightly larger typical size due to surface tension effects. 
This subtle effect must be taken into account in order to obtain the clean power law distributions 
observed in Fig. \ref{power-tam}}. 
This process yields the distributions $P^*(\Delta_m)$ for large avalanches shown in Fig. 
\ref{power-tam}. A power law behavior, followed by an (exponential) cutoff for very large 
avalanches, is thus clearly observed. 

In particular, Fig. \ref{power-tam} shows large avalanche size
distributions $P^*(\Delta_m)$ for $5$ different sizes of the magnetic nanoparticle, namely 
$R=30$, $42$, $60$, $84$ and $120$ spins.
The measured power law exponents, $P^*(\Delta_m) \sim \Delta_m^{-\tau (R)}$, are shown in table \ref{tabla_R}. 
As observed, we find size-dependent corrections to scaling for the exponent $\tau$. Similar corrections have been 
also found in real experimental systems. \cite{Frette} The observed finite$-$size corrections are compatible with a 
functional dependence of the form $\tau (R) = \tau_{\infty} + a/R^2$, where  
$\tau_{\infty} = 1.71(4)$.\footnote{The fact that $\tau_{\infty} < 2$ involves that the mean value of the size power 
law distribution does not exist in the Thermodynamic Limit. \cite{Jensen}}
Analogously, we can perform a similar analysis for the avalanche lifetime distributions, 
$P^*(\Delta_t) \sim \Delta_t^{-\alpha (R)}$. Fig. \ref{power-dur} shows 
$P^*(\Delta_t)$ for large avalanches (once the extrinsic noise has been subtracted) and for the system sizes 
reported above. The measured exponents $\alpha (R)$ are also shown in table \ref{tabla_R}.
Again, these are compatible with a law 
$\alpha (R) = \alpha _{\infty} + a'/R^2$, where $\alpha _{\infty} = 2.25(3)$. 
Hence we expect avalanche power law distributions $P^*(\Delta_m) \sim \Delta_m^{-\tau _{\infty}}$ and 
$P^*(\Delta_t) \sim \Delta_t^{-\alpha _{\infty}}$, with $\tau_{\infty} = 1.71(4)$ and $\alpha _{\infty} = 2.25(3)$, in 
the Thermodynamic Limit. At a first glance this result could seem unphysical, because we have proven before that 
large avalanches are due to the presence of free boundaries, whose importance diminish as the system grows (i.e. as the 
surface/volume ratio goes to zero). However we will prove later in this chapter that the mechanism responsible of 
large avalanches (which appears due to  free boundaries) remains relevant in the Thermodynamic Limit.%
\footnote{In fact, the importance of boundary conditions in the Thermodynamic Limit is not a new idea in Physics. 
For instance, for problems related with (thermal) conductivity, the boundary heat reservoirs play a fundamental role
in order to understand the nonequilibrium physics in the Thermodynamic Limit.}

\begin{figure}[t]
\centerline{
\psfig{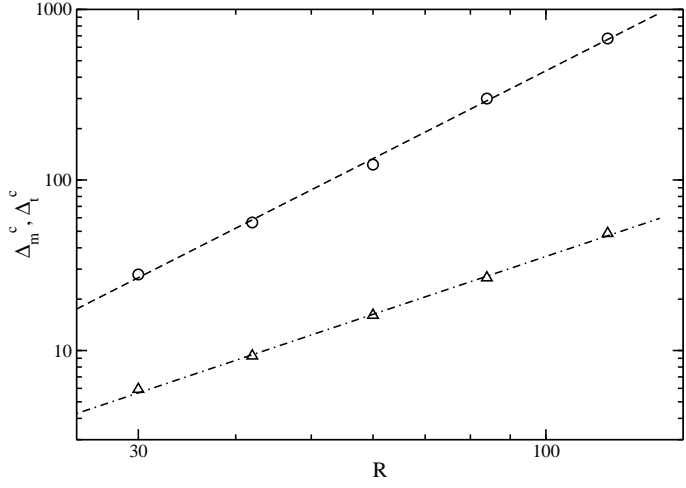}}
\caption[Cutoff dependence on system size.]
{\small Log--log plot of the cutoff dependence with system size. In particular, we observe that 
$\Delta^c_m \sim R^{\beta_m}$ ($\bigcirc$) and $\Delta^c_t \sim R^{\beta_t}$ ($\bigtriangleup$), with $\beta_m=2.32(6)$ 
and $\beta_t=1.53(3)$.
}
\label{cutoff}
\end{figure}

On the other hand, as stated previously, the power law behavior of both $P^*(\Delta _m)$ and $P^*(\Delta _t)$ lasts 
up to an exponential cutoff $\Delta_m^c$ and $\Delta_t^c$, respectively, which depends on system size. 
We measure these cutoff values  (see table \ref{tabla_R}) fitting an exponential function of the form 
$exp[-\Delta_{m(t)}/\Delta_{m(t)}^c]$ to the cutoff tails, and find a power law dependence with $R$, i.e. 
$\Delta_m^c \sim R^{\beta_m}$ and $\Delta_t^c \sim R^{\beta_t}$, where 
$\beta_m = 2.32(6)$ and $\beta_t = 1.53(3)$ (see Fig. \ref{cutoff}). Analogous power law dependences of cutoff 
with system size have been found in real magnetic materials. \cite{Bahiana} Typically, this power law dependence of 
cutoff with system size has been identified as a sign of genuine critical behavior. \cite{Jensen} However, as we will 
prove below, this is not the case here. 

\begin{table}[t]
\centerline{
\begin{tabular}{|c||c|c|c|c|c|}
\hline
R & 30 & 42 & 60 & 84 & 120 \\
\hline \hline
$\tau$ & 2.76(2) & 2.24(2) & 2.06(2) & 1.77(2) & 1.77(2) \\
\hline
$\Delta_m^c$ & 27.9(7) & 56.3(14) & 123(3) & 300(9) & 675(16) \\
\hline
$\alpha$ & 3.70(2) & 3.52(2) & 2.85(2) & 2.61(2) & 2.39(2) \\
\hline
$\Delta_t^c$ & 5.92(20) & 9.3(3) & 16.1(6) & 26.7(7) & 48.6(12) \\
\hline
\end{tabular}
}
\caption[Size dependence of both the power law exponents and cutoffs.]
{\small Here we show the size dependence of both the power law exponents, $P^*(\Delta_m) \sim \Delta_m^{-\tau (R)}$ 
and $P^*(\Delta_t) \sim \Delta_t^{-\alpha (R)}$, and the observed exponential cutoffs, $\Delta^c_m (R)$ and $\Delta^c_t (R)$.
}
\label{tabla_R}
\end{table}

\begin{figure}[t]
\centerline{
\psfig{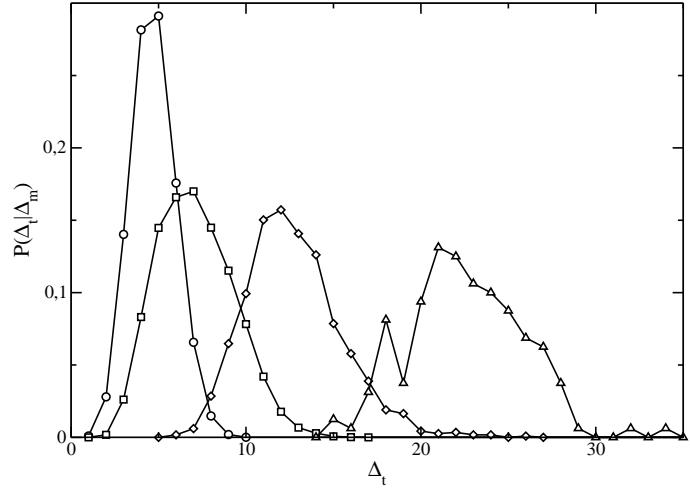}}
\caption[Marginal distribution $P(\Delta_t|\Delta_m)$.]
{\small Normalized marginal distribution $P(\Delta_t|\Delta_m)$ for a circular nanoparticle of radius $R=30$,
after performing 15000 different experiments. In particular, we show $P(\Delta_t|\Delta_m)$ for $\Delta_m=10$ 
($\bigcirc$), $20$ ($\square$), $50$ ($\Diamond$) and $100$ ($\bigtriangleup$). 
Notice  the peaked form of the distribution, and that larger avalanche sizes involve longer lifetimes.
}
\label{doble-hist}
\end{figure}

We can now wonder about the relation between the size and the lifetime of an avalanche. Moreover, we would 
like to know if there is any relation between them at all. With this aim we study the histogram 
$P(\Delta_t|\Delta_m)$, i.e. the probability of measuring an avalanche with lifetime $\Delta_t$ when its size is $\Delta_m$. 
Fig. \ref{doble-hist} shows $P(\Delta_t|\Delta_m)$ for a circular magnetic nanoparticle 
of radius $R=30$. The first relevant conclusion derived from this plot is that the distribution $P(\Delta_t|\Delta_m)$
is peaked. That is, for each value of $\Delta_m$, the marginal 
distribution $P(\Delta_t | \Delta_m)$ shows a narrow peak around certain typical value $\langle \Delta_t\rangle_{\Delta_m}$.
This means that the relation between the lifetime and the size of an avalanche is rather deterministic in our model 
system, for both large and small events. Let us assume that the relation is completely deterministic, and that for large 
enough avalanches it can be written as $\Delta_m \sim \Delta_t^{\gamma}$. This equation defines the exponent $\gamma$.%
\footnote{Many references about Barkhausen Noise \cite{Kuntz,Durin,Mehta,Travesset} refer to $\gamma$ as 
$1/\sigma \nu z$, where $\sigma$, $\nu$ and $z$ are critical exponents. These references assume the existence 
of an underlying critical point in order to explain Barkhausen noise.}
Hence, taking into account the conservation of probability, i.e. 
$P^*(\Delta_m) \text{d}\Delta_m = P^*(\Delta_t) \text{d}\Delta_t$, and assuming that both $P^*(\Delta_m)$ and 
$P^*(\Delta_t)$ are pure power law distributions, we can write a scaling relation among the different avalanche 
exponents,
\begin{equation}
\gamma = \frac{1-\alpha}{1-\tau}
\label{exponentes}
\end{equation}
This scaling relation predicts a value $\gamma = 1.76(8)$ in our system. In order to corroborate such prediction, 
we should calculate from the marginal distribution $P(\Delta_t|\Delta_m)$ the mean value 
$\langle \Delta_t \rangle_{\Delta_m}$ for each $\Delta_m$ in the scaling region, 
thus obtaining a relation $\langle \Delta_t \rangle_{\Delta_m} = {\cal G} (\Delta_m)$ 
which should yield the value of $\gamma$. 
However, we do not know the distribution $P_{small}(\Delta_t|\Delta_m)$ for the extrinsic noise, and it soils 
the scaling region, so this way of measuring $\gamma$ does not yield any definite conclusion. Instead, we are 
able to measure $\gamma$ in an indirect way using the cutoff dependence with $R$. As we previously pointed out, 
$\Delta_m^c \sim R^{\beta_m}$ and $\Delta_t^c \sim R^{\beta_t}$. On the other hand, using the relation
between $\Delta_m$ and $\Delta_t$ assumed above, we can write $\Delta_m^c \sim (\Delta_t^c)^
{\gamma}$, and thus
\begin{equation}
\gamma = \frac{\beta_m}{\beta_t}
\end{equation}
which yields $\gamma = 1.52(5)$. This indirectly measured value is similar, although not compatible, with the one predicted by 
the scaling relation (\ref{exponentes}). This is not a surprise, since the scaling relation (\ref{exponentes}) is
based on the assumption that both $P^*(\Delta_m)$ and $P^*(\Delta_t)$ are pure power law distributions. However,
as we will see later on in this chapter, these probability functions are not purely power-law distributed, but instead 
come from the superposition of many different exponential distributions. This observation breaks the above scaling relation,
eq. (\ref{exponentes}); in fact we can prove (see Fig. \ref{avalteoria} and complementary discussion) that the three
measured exponents, $\tau$, $\alpha$ and $\gamma$, are related in the framework of a superposition of typical scales.


Studying the power spectrum of the avalanches time signal constitutes another way of measuring 
$\gamma$. Many avalanche systems have a power spectrum which shows two different power law regions, one for 
low frequencies and another one for high frequencies. The low frequency region reflects the correlations between 
avalanches, while the high frequency region reflects the dynamics within avalanches. For avalanche systems 
with $\tau _{\infty} < 2$ it has been shown that the high frequency power spectrum decays as 
${\cal P}(\omega) \sim \omega^{-\gamma}$.\cite{Kuntz} Hence we could also obtain the value of $\gamma$ in 
this way (this analysis will be done in a forthcoming work).

\begin{figure}[t]
\centerline{
\psfig{file=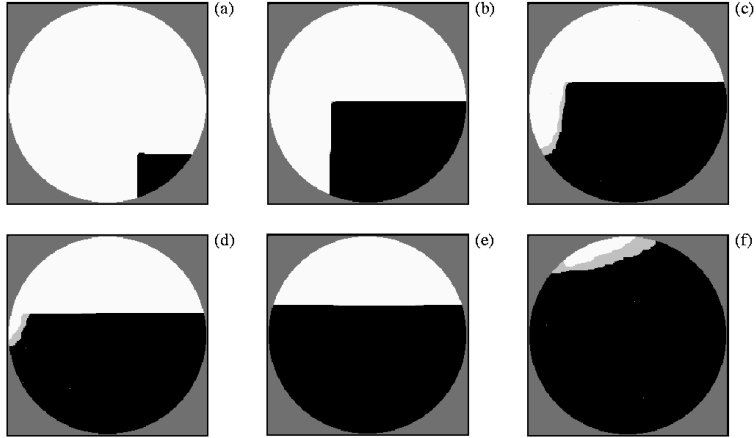,width=10cm,angle=0}}
\caption[Snapshots of the decay of a circular nanoparticle.]
{\small Some snapshots of a particular decay of a circular nanoparticle of radius $R=120$. Avalanches are plotted 
in grey. Notice that large avalanches appear only for curved domain walls.
}
\label{snap-circle}
\end{figure}

\markboth{Scale Free Avalanches during Decay from Metastable States}{\ref{capAval_apOrig} The Physical Origin of 
Scale Invariant Noise}
\section{The Physical Origin of Scale Invariant Noise}
\label{capAval_apOrig}
\markboth{Scale Free Avalanches during Decay from Metastable States}{\ref{capAval_apOrig} The Physical Origin of 
Scale Invariant Noise}

Let us summarize our results up to now. We have observed that the presence of free boundaries and impurities
in a magnetic system induces large avalanches in the demagnetization process from a metastable state. These large 
avalanches follow power law distributions for their sizes and lifetimes up to certain exponential cutoff, which depends 
algebraically on system size. Moreover, the power law distributions converge in the Thermodynamic Limit to size 
independent power laws, $P^*(\Delta_m) \sim \Delta_m^{-\tau _{\infty}}$ and 
$P^*(\Delta_t) \sim \Delta_t^{-\alpha _{\infty}}$ with $\tau _{\infty} = 1.71(4)$ and $\alpha _{\infty} = 2.25(3)$.
We have also found that the relation between the size and the lifetime of an avalanche is rather deterministic. In 
fact, this relation can be quantified with an exponent $\gamma=1.52(5)$.

A physicist trained in critical phenomena would say that these results stron-gly support the existence of an underlying 
continuous phase transition or second order critical point responsible of the observed scale invariance. This critical 
point would induce macroscopically large correlation lengths, and thus all scales in the system should be equally 
relevant. Moreover, the response to a small (microscopic) perturbation should appear at any scale (even at 
macroscopic ones). The characterization of this critical point should depend only on the symmetries and conservation 
laws present in the system, and not on the specific microscopic details.
However, as we will see below, the observed power law  behavior is \textit{effective}, in the sense that the system is 
not really critical (i.e. the system does not present any singularity for our parameters). 
Instead, we will show that a finite (but large) number of different, 
gap$-$separated typical scales appear superposed in such a way that the global distributions exhibit several decades 
of power law behavior, as shown in Figs. \ref{power-tam} and \ref{power-dur}.

\subsection{Avalanches and Domain Wall Curvature}
\label{capAval_apOrig_subDom}

First of all, let us understand the origin of large avalanches in the circular magnetic nanoparticle. Fig. 
\ref{snap-circle} shows some snapshots of the temporal evolution of a magnetic particle of radius $R=120$.
Due to the low temperature of simulations, the dynamics is restricted to the interface, the bulk being almost 
completely frozen. Hence the evolution of the system will be determined by the interface and its interplay
with the open boundaries. 
It can be also shown that for our parameter values, configurations with only flat domain walls are 
metastable\footnote{The stable phase shrinkage rate is larger than the stable phase growth rate for these
configurations (see Chapter \ref{capMedio}).}. 
In fact, there are free energy barriers which impede flat domain walls to advance. In this way, the system 
presents many different metastable states. On the other hand, in Fig. \ref{snap-circle} we observe again two 
different types of avalanches (plotted in grey): small, local avalanches associated with a 
flat domain wall, and large, global avalanches involving curved regions of the domain wall. In this latter case large 
avalanches appear because curvature increases domain wall interfacial energy, thus giving rise to large events 
towards configurations with less interfacial energy (i.e. less curvature). 
\begin{figure}[t]
\centerline{
\psfig{file=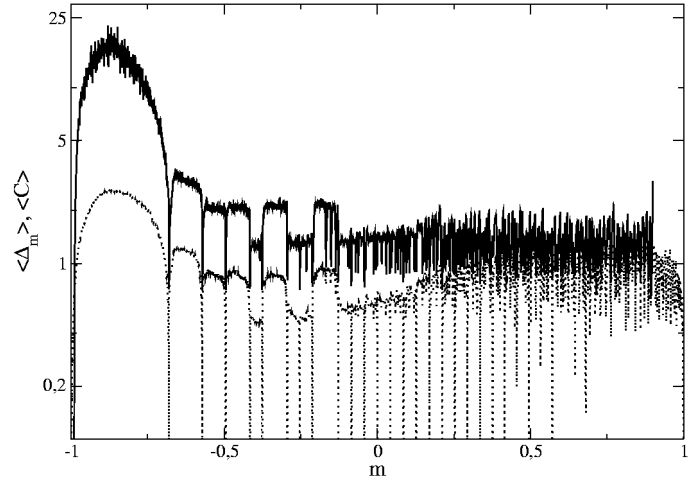,width=9cm,angle=0}}
\caption[Mean avalanche size and curvature as a function of $m$.]
{\small Semilog plot of the mean avalanche size (solid line) and the mean curvature (dashed line), $\langle\Delta_m\rangle$ 
and $\langle{\cal C}\rangle$ respectively, as functions of magnetization, after averaging over 3500 different runs. It is 
clear from this plot that, in average, large avalanches appear in the final part of the evolution. There are also some values 
of magnetization characterized by a very small mean avalanche size. In any case, this function shows clearly non--trivial 
structure. The same is true for $\langle {\cal C}\rangle (m)$. Also noticiable is the high degree of correlation between 
$\langle\Delta_m\rangle(m)$ and $\langle{\cal C}\rangle(m)$
}
\label{reprod1}
\end{figure}

\begin{figure}[t]
\centerline{
\psfig{file=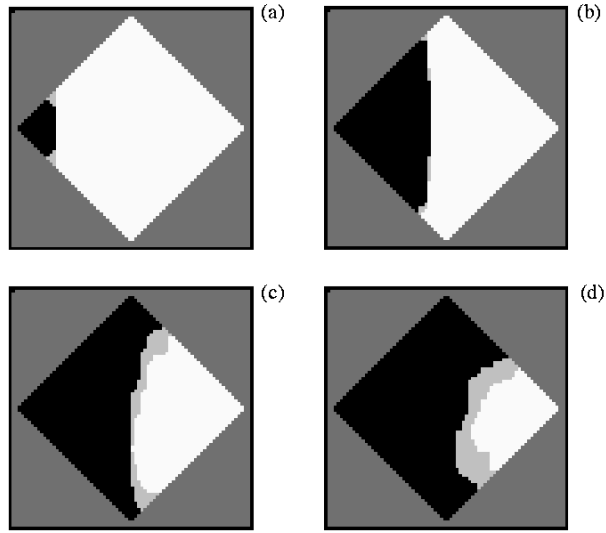,width=8cm,angle=0}}
\caption[Snapshots of the rhombus demagnetization process]
{\small Some snapshots of the rhombus demagnetization process. Again, avalanches are plotted in grey. Notice that the 
domain wall remains flat when the contact angle with the open boundary is greater that $90^o$, and then only small avalanches 
appear. On the other hand, the domain wall gets curved when the contact angle is less that $90^o$, and large avalanches do 
appear.
}
\label{snap-rombo}
\end{figure}
In order to understand more deeply the relation between avalanches and domain wall curvature, let us study both the
mean avalanche size and the mean domain wall curvature as a function of system magnetization, $\langle \Delta_m \rangle (m)$
and $\langle{\cal C}\rangle (m)$ respectively. We will take as a measure of curvature ${\cal C}$ the number of step--like up 
spins in the interface between the stable phase and the metastable one, where step--like up spins are interfacial spins in 
class 3 (see Table \ref{tabclasesbis}), i.e. up spins with two up neighbors and two down neighbors%
\footnote{This way of measuring curvature is valid only for low temperatures, since in this case clusters 
are very compact.}. Fig. \ref{reprod1} shows $\langle \Delta_m \rangle$ and $\langle{\cal C}\rangle$ as a function
of magnetization $m$ for a particle of radius $R=30$, after averaging over 3500 different demagnetization experiments.
Here we observe that both $\langle \Delta_m \rangle (m)$ and $\langle{\cal C}\rangle (m)$ are highly non-homogeneous
functions of magnetization, showing a non-trivial structure. In particular, we observe that there are certain well-defined 
magnetization values for which avalanches are typically very small. These magnetization values correspond to configurations
where domain walls are flat, as can be deduced from the curvature plot. On the other hand, we observe that there are other 
magnetization values where typical avalanches involve many spins. In particular, we observe in Fig. \ref{reprod1} that 
for magnetizations in the range $m\in [-1,-0.75]$ the avalanche mean size is much larger than for other magnetizations
(compare these results with the snapshots shown in Fig. \ref{snap-circle}).
A main conclusion derived from Fig. \ref{reprod1} is the high correlation existing between the typical avalanche size and
the domain wall curvature, as defined above. Large curvature implies large avalanches and reversely. Hence the size of an 
avalanches is perfectly determined by the curvature of the interface when this avalanche starts.

\begin{figure}[t]
\centerline{
\psfig{file=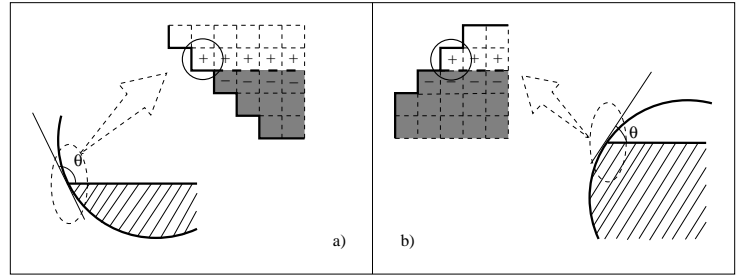,width=10cm}}
\caption[Sketch of the contact angle effect.]
{\small Sketch of the contact angle effect described in the text. In $(a)$ the contact angle $\theta$ is larger 
than $90^{\text{o}}$, and microscopically border up spins near the interface have a low probability of being flipped if 
selected. On the other hand, in $(b)$ the contact angle $\theta$ is smaller than $90^o$, and border up spins near the 
interface have probability one of being flipped if selected.
}
\label{sketch-curvatura}
\end{figure}

Our next question concerns the origin of 
domain wall curvature. As it can be easily guessed, its origin underlies on the interplay between the interface and the open 
boundaries. However, in order to understand better the mechanism which gives 
rise to curved interfaces as the critical droplet grows, we study now demagnetization from a metastable state in
our model system defined on a rhombus with free boundary conditions. Fig. \ref{snap-rombo} shows some snapshots
of the rhombus demagnetization process. For our parameter values this system demagnetizes through the nucleation of a 
single critical droplet. As predicted in recent theoretical studies, \cite{Cirillo} this critical droplet always appears 
in one of the four corners of the rhombus geometry. It is clear from Fig. \ref{snap-rombo} that while the domain wall 
advances forming a convex angle ($\theta >90^{\text{o}}$) with the open boundary the interface remains flat, and only small 
avalanches appear. However, when the contact angle is concave ($\theta <90^{\text{o}}$), the interface gets curved, as a 
consequence of the faster growth of the domain wall near the concave open borders, and large avalanches towards 
configurations with less interfacial energy develop. Microscopically, the interface gets curved because for concave 
contact angles the unflipped border spins near the interface have a probability one of being flipped if selected. This is 
shown in Fig. \ref{sketch-curvatura}, where this mechanism is sketched for the circular particle. In this way, as the 
interface advances subject to concave boundary conditions, it gets curved.

\begin{figure}[t]
\centerline{
\psfig{file=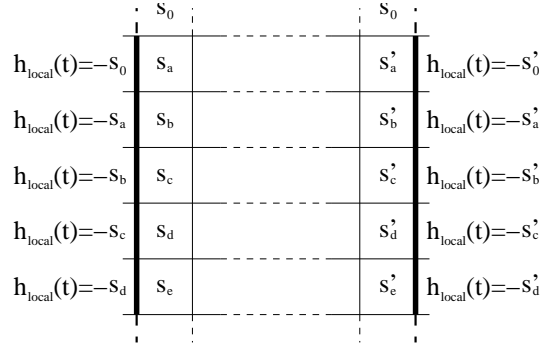,width=7cm}}
\caption[Sketch of the (concave) dynamic boundary conditions.]
{\small Sketch of the dynamic boundary conditions in the $\hat{x}$--direction used in the new system explained in 
the text. Here we represent the first and last columns of the system, and the local, dynamic magnetic field that each spin in these
columns suffers as the system evolves in time. The effect of these dynamic boundary conditions is equivalent to eliminate 
the interaction of each border spin with its up neighbor, i.e. concave boundary conditions. 
}
\label{sketch-concavo}
\end{figure}

\subsection{Avalanche Statistics for Constant Domain Wall Curvature}
\label{capAval_apOrig_subCurv}

We have shown that large avalanches originate due to domain wall curvature, and that the size of the avalanche is intimately
related to the specific curvature of the interface. 
Now we want to know the distribution of avalanches for a 
given curved domain wall. That is, we wonder what is the probability of finding an avalanche of a given size $\Delta_m$
if the domain wall has certain constant curvature.
In order to find out which is this relationship we have designed an (unrealistic) modification of 
our basic system, where an interface with constant (up to small fluctuations) non-zero curvature develops, evolving
via avalanches. Let us define again our system in a $L_x \times \infty$ lattice, with concave open boundary conditions in 
the $\hat{x}$ direction. This is done in practice in the following way. 
The system is set on a $L_x \times L_y$ square 
lattice, with very particular boundary conditions. The lattice is open in the $\hat{y}$ direction. Spins in the upper row 
are fixed to $+1$, and spins in the lower row are fixed to $-1$. On the other hand, boundary conditions in the $\hat{x}$ 
direction are \textit{dynamic}: the lattice is also open in the $\hat{x}$ direction, although each spin in the first and 
last column suffers an additional dynamic magnetic field, equal at any time to the negative value of its up neighbor. This 
is sketched in Fig. \ref{sketch-concavo}. For each spin in the first and the last columns, the effect of these dynamic 
boundary conditions is to effectively decouple this spin with its up neighbor. In this way we emulate a concave, step--like 
border (as the one found by the interface in the rhombus system in the second part of its evolution), with a fixed
distance between both concave borders. The initial configuration is identical to the one exposed for the semi--infinite 
system described previously, and we shift the system in a similar way in order to get an infinitely evolving domain wall. 
Thus, we simulate a semi--infinite system subject to open concave boundary conditions with a fixed distance $L_x$ between 
the concave borders.


\begin{figure}[t]
\centerline{
\psfig{file=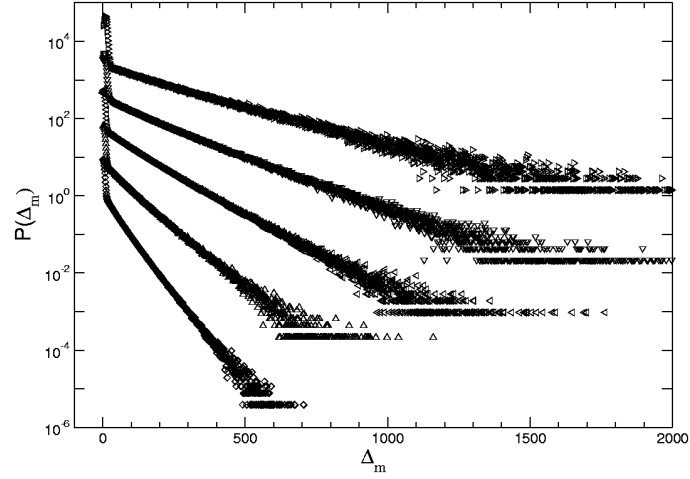,width=9cm,angle=0}}
\caption[Avalanche size distribution for the concave semi-infinite system.]
{\small Avalanche size distribution $P(\Delta_m)$ for the semi--infinite system with concave open boundary 
conditions described in text. Here we show $P(\Delta_m)$ for different system sizes, namely (from bottom to top) 
$L_x=20$, $25$, $30$, $35$ and $40$. The large avalanche tails are stretched exponentials, with well defined typical size. 
Distributions have been shifted in the vertical direction for visual convenience.
}
\label{aval-concavo}
\end{figure}

After a short transient, the initially flat domain wall reaches an stationary state, with an almost constant 
(up to small fluctuations) non--zero curvature, which depends on the size $L_x$. Measuring curvature as stated previously, 
we notice that the steady domain wall curvature in this system is proportional to the 
system size $L_x$. On the other hand, this system evolves through avalanches, whose distribution we can measure. 
Fig. \ref{aval-concavo} shows the avalanche size distribution $P(\Delta_m)$ for this system and several values of $L_x$ 
(i.e. several different curvatures). As for the circular nanoparticle, we observe that $P(\Delta_m)$ shows two different 
well defined regions, the first one being related with the exponentially distributed extrinsic noise, explained before. 
The second regime is compatible with a stretched exponential function of the form
\begin{equation}
P_2(\Delta_m) = \text{e}^{-(\Delta_m / \bar{\Delta}_2)^{\eta}}
\end{equation}
where we measure $\eta \approx 0.89$, independent of system size for large enough values of $L_x$. This stretched exponential 
function is characterized by a \textit{typical} size $\bar{\Delta}_2$, which depends on $L_x$ (i.e., on curvature) 
in an exponential fashion.
These results clarify what is going on in the circular magnetic 
nanoparticle. First, we observe that an interface with a given curvature evolves through large avalanches of a well 
defined typical size. Moreover, this typical size $\bar{\Delta_2}$ strongly (i.e. exponentially) 
depends on domain wall curvature. On the other 
hand, as we observed in Fig. \ref{snap-circle}, the interface of the growing critical droplet in the circular magnetic 
nanoparticle gets curved as the droplet grows. Furthermore, interfacial curvature, as defined above, takes a wide 
range of different values as the particle demagnetizes. Thus, the large avalanche distributions observed in the circular 
particle are just superpositions of distributions with well defined typical sizes.

The above results have another interesting implication. We have observed that the large avalanche typical size grows 
\textit{exponentially fast} with the system size $L_x$. 
Hence, although the origin of large avalanches is connected 
to the presence of an open boundary (which is a surface effect), the mechanism which gives rise to large avalanches will 
be relevant in the Thermodynamic Limit, and large (infinite) avalanches will appear in this limit.

\begin{figure}[t]
\centerline{
\psfig{file=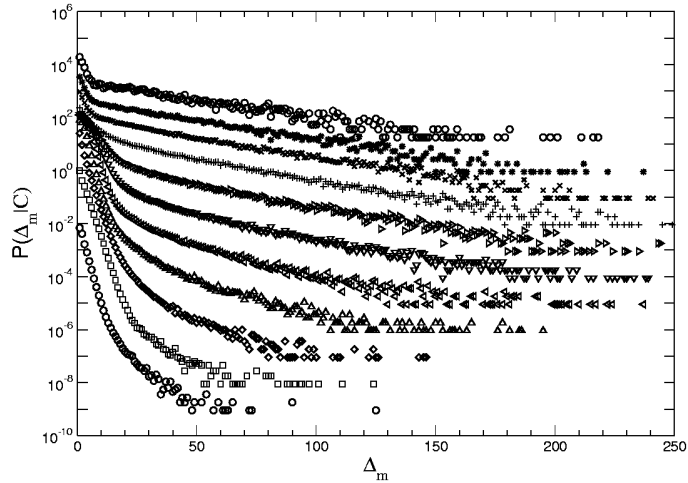,width=9cm,angle=0}}
\caption[Avalanche size distribution for several fixed curvatures.]
{\small Avalanche size distribution for several fixed curvatures (measured as explained in the text) for a 
circular nanoparticle of radius $R=30$, after performing 15000 different demagnetization experiments. 
Curvature grows from bottom to top. Notice that, as curvature increases, the large 
avalanche distribution for this curvature, which is an stretched exponential, increases its typical size.
Curves have been shifted in the vertical direction for visual convenience.
}
\label{aval-circle-curva}
\end{figure}

The above results suggest us to study avalanche distributions in the circular nanoparticle as a function of domain wall 
curvature (as defined previously). Fig. \ref{aval-circle-curva} shows $P(\Delta_m |{\cal C})$ for the circular
nanoparticle, which is the probability of 
measuring an avalanche of size $\Delta_m$ when the interfacial curvature is ${\cal C}$. Here we note that 
the domain wall takes many different curvature values as demagnetization proceeds. For each constant curvature, the tail 
of the avalanche size distribution follows again an stretched exponential law, whose typical size grows with curvature.

This confirms what we  previously stated, i.e. that large avalanches in the circular magnetic nanoparticle have a large 
(but finite) number of different, gap--separated typical sizes. Hence, the power law observed initially for the 
avalanche size distribution (see Fig. \ref{power-tam}) is \textit{effective}: our system is not critical. Although we 
observe an effective scale invariance, the system's correlation lengths are non--divergent, but finite. Moreover, if we 
slightly perturb the system, its response will be equally small.

\subsection{Power Law Behavior as a Consequence of Superposition of Many Different Typical Scales}
\label{capAval_apOrig_subMath}

There is still an annoying remaining question: how can a finite superposition of distributions with well defined 
different typical scales give rise to a scale invariant power law global distribution ?. In order to answer this question 
we solve a mathematical exercise. Let us assume that we have a system that evolves via burst-like events of magnitude $x$.
The state of this system is characterized by an observable $A$, in such a way that the probability of finding an event
of magnitude $x$ when the system is in a state characterized by $A$ is ${\cal P}(x|A)=A\text{exp}(-Ax)$. 
Therefore the observable $A$ fixes the typical scale of burst-like events. If ${\cal Q}(A)$ is the
probability of finding the system in a state characterized by $A$, the probability of finding an event of size $x$ is,
\begin{equation}
{\cal P}(x)=\int_0^{\infty} {\cal Q}(A){\cal P}(x|A) \text{d}A
\label{superposexp}
\end{equation}
where we have assumed that the observable $A$ varies continuously from $0$ up to $\infty$. Hence, due to the exponential 
form of ${\cal P}(x|A)$, ${\cal P}(x)$ is just the Laplace transform of $A{\cal Q}(A)$. If we now assume the simplest
case, namely that the distribution ${\cal Q}(A)$ is flat so all possible values of $A$ are equally probable, we obtain
${\cal P}(x)\sim x^{-2}$. Hence, an {\it infinite} superposition of exponential distributions with well-defined typical scales
(defined by $A$) yields a power law (scale free) probability function with exponent $-2$. This result is not surprising. 
Power law behavior is a consequence of the lack of any typical scale in the system.
However, we can interpret this lack of typical scale in the opposite way, saying that in a system exhibiting power law behavior
all possible scales are present, so there is no {\it typical} scale. 
This is what eq. (\ref{superposexp}) says.
\begin{figure}[t]
\centerline{
\psfig{file=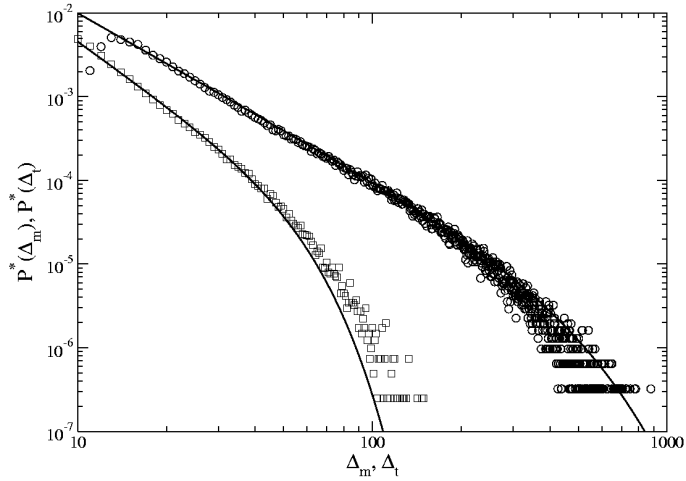,width=9cm,angle=0}}
\caption[Superposition of exponential distributions.]
{\small Large avalanche size ($\bigcirc$) and lifetime ($\square$) distributions for a circular nanoparticle
with radius $R=60$ (these results are the same than those plotted in Figs. \ref{power-tam} and \ref{power-dur}). The upper line
corresponds to the prediction derived from eq. \ref{powerdiscreta} for $N=200$, $A_{min}=0.007$ and $A_{max}=1$. The lower
line is derived from eq. \ref{powerdiscreta} assuming that $\Delta_m \sim \Delta_t^{\gamma}$, where we use $\gamma=1.52(5)$,
the value we previously measured.}
\label{avalteoria}
\end{figure}

However, avalanches in our magnetic nanoparticle show a {\it finite} set of well-defined typical scales\footnote{We assume
for the following calculation that these typical scales are characterized by exponential distributions, instead of
stretched exponentials, as we found in previous sections. We think that this simplification, which makes easy the calculation,
does not involve any fundamental difference in the consequent physical discussion.}. In this case the
continuous sum in eq. (\ref{superposexp}) must be substituted by a discrete sum over the the typical scales present in the 
system. In order to explicitely perform the calculation, let us assume that we have $N+1$ different typical scales $A_n$, with
$n\in [0,N]$, all of them in a finite interval, $A_n \in [A_{min},A_{max}]$ $\forall n\in [0,N]$. We further assume that 
these typical scales are equally spaced in this interval, so $A_n = A_{min} + n\Delta$, where $\Delta = (A_{max}-A_{min})/N$.
Therefore, assuming again that ${\cal Q}(A)$ is a constant function, we now write,
\begin{eqnarray}
{\cal P}(x) & = & \sum_{n=0}^N A_n \text{e}^{-A_nx} \Delta = 
\frac{\Delta \text{e}^{-A_{min}x}}{1-\text{e}^{-\Delta x}}
\Big[A_{min} - A_{max} \text{e}^{-(N+1)\Delta x} \nonumber \\
& - & \Delta\frac{1-\text{e}^{-N\Delta x}}
{1-\text{e}^{\Delta x}}\Big]
\label{powerdiscreta}
\end{eqnarray}
where we have used eqs. (\ref{sum2}) and (\ref{sum6}). Fig. \ref{avalteoria} shows the event size distribution ${\cal P}(x)$
obtained from eq. (\ref{powerdiscreta}) for $N=200$, $A_{min}=0.007$ and $A_{max}=1$. 
As we can observe in this figure, the curve follows power law behavior up to an exponential cutoff given by 
$\text{exp}(-A_{min}x)$.
Fig. \ref{avalteoria} also shows for comparison the avalanche size distribution measured for a circular particle with $R=60$.
The agreement between the theoretical curve based on a finite superposition of different typical scales and the measured
distribution is very good. Moreover, assuming that eq. (\ref{powerdiscreta}) represents the avalanche size distribution, and 
using the relation $\Delta_m \sim \Delta_t^{\gamma}$ between the size and the lifetime of an avalanche, we can obtain the
avalanche lifetime distribution via the conservation of probability. Thereby, if ${\cal P}(\Delta_m)$ is the probability of 
finding an avalanche with size $\Delta_m$, the probability of finding an avalanche with lifetime $\Delta_t$ is
$\gamma \Delta_t^{\gamma-1} {\cal P}(\Delta_t^{\gamma})$. This curve, also shown in Fig. \ref{avalteoria} for the same parameters
described above, agrees with the measured avalanche lifetime distribution for the $R=60$ magnetic particle when we use 
the previously measured value $\gamma=1.52(5)$. This agreement confirms the measured value for the exponent 
$\gamma$, and on the other hand it also strengthens our conclusion about the origin of the scale invariance in this problem.
Hence the superposition of a finite (but large) number of exponential distributions with different 
typical rates results in a global distribution which shows several decades of power law behavior, together with an 
exponential cutoff corresponding to the slowest exponential typical rate.

\markboth{Scale Free Avalanches during Decay from Metastable States}{\ref{capAval_apConc} Conclusions and Outlook}
\section{Conclusions and Outlook}
\label{capAval_apConc}
\markboth{Scale Free Avalanches during Decay from Metastable States}{\ref{capAval_apConc} Conclusions and Outlook}

In this chapter we have studied how an impure ferromagnetic nanoparticle evolves from the metastable phase towards the
stable one. Under the combined action of both impurities (which involve nonequilibrium conditions) and free borders, 
the formation of a nucleus of the stable phase turns out to proceed by avalanches. This burst-like evolution
characterizes the dynamics of many nonequilibrium system.\cite{Jensen} In addition to small events, which show 
up as a completely random, thermal effect (extrinsic noise\cite{Spaso}; see Appendix \ref{apendAval}), we have described 
{\it critical avalanches}. These are typically much larger than the extrinsic noise, while they {\it apparently} show
no temporal and spatial scale. We find for these large avalanches size and lifetime distributions which follow
power laws, $P(  \Delta_m)  \sim \Delta_m^{- \tau}$ and $P(\Delta_t) \sim \Delta_t^{- \alpha}$ with 
$\tau \rightarrow 1.71(4)$ and $\alpha \rightarrow 2.25(3)$ for large enough systems. This scale free behavior holds up to
exponential cutoffs, which grow as a power law of the system size, $\Delta_m^c \sim R^{\beta_m}$ and 
$\Delta_t^c \sim R^{\beta_t}$, with $\beta_m = 2.32(6)$ and $\beta_t = 1.53(3)$ respectively. In addition, the size
and lifetime of an avalanche are related via $\Delta_m \sim \Delta_t^{\gamma}$, with $\gamma=1.52(5)$. A detailed analysis 
of these scale free avalanches reveals that they are in fact the combined result of many avalanches of different 
well-defined \textit{typical} size and duration. That is, the simplicity and versatility of our 
model system allows us to identify many different types of avalanches, each type characterized by a 
probability distribution with well defined typical size and duration, associated with a particular 
\textit{curvature} of the domain wall. Due to free borders and the microscopic impurity the system 
visits a broad range of domain wall configurations, and thus the combination of these avalanches 
generally results (see section \ref{capAval_apOrig_subMath}) in a distribution 
which exhibits several decades of power law behavior and an exponential 
cutoff. However, this apparent scale-free behavior does not mean that avalanches are critical, in the sense of a 
second order phase transition where diverging correlation lengths appear. Instead, we find that avalanches in the
magnetic nanoparticle have a large (but finite) number of different, gap-separated typical sizes and durations.
In this way we have proposed in this chapter a new mechanism to obtain power law distributions not related to any
underlying critical dynamics.\cite{Sornette}

However, the proposal of this new mechanism is not so important on its own. The deep insight derived from this analysis
comes when we extrapolate the conceptual framework here developed to the understanding of Barkhausen Noise in particular
and $1/f$ Noise in general. As we previously stated, Barkhau-sen Noise is the noise by which an impure ferromagnet responds
to a slowly varying external magnetic field. This response is not continuous, but burst-like. In particular, magnetization
jumps are observed as a function of the applied field which are called {\it avalanches}. Experimentalists are able to
measure the size and lifetime of these avalanches, finding that both magnitudes follow power law distributions with well-defined
exponents. In order to obtain these exponents they must filter out the extrinsic noise which soils the main signal. A
main feature characterizing Barkhausen Noise is that the scale-free behavior codified by the observed power law avalanche 
distributions appears without any need of fine tuning. For many years\footnote{In fact, Barkhausen Noise was discovered a
century ago, and since then its origin remains mysterious.} theoretical physicist have been wondering about the 
origin of this {\it spontaneous} or {\it self-organized} scale invariant behavior. Different theoretical approaches have been
proposed as explanation of Barkhausen Effect (most of these approaches are incompatible among them\cite{Spaso}). 
All these theories are based on assuming the existence of an underlying critical point, responsible of the observed scale 
invariance. The main theoretical approaches are divided in three different branches:
\begin{itemize}
\item The Random Field Ising Model (RFIM) has been proposed as a possible theoretical framework in order to explain Barkhausen 
Noise.\cite{Sethna,Kuntz} In this model a second order phase transition driven by the disorder is observed. The calculated
critical exponents are similar to those measured in some Barkhausen experiments. This phase transition is a
plain old one, in the sense that fine tuning of parameters is needed in order to reach the critical region, as opposed to
observations in Barkhausen materials.

\item Another theoretical approach is based on disordered interfaces and the pinning-depinning 
transition.\cite{Zapperi,Durin,Durin1} As above, here an underlying critical point is assumed. Barkhausen Noise is
analyzed in this case in terms of the domain wall motion in a bulk with quenched disorder. This theoretical approximation is
the most popular in the scientific community.

\item Some SOC models have been also proposed as possible explanation to Barkhausen Noise. Nowadays,
this theoretical approach is almost neglected.
\end{itemize}
\begin{table}[t]
\centerline{
\begin{tabular}{|c||c|c|c|}
\hline
Ref. & $\tau$ & $\alpha$ & $\gamma$ \\
\hline \hline
\cite{Mehta} & $1.46(5)$ & $----$ & $1.70(5)$ \\
\hline
\cite{Durin1}$_1$ & $1.27(3)$ & $1.5(1)$ & $1.77(12)$ \\
\hline
\cite{Durin1}$_2$ & $1.50(5)$ & $2.0(2)$ & $----$ \\
\hline
\cite{Spaso} & $1.77(9)$ & $2.25(8)$ & $1.51(1)$ \\
\hline
Our simulation & $1.71(4)$ & $2.22(3)$ & $1.52(5)$ \\
\hline
\end{tabular}
}
\caption[Exponents for different experiments on Barkhausen Noise.]
{\small In this table we show results on \textit{critical} exponents for different experiments on Barkhausen Noise. 
As it is clear from direct inspection, universality is not observed. The last row shows our results for the magnetic noise 
in the circular nanoparticle.}
\label{tabla_exp}
\end{table}

All these theoretical models reproduce some aspects of Barkhausen experiments. The assumption these approaches make about the 
existence of an underlying critical point responsible of the observed scale invariance involves that Barkhausen experiments
must show {\it universality}. That is, if the scale invariance observed in Barkhausen Noise is derived from an underlying
critical point, it is expected that the critical exponents measured for the avalanches will be universal, i.e. independent
of the microscopic details of the material. These critical exponents should only depend on the symmetries and conservation laws
which characterize the material. However, this is not observed in practice. Table \ref{tabla_exp} compares {\it critical}
exponents measured in different Barkhausen experiments, and the exponents we have measured in our simple model. A comparison
among the different experimental results shown in Table \ref{tabla_exp} clearly proves the lack of universality in
Barkhausen materials. This fact contradicts the assumption about the existence of an underlying critical point in Barkhausen
materials. Furthermore, the power of most practical applications of Barkhausen Noise is based on the sensitivity
of Barkhausen emission to microstructural details in the material.\cite{Sipahi1,Sipahi2} 
Such sensitivity is incompatible with the concept of
universality derived from a critical point, where by definition the behavior of the system is not sensitive to details, but
only to symmetries and conservation laws.
\begin{figure}[t]
\centerline{
\psfig{file=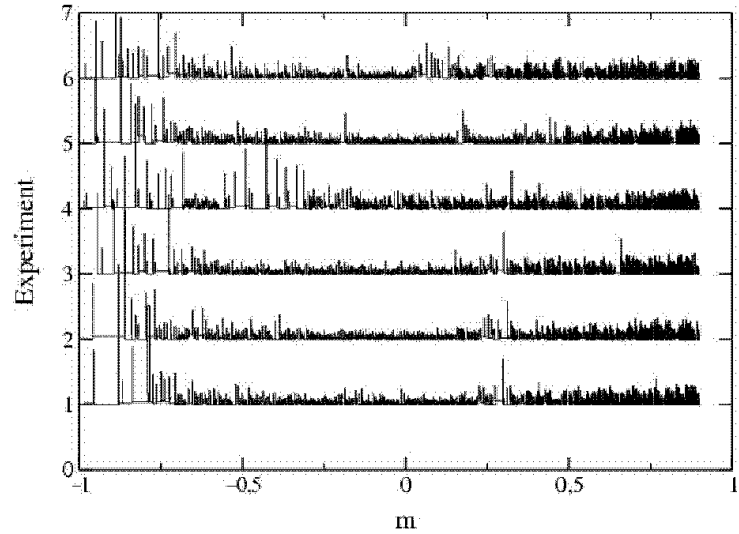,width=7cm,angle=-90}}
\caption[Avalanche sizes as a function of $m$ for different runs.]
{\small Avalanche sizes as a function of magnetization for several different runs. Large avalanches appear in the final
part of the evolution ($m \approx -1$). Notice also that, in general, large avalanches appear for approximately the same 
values of magnetization for each run (see, for instance, the large avalanche appearing for $m \approx 0.3$ in the four first runs).
}
\label{reprod2}
\end{figure}

Let us now analyze the relation between our results for the magnetic nanoparticle and Barkhausen Noise. In Table \ref{tabla_exp}
we observe that the exponents we have obtained for the avalanches in the magnetic nanoparticle are almost equal to
those measured by Spasojevi\'c et al\cite{Spaso} in Barkhausen experiments. In particular, we have obtained exponents
$\tau _{\infty} = 1.71(4)$, $\alpha_{\infty} = 2.25(3)$ and $\gamma = 1.52(5)$, while Spasojevi\'c et al have performed
experiments on quasi-bidimensional VITROVAC, measuring exponents $\tau = 1.77(9)$, $\alpha = 2.22(8)$ and $\gamma = 1.51(1)$.
The results of our simulations are perfectly compatible with these experimental measures. These observations led us to suspect
that Barkhausen Noise might also come from the superposition of more elementary events. In fact, the $1/f$ noise behavior in this 
case is assumed to reflect topological rearrangements of domain walls,\cite{Che_yotros} which result in practice in a 
series of jumps between different metastable states, which is the basic process in our model. 

Our system also reproduces some other phenomena observed in real materials, which support the above hypothesis about the origin
of Barkhausen Noise. For instance, our system shows {\it reproducibility}. It is observed during the demagnetization process
in the ferromagnetic nanoparticle that large avalanches are reproducible from one experiment to another.
This means that large avalanches usually appear at the same stages of evolution, independently of the observed experiment.
Fig. \ref{reprod2} shows the avalanche size as a function of magnetization for several demagnetization experiments in our
magnetic nanoparticle. It is clear from this figure that there are certain magnetizations for which large avalanches usually 
appear (see for instance avalanches for magnetization $m \approx 0.3$ or $m\in [-1,-0.75]$). On the other hand, extrinsic
noise (i.e. small avalanches) shows high variability. The reproducibility phenomenon points out the existence of a
dynamic correlation in the system. The same property has been observed in experiments with Perminvar and a Fe-Ni-Co 
alloy.\cite{Urbach} In these experiments the hysteresis loop associated to the material is studied, and it is observed that 
some (large) avalanches are almost perfectly reproducible from one hysteresis loop to another, 
while other (small) avalanches show no reproducibility. That is, measuring the size of the
avalanches as a function of the applied magnetic field it is found that there are certain values of the external field for 
which very large avalanches develop, and this property is observed for many different hysteresis loops.

In order to investigate the origin of reproducibility in our system, let us come back to Fig. \ref{reprod1}. There
we plotted the mean avalanche size $\langle\Delta_m \rangle (m)$ and the domain wall mean curvature $\langle{\cal C}\rangle (m)$ 
as a function of magnetization for a system with radius $R=30$, after averaging over 3500 different runs. The highly 
inhomogeneous (non-trivial) shape of $\langle\Delta_m \rangle (m)$ confirms the above conclusion, i.e. our magnetic nanoparticle
shows reproducibility. Moreover, the mean curvature $\langle{\cal C}\rangle (m)$ is also reproducible from one experiment to
another. As we previously discussed, the high correlation between $\langle\Delta_m \rangle (m)$ and 
$\langle{\cal C}\rangle (m)$ implies that the typical scale of an avalanche is completely determined by the curvature
of the domain wall when this avalanche starts. Hence the reproducibility observed in avalanches is due to the reproducibility 
of the domain wall curvature. The observed domain wall curvature reproducibility points out that the system evolves
from the metastable phase to the stable one through certain {\it typical} configurations, each one characterized by a typical 
droplet shape and a typical domain wall curvature. Fig. \ref{snap-circle} shows an example of this typical evolution (we came to
this conclusion after looking at many different particular evolutions). Different experiments evolve approximately in the same way
because, on one hand, the system is very efficient selecting the most 
energetically favorable configurations during the evolution due to the low temperature of simulations (in this case
the free energy minima are very deep), and on the other hand the presence of the free boundaries involves that the droplet always 
nucleates at the border. In experiments showing reproducibility it is concluded that the presence of quenched disorder in 
these real systems is essential for the observed reproducibility.\cite{Urbach} 
In our case, the presence of the border (together with the low temperature) is the responsible of the reproducibility phenomenon. 
In this sense we could interpret the open boundary in our system as a quenched disorder distributed in a very particular way.

Summarizing, we observe that the exponents characterizing the avalanche distributions in our system reproduce some experimental
results on Barkhausen Noise. Moreover, our ferromagnetic nanoparticle also shows reproducibility, which is observed in real 
Barkhausen materials. In addition, the avalanche size and lifetime distributions show exponential cutoffs which depend algebraically
on system size. This behavior has been also reported in real materials \cite{Bahiana}. Finally, the measured exponents $\tau$ and
$\alpha$ show finite size corrections similar to those found in experiments with avalanche systems\cite{Frette}. All these
similarities, together with the fact that experimental observations do not support the existence of universality\footnote{Saying 
that universality is not observed is equivalent to affirm that there is no phase transition underlying Barkhausen Noise.} in
Barkhausen Noise, led us to suspect that Barkhausen Noise might also come from the superposition of more elementary events with
well-defined typical scales.

The chances are that our observation that scale invariance originates in a combination of simple events, which we can prove in 
our model cases, is a general feature of similar phenomena in many complex systems \cite{Sethna}. This should explain
why distributions exhibiting power law, exponential or stretched exponential behavior have been identified in different but related
experimental situations and in different regimes of the same experiment.

Finally, let us mention that the analysis method introduced in this chapter in order to identify the origin of
different avalanches and the superposition of different typical scales can be easily exported to many experimental situations,
simplifying the investigation about the origin of Barkhausen emissions in particular, and $1/f$ Noise in general.

\part{Nonequilibrium Phase Separation, Absorbing States and Heat Conduction}

\chapter{Kinetics of Phase Separation in the Driven Lattice Gas: Self-Similar Pattern
Growth under Anisotropic Nonequilibrium Conditions}
\label{capDLG}

\markboth{Kinetics of Phase Separation in the Driven Lattice Gas}{\ref{capDLG_apIntro} Introduction}
\section{Introduction}
\label{capDLG_apIntro}
\markboth{Kinetics of Phase Separation in the Driven Lattice Gas}{\ref{capDLG_apIntro} Introduction}

The first part of this thesis has been devoted to the study of metastability and avalanches in a 
nonequilibrium magnetic model. There we have investigated the dynamic process by which a nonequilibrium
spin system subject to a nonzero magnetic field evolves from a metastable phase towards the truly stable one.
There are other dynamical phenomena in nonequilibrium systems which involve transformations between different
phases. In particular, a very interesting phenomenon is the segregation process that emerges when a nonequilibrium
(conserved) system evolves from a disordered phase towards the ordered one. The aim of this chapter is to investigate
the effects that nonequilibrium anisotropic conditions induce on this phase separation phenomenon.

Many alloys such as Al-Zn, which are homogeneous at high temperature, undergo
phase separation after a sudden quench into the miscibility gap (for details,
see the reviews \cite{rev1}-\cite{rev5}, for instance). One first observes
nucleation in which small localized regions (\textit{grains}) form. This is
followed by \textquotedblleft spinodal decomposition\textquotedblright. That
is, some grains grow at the expense of smaller ones, and eventually coarsen,
while their composition evolves with time. In addition to theoretically
challenging, the details are of great practical importance. For example,
hardness and conductivities are determined by the spatial pattern finally
resulting in the alloy, and this depends on how phase separation competes with
the progress of solidification from the melt.

A complete kinetic description of these highly non-linear processes is
lacking. \cite{rev5} Nevertheless, the essential physics for some special
situations is now quite well understood. This is the case when nothing
prevents the system from reaching the equilibrium state, namely, coexistence
of two thermodynamic phases. The simplest example of this is the (standard)
lattice gas evolving from a fully disordered state to segregation into
\textit{liquid} (particle-rich phase) and \textit{gas} (particle-poor phase).
(Alternatively, using the language of the isomorphic lattice binary alloy,
\cite{ma} the segregation is into, say Al-rich and Zn-rich phases.) As first
demonstrated by means of computer simulations,\cite{prl00,rev1,rev2} this
segregation, as well as similar processes in actual mixtures exhibit time
\textit{self-similarity}. This property is better defined at sufficiently low
temperature, when the thermal correlation length is small. The system then
exhibits a \textit{single} relevant length, the size $\ell\left(  t\right)  $
of typical grains growing algebraically with time. Consequently, any of the
system properties (including the spatial pattern) look alike, except for a
change of scale, at different \textit{times}.

This interesting property is revealed, for example, by the sphericalized
structure factor $S\left(  k,t\right)  $ as observed in scattering
experiments. After a relatively short transient time, one observes that
$S\left(  k,t\right)  \sim J\left(  t\right)  \cdot F\left[  k\ \ell\left(
t\right)  \right]  .$ Taking this as a hypothesis, one may interpret $J$ and
$\ell$ as phenomenological parameters to scale along the $S$ and $k$ axes,
respectively. The hypothesis is then widely confirmed, and it follows that
$J\left(  t\right)  \sim\ell\left(  t\right)  ^{d}$ where $d$ is the system
dimension. It also follows that $F(\varkappa)=\Phi(\varkappa)\cdot\Psi\left(
\sigma\varkappa\right)  $ where $\Phi$ and $\Psi$ are universal functions. In
fact, $\Phi$ describes the diffraction by a single grain, $\Psi$ is a grain
interference function, and $\sigma$ characterizes the point in the
(density$-$temperature) phase diagram where the sample is quenched. It then
ensues that $\Psi\approx1$ except at small values of $k$, so that, for large
$\varkappa$, $F(\varkappa)$ becomes almost independent of density and 
temperature, and even the substance investigated.\cite{prl00,zahra,rev5}

The grain distribution may also be directly monitored. A detailed study of
grains in both microscopy experiments and computer simulations confirms time
scale invariance. More specifically, one observes that the relevant length
grows according to a simple power law, $\ell\left(  t\right)  \sim t^{a},$ and
one typically measures $a=1/3$ at late times. This is understood as a
consequence of diffusion of monomers that, in order to minimize surface
tension, evaporate from small grains of high curvature and condensate onto
larger ones (\textit{Ostwald ripening}). In fact, Lifshitz and Slyozov, and
Wagner independently predicted $\ell\sim t^{1/3},$\cite{LSW} which is often
observed, even outside the domain of validity of the involved
approximations.\cite{LSWbis} In some circumstances, one should expect other,
non-dominant mechanisms inducing corrections to the Lifshitz-Slyozov-Wagner
one. \cite{rev1,rev3,rev5} For instance, effective diffusion of grains
(\textit{Smoluchowski coagulation}) leads to $a=1/6,$ which may occur at early
times;\cite{toral} interfacial conduction leads to $a=1/4;$\cite{huse,puri0}
and, depending on density and viscosity, a fluid capable of hydrodynamic
interactions may exhibit crossover with time to viscous ($a=1$) and then
inertial ($a=2/3$) regimes.\cite{rev4}

Extending the above interesting picture to more realistic situations is an
open question. The assumption that the system asymptotically tends to the
coexistence of two \textit{thermodynamic} (equilibrium) phases is often
unjustified in Nature. This is the case, for example, for mixtures under a
shear flow, whose study has attracted considerable attention,
e.g.\cite{critic}-\cite{corberi}. The problem is that sheared flows
asymptotically evolve towards a \textit{nonequilibrium} steady state and that
this is highly anisotropic. Studying the consequences of anisotropy in the
behavior of complex systems is in fact an important challenge (see, for
instance, \cite{MarroDickman,mandel,katz}). Another important example is that of
binary granular mixtures under horizontal shaking. The periodic forcing 
causes in this case phase separation and highly anisotropic clustering.\cite{Mullin}

In this chapter, we study in detail the kinetics of the driven lattice gas (DLG)
\cite{katz} following a deep quench. Our motivation is twofold. On one hand,
the DLG is recognized to be an excellent microscopic model for nonequilibrium
anisotropic phenomena.\cite{MarroDickman} On the other, the DLG is not affected by
hydrodynamic interactions, which makes physics simpler. Our goal is timely
given that the asymptotic state of the DLG is now rather well understood, and
previous studies of kinetics altogether reveal an intriguing
situation.\cite{katz}-\cite{albano} Following this pioneering effort, we here
present a new theoretical description of the essential physics during
anisotropic, nonequilibrium pattern growth. This is compared with new
extensive computer simulations.\cite{new}

\section{Model and Simulation Details}
\label{capDLG_apMod}

The DLG consists in a $d-$dimensional, e.g., simple-cubic lattice with
configurations $\mathbf{n=}\left\{  n_{i};i=1,...,N\right\}  $. The variable
at each lattice site has two possible states, $n_{i}=1$ (\textit{particle}) or
$0$ (\textit{hole}). As for the standard lattice gas, dynamics is a stochastic
process at temperature $T$ consisting in nearest-neighbor (NN) particle/hole
exchanges. This conserves the particle density, $\rho=N^{-1}\sum_{i}n_{i},$
and depends on $\mathbf{n}.$

A distinguishing feature of the DLG is that exchanges are favored in one of
the principal lattice directions, say $\vec{x}.$ Therefore, assuming periodic
(toroidal) boundary conditions, a net current of particles is expected to set
in along $\vec{x}.$ This is accomplished in practice by defining a biased
transition rate. We shall refer here to the \textit{energy }function
$H=-4J\sum_{NN}n_{i}n_{j},$ which describes attractive interactions between
particles at NN sites, and to the transition rate (per unit time):\cite{MarroDickman}
\begin{equation}
\omega(\mathbf{n}\rightarrow\mathbf{n}^{\ast})=\min\left\{  1,\operatorname{e}%
^{-\left(  \Delta H+E\delta\right)  /T}\right\}  . \label{rateDLG}%
\end{equation}
$\mathbf{n}^{\ast}$ stands for configuration $\mathbf{n}$ after jumping of a
particle to a NN hole; $\Delta H=H\left(  \mathbf{n}^{\ast}\right)  -H\left(
\mathbf{n}\right)  $ is the \textit{energy} change brought about by the jump;
and units are such that both the coupling strength $J$ and the Boltzmann
constant are set to unity. One further defines $\delta=\left(  \mp1,0\right)
$ for NN jumps along $\pm\vec{x}$ or along any of the transverse directions,
say $\vec{y},$ respectively. Consistent with this, $\vec{E}=E\vec{x}$ may be
interpreted as a field driving particles, e.g., an electric field if one
assumes that particles are charged. (One may adopt other interpretations,
e.g., the binary alloy one.\cite{ma} Dynamics then consists in interchanges
between particles of different species, one of them favored along $\vec{x}.$)

The DLG was described as modeling surface growth, fast ionic conduction and
traffic flow, among a number of actual situations of practical
interest.\cite{MarroDickman} A common feature in these situations is anisotropy, and
that steady states are out of equilibrium. Both are essential features of the
DLG induced by the rate (\ref{rateDLG}). The only trivial case is for $E=0,$
which reduces (\ref{rateDLG}) to the Metropolis algorithm. In this case, detailed
balance holds, and one simply has the familiar lattice gas with a unique
(equilibrium) steady state. For any, even small $E,$ qualitatively new
behavior emerges. In fact, detailed balance breaks down for $E>0$ and,
consequently, the steady state depends on $\omega(\mathbf{n}\rightarrow
\mathbf{n}^{\ast}).$ Increasing $E,$ one eventually reaches
\textit{saturation.} That is, particles cannot jump backwards, i.e., $-\vec
{x},$ which formally corresponds to an \textit{infinite field} $(E=\infty).$

The way in which the microscopic anisotropy (\ref{rateDLG}) conveys into
macroscopic behavior is amazing.\cite{MarroDickman} Consider, for simplicity, $d=2,$
$\rho=%
\frac12
$ and $E=\infty.$ The system then exhibits a critical point at $T=T_{C}%
^{\infty}\simeq1.4T_{C}\left(  E=0\right)  $, where $T_c(E=0)\approx 2.2691 \equiv T_{ons}$, 
with novel critical
behavior.\cite{prl01,albano} Furthermore, the asymptotic, steady states below
$T_{C}^{\infty}$ do not comprise equilibrium phases. Instead, one observes a
particle \textit{current} and fully \textit{anisotropic} phases; both are
nonequilibrium features. The intensity of the current increases with $T,$ and
suddenly changes slope at $T_{C}^{\infty}$ (in fact, this property may serve
to accurately locate the critical point). The stable ordered configurations
consist of one stripe, to be interpreted as a \textit{liquid }(rich-particle)
phase of density $\rho_{L}\left(  T\right)  .$ The \textit{gas}
(poor-particle) phase of density $\rho_{G}\left(  T\right)  $ fills the
remainder of the system. Except for some microscopic roughness, the interface
is linear and rather flat, in general\footnote{Some partial, inconclusive studies 
about interfaces in the Driven Lattice Gas\cite{interf} have reviewed in 
Ref. \cite{MarroDickman}.}.

The computer evolutions reported here always begin with a completely
disordered state to simulate the system at infinite temperature. We then model
a sudden quench and the subsequent time evolution. With this aim, one proceeds
with rate (\ref{rateDLG}) that involves the temperature $T$ at which the system
is quenched. The run is followed until one stripe is obtained (eventually, in
order to save computer time, the run was sometimes stopped before reaching the
final stationary state). The code involves a list of $\eta\left(  t\right)  $
particle-hole NN pairs from where the next move is drawn. Time is then
increased by $\Delta t=\eta\left(  t\right)  ^{-1},$ so that its unit or
\textit{MC step} involves a visit to all sites on the average\footnote{This 
corresponds to the standard Monte Carlo method only if the 
time increment $\Delta t$ is drawn from a Poisson distribution. Taking constant 
$\Delta t = \eta(t)^{-1}$ involves some approximation. However, if the number
of particle-hole pairs in the system is sufficiently large, the approximation is excellent.
In particular, for a half filled system subject to an {\it infinite} drive (our model case), 
the minimum amount of particle-hole pairs present in the system will be of order 
$2L_{\parallel}$, which is large for large lattices.}.

The lattice is rectangular, $L_{\parallel}\times L_{\perp},$ with sides
ranging from 64 to 256 and, in a few cases, 512. Results concern an average
over around thousand independent runs. Due to the great computational effort
which is consequently involved, this chapter describes simulations concerning a
single point of the two-dimensional DLG phase diagram. That is, most of our
evolutions are for $\rho=%
\frac12
$ and $E=\infty,$ and simulate a quench at $T=0.8T_{C}\left(  E=0\right)
\simeq0.6T_{C}^{\infty}.$ This choice is motivated by the fact that clustering
is then reasonably compact, which helps to obtain good statistics, while it
proceeds fast enough, so that one can observe full relaxation to the steady
state. In spite of this restriction, brief investigation of other points,
together with some of our observations below, led us to believe that the
validity of our results extends to a large domain around the center of the
miscibility gap; in fact, such generality of behavior has been reported for
$E=0.$\cite{rev2,prl00,toral,rev5}

\section{Growth of Order}
\label{capDLG_apGro}
\begin{figure}
\centerline{
\psfig{file=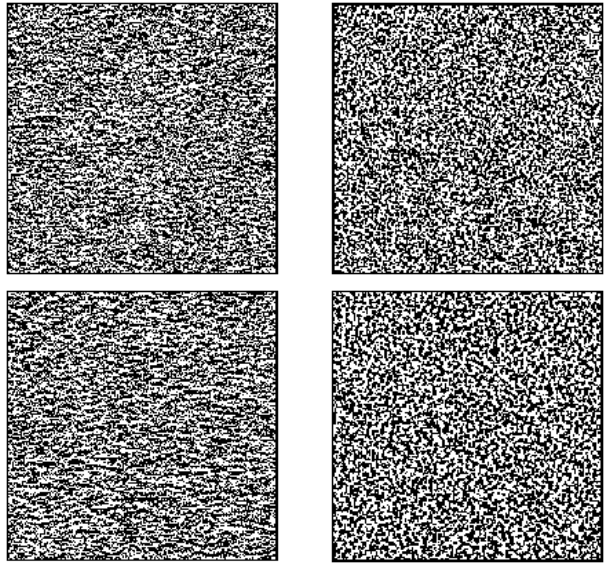,width=8.0cm,angle=0}}
\caption[Snapshots of the early time evolution of DLG.]
{\small A series of MC snapshots comparing very early patterns for
the DLG (with an infinite horizontal field) and for the standard lattice gas,
i.e., zero field (LG) at the same time. This corresponds to a 256x256 lattice
at $T\simeq0.6T_{C}^{\infty}.$ The time (in MC steps) is here $t=$
4 and 10 (from top to bottom) for the DLG (left column) and for the LG
(right column).\medskip}
\label{snap1}
\end{figure}

The DLG exhibits different time regimes during phase separation. Though they
parallel the ones for $E=0,$ the \textit{peculiarities} induced by the
anisotropic condition are essential.

Starting from complete disorder, there is a very short initial regime in which
small grains form. The novelty is that typical grains are now fully
anisotropic, stretched along $\vec{x}.$ The grains then rapidly coarsen to
form macroscopic strings, as illustrated in Fig. \ref{snap1}. Sheared fluids (an
experimentally accessible situation that also involves both nonequilibrium
physics and anisotropy) seem to exhibit similar initial
regimes.\cite{onuki,corberi} That is, during a short time interval, they show
larger growth rate along the flow than in the other directions, which is
assumed to correspond to the initial formation of anisotropic regions.
Afterwards, sheared fluids develop string-like macroscopic domains similar to
the ones in the DLG.
\begin{figure}[t!]
\centerline{
\psfig{file=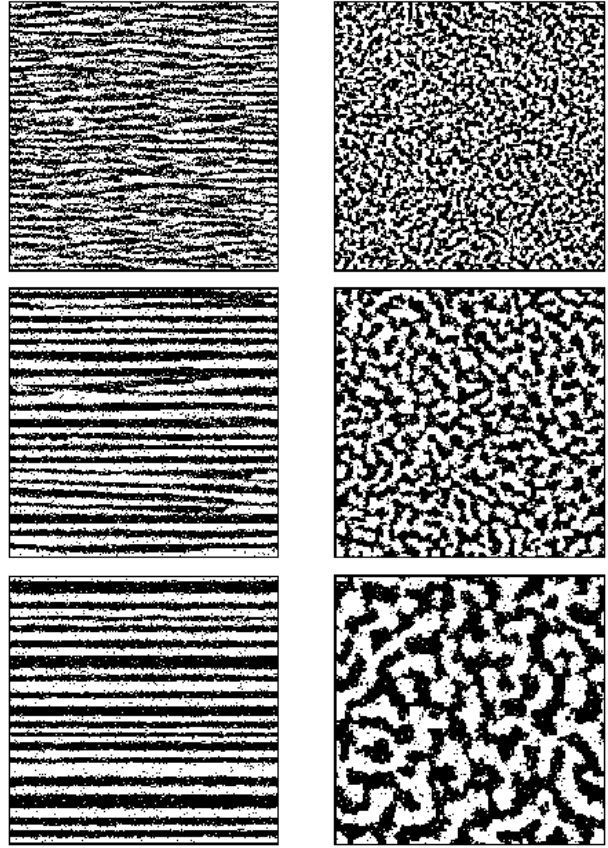,width=8.0cm,angle=0}}
\caption[Snapshots of the time evolution of DLG.]
{\small The same as Fig. \ref{snap1} but at late time, namely, 
$t=$ 100, 1000 and 10000 MC steps (from top to bottom) for the DLG (left
column) and for the LG (right column).\medskip}
\label{snap2}
\end{figure}

Figs. \ref{snap1} and \ref{snap2} include a comparison with the zero-field case, i.e., the standard,
isotropic lattice gas (LG). This clearly illustrates the strong anisotropy of
nucleation and early phase separation for the DLG. Close inspection of these
and similar graphs also seems to indicate relatively small but significant
differences in the degree of segregation between the two cases at a given
time. That is, at small distances, there is a more homogeneous distribution of
particles, both longitudinally and transversely, in the DLG than in the LG.
The latter shows up more segregated at the same time, which is already rather
evident by direct inspection of graphs for $1<t\leq100$ in Figs. \ref{snap1} and \ref{snap2}. We believe
this reveals the different role played by surface tension as the degree of
anisotropy is varied: Typical DLG grains are rather linear except at their
longitudinal ends, where curvature may be even stronger than for the spherical
clusters in the LG at comparable times. This seems to be at the origin of a
smother transverse distribution of particles in the DLG at early times. On the
other hand, the field also tends to smooth things longitudinally.

In order to quantify the aforementioned observation, we evaluated the number 
of broken bonds in the direction of (perpendicular to) the field, $n_{\parallel}(t)$
($n_{\perp}(t)$) as a function of time during the early evolution stage. 
Then $A(t)\equiv [n_{\perp}(t)+n_{\parallel}(t)]/2N$ is
the density of broken bonds. The higher the degree of segregation at time 
$t$, the smaller is $A(t)$. For instance, we observe in a large $256\times 256$ lattice
that $A(t=10)=0.295$ and $A(t=10)=0.38$ for the LG and DLG respectively, confirming the 
above observation. On the other hand, let $B(t)\equiv [n_{\perp}(t)-n_{\parallel}(t)]/2N$. 
One would expect $B(t)\approx 0$ (up to fluctuations) only for the isotropic system.
In fact, we measured $B(t)\approx 0$ for the LG, while $B(t)$ rapidly converges to a 
nonzero value $B(t)\approx 0.05$ for the DLG at early times (again for a large 
$256\times 256$ lattice). We take this number, $B(t)=0.05$, as characterizing the 
anisotropic shape of DLG clusters at early times.

The difference of segregation between the DLG and the LG at early
times merits further study. This will need to take into account the anisotropy
of surface tension. In any case, this concerns a regime very near the initial,
melt state that only bears minor practical importance, given that it extends
extremely shortly on the macroscopic time scale. We are interested in the rest
of this chapter on the subsequent evolution, to be described on the assumption
of a simple flat interface, which holds in Fig. \ref{snap2} for $t>100.$
\begin{figure}
\centerline{
\psfig{file=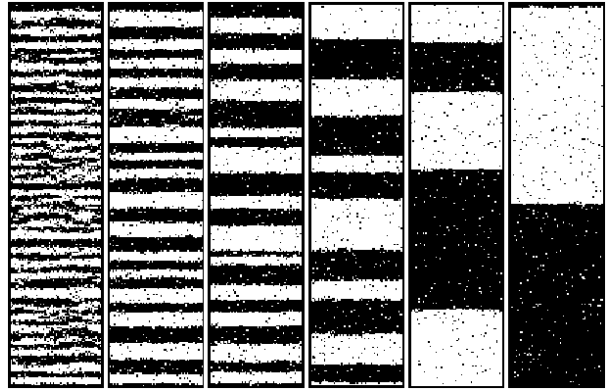,width=8.0cm,angle=0}}
\caption[Snapshots of the whole stripe coarsening process.]
{\small A series of MC snapshots that illustrate (late) growth at
$T\simeq0.6T_{C}^{\infty}.$ This is for a rectangular lattice of
size $L_{\perp}\times L_{\parallel}=256\times64$ and $t=$
10$^{2}$, 10$^{4}$, 10$^{5},$ 10$^{6},$
10$^{7}$ and 1,1x10$^{8}$ MC steps,
respectively, from left to right.\medskip}
\label{snap3}
\end{figure}

The DLG strings coarsen with time until well defined, relatively-narrow
longitudinal (i.e., directed along $\vec{x})$ stripes are formed. (For
periodic boundary conditions, the case of our simulations, each stripe forms a
ring.) This results into a multi-stripe state, as illustrated in Figs. \ref{snap2} and
\ref{snap3}. The ordering time in the DLG, defined 
as the time the system needs to form the stripes, scales with the system size 
in the direction of the field, $L_{\parallel}$, 
since in this case ordered clusters (stripes) 
percolate along the field direction (see below).\cite{mukamel} This is not the case 
for the equilibrium LG, where the ordering time depends exclusively on system 
intensive parameters such as temperature and density.

The multi-stripe states are not stable, however. They are only partially
segregated and, in fact, a definite tendency towards a fully segregated state
with a single stripe is generally observed in computer simulations. One may
also develop simple arguments indicating that, in general, a multi-stripe
state will monotonically evolve until forming a single stripe.\cite{MarroDickman,25bis}
It is true that, in practice, the complete relaxation may take a very long
time. More specifically, a macroscopic system may take to decay into the true
stable state a long, \textit{macroscopic} time interval, namely, a time that
may show up as mathematically \textit{infinite} in some time scales. In fact,
the complete relaxation time is observed to increase with system size, as first
demonstrated in \cite{mukamel}. It should also be remarked that this property
is not a nonequilibrium feature but occurs already in the equilibrium ($E=0$)
case; see, for instance, \cite{rev2,rev4} and references therein. Slow
relaxation is a consequence of the conservation of particle density $\rho$
implied by the particle-hole exchange dynamics; this induces scale invariance,
namely, slow (power-law) evolution of correlations so that, once enough order
sets in, all but very small pattern modifications during a single MC step are
precluded. Consequently, certain individual runs sometimes block for a long
time in a state with several stripes; however, this does not correspond to the
average behavior. As illustrated by Fig. \ref{snap3}, which shows a typical evolution,
and demonstrated below by our averages corresponding to thousand evolutions,
the number of stripes monotonically decreases with time (see also
section \ref{capDLG_apCont}), and the whole relaxation can easily be observed in computer
simulations if one waits long enough.

We next attempt a theoretical description of the relaxation process. Our
interest is on the \textit{anisotropic spinodal decomposition} by which the
earliest state with many well-defined stripes decays into a single stripe.We
shall assume that relaxation is a consequence of monomer events causing
effective diffusion of liquid stripes. (Note that assuming gas stripes here
would be completely equivalent.) That is, due to single particle processes,
liquid stripes move transversely as a whole, and may collide and eventually
coalesce with one of the neighboring stripes; see the late evolution depicted
in Fig. \ref{snap3}. We notice that coalescence implies evaporation of the gas stripe
between the two involved liquid stripes. Therefore, given the particle/hole
symmetry, our assumption is in a sense equivalent to assuming that growth is
due to evaporation of stripes;\cite{mukamel} however, the view adopted here
allows for a more detailed description below.

In order to evaluate the implications of stripe effective diffusion via
monomer events, lets assume that stripes are well defined, compact and exhibit
a (linear) interface which is rather flat. This is perfectly justified at
sufficiently low temperature (the case analyzed in detail here),\cite{MarroDickman} and
it might hold more generally, in a wide region including the center of the
miscibility gap but excluding the critical region. Under this assumption,
consider a stripe of mean width $\ell\left(  t\right)  $ that consists of $M$
particles whose coordinates along the transverse (vertical) direction are
$y_{j}\left(  t\right)  ;$ $j=1,\ldots,M.$ We characterize the stripe position
by its center of masses, $Y_{\text{cm}}\left(  t\right)  \equiv M^{-1}\sum
_{j}y_{j}\left(  t\right)  .$

Let us evaluate the \textit{mobility} coefficient $\mathcal{D}_{\ell}\equiv
N_{\text{me}}\langle(\Delta Y_{\text{cm}})^{2}\rangle$ which depends on the
stripe width $\ell(t).$ Here $N_{\text{me}}$ is the number of monomer events
per unit time, and $\langle(\Delta Y_{\text{cm}})^{2}\rangle$ is the mean
squared displacement of the stripe associated to one of the monomer events. We
think of two possible types of events, each giving a different contribution to
$\mathcal{D}_{\ell}:$

\noindent\ (A) Evaporation-condensation of particles and holes in the stripe surface.
Here particles (holes) at the stripe interface evaporate to the hole (particle) gas,
and condensate later at the same interface.
The evolution of the evaporated particle (hole) in the 
bulk can be seen as a one dimensional random walk with two absorbing
walls, the left and right interfaces, respectively. According to standard
random walk theory,\cite{gambler} the evaporated particle (hole) will go again 
with unit probability to one of the (two) possible interfaces, Moreover, the random 
walker will stick again to its original interface with high probability, so trapping 
a particle (hole) from the opposite interface is unlikely.
Consequently, in this case (A), $N_{\text{me}}$ is
simply the evaporation rate. That is, $N_{\text{me,A}}=\nu\sum_{j}^{\prime
}\exp(-2T^{-1}\Delta_{j}),$ where $\nu$ is the \textit{a priori} frequency,
the sum is over the surface particles and $\Delta_{j}$ is the number of
resulting broken bonds. For a flat linear interface, particles can only jump
transversely away the surface, $\nu$ equals the inverse of the lattice
coordination number, $q,$ and one may write $N_{\text{me,A}}\approx
4q^{-1}L_{\parallel}\exp(-2\bar{\Delta}/T)$ where $\bar{\Delta}$ is the mean
number of broken bonds per evaporation event. We multiplied here by 2 to take
into account evaporation of surface holes that travel within the stripe to
reach the (same) surface again. On the other hand, evaporation processes
induce changes $\Delta Y_{\text{cm}}=M^{-1}\delta y,$ where $\delta y$ is the
net particle (transverse) displacement, and $M\approx L_{\parallel}\times
\ell(t)$ for compact stripes. Therefore,
\begin{equation}
\mathcal{D}_{\ell}^{(\text{A})}\sim4q^{-1}\langle\delta y^{2}\rangle
\operatorname{e}^{-2\bar{\Delta}/T}L_{\parallel}^{-1}\ell^{-2}. \label{dA}%
\end{equation}

\noindent\ (B) A hole jumps one lattice spacing away within the stripe
interior. This induces $\Delta Y_{\text{cm}}=1/M$ or 0, depending on the jump
direction. One may write $N_{\text{me,B}}=2\nu\rho_{h}\left(  T\right)
L_{\parallel}\ell p_{h}\left(  T\right)  ,$ where $\rho_{h}$ is the density of
holes, $L_{\parallel}\ell$ is the \textit{volume} or total number of sites
within the liquid stripe, and $p_{h}$ is the jumping probability per unit
time. The factor 2 here comes from the fact that a hole modifies
$Y_{\text{cm}}$ when jumping to any of the two directions $\pm\vec{y}.$ At low
$T,$ $\rho_{h}$ is small; holes are then rather isolated from each other, so
that jumps do not modify the number of broken bonds, and $p_{h}\approx1.$ It
ensues
\begin{equation}
\mathcal{D}_{\ell}^{(\text{B})}\sim2q^{-1}\rho_{h}L_{\parallel}^{-1}\ell^{-1}.
\label{dB}%
\end{equation}

\noindent Note that a different dependence of (\ref{dA}) and (\ref{dB}) on
$\ell$ is a consequence of the fact that the rates $N_{\text{me,A}}$ and
$N_{\text{me,B}}$ involve processes consisting in evaporation on the interface and
diffusion on the bulk, respectively.

For $\rho=%
\frac12
,$ one has on the average stripes of width $\ell$ that are separated a
distance $\ell$ from each other. Therefore, a given stripe takes a mean time
$\tau_{\ell}=\ell^{2}/\mathcal{D}_{\ell}$ to find (and thus to coalesce with)
another one, and this causes its width to increase by $\Delta\ell=\ell.$
Consequently, $\operatorname{d}\ell/\operatorname{d}t\sim\Delta\ell
\ \tau_{\ell}^{-1}=\mathcal{D}_{\ell}\ell^{-1}.$ Together with (\ref{dA}) and
(\ref{dB}), respectively, this implies that mechanism A is characterized by a
power law $\ell\sim t^{1/4},$ and that mechanism B is to be associated with
$\ell\sim t^{1/3}.$ Furthermore, assuming that pattern growth in the DLG is
the result of competition between the two mechanisms, and that they are
independent of each other, $\mathcal{D}_{\ell}=\mathcal{D}_{\ell}^{(\text{A}%
)}+\mathcal{D}_{\ell}^{(\text{B})},$ it follows that
\begin{equation}
\frac{\operatorname{d}\ell}{\operatorname{d}t}\sim\frac{1}{L_{\parallel}%
}\left(  \frac{\alpha_{\text{A}}}{\ell^{3}}+\frac{\alpha_{\text{B}}}{\ell^{2}%
}\right)  , \label{kineq}%
\end{equation}
where $\alpha_{\text{A}}=4\nu\langle\delta y^{2}\rangle\operatorname{e}%
^{-2\bar{\Delta}/T}\ $and $\alpha_{\text{B}}=2\nu\rho_{h}.$ This is our
general result for the DLG as far as the field $E$ is large, e.g., infinite,
and the temperature $T$ is low enough so that the interfaces, and mechanisms A
and B, are sufficiently simple as assumed. This is to be compared with the
Lifshitz-Slyozov-Wagner behavior $\operatorname{d}\ell/\operatorname{d}%
t\sim\ell^{-2}$ which assumes spatial isotropy and diffusion directly governed
by surface tension. Formally, (\ref{kineq}) is similar to an equation obtained
before by assuming isotropic conditions; see section \ref{capDLG_apIntro}.\cite{huse}

The consequences of (\ref{dA})--(\ref{kineq}) are as follows. Both (\ref{dA})
and (\ref{dB}) imply independently that
\begin{equation}
\ell\sim\left(  \varphi\theta\right)  ^{1/\varphi}\left(  t/L_{\parallel
}\right)  ^{1/\varphi}. \label{t1314}%
\end{equation}
The difference is that $\theta=\alpha_{A}$ and $\varphi=4$ from (\ref{dA})
while one obtains $\theta=\alpha_{B}$ and $\varphi=3$ from (\ref{dB}). On the
other hand, for sufficiently late times, $\ell$ becomes large and equation
(\ref{kineq}) simply solves into
\begin{equation}
\ell(t)\sim\alpha t^{1/3}+\zeta, \label{t13}
\end{equation}
where $\alpha^{3}=3\alpha_{\text{B}}L_{\parallel}^{-1}$ and $\zeta
=\alpha_{\text{A}}/2\alpha_{\text{B}}.$ That is, the prediction is that hole
diffusion within the stripe (mechanism B) will be dominant at late times. A
different hypothesis, based on the stripe evaporation picture, was shown in \cite{mukamel} 
to imply $\ell\sim\left(t/L_{\parallel}\right)  ^{1/3}$. 
This coincidence is not surprising since, as argued above, the coalescence of two 
particles stripes implies the evaporation of the intermediate hole stripe and due to the 
particle/hole symmetry in our system, both mechanisms (stripe diffusion/coalescence and stripe 
evaporation) correspond to the same physical process yielding the same behavior. 
In order to uncover the close analogy between the two pictures, one may notice that,
to evaporate a particle stripe, many of its particles must cross the 
surrounding hole stripes and stick on the neighboring particle stripes (this
is so since the particle density in the gas phase remains almost constant).
This particle migration process through the surrounding hole stripes is
in fact what we have called 'hole diffusion within the stripe' in the presence of
particle/hole symmetry. Hence the fundamental mechanism involved in a
stripe evaporation is the diffusion of its constituents through the neighboring
stripes. This observation is a key one to understand the relation between 
stripe's evaporation and hole (particle) diffusion.

The effect of mechanism A ---surface evaporation and subsequent
conden-sation--- on growth is more subtle. In fact, our theory predicts a
crossover from the $t^{1/4}$ regime to the $t^{1/3}$ regime as time is
increased. That is, the two mechanisms will have a comparable influence at
$t\sim\tau_{\text{cross}}$ with
\begin{equation}
\tau_{\text{cross}}=\frac{\left(  4\alpha_{\text{A}}\right)  ^{3}}{\left(
3\alpha_{\text{B}}\right)  ^{4}}L_{\parallel}. \label{time1}%
\end{equation}
For times $t<\tau_{\text{cross}}$, mechanism A is dominant and the $t^{1/4}$ behavior is 
expected, while mechanism B is dominant for $t>\tau_{\text{cross}}$ and the asymptotic
$t^{1/3}$ growth law is then observed.
The crossover time $\tau_{\text{cross}}$ is a macroscopic, observable time. Further, we may 
define the time $\tau_{\text{ss}}$ at which a single stripe is reached by the condition that
$\ell\left(  t\right)  \approx%
\frac12
L_{\perp}.$ One obtains%
\begin{equation}
\tau_{\text{ss}}=\frac{L_{\parallel}}{\alpha_{\text{B}}}\left\{
\frac{L_{\perp}^{3}}{24}-\frac{\zeta L_{\perp}^{2}}{4}+2\zeta^{2}L_{\perp
}-8\zeta^{3}\left[  \ln\frac{\alpha_{\text{B}}\left(  2\zeta+%
\frac12
L_{\perp}\right)  }{L_{\parallel}}-\ln\frac{2\zeta\alpha_{\text{B}}%
}{L_{\parallel}}\right]  \right\}  . \label{time2}%
\end{equation}
Hence our system is characterized by two different time scales, namely, $\tau_{\text{cross}}$
and $\tau_{\text{ss}}$. They depend on system size in a different way. 
For large systems one generally obtains $\tau_{\text{ss}} \gg \tau_{\text{cross}}$, so that
the system converges,
after a short, perhaps unobservable transient time, to the relevant $t^{1/3}$ 
behavior. However, there are small systems for which $\tau_{\text{ss}}<\tau_{\text{cross}}$. 
These systems will reach the stationary (single-striped) state 
before having time to enter into the asymptotic $t^{1/3}$ regime. For
these small systems, the only relevant behavior is the $t^{1/4}$ one.
Therefore, there is  a {\it size} crossover between $t^{1/4}$ asymptotic 
behavior for small systems and 
$t^{1/3}$ asymptotic behavior for large ones. The condition 
$\tau_{\text{cross}}(T,L_{\parallel}) = \tau_{\text{ss}}(T,L_{\parallel},L_{\perp})$ 
defines the crossover size.

Consider now the parameter $\gamma\equiv\tau_{\text{cross}}\left(  T,L_{\parallel
}\right)  /\tau_{\text{ss}}\left(  T,L_{\perp},L_{\parallel}\right)  .$ It
follows that the $t^{1/3}$ behavior is dominant for $\gamma\ll1.$ However, one
also has that $\gamma\left(  T,L_{\perp},L_{\parallel}\right)  \rightarrow0$
for finite $T$ in the thermodynamic limit ($L_{\perp},L_{\parallel}%
\rightarrow\infty,\ L_{\perp}/L_{\parallel}=\operatorname{const}.).$
Consequently, the $t^{1/3}$ growth law is the general one, namely, the only
one we should expect to observe in a macroscopic system. Corrections to this
should only occur at early times in small systems. This is fully confirmed below.

One may also define a longitudinal length,\cite{prl01,mukamel} say
$\ell_{\parallel}\sim t^{a_{\parallel}},$ where one expects $a_{\parallel
}>1/3$ (given that the growth is more rapid longitudinally than transversely).
This length is only relevant during the initial regime, until stripes become
well-defined, all of them extending the whole length $L_{\parallel}$. This
condition may be taken as defining the onset of the multi-stripe state, which
may be characterized by $\ell_{\parallel}\left(  \tau_{\text{ms}}\right)
=L_{\parallel},$ from where it follows that $\tau_{\text{ms}}\sim
L_{\parallel}^{1/a_{\parallel}}.$ Interesting enough, this is on the
macroscopic time scale, as for both $\tau_{\text{cross}}$ and $\tau
_{\text{ss}}$ (more precisely, $\tau_{\text{cross}}\sim L_{\parallel}$ and
$\tau_{\text{ss}}\sim L_{\parallel}L_{\perp}^{3}).$ The fact that all these
relevant times are on the macroscopic, observable time scale confirms that, as
argued above, the single-stripe (and not the multi-stripe) state is the only
stable one in general. It is also to be remarked that, once the multi-stripe
state sets in, the only relevant length is the transverse one, $\ell.$ Of
course, this is compatible with the possible existence of two correlation
lengths describing thermal fluctuations at criticality. 
\begin{figure}
\centerline{
\psfig{file=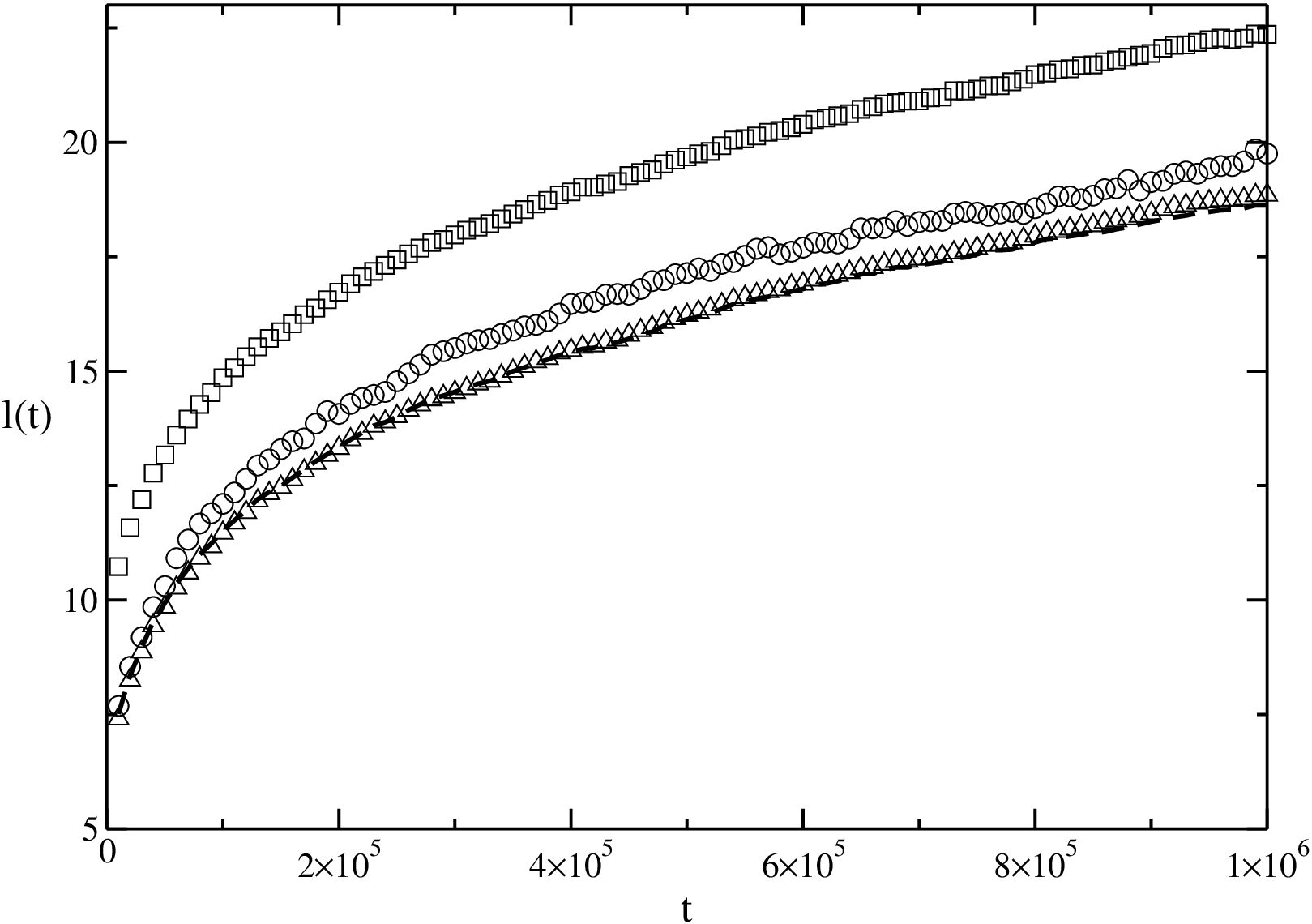,width=9.0cm}}
\caption[Time evolution of the relevant length scale.]
{\small Time evolution of the relevant length, $\ell\left(
t\right)$, as obtained by different methods, namely, from the
number $N_{s}$ of stripes (dashed line), from the maximum width,
$\ell_{\text{{\small max}}}$ $\left(  \Box\right)$ from the
mass, $\ell_{M}\left(  {\large \bigtriangleup}\right)$, and from
the peak of the structure function, $\ell_{S}\equiv2\pi/k_{\perp,\text{max}
}\left(  \bigcirc\right)$; these quantities are defined in the main
text. The graphs here correspond to an average over 600 independent runs for
the 128x128 lattice.\medskip}
\label{length}
\end{figure}

In order to test our predictions, several measures of the relevant length in
computer simulations were monitored, namely:

\begin{itemize}
\item the maximum width of the stripe, $\ell_{\text{max}},$ averaged over all
stripes in the configuration. This maximum width is defined as the distance in the
direction perpendicular to the field between the leftmost and the rightmost particles 
within the stripe;

\item $\ell_{M}\equiv M/L_{\parallel},$ where $M=M\left(  t\right)  $ is the
\textit{mass,} or number of particles belonging to the stripe, averaged over
all stripes in the configuration. This mass width is defined as the width of a 
perfectly dense stripe with $M$ particles.

\item $\ell_{s}\equiv L_{\perp}/2N_{s},$ where $N_{s}$ is the number of
stripes in the configuration.
\end{itemize}

\noindent After averaging over many independent evolutions, all these
quantities happen to behave similarly with time. Further measures of the
relevant length that we define in the next section behave in the same way. We
shall refer to this common behavior, which is illustrated in Fig. \ref{length}, as
$\ell\left(  t\right)  .$ (It is noticeable that, before showing a common
behavior, Fig. \ref{length} reveals some significant differences between our measures of
$\ell\left(  t\right)  $ at early times. This confirms the more difficult
description ---not attempted here--- which is required by the initial regime.) 
\begin{figure}[t]
\centerline{
\psfig{file=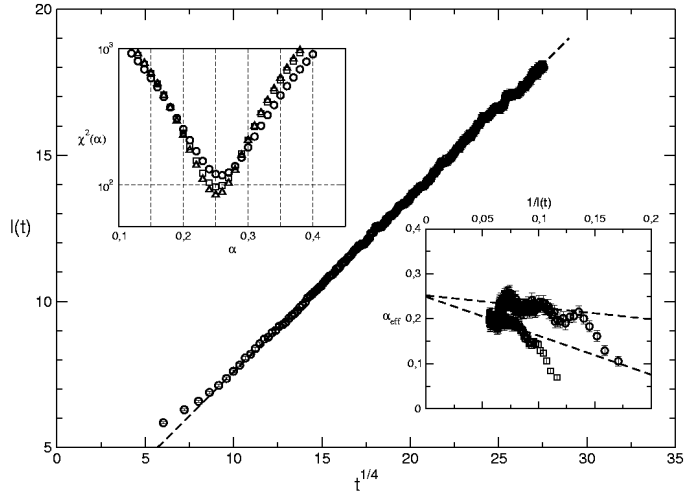,width=9.0cm}}
\caption[Growth law for small systems.] 
{\small The main graph shows $\ell\left(  t\right)  =\ell
_{S}\left(  t\right)$ versus $t^{a_{\perp}}$ for
$a_{\perp}=1/4$ in the case of the \textquotedblleft
small\textquotedblright\ 64x64 lattice. A similar behavior is obtained for any
of the studied measures of $\ell$ (see the main text for
definitions), which are represented in the insets by different symbols,
namely, $\ell_{\text{{\small max}}}$ ($\Box$), $\ell_{S}$
($\bigcirc$), and $\ell_{M}$ ($\triangle$).
The upper inset shows the chi square function for varying $a_{\perp}%
$ as obtained from a series of fits; a well-defined minimum is
exhibited indicating that $a_{\perp}\simeq1/4$ in this case. The
lower inset shows the \textit{effective exponent,} $\text{d}\log
_{2}\ell\ /\ \text{d}\log_{2}t,$ as a function of 
$1/\ell\left(  t\right)  $; this extrapolates to the same value of
$a_{\perp}$. \medskip}
\label{uncuarto}
\end{figure}

\begin{figure}
\centerline{
\psfig{file=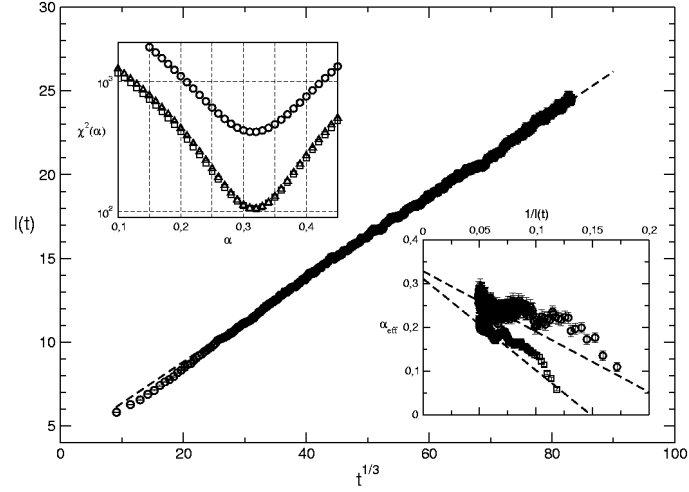,width=9.0cm}}
\caption[Growth law for large systems.]
{\small Same as Fig. \ref{uncuarto} but demonstrating that $a_{\perp}=1/3$
for the \textquotedblleft large\textquotedblright\ $L_{\perp}\times L_{\parallel} = 256x64$ 
lattice (one obtains a similar result for larger $L_{\parallel}$).\medskip}
\label{untercio}
\end{figure}

In Figs. \ref{uncuarto} and \ref{untercio} 
we illustrate our analysis and main results concerning the (late)
time evolution of $\ell\left(  t\right)  .$ The predictions above are
confirmed and, in particular, \textquotedblleft small\textquotedblright%
\ lattices ---Fig. \ref{uncuarto}--- happen to behave differently than \textquotedblleft
large\textquotedblright\ lattices ---Fig. \ref{untercio}. In both cases we plotted
$\ell\left(  t\right)  $ versus $t^{a_{\perp}}$ for varying $a_{\perp},$
looking for the best linear fit $\ell(t)=\alpha t^{a_{\perp}}+\zeta,$
excluding the initial time regime. The upper insets in the figures show the
chi square function associated to each fit, namely,
\begin{equation}
\chi^{2}\left(  a_{\perp}\right)  =\sum_{i=1}^{\eta}\frac{\left[  \ell
(t_{i})-\left(  \alpha t_{i}^{a_{\perp}}+\zeta\right)  \right]  ^{2}}{\alpha
t_{i}^{a_{\perp}}+\zeta},
\end{equation}
for a least-squares fit to $\eta$ data points using parameters $a_{\perp},$
$\alpha$ and $\zeta$. The graphs confirm the existence of a common
behavior for all the monitored measures of $\ell\left(  t\right)  $ (indicated
by different symbols). These graphs also demonstrate that $\ell(t)=\alpha
t^{a_{\perp}}+\zeta,$ with small $\zeta,$ during the whole time regime of
consideration. On the other hand, the upper insets indicate that $a_{\perp}$
is very close to $\frac{1}{4}$ for \textquotedblleft small\textquotedblright%
\ systems (in fact, for $L_{\perp}\leq128$) while $a_{\perp}\simeq\frac{1}{3}$
as the system becomes larger, say $L_{\perp}\geq256$ that corresponds to a
\textquotedblleft large\textquotedblright\ lattice according to familiar MC
standards. As an alternative method to analyze $\ell\left(  t\right)  ,$ one
may evaluate
\begin{equation}
\overline{a}\left(  t\right)  \equiv\frac{\operatorname{d}\log_{n}\ell\left(
t\right)  }{\operatorname{d}\log_{n}t}.
\end{equation}
Our prediction is that $\overline{a}\left(  t\right)  =a_{\perp}-\zeta
a_{\perp}/\ell\left(  t\right)  ,$ i.e., this should provide the exponent
$a_{\perp}$ by extrapolating to large $\ell\left(  t\right)  $ (late time).
The insets at the bottom of Figs. \ref{uncuarto} and \ref{untercio} 
show the results for $n=2.$ They are in
agreement with the other method, and again confirm our predictions.

\begin{figure}
\centerline{
\psfig{file=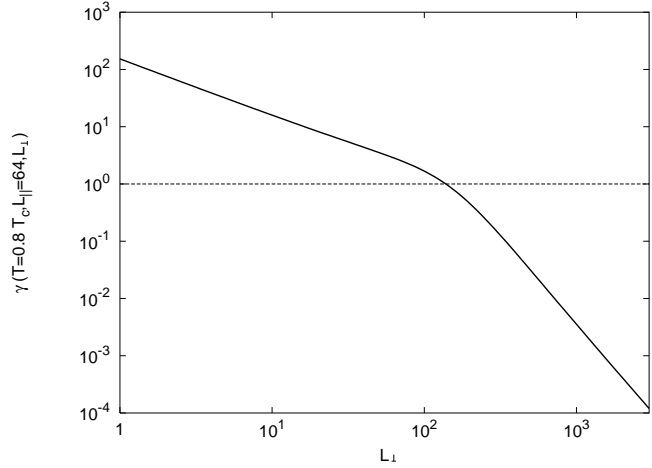,width=9.0cm}}
\caption[The parameter $\gamma(T,L_{\perp},L_{\parallel})$ signaling the size crossover.]
{\small The parameter $\gamma=\tau_{\text{{cross}}
}(T,L_{\parallel})/\tau_{\text{{ss}}}(T,L_{\perp},L_{\parallel}),$
with the characteristic times $\tau_{\text{cross}}$ and
$\tau_{\text{ss}}$ defined in the main text, as a function of
$L_{\perp}$ for $L_{\parallel}=64$, using our estimates for the amplitudes $\alpha_A$ and $\alpha_B$. 
This confirms our distinction between \textquotedblleft small\textquotedblright\ and
\textquotedblleft large\textquotedblright\ lattices, as explained in the main
text.\medskip}
\label{gamma}
\end{figure}

As indicated above, the size crossover between the $t^{1/4}$ and $t^{1/3}$ asymptotic regimes
is expected for a system size $(L_{\parallel},L_{\perp})$ such that 
$\tau_{\text{cross}}\left(  T,L_{\parallel}\right)
=\tau_{\text{ss}}\left(  T,L_{\perp},L_{\parallel}\right)  .$ 
In order to make
this condition explicit, we need to estimate the amplitudes $\alpha_{\text{A}%
}\ $and $\alpha_{\text{B}}$ in (\ref{kineq}); see equations (\ref{time1}) and
(\ref{time2}). These amplitudes, which state the relative importance of
surface evaporation/condensation versus bulk hole-diffusion, are given
respectively by $\alpha_{\text{A}}=4q^{-1}\langle\delta y^{2}\rangle
\operatorname{e}^{-2\bar{\Delta}/T}\ $and $\alpha_{\text{B}}=2q^{-1}\rho_{h}.$
We note that, for a sufficiently flat interface (i.e., one that involves
microscopic --but not macroscopic-- roughness), $\langle\delta y^{2}%
\rangle\sim\mathcal{O}\left(  1\right)  $ and $\bar{\Delta}\simeq5.$ On the
other hand, the excess energy associated to an isolated hole is 16, so that
$\rho_{h}\sim\exp\left(  -16/T\right)  $ is a rough estimate of the hole
density. As depicted in Fig. \ref{gamma}, it follows numerically, in full agreement with
our observations, that $a_{\perp}=\frac{1}{4}$ is to be observed only at early
times, earlier for larger systems; to be more specific, the crossover for
$L_{\parallel}=64,$ for instance, is predicted for $L_{\perp}\sim140,$ which
confirms the above; see also Figs. \ref{uncuarto} and \ref{untercio}.

This behavior may be understood on simple grounds. The surface/volume ratio is
large initially and, consequently, mechanism A (based on surface events) is
then dominant. This is more dramatic the smaller the system is. That is, the
surface is negligible for macroscopic systems, in general, and, as illustrated
in Fig. \ref{snap3}, even if the surface is relevant at very early times, its ratio to
the volume will monotonically decrease with time. This causes hole diffusion
in the bulk (mechanism B) to become dominant, more rapidly for larger systems,
as the liquid phase is trying to exhibit only two surfaces.
On the other hand, ref. \cite{mukamel} 
studies the stripe coarsening process in the infinitely 
driven lattice gas. Pure $t^{1/3}$ behavior is reported assuming 
the stripe evaporation mechanism.
This result is perfectly compatible with our results, given that the systems in ref. 
\cite{mukamel} correspond to very large values of $L_{\perp}$ ($800$ and $960$) and small 
values of $L_{\parallel}$ ($8$, $16$ and $32$). For these shapes our theory also predicts
the (simple) $t^{1/3}$ asymptotic behavior.

\section{Correlations and the Structure Factor}
\label{capDLG_apCor}

\begin{figure}
\centerline{
\psfig{file=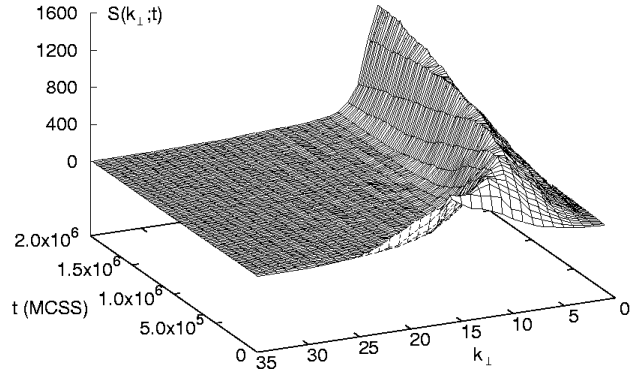,width=10.0cm}}
\caption[Time development of the structure factor.]
{\small Time development of the structure factor $S(k_{\perp};t)$, as defined in the
main text, for a \textquotedblleft large\textquotedblright\ lattice 
$L_{\perp}\times L_{\parallel}= 256\times 256$ during
early and intermediate phase segregation. A peak grows with time as it shifts
towards the small values of $k_{\perp}.\medskip$}
\label{sfactor}
\end{figure}

Consider now the Fourier transform of the pair correlation function
$C(x,y;t)=\left\langle n_{0,0}\left(  t\right)  \ n_{x,y}\left(  t\right)
\right\rangle ,$ where $n_{x,y}$ stands for the occupation variable at lattice
site $\vec{r}=\left(  x,y\right)  .$ This is the so-called structure factor,
$S(\vec{k},t),$ where $\vec{k}=(k_{\Vert},k_{\perp}).$ Given that the
$k_{\parallel}$ dependence is only relevant at early times, before the
multi-stripe state sets in, i.e., for $t<\tau_{\text{ms}}$, we shall set
$k_{\parallel}=0$ in the following. That is, our interest here is on%
\begin{equation}
S\left(  k_{\perp};t\right)  =\frac{1}{L_{\parallel}L_{\perp}}\left\vert
\sum_{x,y}n_{x,y}\left(  t\right)  \exp\left[  \text{i}k_{\perp}y\right]
\right\vert ^{2}.
\end{equation}
As illustrated in Fig. \ref{sfactor}, this function develops a peak at $k_{\perp
}=k_{\text{max}}\left(  t\right)  $ immediately after quenching. The peak then
monotonically shifts towards smaller wave numbers with increasing $t;$ in
fact, one expects $k_{\perp}\rightarrow0$ as $t\rightarrow\infty$ in a
macroscopic system. The wave length $\ell_{S}\equiv2\pi/k_{\text{max}}$ turns
out to be an excellent characterization of the relevant order, namely, it
measures both the stripe width and the stripe separation during phase
segregation. In particular, we confirm that $\ell_{S}\left(  t\right)  $ has
the common behavior discussed above for length $\ell\left(  t\right)  ;$ see
Figs. \ref{length}, \ref{uncuarto} and \ref{untercio}.

\begin{figure}
\centerline{
\psfig{file=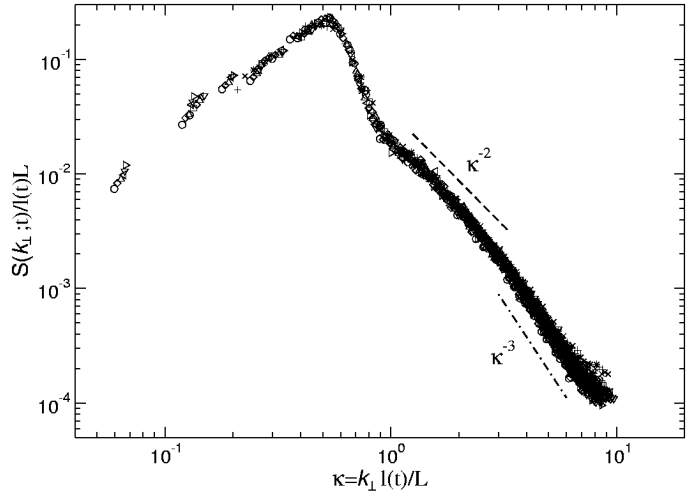,width=9.0cm,angle=0}}
\caption[Scaling with both time and size of the structure function.]
{\small The scaling with both time and size of the structure
function to show that $\Phi(\kappa)\equiv S\left(  k_{\perp};t\right)
/\ell L,$ with $\kappa=k_{\perp}\ell L^{-1}$, is
well-defined and universal, i.e., the same at any time (excluding some early
evolution) and for any square lattice of side $L.$ This plot
includes all data for $t\geq$ 10$^{4}$ MC steps and 64x64,
128x128, and 256x256 lattices. The broken lines illustrate the different kinds
of behavior of $\Phi\left(  \kappa\right)$ that are discussed in
the main text.$\medskip$}
\label{scalingfactor}
\end{figure}

The fact that the DLG shows a unique \textit{time-dependent} relevant length,
$\ell_{\perp}=\ell\left(  t\right)  ,$ has some important consequences. For
example, extrapolating from the equilibrium case (see section \ref{capDLG_apIntro}),\cite{prl00} one
should probably expect dynamical scaling, i.e.
\begin{equation}
S\left(  k_{\perp};t\right)  \propto\ell\left(  t\right)  F\left[  k_{\perp
}\ell\left(  t\right)  \right]
\end{equation}
for the anisotropic DLG in two dimensions. This is indeed observed to hold
during most of the relaxation and, in particular, during all the segregation
process after formation of well-defined stripes. This is illustrated in Fig. \ref{scalingfactor}
depicting the scaling function $F.$ A time-dependent mean-field model of a
binary mixture in shear flow has recently been demonstrated to exhibit a
similar property, though involving two lengths both behaving differently from
$\ell\left(  t\right)  $ above.\cite{corberi} 

The structure factor may be obtained by scattering, which makes it an
important tool in many studies. Analyzing further the details of functions
$S\left(  k_{\perp};t\right)  $ and $F\left(  \varkappa\right)  $ or,
alternatively, the universal function $\Phi\left(  \varkappa\right)  \equiv
S/\ell L$, as observed in computer simulations is therefore of great interest
(the extra $L$ factor in the definition of $\Phi\left(  \varkappa\right)$ is
our finite size scaling ansatz).
Experimental studies often refer to the mean `radius of gyration' of the
grains as the slope of the straight portion in a plot of $\ln\left[  S\left(
k,t\right)  \right]  $ \textit{versus} $k^{2}.$\cite{guin} We checked the
validity under anisotropic conditions of this concept, which is in fact quite
useful in equilibrium even outside the domain of validity of its
approximations.\cite{prl00} We confirm that $S\left(  k_{\perp};t\right)  $
exhibits the Guinier Gaussian peak, namely,
\begin{equation}
\Phi\left(  \varkappa\right)  \sim\exp\left[  -\text{const.}\left(
\varkappa-\varkappa_{\text{max}}\right)  ^{2}\right]
\end{equation}
around the maximum $\varkappa_{\text{max}}.$ More intriguing is the behavior
of $\Phi\left(  \varkappa\right)  $ before the peak, $\varkappa<\varkappa
_{\text{max}}.$ Fig. \ref{scalingfactor} indicates that scaling does not hold in this region even
at the end of our (otherwise long enough) simulations. This is so because
$\Phi\left(  \varkappa\right)  $ goes as $\rho^{2}L/\ell\left(  t\right)  $ at
$k_{\perp}=0$, 
and thus depends on time for very small values of $\kappa$, breaking the
scaling observed for larger values of $\kappa$.
However, a detailed study of data reveals that the scaling
function near the origin tends with time towards a common envelope
$\Phi\left(  \varkappa\right)  \sim\varkappa^{1+1/3}$ for $\varkappa
_{0}<\varkappa<\varkappa_{\text{max}};$ we do not have a simple explanation of
this. In any case, this behavior breaks down close to the origin,
$\varkappa\lesssim\varkappa_{0},$ where $\Phi\left(  \varkappa\right)
\rightarrow0$ as $\varkappa\rightarrow0$ and $t\rightarrow\infty$ for the
infinite system.

The behavior after the peak, $\varkappa>\varkappa_{\text{max}},$ may be
predicted on simple grounds. The (sphericalized) structure factor for
(equilibrium) isotropic binary mixtures is known to satisfy the Porod's law,
$S\sim k^{-(d+1)}$ at large enough $k,$ where $d$ is the system
dimension,\cite{prl00} i.e., $S\sim k^{-3}$ in two dimensions. The main
contribution to the large-$k$ tail comes from the short-distance behavior of
$C(x,y;t).$ That is, the Porod's region for the DLG may be taken to correspond
to $\lambda_{\bot}\ll k_{\bot}^{-1}\ll\ell\left(  t\right)  ,$ where
$\lambda_{\bot}$ stands for a (transverse) thermal length that characterizes
the smallest, thermal fluctuations. Let two points, $\vec{r}_{0}$ and $\vec
{r}_{0}+\vec{r},$ $\vec{r}=(x,y).$ For any $x$ such that $\lambda_{\perp}\ll
x\ll\ell(t),$ one roughly has that the product $n_{\vec{r}_{0}}\left(
t\right)  \ n_{\vec{r}}\left(  t\right)  $ equals $+1$ if the two points are
on the stripe, and $0$ otherwise, i.e., if either an interface exists between
them or else the two points belong to the gas between stripes. Since $x\ll
\ell(t)$, the probability that $\vec{r}$ crosses more than one interface is
negligible. For a half-filled system, the probability that $\vec{r}_{0}$ lies
at a particle stripe is $%
\frac12
,$ and the probability that both $\vec{r}_{0}$ and $\vec{r}_{0}+\vec{r}$
belong to the same stripe is roughly $%
\frac12
\left(  \ell\left(  t\right)  -x\right)  /\ell\left(  t\right)  .$ Hence,
\begin{equation}
C\left(  x,y;t\right)  \simeq\frac{1}{2}\left(  1-\frac{x}{\ell\left(
t\right)  }\right)  ,\qquad{\small x\ll\ell\left(  t\right)  .}%
\end{equation}
By power counting, this implies the \textit{anisotropic Porod law} (in two
dimensions):%
\begin{equation}
S\left(  k_{\perp};t\right)  \sim\frac{1}{\ell\left(  t\right)  \ k_{\bot}%
^{2}},\qquad{\small \lambda_{\perp}\ll k_{\bot}^{-1}\ll\ell\left(  t\right)
.} \label{Porod}%
\end{equation}
Therefore, $\Phi\left(  \varkappa\right)  \sim\varkappa^{-2}L^{-1},$ which is
confirmed in Fig. \ref{scalingfactor}. This is in contrast with the (isotropic) Porod's result.
The difference is a consequence of the fact that the DLG clusters are stripes
that percolate in the direction of the field, instead of the isotropic
clusters of the LG. The short-distance pair correlation function for the
latter is $C(\vec{r};t)\simeq%
\frac12
\left(  1-|r|/\ell\left(  t\right)  \right)  $, from which one has that
$\Phi\left(  \varkappa\right)  \sim\varkappa^{-3}.$ It follows that anisotropy
may easily be detected by looking at the tail of the structure factor.

The detailed analysis of $S\left(  k_{\perp};t\right)  $ also reveals that, as
$L_{\parallel}$ is increased in computer simulations, the anisotropic behavior
$\Phi\sim\varkappa^{-2}$ crosses over to $\Phi\sim\varkappa^{-3}$ for larger
$\varkappa$; see Fig. \ref{scalingfactor}. We believe this reflects the existence of standard
thermal fluctuations. That is, very small clusters of particles occur in the
gas in the asymptotic regime whose typical size in the direction perpendicular
to the field is of order $\lambda_{\perp}$. These very-small asymptotic
clusters are rather isotropic, namely, they do not differ essentially from the
corresponding ones in equilibrium binary mixtures. More specifically, for
$x\sim\lambda_{\perp},$ one may approximate $C(\vec{r};t)\sim1-|r|/\lambda
_{\perp}(t),$ which implies the $\varkappa^{-3}$ power-law tail for large
$\varkappa.$ On the other hand, according to (\ref{t1314}), the mean stripe
width grows as $\ell(t)\sim(t/L_{\parallel})^{a}$ with $a=1/4$ or $a=1/3,$
depending on the value of $L_{\perp}$. Therefore, the number of stripes at
time $t$ is proportional to $L_{\perp}L_{\parallel}^{a}/t^{a}$ and, for a
given time, the number of stripes increases with $L_{\parallel}$ as
$L_{\parallel}^{a}$. We also know that, at a given time, the number of small,
fluctuating clusters is proportional to $L_{\parallel}$. Hence the relative
importance of small clusters due to thermal fluctuations as compared to
stripes is proportional to $L_{\parallel}.$ In fact, the $\varkappa^{-3}$ tail
is observed for large enough values of $L_{\parallel}$ but not for small lattices.

\section{A Continuum Description}
\label{capDLG_apCont}

The rigorous derivation of a general continuum analog of the driven lattice
gas is an open problem.\cite{MarroDickman} Recent studies led to the following proposal
for a coarse-grained density, $\phi(\mathbf{r},t):$\cite{Rbien}
\begin{equation}
\partial_{t}\phi(\mathbf{r},t)=\tau_{\perp}\nabla_{\perp}^{2}\phi
-\nabla_{\perp}^{4}\phi+\frac{\lambda}{6}\nabla_{\perp}^2\phi^{3}+\tau
_{\parallel}\nabla_{\parallel}^{2}\phi+\nabla_{\perp}\xi(\mathbf{r}%
,t).\label{bien}%
\end{equation}
Here, the last term stands for a conserved Gaussian noise representing the
fast degrees of freedom, and $\tau_{\perp}$, $\tau_{\parallel}$ and $\lambda$
are model parameters. Compared to previous proposals,\cite{Rmal,zia} this
Langevin type of equation amounts to neglect a non-linear current term,
$-\alpha\nabla_{\parallel}\phi^{2},$ that was believed to be essential
(\textit{relevant}) at criticality. However, one may show that, at least in
the limit $E\rightarrow\infty,$ the coefficient $\alpha$ cancels out (due in
this case to a subtle saturation effect).\cite{Rbien} In fact, recent scaling
analysis has unambiguously confirmed that a particle current is not relevant
and that equation (\ref{bien}) captures the correct critical behavior of the
DLG.\cite{prl01,albano} Consequently, an important question is now whether
(\ref{bien}) reproduces also the kinetic behavior of the DLG as described in
previous sections. We present here a first confirmation that, as
compared with other approaches,\cite{zia} (\ref{bien}) is indeed a proper
continuum description of the DLG kinetic relaxation. 
\begin{figure}
\centerline{
\psfig{file=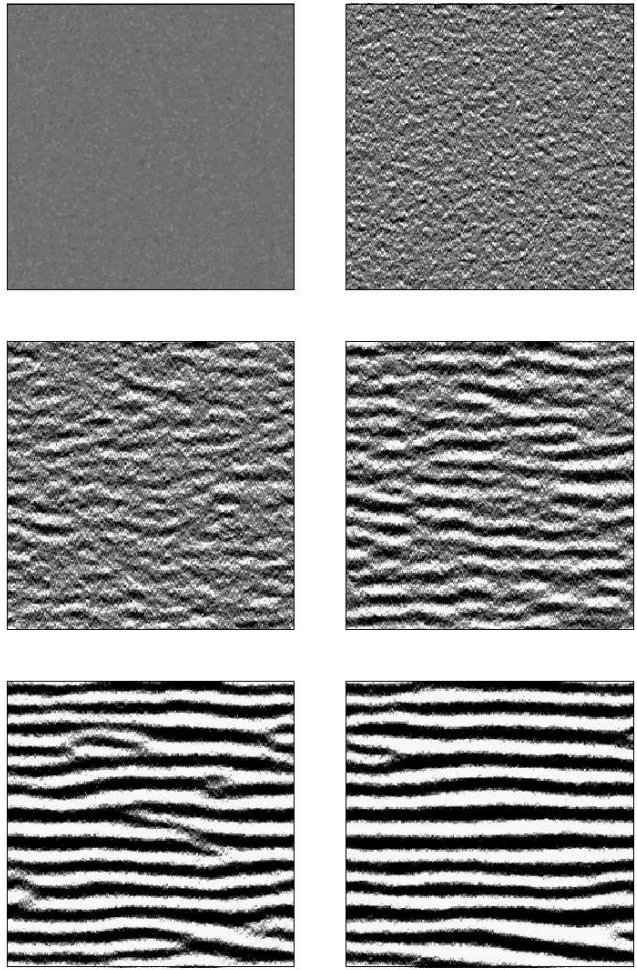,width=10.0cm,angle=0}}
\caption[Snapshots of the time evolution of the continuous equation.]
{\small Series of snapshots as obtained from equation (\ref{discrte}) 
for the $256\times256$ lattice with parameters as given in the main text.
Time (arbitrary units) is $t=0,10,100,200,500$, and $1000$ respectively, from
left to right and from top to bottom.$\medskip$}
\label{snapcont}
\end{figure}

In order to numerically integrate (\ref{bien}), let us introduce the indexes
$i,j=1,\dots,N$ to represent, respectively, the two components of
$\mathbf{r\equiv}\left(  x_{\perp},x_{\parallel}\right)  .$ One thus makes a
trivial discretization of the space, and then of the time by Cauchy-Euler
method.\cite{CEmethod} The result is a set of $N^{2}-1$ coupled nonlinear
equations, namely,%
\begin{align}
\phi(i,j;t+\Delta t) &  =\phi(i,j,t)\label{discrte}\\
&  +\Delta t\left[  \tau_{\perp}\widetilde{\nabla}_{\perp}^{2}\phi
-\widetilde{\nabla}_{\perp}^{4}\phi+\frac{\lambda}{6}\widetilde{\nabla}%
_{\perp}^{2}\phi^{3}+\tau_{\parallel}\widetilde{\nabla}_{\parallel}^{2}%
\phi\right]  +\sqrt{\Delta t}\widetilde{\nabla}_{\perp}\xi(i,j;t).\nonumber
\end{align}

This equation is to be solved by the computer. With this aim, we may write
$\widetilde{\nabla}_{\perp}\xi(i,j;t)=\left[  \xi(i+1,j;t)-\xi
(i-1,j;t)\right]  /2\Delta x_{\perp}$ and $\phi(i,j;t)\equiv\phi(i\Delta
x_{\perp},j\Delta x_{\parallel};t)$ where $\Delta x_{\perp}=L_{\perp}/N$ and
$\Delta x_{\parallel}=L_{\parallel}/N.$ The maximum value of $\Delta x_{\perp
}$ is thus limited by the interface width. For Fig. \ref{snapcont}, which concerns a
$256\times256$ lattice $(N=256)$ we ---rather arbitrarily--- used $\Delta
x_{\perp}=\Delta x_{\parallel}=1.7,$ and $\Delta t=0.05,$ which produce a
locally stable solution. The parameters $\tau_{\perp}$, $\tau_{\parallel}$ and
$\lambda$, are fixed on the basis of its physical meaning. The \textit{mass}
terms $\tau_{\parallel}$ and $\tau_{\perp}$ represent temperatures along the
longitudinal and transverse directions, respectively, relative to the critical
temperature, i.e., $\tau_{\perp}\sim(T_{\perp}-T_{C}^{\infty}).$ Given the
anisotropy of phase segregation, with longitudinal interfaces only,
$\tau_{\perp}<0$ and $\tau_{\parallel}>0.$ On the other hand, $\left\vert
\tau_{\perp}\right\vert $ should be small enough to allow for a relatively
fast evolution. Our choices for Fig. \ref{snapcont} are $\tau_{\perp}=-0.25$, $\tau
_{\parallel}=0.5$ and $\lambda=1$.

It is remarkable that, in spite of some apparent similarity, the
problem here differs from the one in the study of (standard) spinodal
decomposition by means of the isotropic ($E=0$) Cahn-Hilliard equation. In
equilibrium,\cite{CHeq} one usually assumes that the influence of noise on
growth, which is then assumed to be directly driven by surface tension, is
negligible far from criticality. The noise term in (\ref{discrte}) may be
expected to be important in a more general context, however. That is, as
described in section \ref{capDLG_apGro}, the DLG develops striped patterns in which
surface tension smooths the interfaces but has no other dominant role on the
basic kinetic events. Consequently, neglecting the noise in (\ref{discrte})
would turn metastable any striped geometry after coarsening of strings, which
is not acceptable (see section \ref{capDLG_apGro}).

Finally, it  is  interesting  to notice that  if a one-dimensional structure is assumed, and 
the gradient in the direction parallel to the field  in  Eq. \ref{bien} is  eliminated,
then  this equation reduces to the one-dimensional time-dependent Ginzburg-Landau model 
in \cite{kawakatsu}. 
There  it was found a $\ln(t)$ growth at zero temperature  and a crossover
from $\ln(t)$ to $t^{1/3}$ at finite temperatures.

\section{Conclusion}
\label{capDLG_apConc}

This chapter presents a theoretical description of spinodal decomposition in the
DLG, and compares it with new data from a kinetic Monte Carlo study. This is
also compared with the kinetic implications of a Langevin, continuum equation
that had previously been shown to capture correctly the critical behavior of
the DLG. The resulting picture from these three approaches, which is
summarized below, should probably hold for a class of highly-anisotropic phase
segregation phenomena. In fact, our results provide a method for analyzing
experiments that could be checked against laboratory realizations of the DLG,
i.e., the case of phase segregation under biased fields or other influences
such as electric fields, gravity and elastic stresses.

Immediately after a deep quench, there is an early regime in which anisotro-pic
grains develop. They tend to coarsen to form small strings that then combine
into well-defined thin stripes. Such nucleation and early coarsening (Figs. \ref{snap1} and \ref{snap2})
seem governed by surface tension at the string ends competing with other both
surface and bulk processes. This complicated situation typically extends less
than 10$^{3}$ MC steps in computer simulations, which corresponds to a very
short macroscopic time, so that it would be hardly observable in experiments.
As a matter of fact, most of the system relaxation proceeds by coarsening of
stripes until full segregation (Fig. \ref{snap3}). Surprisingly enough, this regime,
which has been studied for more than a decade now,\cite{aurora}-\cite{mukamel}
happens to be theoretically simpler than the corresponding one for the
isotropic case.\cite{rev1}-\cite{toral}

The evolution from many stripes to a single one mainly proceeds by competition
of two mechanisms: (A) evaporation of a particle (hole) from one stripe surface and
subsequent deposition at the same surface, and (B) diffusion of a hole within
the bulk of the stripe. The first one dominates initially (and lasts more for
smaller systems), when the surface/volume ratio is relatively large. Mechanism
A implies that the relevant length (as defined in Fig. \ref{length}) increases with time
according to $\ell\left(  t\right)  \sim t^{1/4}.$ The surface/volume ratio
decreases with time, however, and mechanism B soon becomes dominant. This
implies $\ell\left(  t\right)  \sim t^{1/3}$ which is the general prediction
for a macroscopic system (cf. Figs. \ref{uncuarto}, \ref{untercio} and 
\ref{gamma})\footnote{The $t^{1/3}$ behavior 
here is in contrast with the logarithmic growth that is assumed to govern a class 
of lattice models in which coarsening is not a direct consequence of surface 
tension.\cite{voter}}. This was obtained
before by assuming coarsening of two (liquid) stripes by evaporation of the
gas stripe placed between them;\cite{mukamel} see also \cite{aurora}. Note that
the $t^{1/3}$ law is precisely the behavior which is acknowledged to be
dominant under isotropy, but this has a different origin in the equilibrium
case.\cite{rev4,rev5} Note also that surface tension determines evaporation
rates but has no other influence on mechanisms A and B.

The $t^{1/3}$ growth law, (\ref{t13}), is perfectly confirmed by the DLG data
(Fig. \ref{untercio}). This indicates time-scale invariance. In fact, such invariance was
demonstrated for the isotropic case, in which the situation is somewhat more
involved (section \ref{capDLG_apIntro}). 
The invariance property may be better analyzed by looking at
the structure factor transversely to the drive, $S\left(  k_{\perp},t\right)
$ (Fig. \ref{sfactor}). This exhibits \textit{dynamic scaling}, i.e., it remains
self-similar during phase segregation\footnote{More properly, one should speak 
here of \textquotedblleft self-affinity\textquotedblright, given the underlying 
anisotropy.\cite{mandel}}. More specifically,
$\Phi(\varkappa)\equiv S\left(  k_{\perp};t\right)  /\ell L,$ with
$\varkappa=k_{\perp}\ell L^{-1}$, is universal, namely, the same at any
(sufficiently late) time $t$ and for any square lattice of side $L.$
Furthermore, the function $\Phi(\varkappa)$ has a well-defined shape. In
particular, it exhibits the Guinier Gaussian peak, and this is followed by the
\textit{anisotropic Porod} decay, $\Phi\left(  \varkappa\right)  \sim
\varkappa^{-2}$ and then by a \textit{thermal tail} $\Phi\left(
\varkappa\right)  \sim\varkappa^{-3}$ (Fig. \ref{scalingfactor}). Also noticeable is the fact
that the the parameter to scale along the $S$ axis is $J(t)=\ell$ and not
$\ell^{2}$ as under isotropy.

Our results in this chapter have two main restrictions, both due to the great
computational effort required by this problem.\cite{MarroDickman} Firstly, they follow
from an extensive analysis of only one phase-diagram point, namely, $\rho=%
\frac12
,$ $E=\infty,$ and $T=0.8T_{C}^{0}.$ However, our own observations (including
brief investigation of other points), together with an extrapolation of the
many results known for the isotropic case, strongly suggest that the picture
in this chapter holds within a large domain around the center of the miscibility
gap\footnote{Extending some of our arguments to $\rho\neq\frac{1}{2}$ needs
some care but the whole picture should still be valid, at least not far from
$\rho=\frac{1}{2}.$}. In fact, the scaled structure factor for isotropic systems
was shown to be almost independent of density and temperature, and even the
substance investigated, in a wide region below the coexistence line.\cite{LSWbis} 
Our consideration of only a two-dimensional system
does not seem a real restriction neither. That is, adding an extra
(transverse) dimension should not essentially modify the picture
here\footnote{Note that the Lifshitz-Slyozov-Wagner behavior is known to be
valid in some cases in both $d=2$ and $d=3$.\cite{rev5}}.

It would be interesting to look next in the laboratory for both time-scale
invariance and $t^{1/3}$ growth under highly anisotropic conditions. In fact,
there are some evidences of such behavior in sheared fluids (section \ref{capDLG_apIntro}), and one
may think of some more direct experimental realizations of the driven lattice
gas. In particular, coarsening striped patterns very similar to those observed
in our system are found in some intriguing experiments on granular binary mixtures 
under shaking.\cite{Mullin} We think that the mechanisms we propose in this chapter should help
the understanding of such experimental results. In general,
we hope our observations will motivate both experiments and future more
complete theories.

\chapter{Dynamic Phase Transitions in Systems with Superabsorbing States}
\label{capLipo}

\section{Introduction}
\label{capLipo_apIntro}

Dynamic phase transitions separating active from  fluctuation-free
 absorbing phases appear in a vast group of physical phenomena
and models
as, for instance, directed percolation
\cite{reviews,Hinrichsen},  catalytic reactions \cite{catal}, 
 the pining of surfaces by disorder \cite{Barabasi},
 the contact process
\cite{CP}, damage spreading transitions \cite{damage}, 
nonequilibrium wetting \cite{firstorder2}, or  sandpiles
\cite{soc,soc2}. See \cite{reviews} and  \cite{Hinrichsen} for recent
reviews. 
Classifying these transitions into universality
 classes is a first
priority theoretical task. As conjectured by Janssen and Grassberger
\cite{conjecture} some
time ago and corroborated by a huge number of theoretical studies
 and computer simulations, systems exhibiting a continuous transition into a 
unique absorbing state with no extra symmetry or conservation law 
belong to one and the same universality class, namely that of 
directed percolation (DP). At a field theoretical level this class 
is represented by the Reggeon field theory (RFT) \cite{RFT}.

 This universality conjecture has been extended to include
multicomponent systems \cite{Lai}
 and systems with infinitely many absorbing states
\cite{Muchos,many}. On the other hand some other, 
less broad, universality classes
of systems with absorbing states
have been identified in recent years. They all 
include some extra symmetry or conservation law, 
foreign to the DP class.
 For example, if two symmetric absorbing states exist 
(which in many cases is equivalent to having activity parity-conservation 
 \cite{baw}), the universality class is other than DP, and the corresponding
 field theory differs from RFT  \cite{CT}.         
A second example is constituted by systems with absorbing states 
in which fluctuations occur only at the interfaces separating active 
from absorbing regions, but not in the bulk of compact active regions
(examples of this are the {\it voter model} or compact directed percolation
\cite{CDP}).
In this case the exponents are also non DP. A third and last example
is that of systems with many absorbing states
 in which the activity field is coupled
 to an extra conserved field.
This type of situation appears, for example, 
in conserved sandpile models, and 
has been recently shown to define a
new universality class \cite{soc,Conserved}.
   Apart from these and some few other well
 known examples\footnote{For example, long range interactions do also change
the universality.\cite{other}}, systems
with absorbing states belong generically into the DP universality class.

  Recently, it  has been proposed a very simple, biologically motivated 
model, exhibiting a continuous transition into an absorbing phase,\cite{Lip1} and
claimed that this model shows a sort of ``superuniversality'', i.e.
in both 
one and two dimensions the model has the same critical exponents,
namely those of one-dimensional DP. Consequently, the 
system has been hypothesized to show a rather strange 
 {\it dimensional reduction}\footnote{Dimensional reduction is not 
a new concept in statistical physics.
For example, quenched disordered magnetic systems were some time ago claimed
to behave in d dimensions as their corresponding pure counterparts in $d-2$
dimensions \cite{Parisi}. However, this results, is at odds with simple domain 
wall arguments, and
has recently proven to fail \cite{Brezin}.} in two dimensions.
This conclusion, if confirmed, would break the Janssen-Grassberger 
conjecture, since it is not clear that any new symmetry or extra conservation
law is present in this model. In what follows we show what are the
physical reasons why this model does not show directed percolation
behavior: the presence of what we called {\it superabsorbing sites}
is at the basis of this anomalous behavior. We will discuss also how
DP can be restored by changing the geometry of the lattice
on which the model is defined.

\section{The Model}
\label{capLipo_apMod}

 The model (called from now on Lipowski model) is defined operationally as follows:
let's consider a square d-dimensional 
lattice. At a bond linking neighboring sites, $i, j$,
a random variable $w= w_{ij}$ is assigned.
 Different bonds are 
uncorrelated, and $w$ is distributed homogeneously in the interval 
 $[0,1]$.  At each site $i$ one defines $r_i$ as the
sum of the four bonds connecting this site to its four nearest neighbors.
If $r_i$ is larger that a certain threshold, $r$ (that acts as a control 
parameter) the site is declared active,
 otherwise the site is inactive
or absorbing.  Active sites are considered unstable; at each step
one of them is chosen randomly  and its
 four associated $w_{ij}$ bond variables
are replaced by four freshly chosen independent random values
 (extracted from the same 
 homogeneous probability distribution), and time is incremented by
an amount
$\Delta t = 1/n(t)$, where $n(t)$ is the number of active sites at that
time.  Critical exponents are defined as it is customary in the
realm of absorbing phase transitions \cite{reviews}.

   It is clear that for small values of $r$, for instance $r=0$, the
system will always be active, while for large enough values of $r$ 
 an absorbing configuration (with
$r_i < r$ for all sites $i$) will be eventually reached.
 Separating these two regimes we observe
 a critical value of $r$, $r_c$,
 signaling the presence of a continuous phase
transition.  In $d=1$ $r_c \approx 0.4409 $ \cite{Lip1}, while
for $d=2$ we find $r_c=1.38643(3)$.  As bond variables are continuous 
it is obvious that there is a continuous degeneracy of the absorbing
state (i.e. infinitely many absorbing configurations). 

 In the one dimensional case, all the measured critical exponents  
take the expected DP values \cite{Lip1}, compatible with theoretical
 predictions for 
systems with many absorbing states \cite{many,JSP}. The only
discrepancy comes from the fact that the spreading 
exponents $\eta$ and $\delta$ (see section \ref{capLipo_apAna_subSprea} for definitions)
 appear to be non-universal, but
the combination $\eta+\delta$ coincides with the DP expectation. This 
non-universality in the spreading  is however generic of one-dimensional 
systems with an infinite number of absorbing states \cite{PCP,JSP},
and therefore it does not invalidate the conclusion that the
system behaves as DP.

 In two dimensions the only measured critical exponent in \cite{Lip2}
 is the order parameter one,  $\beta$, which
 has been reported to take a value surprisingly close to the
one dimensional DP expectation, $\beta \approx 0.27$  \cite{Lip2}.
Based on this observation it has been claimed that
the system exhibits a sort of dimensional reduction. This possibility
would be very interesting from a theoretical point of view and
elucidating it constitutes the main original motivation of what
follows.         

Finally let us mention that for spreading experiments it was found that,
as happens generically in two-dimensional systems with many absorbing states
\cite{JSP,Chate}, the critical point is shifted, and
 its location depends on the nature of the absorbing
environment the initial seed spreads in.  In particular,
 the annular type of growth described in
\cite{Lip2} in the case of spreading into favorable media is
typical of spreading in two-dimensional systems
 with many absorbing states, and it is well known 
to be described by dynamical percolation \cite{JSP,Chate}.

\section{Model Analysis}
\label{capLipo_apAna}

In order to obtain reliable estimations for $\beta$ and 
 determine other exponents, we have performed extensive
Monte Carlo simulations in $d=2$ combined with finite size scaling analysis,
as well as properly defined spreading experiments.

\subsection{Finite Size Scaling Analysis}
\label{capLipo_apAna_subSca}

We have considered square lattice with linear dimension $L$ 
ranging
 from 32 to 256.  Averages are performed over a number of
independent runs
ranging from $10^2$ to $10^5$ depending on the distance to the critical point 
and on system size.
The first magnitude we measure is the averaged density
of active sites,
$ \rho(L,r,t)$, which for asymptotically large times converges to 
a stationary value $\rho(L,r)$. 
Observe that for small system
sizes the system always reaches an absorbing configuration in finite
time and therefore the only truly
 stationary state is $\rho=0$. 
In order to extrapolate the right asymptotic behavior in the active phase
one has to determine $\rho(L,r)$
averaged over the runs which have not reach an absorbing configuration.
A peculiarity of this system is that its convergence towards  a well
defined 
stationary state is very slow,  fluctuations around mean values
are extremely persistent 
and, therefore, a huge number of runs is needed in order 
to obtain smooth evolution curves. 
Owing to this fact, numerical studies are
rather costly from a computational point of view.
 The reasons underlying such anomalously long lived fluctuations
will be discussed in forthcoming sections.
The maximum times considered are $8 \times 10^5$ Monte Carlo steps
per spin; this is
 one order of magnitude 
larger than simulations presented in \cite{Lip2}.  
Near the critical point the
relaxation times are very large (larger than $10^5$) and, 
in order to compute stationary 
averages, transient effects have been cut off. We observe the 
presence 
of a continuous phase transition separating the active from 
the absorbing phase at a value of $r \approx 1.38$.

\begin{figure}
\centerline{
\psfig{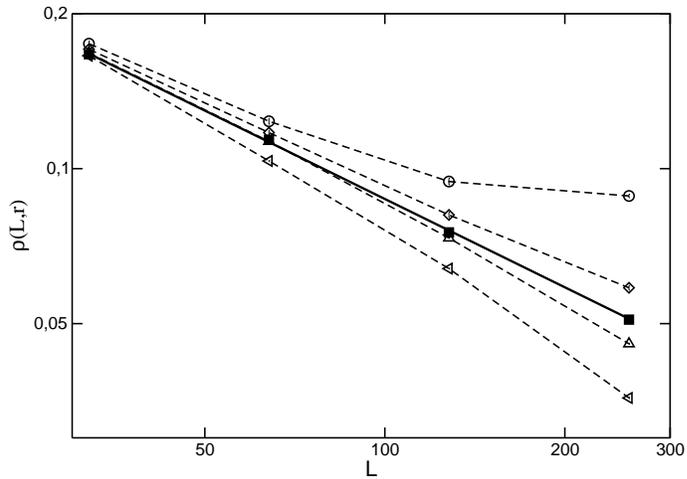}}
\caption[Density of active sites as a function of $L$.]
{\small
Log-log plot of the density of active sites
as a function of $L$ (the linear system size) for 
different values of $r$: from top to bottom, $1.38630$,
$1.38640$, $1.38643$, $1.38645$, and $1.38650$ respectively.
The straight solid line corresponds to the critical point
$r_c=1.38643(3)$.}
\label{beta_nu}
\end{figure}

Assuming that finite size scaling holds \cite{FSS} in the vicinity of
the critical point point $r_c$, we expect
for values of $r < r_c$ (i.e. in the active phase)
\begin{equation}
\rho(L,r) \sim L^{-\beta/\nu_\perp} {\cal{G}}(L/\Delta^{-\nu_\perp})
\label{uno}
\end{equation}
where $\Delta=| r-r_c|$.
Right at the critical point, this corresponds to a straight 
line in a double logarithmic plot 
of $\rho(L,r)$ vs. $L$. 
In Fig. \ref{beta_nu} it can be seen that, in fact, we
observe a straight line as a function
 of $log(L)$ for $r=1.38643(3)$ which
constitutes our best estimation of $r_c$. 
This finite size analysis
allows us to determine $r_c$  with much better precision 
 than in the previous estimations
\cite{Lip2}.  From the slope of the previous log-log plot we measure
$\beta/\nu_\perp = 0.57(2)$ which is quite far from both, the 
one-dimensional DP exponent
$\beta/\nu_\perp= 0.2520(1)$, and the two-dimensional value
$0.795(5)$. 

\begin{figure}
\centerline{
\psfig{file=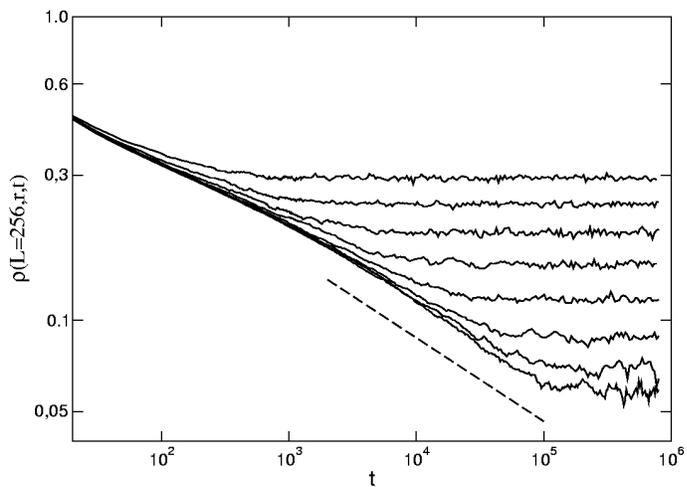,width=9cm,angle=0}}
\caption[Time evolution of the density of active sites.]
{\small
Log-log plot of the time evolution of the density of active sites for $L=256$ and different 
values of $r$ in the active phase, namely, from top to bottom
$r=1.38143$, $1.38402$, $1.38527$, $1.38587$, $1.38616$, $1.38630$,
$1.38637$, and $1.38640$ respectively.
 From the slope of the straight dashed line 
we estimate $\theta=0.275(15)$.
}
\label{theta}
\end{figure}

\begin{figure}
\centerline{
\psfig{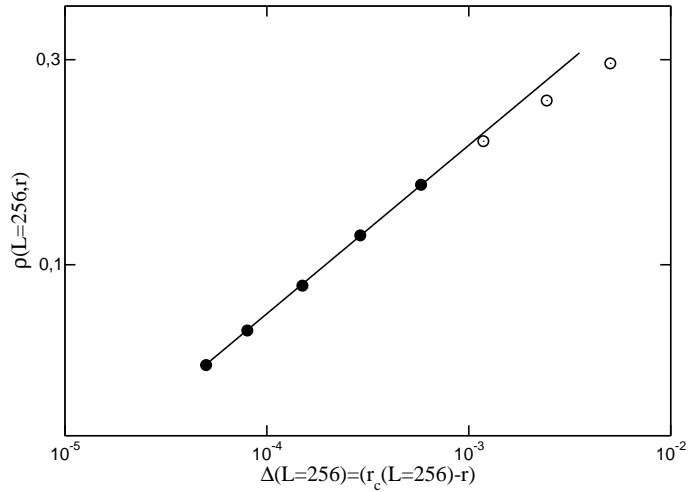}}
\caption[Stationary density of active sites near the critical point.]
{\small
Log-log plot of the stationary density of active sites as a function of the 
distance to the critical point, for $L=256$ and different values of
$r$ in the active phase (the same values reported in Fig \ref{theta}).
The best fit
gives $\beta=0.40(2)$ and $r_c(L=256) \approx 1.38645$. Filled (empty)
 circles are used to represent scaling (not scaling) points.
}
\label{beta}
\end{figure}

We have  considered the larger available system size,
$L=256$, and studied the time decay of a fully
active initial state for values of $r$ close to $r_c$ in the 
active phase (see Fig. \ref{theta}). The stationary values for
large values of $t$ should scale as  $\rho(L,r) \sim \Delta(L)^\beta$.
 From the best fit of our data (see Fig. \ref{beta}) we determine both
 $r_c(L=256) \approx 1.38645 $ 
and $\beta =0.40(2)$.

 At the critical point, $\rho(r=r_c ,t) \sim t^{-\theta}$.
 From the asymptotic slope of the curve for $r_c(L=256)$
  in Fig. \ref{theta}, we measure
$\theta = 0.275(15)$. In this way, we have already determined three 
independent exponents. 
From these, using scaling laws, we can determine others, as for example
 $\nu_\perp= \beta/ (\beta/\nu_\perp) = 0.69(9)$ (to be compared with the 
DP prediction $1.09$ in $d=1$ and $0.733$ in two dimensions \cite{exp}).

\begin{figure}
\centerline{
\psfig{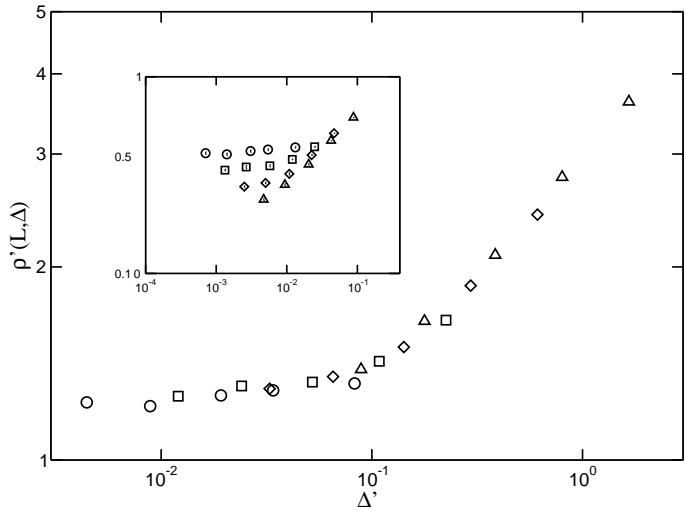}}
\caption[Data collapse for the density of active sites.]
{\small
Log-log data collapse for the density of active sites: $\rho' (L, \Delta) = 
\rho (L, \Delta) L^{\beta/\nu_\perp}$ and $\Delta' =\Delta L^{1/\nu_\perp}$. 
Using the obtained 
exponent values, $\beta/\nu_\perp \approx 0.57$, and $\nu_{\perp} \approx 0.69$,
 a reasonably good data collapse is observed.
In the inset we show an attempt to collapse data
using one-dimensional DP exponent values. There is no evidence of scaling
neither in this case nor using two-dimensional DP exponents.
}
\label{collapse}
\end{figure}

  To further verify the consistency of our results we have  considered
$\rho(L,r)$  computed for different values of $r$ and $L$,
and assumed that $\rho(L,r) L^{\beta/\nu_\perp}$, 
depends on $r$ and $L$ through 
the combination $L^{1/\nu_\perp} \Delta $ \cite{reviews}. 
 In Fig. \ref{collapse}, we show the
corresponding data collapse 
which is  rather good when
the previously reported values of $\beta$ and $\nu_\perp$ are used.
In the inset
we verify that the data points are broadly scattered when one-dimensional 
DP exponent values are considered, 
showing that the dimensional reduction hypothesis
is not valid. Data collapse is neither observed
 using two-dimensional DP exponents; 
this provides a strong evidence that we are in the presence 
 of anomalous (non-DP) scaling behavior.
 Finally, let us remark that the observed scaling does
not extend over many decades for any of the computed steady state
magnitudes. Much better scaling is observed for spreading
exponents as will be shown in the following section.


\subsection{Spreading Experiments and Superabsorbing States}
\label{capLipo_apAna_subSprea}

In order to further verify and 
support our previous conclusion we have performed also
 spreading experiments as it is customarily done
 in systems with absorbing
states \cite{Torre,reviews}.
 These consist in locating a seed of activity at the center
of an otherwise absorbing configuration, and studying how it spreads 
 on average in that medium \cite{reviews}. In the absorbing 
phase the seed dies exponentially fast, propagates indefinitely in
the active phase, while the critical point corresponds to a 
marginal (power law) propagation regime \cite{reviews}. 

As said before, it is well established that two-dimensional
 systems with  
infinitely many absorbing states show some peculiarities
 in studies of the spreading of a localized activity seed.
 The absorbing environment surrounding the seed
can either favor or un-favor the propagation of activity depending on 
its nature  
 (see \cite{JSP,Chate} and references therein).
For the, so called, {\it natural} initial conditions \cite{reviews} the
critical point for spreading coincides with the bulk critical point, and 
standard DP exponents are expected. In order to generate such 
natural configurations one could start the system with 
some highly active configuration
and run the system right at the critical point;
 once it reaches an absorbing configuration
it can be taken as a natural or
 self-generated environment for spreading. 
An alternative, more efficient way of proceeding, inspired in sandpile
systems \cite{soc}, is as follows. 
 One considers an arbitrary absorbing configuration
 and runs a spreading experiment.
 Once the epidemic (or ``avalanche'' in the language of 
self-organized criticality \cite{soc}) is over,
one considers the newly reached absorbing configuration as initial state
for a new spreading experiment avalanche.
 After iterating this process a number of times
the system reaches a statistically stationary absorbing state:
 the natural one
(see \cite{soc} and references therein).  Using this absorbing state for
 spreading leads to DP exponent values
 (and critical point) in systems with many absorbing states as 
for example the pair contact process \cite{PCP,dvz}.

\begin{figure}
\centerline{
\psfig{file=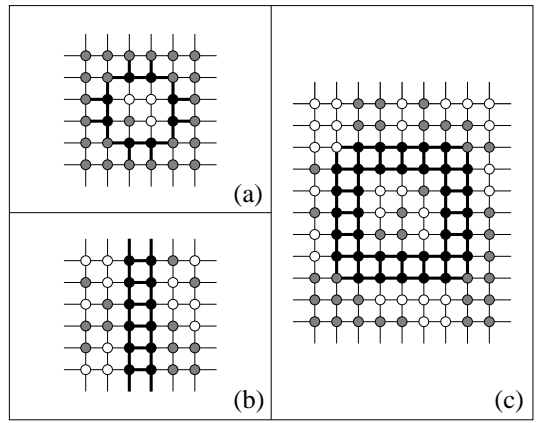,width=7cm,angle=0}}
\caption[Different frozen configurations of superabsorbing sites.]
{\small Different frozen configurations of superabsorbing (black) sites. 
White (grey) color stands for active (absorbing) sites.
(a) Blocking configuration for spreading from the central cluster of
four sites. Black sites 
cannot change their state whatever the state or dynamics inside
the cluster might be. Black bonds remain also frozen.  
(b) Spanning frozen cluster of superabsorbing sites.
(c) Almost-frozen cluster of superabsorbing sites. This, and analogous
structures, can be destabilized from the outside corners.
}
\label{cluster1}
\end{figure}

 By following this procedure we have found a very peculiar
property of this model,
 that we believe to be at the basis of its deviating from DP.
If the initial seed is located for all avalanches in the same 
site (or small group of localized sites), 
as is usually the case, after a relatively 
small number of avalanches
 the system reaches an absorbing configuration such that it is
impossible to propagate activity
for any possible forthcoming avalanche
beyond a certain closed contour.
 For example, 
configurations as the one showed in Fig. \ref{cluster1}.a are generated.
The four sites at the center are the ones at which activity seeds 
are placed in order to start avalanches. White sites are active
and grey ones are absorbing. 
At  each marked-in-black site, the sum of
the three (black) bonds connecting it to sites other than a 
central one is smaller than $r_c-1=0.38643(3)$. In this way,
regardless the value of the bond connecting the site
to the central region the site remains inactive: it is a {\it superabsorbing
site}.  The existence of ``inactive forever'' sites have been 
already pointed out in \cite{Lip2,Lip3}. 
 In the configuration showed in Fig. \ref{cluster1}.a 
activity cannot propagate out of the ``fence'' of superabsorbing
sites: the cluster of superabsorbing sites will remain frozen
indefinitely, and activity cannot possibly spread out. 
All avalanches will necessarily die after a few time steps. 
This type of blocking structure is quite generic,
 and appears in all experiments after some relatively short 
transient.

   In conclusion, this way of iterating spreading experiments leads always
to blocking closed configurations of superabsorbing sites
 instead of driving the system to
a natural absorbing configuration. 

  Observe that some activity put out of a blocking fence of sites
in Fig. \ref{cluster1}.a
could well affect any of the external bonds
 of the superabsorbing  sites (the dangling black bonds in Fig. \ref{cluster1}.a),
 converting the corresponding site 
to an absorbing or even an active one. 
Therefore, in order to overcome this  
difficulty of the frozen blocking 
configurations and be able to perform
 spreading experiments in some meaningful way,
we iterate avalanches by locating the initial seed 
at randomly chosen sites in the lattice. In this way there is always 
a non-vanishing probability of destroying 
blocking ``fences'' by breaking them from 
outside as previously discussed. Measurements of the different relevant
magnitudes are stopped when the system falls into an  
 absorbing configuration or alternatively whenever a linear
 distance $L/2$ from the avalanche origin is reached.
 Observe that in the second case
the dynamics has to be run further
in order to reach a new absorbing configuration at which 
launching the next avalanche.

\begin{figure}
\centerline{
\psfig{file=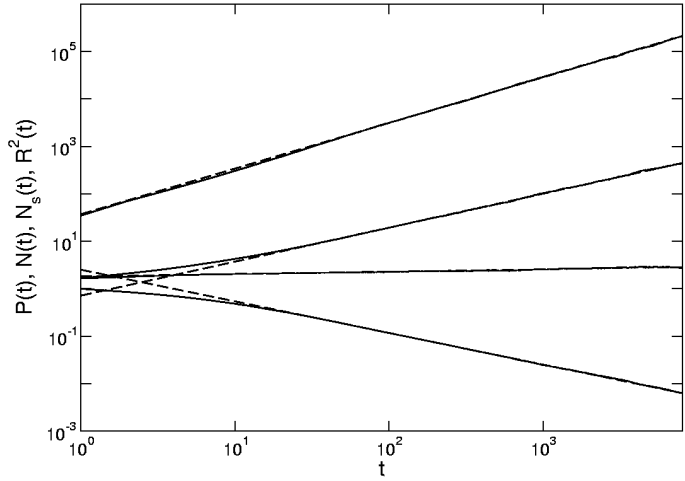,width=9cm,angle=0}}
\caption[Numerical results for spreading experiments]
{\small Numerical results for spreading experiments. 
$R^2(t)$ (topmost curve), $N_s(t)$ 
 (second curve from above),
$N(t)$ (third curve from above),
and $P(t)$ (bottom curve).
From the slopes we estimate
$z=0.96(1)$ and $\eta+\delta=0.71(1)$, $\eta=0.05(1)$ and $\delta=0.66(1)$
respectively.
}
\label{spreading}
\end{figure}

We monitor the following magnitudes:
the total number of active sites in all the runs as a function of
time $N(t)$  (we also estimate $N_s(t)$ defined as the average 
number of active sites restricted to surviving runs),
the surviving probability  $P(t)$, and the average square 
distance from the origin, $R^2(t)$. At the critical point
these are expected to scale as $N(t) \sim t^\eta$, $P(t) \sim t^{-\delta}$
and $R^2(t) \sim t^z$. Results for this type of measurements
are reported in Fig. \ref{spreading}.

 We obtain rather good algebraic behaviors at the previously
estimated critical point, $r_c$, confirming that the iteration-of-avalanches
procedure leads the system to a natural absorbing environment. Slightly
subcritical (supercritical) values of $r$ generate downward (upward)
 curvatures in this plot for all the four magnitudes.
Our best estimates for the exponents at criticality  
are: $z= 0.96(1)$, $\eta=0.05(1)$, $\delta=0.66(1)$ (see Table \ref{tabla_abs}).
To double check our results we also plot $N_s(t)$, which
 is expected to scale with an exponent 
$\eta +\delta$. An independent measurement of its slope in the log-log
plot gives $\eta+\delta=0.71(1)$, in perfect agreement with the
previously obtained results.

We can use these values to verify the hyperscaling relation \cite{Mendes,JSP}
\begin{equation}
\eta+\delta+\theta ={d \frac{z}{2}}.
\label{hyper}
\end{equation}
Substituting the found values for $z$ and $\eta+\delta$ we obtain 
$\theta \approx 0.25(2)$, compatible within error bars with the previously 
determined value $\theta =0.275(15)$. 
 
 One more check of the consistency of our 
results by using scaling laws is the following.
As $z=2 \nu_\perp / \nu_\parallel$ \cite{exp},
 we can estimate $\nu_\parallel$ from $z$ and
$\nu_\perp$. Then,  using $\nu_\parallel$ 
and the fact that $\theta=\beta/\nu_\parallel$ we obtain $\theta = 0.27(1)$, again
in excellent agreement with the directly measured value.
\begin{table}
\begin{center}
\begin{tabular}{|c||c|c|c|c|c|c|}
\hline
Model &  $\beta $ & $\beta/\nu_\perp$ & $\theta$ & $\eta$ & $ \delta$  & $ z$  \\
\hline
\hline
Lipowski &  $0.40(2)$  & $0.57(2)$  & $0.275(15)$  &$0.05(1)$&0.66(1)& $0.96(1)$   \\
\hline
 DP, $d=1$  &  $0.276 $   &  $0.252$   & $0.159$  &  $0.313$ & $0.159$   & $1.265$ \\
\hline
 DP, $d=2$  &  $0.583$    &  $0.795$   & $0.450$  &  $0.229$ & $0.450$   & $1.132$ \\
\hline
\end{tabular}
\end{center}
\caption[Critical exponents for Lipowski model and DP.]
{\small Exponent values for the two dimensional Lipowski model
and directed percolation in both one and two dimensions. 
Figures in parenthesis denote
statistical uncertainty (note that error-bars are statistical errors
coming from power-law fittings, and therefore do not include eventual 
systematic corrections to scaling).}
\label{tabla_abs}
\end{table}

  In Table \ref{tabla_abs}, we present the collection of exponents and compare them
with DP values in both one and two dimensions \cite{exp}. There is no trace
of dimensional reduction: this model does not behave, at least up to the
scales we have analyzed, as any other known universality class.

\subsection{More about Superabsorbing States}
\label{capLipo_apAna_subMore}

 Let us recall our definition of superabsorbing states.
 A site, three of whose associated bonds take values
such that the sum of them is smaller that $r-1$, 
cannot be activated from the remaining direction by neighboring activity.
We say that this site is superabsorbing in that direction (or
it is in a superabsorbing state).  A site can
be superabsorbing in one or more than one directions.
Still a site in a superabsorbing state
can obviously be activated by neighboring
activity in any of the remaining directions (if any).
                                              
  Having stated the existence of frozen clusters in standard spreading
 experiments (when initialized from a  fixed localized set of sites),
one may wonder whether there are similar frozen structures in simulations
 started with an homogeneous initial distribution of activity, or in the
 modified type of spreading experiments we have just used (i.e. allowing
the initial seed to land at a randomly chosen site) in the neighborhood 
of the critical point. 

In principle, for any finite lattice, the answer to that question is 
affirmative.
In Fig. \ref{cluster1}.b we show the shape of a frozen cluster of superabsorbing
sites: any of the sites in it is superabsorbing with respect to the
corresponding outward direction, and it cannot be ``infected'' from
any of the other directions as neighboring sites are
 similarly superabsorbing.
If a cluster like that is formed (or put by hand on the initial state)
it will remain superabsorbing forever. However the probability to form 
such a perfectly regular chain is extremely small for large system sizes.
Observe also that in order to have a completely frozen two-site 
broad band structure it has to be unlimitedly long (or closed if periodic
boundary conditions are employed). If instead it was finite,
 then sites at the corners
 would be linked to two external susceptible-to-change bonds and, therefore,
 themselves would be susceptible to become active: they would not be blocked
forever.
 In this way any finite structure of 
superabsorbing sites in the square lattice is unstable:
 it can be eaten up (though very slowly)
 by the dynamics, and is therefore not fully frozen.
   For instance, the cluster of superabsorbing 
sites represented in Fig. \ref{cluster1}.c is almost-frozen, but not really
frozen as it may lose its superabsorbing character from the 
outside corners
as previously described.  Analogously, any other cluster shape of 
superabsorbing sites may be destabilized from its outside corners.

\begin{figure}
\centerline{
\psfig{file=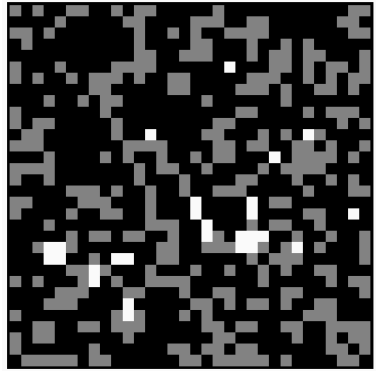,width=5cm,angle=0}}
\caption[Snapshot of a critical configuration with superabsorbing sites.]
{\small Snapshot of a configuration in a $32 \times 32$
lattice in the stationary regime for a value of $r$ close to 
the critical point.
 White color denotes activity, black corresponds
to superabsorbing sites, while grey  stands for absorbing sites. 
Observe that superabsorbing sites percolate through the lattice.}
\label{snapshot}
\end{figure}

  In conclusion, frozen clusters of superabsorbing sites do not appear
spontaneously. However, almost-frozen regions do appear and
may have extremely long life spans, specially 
 close to the critical point where activity is scarce, and therefore
the possibility of destabilizing them is small.  
In order to give an idea of how frequently superabsorbing sites appear
 we present in Fig. \ref{snapshot} a snapshot of a typical system-state
near the critical point.

 White corresponds to active sites, while the remaining sites are absorbing: 
in black we represent superabsorbing (in one or more than one directions)
 sites,
 while simple absorbing (not-superabsorbing) sites are marked in grey
color.
 Observe that superabsorbing sites are ubiquitous; in fact they percolate
 through the system.  Among them, about
one forth are superabsorbing in all four directions.
  
   Even though none of the clusters of superabsorbing sites is 
completely 
frozen, and in principle, activity could reach any lattice site, the 
dynamics is {\it glassy} \cite{glass} in some sense. For instance,
imagine an active region separated 
from an absorbing region by a line of
superabsorbing-in-the-direction-of-the-activity sites. In order to reach
the absorbing region, activity has to circumvent the superabsorbing barrier.
But near the critical point, where activity is scarce, barriers
of superabsorbing sites are intertwisted among them
 forming structures that,
even if not completely frozen, are very unlikely to be infected: activity
has to overcome them progressively in order to reach the interior
of superabsorbing regions.
This resembles  some aspects of
glassy systems for which degrees of freedom are hierarchically 
coupled and, at observable timescales, they may appear
 effectively frozen   \cite{glass}.    
 
 This phenomenology is certainly very different
from DP, and it is the reason why the relaxation towards
stationary states is so slow, and why deviations
from mean values are so persistent in numerical simulations.
 In particular, as superabsorbing regions are
long lived, the time required for the system to self-average is very 
large, and as near the critical point the probability of reaching an absorbing
state is large, in practice, 
the system does not have the time to self-average. 
Consequently, a huge amount of  
independent initial states and runs have to be considered in order
 to measure smooth  well behaved physical magnitudes \cite{SA}.
We strongly believe that this
 type of pathological dynamics is responsible 
for the departure of the Lipowski model from the DP universality class
in two dimensions.
  
 At this point one might wonder whether
the one-dimensional version of this model is essentially different.
Or in other words, why (one-dimensional) DP exponents are
 observed in $d=1$ \cite{Lip1}?.
 The answer to this
question is not difficult if one argues in terms of superabsorbing sites.
First of all notice that in $d=2$, $r_c > 1$. This means that just
by changing one bond, whatever the value of the output is, the site can stay
below threshold if the other three bonds sum less than $r_c-1$; this is
to say
superabsorbing states do exist at criticality.
  However in $d=1$,
$r_c=0.4409 < 1$. In this case by changing one bond value
it is always possible to activate the corresponding site:  superabsorbing
sites do not exist in $d=1$ at the critical point\footnote{Observe that for 
values of the control parameter $r > 1$, well
into the absorbing phase, superabsorbing sites show up also in $d=1$; but
they do not affect the critical region.}. 
Once the ``disturbing'' ingredient is removed from the model, we
are back to the DP class
as general principles dictate.

\subsection{The Honeycomb Lattice}
\label{capLipo_apAna_subHon}

\begin{figure}
\centerline{
\psfig{file=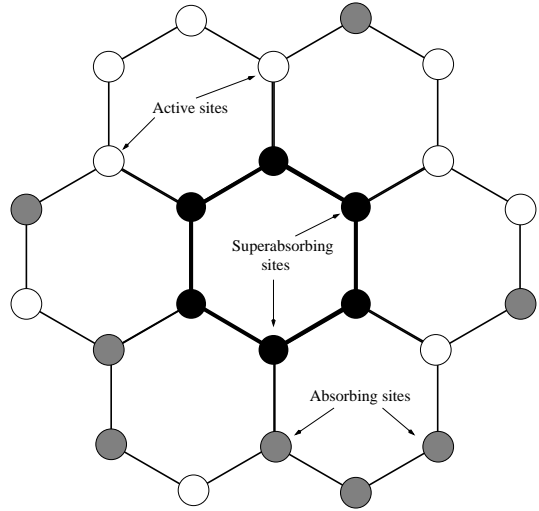,width=7cm,angle=0}}
\caption[Frozen cluster in the honeycomb lattice.]
{\small Frozen cluster in the honeycomb lattice. This type
of frozen structure of superabsorbing sites remains indefinitely 
superabsorbing at the critical point. Black: superabsorbing sites. 
Grey: absorbing sites. White: active sites.
}
\label{cluster2}
\end{figure}

 In order to further test our statement that superabsorbing states are
responsible for the anomalous scaling of the two-dimensional Lipowski model,
we have studied the following variation of it.
 We have considered the model defined
on a honeycomb lattice (with three bonds per site), and performed Monte Carlo
simulations.  In this case there is the (geometrical) 
possibility of having completely frozen
clusters of superabsorbing sites (see Fig. \ref{cluster2}).
                                         
 The main geometrical difference from the
previous case comes from the fact that here cluster-corners are linked only
to one external bond, and therefore are more prone to form
frozen clusters.
 In principle, before performing any numerical analysis,
 there are two alternative possibilities:
either the critical point is located at a value of $r$ smaller than $1$ or
 larger than $1$.            
 In the first case, there would be no superabsorbing sites (in analogy
with the one-dimensional case); in the second case pathologies associated 
with superabsorbing sites should be observed. 
The case $r_c=1$ would be marginal.
Finite size scaling analysis 
indicate the presence of a continuous phase transition
located at $r \approx 1.0092$ (very nearby the marginal case,
but significatively larger than $r=1$).

For Monte Carlo simulations, we have employed lattices
 of up to a maximum of
$256 \times 256 $ sites. All the observed phenomenology is
perfectly compatible with two-dimensional DP behavior. The dynamics
does  not show any of the
anomalies described for the square lattice case.  
 In particular, from the dependence of the stationary activity density
on system size  we evaluate  $\beta/\nu_{\perp} = 0.80(1)$;
 from the time decay at criticality
 $\theta=0.45(1)$, and finally $\beta=0.57(2)$;
 fully confirming consistency
with two-dimensional DP behavior. 
This result seems to be in contradiction with the two alternative 
possibilities presented above.
Let us now discuss why this is the case.  

   As the coordination number is $3$ in this case, 
the sum of two bond-values has to
be smaller than  $r_c -1 \approx 0.0092$ in order to have a superabsorbing 
site in the direction of the remaining bond at criticality.
As the
two bonds are independent random variables, the probability of creating a
superabsorbing site if the two of them are changed, is fewer
than $ 0.5 \%$, and the probability to generate frozen clusters 
(composed by six neighboring superabsorbing sites as shown in
 Fig. \ref{cluster2}), is 
negligible at the critical point. In fact, we have not been able to observe
any of them in our simulations.
 This means that one should study extremely large system sizes
and extraordinarily  long simulations
in order to see  anomalies associated with superabsorbing sites,
otherwise, for any feasible simulation the behavior is expected to 
be DP-like. 
  The observation of DP exponents in this case strongly supports the 
hypothesis that superabsorbing states are at the basis of the 
anomalous behavior of the model on the square lattice.

 However, strictly speaking, the system should exhibit a (unobservable)
 first-order phase transition at $r=1$ in the thermodynamic limit.
Indeed, for values of $r$ larger than $1$ there is a finite,
 though extremely small, probability of   
creating frozen clusters of superabsorbing sites 
(as the one in Fig. \ref{cluster2}). As
this is an irreversible process, 
after some (divergently long) transient there would be a
percolating network of frozen clusters of superabsorbing 
sites, and the only possible stationary state would be an absorbing one
with zero activity.
On the other hand, for values of $r$ smaller than unity, the probability of
creating superabsorbing sites is strictly zero, 
and there will be a non-vanishing density
of activity. As the density at $r=1$, almost independent of system size,
is $\rho  \approx 0.18$, the transition is expected to be discontinuous,
and therefore the DP transition observed in our simulations 
is merely a finite size effect, and should disappear for large
enough sizes and long times.
  In any case,
this first order transition is unobservable computationally. 

\section{Conclusions and Outlook}
\label{capLipo_apConc}

Summing up, we have shown that the two-dimensional 
Lipowski model does not belong to any known universality class.     
 We have measured different critical 
exponents by running Monte Carlo simulations started
 from homogeneous initial states and also by performing spreading
experiments.
 In any case, we find absolutely no trace of dimensional reduction,
 and there
is neither evidence for the system to behave as
two-dimensional DP. Instead, a novel scaling behavior
 is observed.      
The main relevant physical ingredient of this class is 
the presence of superabsorbing sites, and almost-frozen clusters
of superabsorbing sites which slow down enormously the dynamics.   

 The previous conclusion is strongly supported by two other observations: (i) 
the regular DP behavior observed in the one-dimensional version of the
model for which superabsorbing states do not appear at criticality, and 
(ii) the two-dimensional DP behavior observed for the two-dimensional model 
defined on a honeycomb lattice, for which the probability
of generating superabsorbing sites at criticality is almost negligible. 

   In general, superabsorbing sites can either arrange into completely
 frozen clusters or not depending on dimensionality, coordination
number and other system details. Let us distinguish three main cases:
\begin{itemize}
\item When completely 
frozen clusters of superabsorbing
sites appear below (or above) a certain value of the control parameter
but not above (below), first order
transitions are expected 
(as occurs in the multiplicative model discussed in Appendix \ref{apendLipo}
\cite{Lip3}).

\item If completely frozen clusters do not appear at criticality,
but instead almost-frozen clusters are present, 
we expect anomalous behavior (as 
occurs in the original Lipowski model \cite{Lip2}). 
   
\item If neither frozen nor almost-frozen clusters are
 observed at criticality 
(as is the case for the one dimensional version of the model \cite{Lip1})
we expect standard directed percolation behavior. 
\end{itemize}

There are two possible follow-ups of this work. First, 
it would be worth studying in more realistic situations
as, for instance,
 in surface catalysis (dimer-dimer or dimer-trimer)  models
 \cite{Muchos}
whether effects similar to those described in this chapter 
play any relevant role. In particular, for those models 
depending upon lattice and particle geometry there are
cases in which activity cannot propagate to neighboring regions,
but is constrained to evolve following certain directions or paths.
 It would be rather interesting to sort out whether
anomalies reported for those models \cite{Muchos} 
are related to the existence 
of superabsorbing states.
     
On the other hand, from a more theoretical point of view, an interesting question is:
 what is the field theory or Langevin 
equation capturing the previously described
 phase transition with superabsorbing states? and, how does it 
 change with respect to Reggeon field theory?.
Establishing what this theory looks like, would clarify greatly at a 
field theoretical level the effect of superabsorbing states on phase 
transitions, and would permit to shed some light on 
the degree of universality of this anomalous phenomenology.
Our guess is that a
Reggeon field theory \cite{RFT,conjecture} 
with a spatio-temporal dependent anisotropic Laplacian term
(which, for example, would enhance, un-favor or forbid 
diffusion from certain sites in certain directions) could be a good
candidate to describe this new phenomenology.
  Analogously to what 
happens in field theoretical descriptions of other systems with many a
absorbing states \cite{Muchos,JSP}, the inhomogeneous Laplacian-term 
 coefficient should be described by a second physical field
 coupled to the activity field in such
a way that its fluctuations would vanish upon local
 absence of activity. 
 Further pursuing this
line of reasoning is beyond the scope of the present chapter. 
 As long as this  program has not been completed, is not safe
to  conclude 
unambiguously that the anomalies described in this chapter are
relevant in the limit of extremely large times and system sizes.

\chapter{Heat Conduction and Fourier's Law in One-Dimensional Systems}
\label{capFou}

\section{Introduction}
\label{capFou_apIntro}

A research on dynamics of nonequilibrium systems must include a reference to 
\textit{transport phenomena}. These dynamic processes appear ubiquitously 
in Nature. Some classical examples are heat and mass transport in fluids (eg. boiling water), 
diffusion, electric conduction, etc.  
Many natural systems can be understood and analyzed using the transport phenomena jargon, 
as for instance atmospheric dynamics, oceanic currents, ion currents between cells, stellar convection, 
traffic flow, social migration, diffusion of information, etc.
These initially so different systems share certain common features. In general, they 
are usually inhomogeneous, and typically show nonzero gradients of several 
magnitudes, together with non--zero net fluxes carrying energy, mass and/or momentum through the 
system. 

Classically, transport phenomena have 
been studied by Irreversible Thermodynamics, where these processes are analyzed in terms of 
conservation laws, local entropy balance equations and the maximum entropy production postulate. 
\cite{deGroot} This theory builds up using a series of 
phenomenological macroscopic laws based upon the 
proportionality among the fluxes and the thermodynamical forces. Fick's law of diffusion, 
Fourier's law of heat conduction and Ohm's law of electric conduction, for instance, belong 
to this class of linear phenomenological laws. Such laws are completely defined once 
certain proportionality factors, called \textit{transport coefficients}, are specified. Some of 
these transport coefficients are the diffusivity, and the thermal and electrical conductivity 
(related, respectively, to the above mentioned laws). It has been 
shown that the phenomenological laws of Irreversible Thermodynamics remain valid (i.e. 
reproduce experimental results) as far as the system under study stays close enough to 
thermodynamic equilibrium, although the notion of ``close enough to equilibrium'' is not clear.

On the other hand, nonequilibrium Statistical Mechanics is a theory 
which aspire to describe macroscopic 
nonequilibrium phenomena (and, as a limiting case, the equilibrium ones) starting from the 
microscopic description of Nature. 
Therefore, a successful nonequilibrium statistical--mechanical theory should 
be able to calculate microscopically, among other things, the transport coefficients associated to 
the linear laws of Irreversible Thermodynamics, commonly observed in Nature. Moreover, 
this theory should be capable to define precisely the notion of ``close enough to equilibrium''
for a general system. Although great advances have been made in the field, up to now there is 
no closed theory for the mechanical--statistical description of nonequilibrium systems.

In order to further advance in the field, we must understand deeper the microscopic mechanisms 
underlying irreversible processes. In particular, in this chapter we want to investigate the 
microscopic basis of heat transport. The corresponding linear phenomenological law of 
Irreversible Thermodynamics associated to heat transport is Fourier's law, which in its more 
general form reads,
\begin{equation}
c_v(T)\frac{\partial}{\partial t} T(\vec{r},t) = 
\vec{\nabla} \cdot [\kappa \vec{\nabla}T(\vec{r},t)]
\label{ecfou}
\end{equation}
where $T(\vec{r},t)$ is the temperature measured by a probe at position $\vec{r}$ at time $t$, 
$c_v(T)$ is the specific heat per unit volume, and $\kappa$ is the thermal conductivity. Notice 
that, in order to write down this equation, one should assume that no mass transport and/or 
other mechanism different from heat conduction appears in the system. This diffusion--like 
equation describes the time evolution of a macroscopic system whose initial temperature profile
$T(\vec{r},t=0) \equiv T_0(\vec{r})$ should be inhomogeneous. Alternatively, Fourier's law 
(eq. \ref{ecfou}) can be applied to a homogeneous system in contact with heat reservoirs at 
time invariant temperatures $T_{\alpha}$. In the stationary state, the temperature profile 
should be solution of the equation,
\begin{equation}
\vec{\nabla} \cdot [\kappa \vec{\nabla}T(\vec{r},t)] = 0
\label{ecflujo}
\end{equation}
where $J \equiv -\kappa \vec{\nabla}T(\vec{r},t)$ is the stationary heat flux through the system.
This law has been extensively tested in experiments in fluids and crystals. However, we do not
understand yet many of its fundamental aspects. \cite{Bonetto, Lepri}

In particular, derivation of Fourier's law from a microscopic Hamiltonian dynamics is still an 
open question. Moreover, writing eq. \ref{ecfou} we assume that the state of the system is 
completely defined, from the macroscopic point of view, by the local temperature field 
$T(\vec{r},t)$ at any time $t$. Such assumption implicitly involves that the Local Thermodynamic 
Equilibrium (LTE) property holds for the investigated system, which usually is far from clear. 
As a simple picture of what LTE means, let us imagine our 
system divided up into many \textit{small} cells, large enough 
so each one contains a large (\textit{quasi}--macroscopic) number of atoms, and at the same time 
small enough in order to be accurately described by an (equilibrium) Gibbs measure at temperature 
$T(\vec{r},t)$ for each cell at position $\vec{r}$. Thus, although the system is macroscopically 
inhomogeneous, locally thermodynamic equilibrium holds (in the sense of Gibbs measures). This 
concept can be precisely defined using the Hydrodynamic Scaling Limit, where the ratio of 
micro to macro scales goes to zero. \cite{Bonetto} 

Thus, many fundamental questions arise related to the microscopic understanding of heat conduction, 
as for instance: which are the necessary and sufficient conditions in order to observe LTE in a 
system?, which is the interplay among LTE, 
energy equipartition and heat transport?, which are the minimum 
requirements a system must fulfill in order to obey  Fourier's law?, etc. In order to 
answer all these question (and many other, equally interesting questions) we must study simplified 
mathematical models of real systems. The study of these simple models will give us a firm basis in 
order to understand the microscopic origin of heat conduction and the physical hypothesis underlying 
Fourier's law. As usual in Theoretical Physics, mathematical and computational simplicity drive us 
to consider low dimensional systems, namely one and two--dimensional systems, which are comparatively
much easier to handle with than \textit{real} three--dimensional ones.%
\footnote{In spite of their apparent simplicity, low dimensional systems present some unusual features which 
usually do not appear in their three--dimensional counterparts. For instance, transport coefficients 
in low dimensional systems may not even exist.}
On the other hand, there is 
also experimental motivation for studying heat conduction in low dimensional systems. As an example,
let us mention anisotropic crystals, solid polymers, single walled nanotubes, quantum wires, etc. 
\cite{Lepri}

In particular, heat transport in one--dimensional systems is nowadays a highly
interesting problem in the context of both
non-linear dynamics and non--equilibrium statistical physics. Its study has added new insights to 
the understanding of the microscopic origin of normal heat conduction, as we will see below.
Long ago, Peierls was the first one to 
identify a mechanism which gives rise to a finite thermal conductivity. He proposed a successful
perturbative theory, based on a phonon scattering mechanism, in order to explain thermal conductivity 
in solids. \cite{Peierls} In electrically insulating solids, heat is transmitted by lattice 
vibrations. In that case, it is useful to visualize the solid as a gas of interacting phonons. 
These phonons (elementary lattice excitations) store and transport energy through the system.
In a perfect harmonic crystal, phonons behave as a gas of non--interacting particles. Hence, energy 
flow through the system without any loss, so the energy current (assuming that the system is 
subject to a temperature gradient) does not decrease with crystal length. Therefore, a perfect 
harmonic crystal should have an infinite thermal conductivity. However, a real crystal presents 
anharmonicities which give rise to phonon interactions, i.e. phonons scatter among them. In these 
collisions momentum is conserved modulus a vector of the reciprocal lattice. We thus can classify 
phonon collisions in two distinct classes: those which perfectly conserved momentum (\textit{normal 
process}), and those where the difference between the initial and the final momentum is a vector 
$\vec{k}$ of the reciprocal lattice (\textit{umklapp process}). 
Peierls theory shows that, in absence of 
umklapp processes, the thermal conductivity of a solid is infinite. Consequently, this theory 
predicts that we do not expect a finite thermal conductivity in monoatomic one--dimensional 
lattices with nearest neighbor interactions. However, as we will see below, other mechanisms are 
possible which give rise to normal heat conduction in these one--dimensional chains.
More generally, it has been shown that any integrable 
Hamiltonian system must have a divergent thermal conductivity, since its associated normal modes 
behave as a gas of non--interacting phonons, carrying energy from the hot to the cold sources 
without any loss.\cite{Lepri} On the other hand, there are one--dimensional non--integrable systems, to which 
Peierls theory does not apply directly, which also show a divergent thermal conductivity, as for 
instance the Fermi--Pasta--Ulam--$\beta$ model \cite{Lepri2}.

Many recent studies have focused their attention on heat transport in several one--dimensional 
systems, with the hope of identifying the relevant mechanisms underlying normal (finite) heat 
conduction. Some of these models yield a finite thermal conductivity, while others yield an 
infinite $\kappa$. Nowadays, the general belief is that integrability,%
\footnote{A system with $N$ degrees of freedom is integrable if there exists a canonical 
transformation such that the system can be described by $N$ conjugated action--angle canonical 
coordinates. Hence, an integrable system with $N$ degrees of freedom will have $N$ constants of motion.}
total momentum conservation and 
total pressure are the relevant ingredients which define whether a systems presents normal heat transport 
or, instead, anomalous thermal conductivity. In particular, it has 
been shown that the non--integrability is a sufficient condition in order to obtain a non--trivial 
temperature profile, although this property is not sufficient to guarantee normal heat conduction. 
\cite{Hu} Furthermore, there are one--dimensional systems with zero total pressure and translational 
invariant Hamiltonian which show normal heat conduction. On the other hand, it has been shown that the 
effect of local potentials, which break the Hamiltonian translational invariance and simulate 
interactions of the one--dimensional system with its embedding higher--dimensional space, is crucial 
in order to guarantee normal heat transport. Local potentials break total momentum conservation, thus 
identifying this symmetry as a relevant one in the heat transport problem. 
It is actually believed 
that one should not expect in general a finite thermal conductivity in one--dimensional systems with 
momentum conserving interactions and non--zero pressure. \cite{Gendelman} 

The last statement has been formally established in a recent theorem due to Prosen and 
Campbell \cite{Prosen}, which affirms that ``in 1D systems, conservation of total momentum implies 
anomalous conductivity provided only that the average pressure is non-vanishing in thermodynamic limit''.
The goal of this chapter is to show a counterexample to the above theorem. We introduce a system that,
although its particle interaction conserves momentum and it exhibits a nonzero pressure, the energy 
behavior has a diffusive character and Fourier's law holds, thus implying a finite thermal conductivity.
Therefore, we think that in one dimensional systems with nonzero pressure, the conservation of momentum 
does not seem to be a key factor to find anomalous heat transport. We think that there are other 
cooperative mechanisms that can do the job of the local potentials. As we will see below, maybe 
systems having  degrees of freedom that acquire easily energy but release it in a very long times 
scale have, in general, normal thermal conductivity.
On the other hand, and supporting our results, there are strong evidences (which we will describe 
later) pointing out that Prosen and Campbell's theorem is empty, in the sense that all the 
calculations are correct, but they do not predict anything about system's thermal conductivity.

The structure of this chapter is as follows. In section \ref{capFou_apMod} we present our one--dimensional model, 
together with the boundary heat baths used, explaining carefully their properties and the reasons 
underlying our choices. In Section \ref{capFou_apAna} we describe the numerical results obtained. There we study the system 
from several perspectives, all of them pointing out the finiteness of thermal conductivity in the 
model in thermodynamic limit, and thus the validity of Fourier's law in this system. 
Finally, in Section \ref{capFou_apConc} we summarize our 
results and present the conclusions, paying special attention to the fundamental implications that our 
observations have on the microscopic understanding of heat conduction. 

\section{One Dimensional Model of Heat Conduction}
\label{capFou_apMod}

Our model consists in a one--dimensional chain of interacting particles subject to a 
temperature gradient, which induces a heat flux from the hot extreme to the cold one.
In a line of length $L$, there are $N$ point particles of different masses interacting
exclusively via elastic collisions. In order to minimize the finite size effects, the particles
have only two different masses and they alternate along the line, {\it i.e.} 
$m_{2l-1}=1$ and $m_{2l}=(1+\sqrt{5})/2$ with $l=1,...,N/2$. We have chosen 
the masses of the even particles to be the most irrational number (golden number) in order to minimize 
possible periodicities, resonances or non-ergodic behaviors. 
At the extremes of the line there are thermal reservoirs at fixed (time invariant) temperatures 
$T_1=1$  and $T_2=2$ at $x=0$ and $x=L$, respectively. 

\begin{figure}
\centerline{
\psfig{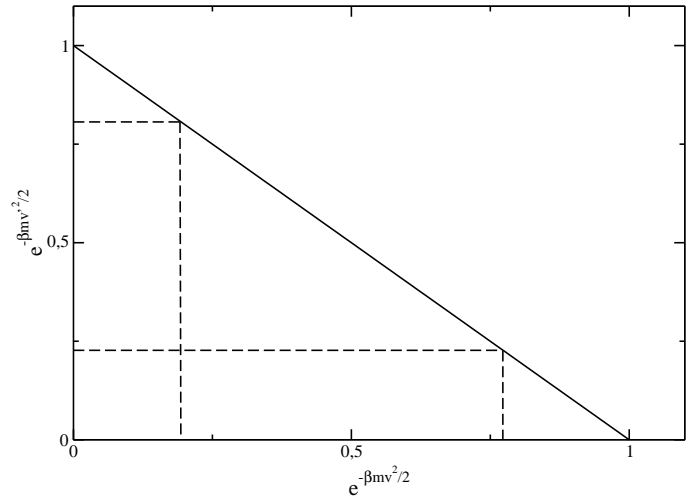}}
\caption[Representation of the deterministic heat bath scheme.]
{\small Graphic representation of the deterministic heat bath scheme. Notice that when the particle 
reaches the end wall with a large velocity, it is reflected with low velocity, and reciprocally. This 
thermal reservoir generates thermal equilibrium.
}
\label{bath}
\end{figure}

We simulate the reservoirs 
by using the following deterministic process: each time particle $1$ ($L$) hits the boundary 
at $x=0$ ($x=L$) with velocity $v$, the particle is reflected with the velocity modulus
\begin{equation}
v'=\left[ {{-2}\over{m_{1(L)}\beta_{1(2)}}}\ln\left(1-e^{-{{\beta_{1(2)}}\over{2}}m_{1(L)}v^2}
\right)\right]^{1/2}
\end{equation}
where $\beta_{1(2)}=1/T_{1(2)}$ (see Fig. \ref{bath}). This reversible and deterministic heat bath is due to 
H. van Beijeren (private communication). This map emulates a true heat bath. Other mechanisms exist that 
emulate the presence of true heat baths, as for example stochastic thermal reservoirs, where the 
particle colliding with the end wall returns with a velocity modulus randomly extracted from a 
Maxwellian distribution for the corresponding temperature. In general the correct, rigorous 
procedure should require studying infinite systems, but this is usually not feasible, and hence one 
must look for alternative, effective implementations of a thermal reservoir, as the one chosen. These 
effective heat baths must thermalize the system exactly as a 
real semi--infinite heat bath should do, since it has 
been shown \cite{Dhar1} that the coupling between the system and the thermal reservoirs dramatically 
affects the physical properties of the system, even in the Thermodynamic Limit. 
Needless to say that we tested, as a preliminary step, 
that our deterministic heat bath generates an equilibrium distribution (Gibbs measure) starting from 
a completely random state, and thus correctly simulates a true thermal source. On the other hand, the
deterministic thermal reservoir shows a very interesting property: 
it is time reversible. Hence, dissipation 
and irreversibility appear intrinsically in the system, and not as a consequence of the randomness 
introduced by stochastic thermal reservoirs. Furthermore, this type of heat bath guarantees that our 
system is completely deterministic, and thus the tools from non-linear system analysis can be used. 
Let us mention that, in order to check the influence of the type 
of reservoir into the measured system properties, we have also used the more conventional stochastic 
boundary conditions described above, but only different finite size effects and no other relevant 
behavior has been observed.

For $T_1 \neq T_2$, due to the temperature gradient, there is a flow of energy from the high 
temperature reservoir to the low one, and the system then evolves to a nonequilibrium stationary 
state. This properties (nonzero temperature gradient and nonzero net heat flux) are typical of 
nonequilibrium systems. On the other hand, numerical studies of our model 
are rather complex, since there relaxation times are very long.%
\footnote{Very often low dimensional systems show very large relaxation times. This is due to the 
restrictions induced by its low dimensionality in phase space flow.}
This problem restricts simulations to relatively small systems, where finite size effects are very 
important. We will notice this problem in the forthcoming analysis. In spite of these difficulties, 
several authors have studied this model before. \cite{Garrido,Dhar2,Hatano} For instance, in 
\cite{Dhar2} special attention is payed to the existence of a non--trivial temperature profile. There 
it is observed that the temperature profile does not change under rescaling of masses. Moreover, it is 
found that $T(x;\nu T_1, \nu T_2) = \nu T(x;T_1,T_2)$. This involves that the independent variables of 
the system are $M/m$, $T_1/T_2$ and $N$. Furthermore, as the number of particles $N$ increases, the 
temperature profile get closer to a limiting profile, similar to that predicted by kinetic theory.
A version of this model in which the masses are randomly placed was already studied
in \cite{Garrido}.  In this work, the system Thermodynamic Limit behavior was not considered but
the Local Thermodynamic Equilibrium (LTE) property was demonstrated. This property guarantees the 
existence of a well--defined local temperature. Moreover, LTE has been numerically observed in the 
alternating masses model for large enough system sizes (otherwise, we couldn't 
define local temperatures). \cite{Dhar2} Some of these works 
\cite{Dhar2,Hatano} have also studied the thermal 
conductivity as a function of system size, arriving to the conclusion that $\kappa$ 
slowly diverges in the Thermodynamic Limit. As we will see below, this type of analysis is strongly 
affected by finite size effects, and thus any conclusion derived from it about thermal conductivity 
could be wrong. In fact, as we will demonstrate, the thermal conductivity $\kappa$ is finite in the 
Thermodynamic Limit for this system, and hence Fourier's law holds here.

\section{Numerical Analysis}
\label{capFou_apAna}

Our goal in this chapter is to check whether the system has a finite thermal conductivity in the 
Thermodynamic Limit, $N,L\rightarrow\infty$ with $N/L=1$, or instead it exhibits a divergent $\kappa$.
With this aim we performed a detailed numerical analysis along several different, complementary lines.

\subsection{The Existence of a Non-trivial Thermal Profile}
\label{capFou_apAna_subEx}

Before going on the analysis of conductivity, we have to check whether this system shows a nontrivial temperature 
profile. This would indicate that LTE holds in the system, and then it makes sense to wonder whether Fourier's law 
holds or not. As a first step, we must give a working definition of local temperature. We define the local 
temperature by measuring the mean kinetic energy of each particle and its mean position $\bar{x}_i$ at the 
stationary state. Assuming that energy equipartition holds at least locally (we checked this point), this mean 
kinetic energy is proportional to the local temperature at position $\bar{x}_i$. There are many other methods to 
measure local temperature, almost all of them based on the Virial theorem. Strictly speaking, the only correct 
method to measure temperature profiles consists in dividing the system in cells, each one with a large enough number 
of particles, and measuring the velocity distribution in every cell. If local equilibrium holds, one expects a 
Maxwellian distribution for each cell, with local temperatures proportional to the variance of this distribution 
\cite{Garrido}. The method we use, and those based on Virial theorem, are just (numerically efficient) 
approximations. These approximations are effective as far as they confirm the existence of a non--trivial 
temperature profile in the Thermodynamic Limit. This is the important conclusion here, and not 
the exact shape of the temperature profile, which of course depends somehow on the thermal baths used and the 
definition of local temperature.

\begin{figure}
\centerline{
\psfig{file=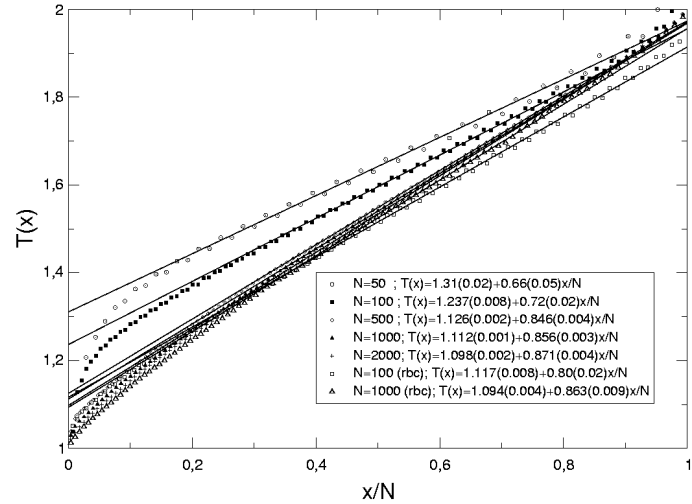,width=9cm}}
\caption[Temperature profile at the stationary state.]
{\small Temperature profile at the stationary state for $N$ particles. Lines are the
best fits of the data in the interval $x/N\in[0.4, 0.6]$. The corresponding equations are 
shown in the box. Errors in the coefficients are in brackets. We have also included, for direct comparison, 
temperature profiles obtained using stochastic boundary conditions (noted as rbc).}
\label{temp}
\end{figure}

We computed the profiles for $N=50$, $100$, $500$, 
$1000$, $2000$ particles, with fixed $N/L=1$, $T_1=1$ and $T_2=2$. Fig.\ref{temp}  
shows the local temperature as a function of $x/N$
(by seeking clearer figures, we have performed local averaging of the temperatures and 
positions to draw only $100$ 
points; no difference is found by drawing all the points). 
We see in Fig.\ref{temp} that the temperatures follow linear profiles in the
interval $x/N\in[0.4, 0.6]$ with slopes 
depending on the system size. This slope apparently tend to converge to unity but we 
find that the convergence is very slow. On the other hand, the temperature profile get curved
near the boundary heat baths. This \textit{surface resistance} \cite{Aoki}, which is consistent 
with previous studies \cite{Dhar2,Aoki}, tends to diminish as the system grows. 
In any case, a non--flat profile is clearly expected in the Thermodynamic Limit.


\subsection{The Averaged Heat Current}
\label{capFou_apAna_subCur}

We now pay attention to the heat flux through the system. If Fourier law holds and the heat conductivity is 
finite, the mean heat current, defined as $J=N^{-1}\sum_{i=1}^N m_iv_i^3/2$, should go to zero as 
$1/N$ whenever $T_{1,2}$ and $N/L$ are kept fix. This is so because in the steady state we have
$J = -\kappa \nabla T(\vec{r},t) \approx \kappa (T_2 - T_1)/N$. If instead $J \sim N^{-\alpha}$ with 
$\alpha < 1$ then we should have a divergent thermal conductivity, $\kappa = J N /(T_2 - T_1)$. \cite{Lepri2}
In our case, the data does not give us a conclusive answer. In fact we fitted our experimental
points ($J$ corresponding up to seven different $N$'s) to behaviors like $J=aN^{-0.71}$, 
$J=aN^{-1}(1+b\ln N)$, both yielding a divergent conductivity, and $J=aN^{-1}(1+bN^{-1})$ 
and $J=aN^{-1}(1+b/\ln N)$, both yielding a finite conductivity. All these four fits had
regression parameters of order $0.999$. Thus we cannot conclude nor convergence 
nor divergence of thermal conductivity from this point of view. Moreover, our fits reflect 
that finite size corrections to the leading order are dominant and that we are far from the asymptotic 
regime for the observable heat current. 
Therefore, the direct use of the Fourier's law $\kappa=J N/(T_2-T_1)$ does not clarifies (from the
numerical point of view) the existence of a finite heat conductivity 
in the Thermodynamic Limit. Some authors find for the same model that the heat current 
goes like $N^{-0.83}$\cite{Dhar2} or $N^{-0.65}$\cite{Hatano}. These results are similar to our direct 
fit to power law behavior. However, in contrast with them, we conclude that such fits are done in a 
non-asymptotic regime.

\subsection{The Current-current Self Correlation Function} 
\label{capFou_apAna_subCor}

It is clear from the previous section that direct use of Fourier's law does not yield any definite 
conclusion about system's thermal conductivity in the Thermodynamic Limit, due to the strong finite 
size effects affecting our data. Hence we must look for other 
different methods in order to conclude about $\kappa$. One of these methods is based upon the Green--Kubo 
formulae. The heat conductivity is connected to the 
total energy current--current time correlation function evaluated 
at equilibrium via its time integral (Green-Kubo formula\cite{deGroot}). 
It can be written as,
\begin{equation}
\kappa = \lim_{\tau \rightarrow \infty} \lim_{L \rightarrow \infty} \int_{-\tau}^{\tau} 
<J(0)J(t)>_{\beta} \text{d}t
\label{Kubo}
\end{equation}
where we can write,
\begin{equation}
<J(0)J(t)>_{\beta} \equiv C(t) =\frac{\int \prod_n \text{d}p_n \text{d}q_n J(0) J(t) exp(-\beta H)}
{\int \prod_n \text{d}p_n \text{d}q_n exp(-\beta H)}
\end{equation}
Hence $C(t)$ is the equilibrium canonical average of $J(0)J(t)$ at inverse 
temperature $\beta$ for a system with Hamiltonian $H$.
The order of limits in eq. (\ref{Kubo}) is crucial in order to precisely define $\kappa$. 
The integral has some meaning (i.e. yields a finite $\kappa$) whenever the correlation function decays
as $<J(0)J(t)>_{\beta} \sim t^{-1-\Delta}$ with $\Delta>0$.
Thus, we have to measure the long time tail of the total energy 
current--current correlation function $C(t)$ in order to conclude about 
system's thermal conductivity. As Green--Kubo formula states, this autocorrelation function must be measured 
in equilibrium. The fact that the equilibrium average should be taken using the canonical ensemble is related with 
one of the fundamental hypothesis underlying Green-Kubo formula: Local Thermodynamic Equilibrium. As previously 
explained, this hypothesis implies that one can define locally a temperature, in such a way that the system behaves 
locally as an equilibrium system with this temperature. This hypothesis implicitly involves the use of 
the canonical ensemble, since in this ensemble the temperature is defined precisely, being the energy a 
fluctuating observable. In order to measure $C(t)$ one thus should simulate our system with heat reservoirs 
at the borders at equal temperatures $T=(T_1+T_2)/2$. However, this procedure is not practical, due to the 
strong finite size effects affecting $C(t)$ as a consequence of the open boundaries. In practice, this finite 
size corrections impede any definite analysis of the self correlation function. 
In order to avoid such difficulties related to the presence of the 
open borders, we measure $C(t)$ in the microcanonical ensemble with total energy $E$ defined by the temperature 
$T=(T_1+T_2)/2$ via the equipartition theorem. In this way, we substitute open 
boundaries by periodic boundary conditions, recovering translational symmetry. Using the 
microcanonical ensemble, where finite size effects are minimized in some sense, 
we are able to successfully analyze the long time behavior of the self correlation 
function $C(t)$. However, as we will see below, some easily identifiable finite size effects related with the
system's finite length remain. One could wonder whether this ensemble change is plausible. The equivalence of 
equilibrium ensembles is well known fact in Equilibrium Statistical Mechanics, although it has not been 
rigorously proved in all cases. Hence one should interchange microcanonical and canonical averages arbitrarily, 
expecting invariant results (up to very small --logarithmic-- finite size corrections) under such modifications.
However, in order to gain confidence in our results, we will also present evidences pointing out the validity 
of this interchange.

Writing eq. (\ref{Kubo}) we have assumed that the total momentum is set to zero 
in the system. Otherwise, if a non--zero net momentum exists, one should use the connected part of the 
autocorrelation function $C(t)$  in eq. (\ref{Kubo}), defined as,
\begin{equation}
C_c(t) \equiv <J(0)J(t)>_{\beta} -<J>_{\beta}^2
\label{connected}
\end{equation}
in order to precisely calculate the thermal conductivity. 
\cite{Bonetto} This is an important point, because if there is a net momentum in the system and one uses the 
non--connected self correlation function $C(t)$ in the Green--Kubo formula the result won't have 
nothing to do with the real thermal conductivity. This subtle technicality has not been taken into account 
in a recent theoretical work \cite{Prosen}, which in principle proves that ``\dots for classical many--body 
lattice Hamiltonians in one dimension, total momentum conservation implies anomalous conductivity in the 
sense of divergent Kubo expression \dots``. However, in their derivation, which is \textit{not} restricted to a 
zero total momentum ensemble, these authors have used the non--connected self correlation function $C(t)$, 
and thus the divergence of the Green--Kubo formula does not implies anything on the system's thermal 
conductivity, contrary to author's claim. Moreover, in this chapter we show a counterexample to the above 
affirmation, i.e. we show a one--dimensional system, which conserves momentum and has non--zero pressure, whose 
thermal conductivity is finite in the Thermodynamic Limit, as we will see below.

\begin{figure}
\centerline{
\psfig{file=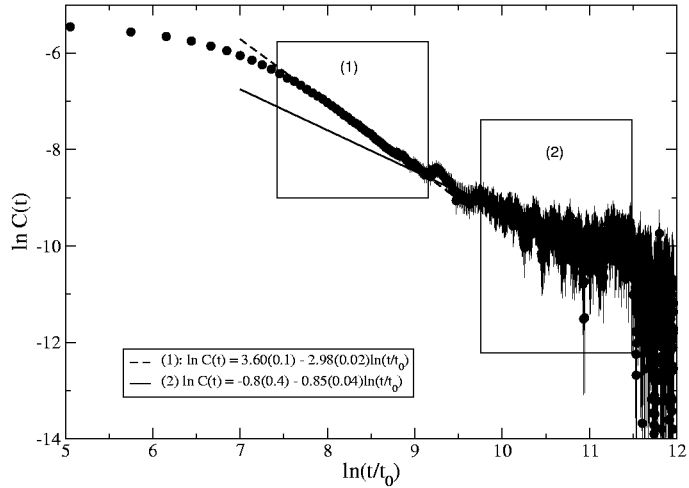,width=9.0cm}}
\caption[Total energy current--current time correlation function.]
{ \small Total energy current--current time correlation function $C(t)$ for the 
different masses system with $N=1000$. The inset shows the results of a power law fit for the 
time decay in regions ($1$) and ($2$). In region $(1)$ we observe $C(t) \sim t^{-1.3}$. Notice 
large error-bars and strong fluctuations for $\ln (t/t_0) > 10$. $t_0$ is the mean collision time.
}
\label{C-dif}
\end{figure}

We hence have measured $C(t)$ using periodic boundary conditions and setting total momentum equal to 
zero, with total energy defined through the equipartition relation, $E=NkT/2$, where $T$ is the arithmetic 
average of $T_1$ and $T_2$. In Fig. \ref{C-dif} we show our results on $C(t)$ for a system with $N=1000$ 
particles. Here we can study two different regions:
\begin{enumerate}
\item Region $(1)$, defined for $\ln (t/t_0) \in [8,9]$, where a power law fit yields $C(t) \sim t^{-1.3}$.
\item Region $(2)$, defined for $\ln (t/t_0) > 10$, where the same algebraic fit yields $C(t) \sim t^{-0.85}$.
\end{enumerate}
The observed slight difference in the time decay exponent of $C(t)$ depending on the fitting region 
is crucial in order to conclude about the 
convergence or divergence of thermal conductivity of our model in the Thermodynamic Limit. Thus we must 
develop a set of physically well--motivated criteria that help us to 
distinguish the true bulk asymptotic behavior from spurious 
finite size corrections. As we will explain below, we think that only region $(1)$ corresponds to the infinite 
system bulk behavior, which involves a finite thermal conductivity in the Thermodynamic Limit, and thus that 
Fourier's law holds for our total--momentum--conserving one--dimensional model. The bump observed in 
$C(t)$ for very long times (namely $\ln (t/t_0) > 9.5$) is typical of system autocorrelations due to finite size 
effects \footnote{As an example, autocorrelation functions of financial time series show this finite size 
behavior.\cite{Stanley}}.

\begin{figure}
\centerline{
\psfig{file=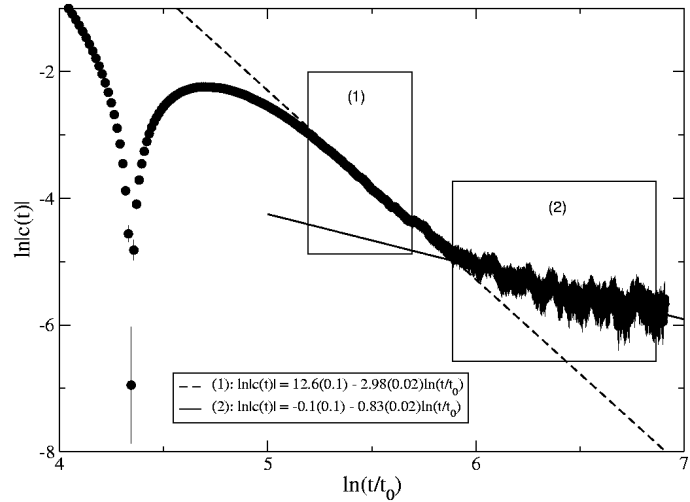,width=9.0cm}}
\caption[Local energy current time correlation function for equal masses.]
{ \small Local energy current--current 
time correlation function $c(t)$ for the equal masses system with $N=500$. The inset shows the 
results of a power law fit for the time decay in regions ($1$) and ($2$). Here $|c(t)| \sim t^{-3}$ 
in region $(1)$, and we observe a $t^{-0.83}$ tail in region $(2)$.}
\label{c-equal}
\end{figure}

As an example of the previous statement, let us study the asymptotic behavior of the local energy current--current 
time correlation function $c(t)=<j_i(0)j_i(j)>_{\beta}$ for the equal masses version of our model, where 
$j_i=m_i v_i^3 /2$. Making equal the particle's masses (i.e. $M=m$) has strong implications on system's 
properties. As a first fact, let us mention that the equal masses version of our model is not ergodic. 
Once we define the set of initial velocities for the particles, these velocities will remain invariant during 
the whole evolution. This is due to the elastic collisional kinematics, which implies that the equal masses 
system effectively behaves as a gas of identical non--interacting particles. Hence the total energy current 
$J$ is a constant of motion for the equal masses case, and thus $C(t)$. Furthermore, this system is integrable
(i.e. it has a macroscopic number of constants of motion), and thus it shows an infinite thermal conductivity.
Long time ago, Jepsen \cite{Jepsen} proved that, for this equal masses system, $<v_i(0)v_i(t)>_{\beta} \sim t^{-3}$,
where the average should be taken over the canonical ensemble. Following the steps stated by Jepsen, it can 
be shown analytically, after a lengthly calculation\footnote{Pedro L. Garrido, private communication.}, that also
$c(t) \propto <v_i(0)^3 v_i(t)^3>_{\beta} \sim t^{-3}$ for this system. In order to learn how system's finiteness 
affect current--current time correlation functions, we have measured $c(t)$ for 
equal masses in a finite system. We have performed such simulation using the canonical ensemble 
and with a Maxwellian initial velocity distribution, in order to reproduce 
the previous exact result. Thus we use our deterministic heat baths at the 
boundaries, with both ends at the same temperature (for completeness, we simulated also this system using 
stochastic thermal reservoirs, obtaining the same results). Notice that, for the equal masses case, we 
cannot interchange freely the canonical and the microcanonical ensembles. The underlying reason is the 
non-ergodic behavior of this system. The presence of the boundary heat reservoirs restores ergodicity, while 
averages in the microcanonical ensemble depend on initial conditions. Hence, since Jepsen calculation is 
done in the canonical ensemble, we must simulate the equal masses system also in this ensemble in order to 
recover the analytical result. Fig. \ref{c-equal} shows the numerical computation of $c(t)$ for $N=500$. 
It is remarkable that we can also define here two different regions: $(1)$ one for 
$\ln (t/t_0) \in [5.1,5.8]$, where a power law fit yields $|c(t)| \sim t^{-3}$, and $(2)$ one for 
$\ln (t/t_0) > 6$, where a power law fit yields $|c(t)| \sim t^{-0.83}$. 
We recover the theoretically predicted asymptotic bulk behavior in region $(1)$, 
while region $(2)$ should be due to finite size 
effects. Moreover, it is intriguing that the finite size time decay exponent 
($\sim 0.83$) is almost the same both in the different masses case and the equal masses one, being these 
systems very different in essence, and for two different time correlation functions 
--namely $C(t)$ and $c(t)$. This fact points out the existence of an underlying common finite size 
mechanism, responsible of this spurious long time decay. We think that autocorrelation effects as those provoked 
by perturbations which travel all around the system and come back to their origin are at the basis of the observed 
long time finite size corrections. In conclusion, coming back to the different masses case, in our opinion the above 
example indicates that only region $(1)$ of Fig. \ref{C-dif} represents the asymptotic bulk behavior. Hence, 
any conclusion about system's conductivity derived from region $(2)$ should be misleading. This result involves a 
finite Green--Kubo thermal conductivity, and hence that Fourier's law holds in our one--dimensional system.

\begin{figure}
\centerline{
\psfig{file=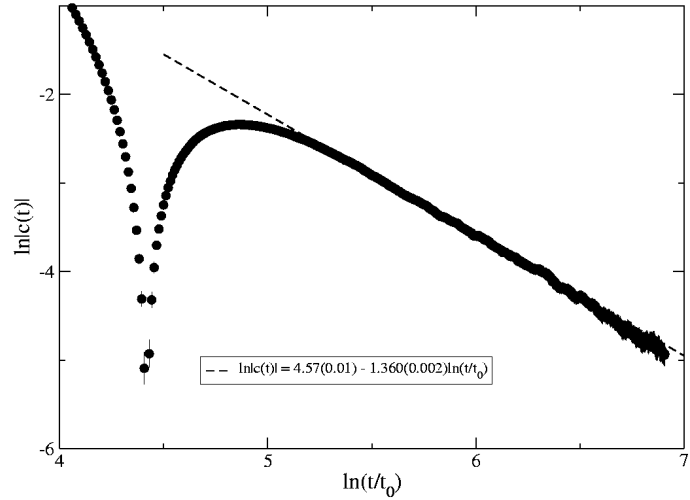,width=9.0cm}}
\caption[Local energy current time correlation for alternating masses.]
{ \small Local energy current--current 
time correlation function $c(t)$ for the alternating masses system with $N=1000$. The inset shows the 
results of a power law fit for the long time decay of $c(t)$. Here $|c(t)| \sim t^{-1.36}$. This result is very 
similar to the observed asymptotic behavior of $C(t)$, and thus reinforces our previous analysis. The number of 
independent averaged histories is of order $10^9$.}
\label{c-diferentes}
\end{figure}

In order to confirm such result we computed the local energy current-current time correlation function $c(t)$ 
at equilibrium for the different masses case. It has much better averaging 
properties than $C(t)$, and thus its asymptotic behavior is 
much easier to distinguish. The obvious question is whether $c(t)$ has something to 
do with the total energy current--current self correlation function $C(t)$ entering Green--Kubo formula, both 
calculated for the different masses model. In general, we can write 
$C(t) \sim \sum_{i,l}<j_i(0)j_l(t)>=\sum_{i}<j_i(0)j_i(t)>+\sum_{i \neq l} <j_i(0) j_l(t)>=
Nc(t)+\sum_{i \neq l}c_{i,l}(t)$, where $c_{i,l}(t) \equiv <j_i(0) j_l(t)>$. Hence, for {\it regular} 
systems, where non--local time correlation functions $c_{i,l}(t)$ decay fast enough with distance, one 
expects a similar time decay for both $C(t)$ and $c(t)$. However, there are {\it anomalous} systems 
for which $c_{i,l}(t)$ decays very slowly, or does not decay at all. One of these anomalous systems 
is the equal masses version of our model. As explained before, this system is not ergodic; $C(t)$ is constant here.
Furthermore, as previously stated, it can be analytically proved that $c(t) \sim t^{-3}$ there. 
Hence, the behaviors of $C(t)$ and $c(t)$ are clearly different for the equal masses case. Moreover, 
since $C(t)$ is the correlation function entering the Kubo formula for conductivity, in this case the fact 
that $c(t) \sim t^{-3}$ implies nothing about the system's conductivity. 
We measure $c(t)$ for the alternating masses case using the canonical ensemble, i.e. we simulate the system 
subject to thermal reservoirs at the borders, both at equal temperature $T_1=T_2=1.5$. This measurement 
yields $c(t) \sim t^{-1-\Delta}$, where $\Delta$ is again very close to $0.3$ (see Fig. \ref{c-diferentes}), which is 
very similar to the observed time decay of $C(t)$. This may be thought as an 
indication that our system is {\it regular}, in the sense stated above. Moreover, since the averaging properties 
of $c(t)$ are much better, the observation of a $t^{-1.3}$ tail in $c(t)$ confirms our analysis for $C(t)$, 
reinforcing our conclusion, i.e. that Fourier's law holds for this unidimensional system. On the other hand, assuming
that our different masses system is regular, the fact that $C(t)$, measured using the microcanonical ensemble, and
$c(t)$, measured using the canonical ensemble, show very similar behavior supports our previous hypothesis about 
the equivalence of microcanonical and canonical ensemble averages for the alternating masses model.

In our opinion the decay of correlations is so slow that it explains the strong finite size effects 
observed in the temperature profile and in the mean heat current. In fact we can argue that 
$J N/(T_2-T_1)=\kappa-A N^{-\Delta}$ which explains why we do not see a clear behavior of $J$ with $N$ with 
system sizes of order $10^3$ (the corrections are of order unity  for those sizes).

\subsection{The Energy Diffusion} 
\label{capFou_apAna_subEdif}

\begin{figure}
\centerline{
\psfig{file=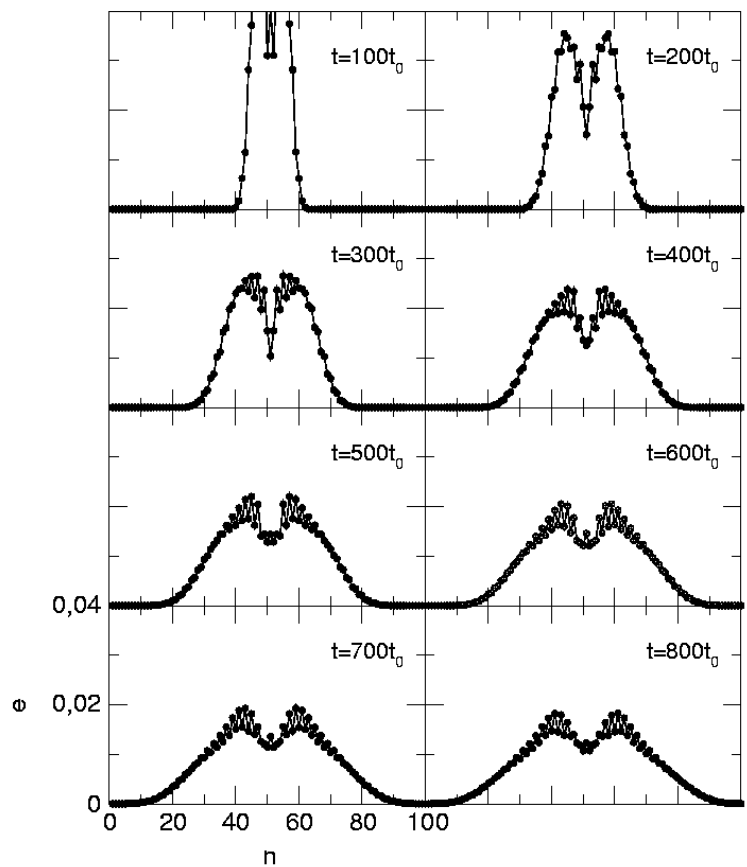,width=12cm,angle=-90}}
\caption[Evolution of the energy distribution.]
{\small Evolution of the energy distribution for an initial condition in which all particles,
$N=100$, are at rest except particle $51$ which has an averaged energy corresponding to temperature
$1.5$. The figure shows averages over $10^7$ independent realizations and $t_0=0.032$.}
\label{edistrib}
\end{figure}

\begin{figure}
\centerline{
\psfig{file=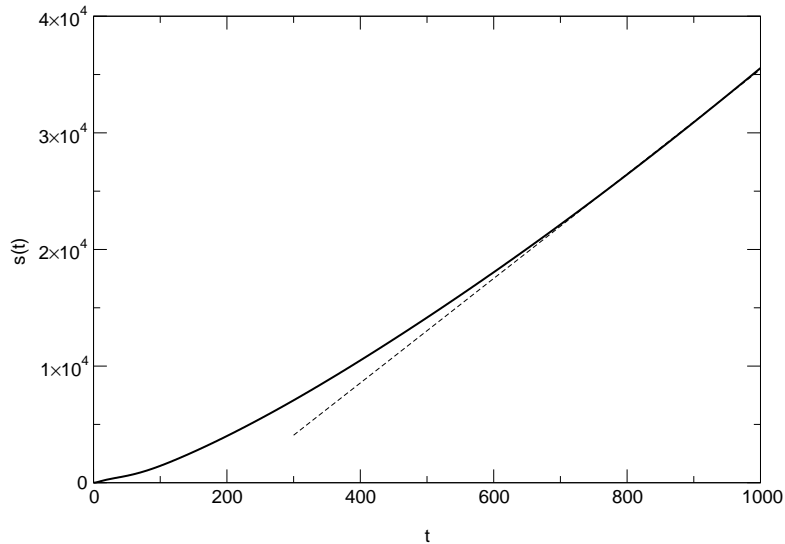,width=8cm,angle=-90}}
\caption[Diffusion of energy.]
{\small Time evolution of $s(t)$ (see the text for definition). A linear behavior 
$s(t) \sim t$ is clear for long enough times, thus indicating a diffusive propagation of energy. 
For shorter times we observe $s(t) \sim t^2$, i.e. ballistic propagation.}
\label{s-t}
\end{figure}

The above facts imply that Fourier's law holds for our one--dimensional alternating masses system, which 
conserves total momentum and has a non--zero pressure. However, this conclusion 
depends critically on data analysis. We have exposed above our analysis, which we think is physically 
well motivated and coherent. In spite of this, there is always a possibility that an 
analysis focused in only one observable may drive to the wrong 
conclusions. Hence, we have attempted to obtain a global, consistent vision of the problem by measuring 
several magnitudes. With this aim in mind we also studied the dynamical aspects of Fourier's law (see eq. 
(\ref{ecfou})). Particularly, we studied the propagation of energy in the system. We prepared the system with 
zero energy (all particles at rest) and positions $x(i)=i-1/2$, $i=1,\ldots, N$. Then, we give to the light particle 
$i=N/2+1$ a velocity chosen from a Maxwellian distribution with temperature $T=1.5$. That is, we introduced an 
energy pulse in an otherwise frozen system, and monitored how the energy flows through the system until any 
boundary particle moves. Finally, we average over many initial conditions 
(changing the initial velocity of the central light particle). If the system follows Fourier's law 
we should see a diffusive type of behavior (if the thermal conductivity is constant). This is due to the Local 
Thermodynamic Equilibrium (LTE) property: if LTE holds in the system, local temperature and local energy are 
proportional (due to local equipartition of energy, involved by LTE), and thus a diffusive behavior of energy 
implies a diffusive behavior of local temperature, which is exactly what Fourier's dynamical law, eq. (\ref{ecfou}), 
states. Figure \ref{edistrib} shows the energy distribution for $N=100$ and different times measured in units 
$t_0=0.032$, where $t_0$ is the mean free time between consecutive particle collisions. 
Let us remark here again that to apply eq. (\ref{ecfou}) the  temperature should have
a smooth variation in the microscopic scale to guarantee that local equilibrium holds.
In Figure \ref{edistrib} we see that, for times larger than $t=200t_0$, the average
variation in the local temperature is of order $0.001$. Therefore, we may assume that
we are in a regime where eq. (\ref{ecfou}) holds. Initially, the energy of the
light particle is transfered to the neighbors very fast and then the particle stays very cold, much 
colder than its neighbors. In fact, in this initial regime, the energy maxima are  moving outwards 
at constant velocity. This behavior ends at around $t=100t_0$. The system then
begins to slow down and, at $t \simeq 300t_0$, the structure of the energy distribution 
changes, and one can then differentiate the behavior corresponding to light particles and heavy ones 
at least around the maxima of the distribution. We measured the mean square displacement of
the energy distribution at each time,
\begin{equation}
s(t)=\sum_n (n-51)^2 e(n,t)
\label{difusion}
\end{equation}
We found that we can fit $\ln s(t)=-6.39(0.04)+2.05(0.01)\ln t$ for $t/t_0\in(30,100)$, which is a ballistic behavior 
that changes smoothly until we reach $t>400t_0$, where we find a diffusive behavior 
$\ln s(t)=-1.00(0.01)+1.005(0.002)\ln t$ (see Fig. \ref{s-t}). This last result confirms that our 
system follows even the dynamical aspects of Fourier's law.

\begin{figure}[t]
\centerline{
\psfig{file=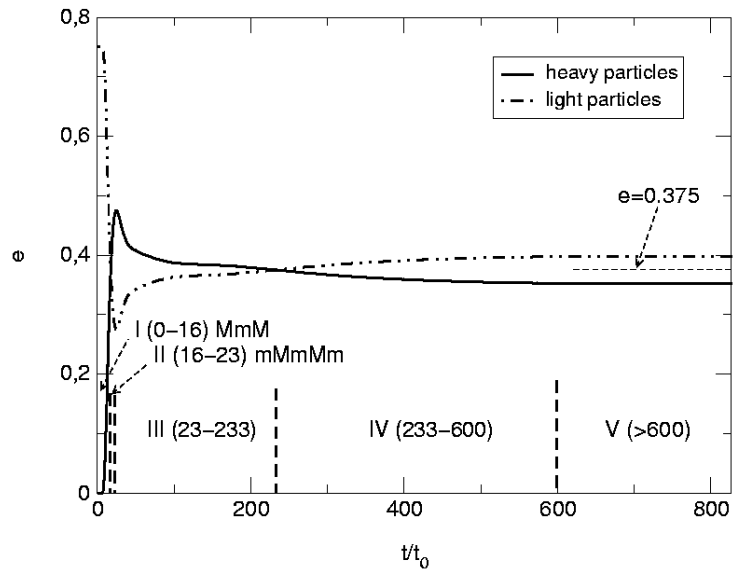,width=9cm,angle=-90}}
\caption[Evolution of the energy stored in the heavy and light particles.]
{\small Evolution of the total energy stored in the heavy and light particles. 
The conditions are the same as in Fig.\ref{edistrib}. $MmM$ indicates that only the central light
particle and the nearest heavy ones are moving in region I.}
\label{etotalmM}
\end{figure}

As we noticed above, in Fig.~\ref{edistrib} we see that the light  and heavy particles seem
to follow different energy distributions, at least for times longer than $t=300t_0$. 
In order to get some more insight about such behavior,
we computed the evolution of the total energy stored in the light (heavy) particles.
The result is shown in Fig.~\ref{etotalmM} where 
we can detect five different time regions:
\begin{enumerate}
\item {\it $t/t_0\in(0,16)$}; only the light particle and the two heavy nearest neighbors have a
nonzero velocity. 
\item {\it $t/t_0\in(16,23)$}; the five central particles (three light and two heavy ones)
are moving. The total energy stored in the light particles reaches a minimum.
\item {\it $t/t_0\in(23,233)$}; the heavy particles begin  to release energy (on the  average) 
until, at $t=233t_0$, both types of particles have the same amount of energy. 
\item {\it $t/t_0\in(233,600)$}; light particles keep getting energy until we reach the next region, 
\item {\it $t>600t_0$}; where the total energy stored in the light  particles reaches a constant 
value  that exceeds to the one corresponding to the heavier ones.
\end{enumerate}
Let us remark  that, in the asymptotic regime $t>600t_0$,
the energy distribution is still evolving and, therefore, this partition of energy between
both degrees of freedom is an asymptotic \textit{dynamical} (non--stationary) property 
of the system, i.e. it does not appear once we reach the steady state.

In order to discard any non--ergodic behavior of our
system we included reflecting boundary conditions at the extremes of the chain and we
did much longer simulations. We saw that the isolated system tends to the equilibrium
in which equipartition of energy between all degrees of freedom holds. 
That is, the total mean energy stored in the light particles is
equal to the one stored in the heavy ones once the stationary state has been reached. 
Moreover, we have checked that the system at any stationary state (equilibrium or non--equilibrium) does not 
present the property of non--equipartition of the energy.  This is not incompatible with the dynamical 
non--equipartition observed above. If due to fluctuations a light particle gains energy over the mean particle 
energy, it will release this energy excess very slowly, thus effectively trapping energy for a long time, 
as a local potential would do, although finally this particle should converge to the mean particle energy.
We think that this dynamical non-equipartition of the energy between degrees of freedom is responsible for the 
normal thermal conductivity. In fact, we see that, around the
distribution maxima, the particles arrange in the form that hot light particles are surrounded
by cold heavy ones. The energy is then trapped and released in a diffusive way. However, we also see
that the release is diffusive when a large enough number of those hot-cold structures
develop. Therefore we think that the mechanism for the thermal resistance 
is somehow cooperative. 

\section{Conclusions}
\label{capFou_apConc}

In this chapter we have studied the microscopic foundations of normal heat conduction, which 
is a dynamic phenomenon very interesting nowadays in the context of nonequilibrium Statistical Mechanics.
In order to do so, we have numerically investigated a simple one-dimensional model where point particles
of alternating masses elastically collide. When this system is subject to a temperature gradient, a net heat 
flux emerges from the hot reservoir to the cold one, together with a non-trivial, non-linear steady 
temperature profile. 

An evaluation of the system thermal conductivity from the observed energy current
does not yield any conclusive answer about its convergence nor divergence, due to the strong finite size effects 
affecting our measurements. We thus use the Green-Kubo formula in order to evaluate $\kappa$ in the Themodynamic 
Limit, for which we measure the total energy current self correlation function, $C(t)$. A careful analysis of
the long time behavior of $C(t)$, together with some other related measurements, allow us to conclude that
our model system, in spite of being a one-dimensional momentum-conserving system with nonzero pressure, exhibits
a finite thermal conductivity in the Thermodynamic Limit, and thus Fourier's law holds in this limit.
We further check this conclusion measuring how energy propagates through the system, finding a diffusive kind
of propagation, compatible with the dynamic version of Fourier's law.

In conclusion, Peierls arguments  have successfully explained the observed thermal
conductivity in solids by applying a perturbative scheme around the lattice harmonic interaction.
The actual belief is that strong anharmonicity is not enough to guarantee a normal
thermal conduction in one dimensional systems. Moreover, it has been proposed that
the key lacking ingredient is that the dynamics of the system 
should  not conserve linear momentum via the existence of local potentials through  the line
(think about particles attached to the one dimensional substrate through some kind of
non-linear springs). In this way, local potentials should
act as local energy reservoirs that slow down the energy flow.
These properties, anharmonicity and non-conservation of
momentum, are in some way the ones used on the original Peierls argument.
We have shown a model that does not follow such clean picture.
Although our one dimensional model is non-linear and it conserves linear momentum 
(with non-zero pressure),
we find that it follows Fourier's law. We think that there are other cooperative mechanisms
that can do the job of the local potentials. Maybe,  systems having  degrees of freedom that
acquire easily energy but release it in a very long times scale have, in general,
normal thermal conductivity.
In any case, we think that it is worth to explore such possibility.

\chapter{Summary, Conclusions and Outlook}
\label{capConc}

In this thesis we have studied the dynamical aspects of some nonequilibrium systems. They are open, hysteretic systems, 
subject to density and/or temperature gradients, energy and/or mass fluxes, under the action of external agents and
different sources of non-thermal noise, etc. Nonequilibrium systems abound in Nature. In fact, they are the rule, being
equilibrium systems an unlikely exception. Examples of out-of-equilibrium systems can be found for instance in biology 
(e.g. living organisms), economy (e.g. traded stocks), geology (e.g. earthquakes), quantum and molecular physics (e.g. 
magnetic nanoparticles), hydrodynamics (e.g. turbulent fluids), astrophysics (e.g. star evolution and structure), 
sociology (e.g. opinion spreading), and so on. Moreover, it seems that nonequilibrium conditions are essential for the 
observed complex structure in Nature.\cite{Kadanoff}

In spite of their importance, up to now nobody has been able to formulate a complete theory connecting the microscopic 
properties of nonequilibrium systems with their macroscopic phenomenology. This connection has been rigorously established
only for equilibrium systems in terms of the partition function.\cite{Balescu} 
The search for a statistical-mechanical 
description of nonequilibrium systems is one of the main aims of modern Physics. Nowadays there is only a set of 
{\it ad hoc} techniques which describe in a partial and approximate way some particular problems in nonequilibrium 
statistical physics. In particular, most of the studies and theoretical developments in nonequilibrium physics have
been centered on nonequilibrium steady states, which constitute the simplest situation in nonequilibrium
phenomena. On the other hand, the dynamical aspects of nonequilibrium systems have been poorly studied, and the aim of this
thesis consists in enlarging a bit our understanding of such processes.

Dynamical phenomena in complex systems are usually related to transformations between different phases. Thus we have 
investigated how nonequilibrium conditions affect such transitions. In particular, we have studied the effects that 
nonequilibrium conditions induce on the dynamic problem of metas-tablity, where a system set in a metastable phase eventually 
evolves towards the stable one. We have also studied how a system under nonequilibrium anisotropic conditions and with 
conserved number of particles evolves from an initially disordered phase via segregation towards an ordered phase. 
In addition to their intrinsic interest due to their ubiquity in Nature, these two examples are
very interesting because they have equilibrium counterparts. That is, both metastability and phase segregation are dynamic 
processes observed in equilibrium systems, and they have been deeply investigated. This fact allows us to deduce, comparing 
both the equilibrium and nonequilibrium cases, the net effects induced by nonequilibrium conditions, which yields many hints
about the relevant ingredients that must be taken into account in order to build up a general formalism for nonequilibrium systems.
On the other hand, there are dynamic processes intrinsic to nonequilibrium systems. This is the case, for example, of dynamic
phase transitions between an active phase, characterized by a non-trivial dynamics, and an absorbing phase, where the system
is frozen. This phase transition is irreversible, and hence it is a pure nonequilibrium phenomenon. In this thesis we have 
investigated the effects that a new, hidden symmetry has on the universality observed in these absorbing phase transitions.
Another dynamic processes with no equilibrium counterparts are transport phenomena in general, and heat conduction 
in particular. We have also investigated the microscopic origins of normal heat conduction and Fourier's law.
Of course, there are many nonequilibrium dynamic phenomena that have not been studied in this thesis. As an example,
just mention all the rich, complex and yet not fully understood phenomenology observed in fluids: convection, turbulent flow, 
etc.\cite{Navier} However, we think that the nonequilibrium dynamic phenomena studied here yield a comprehensive overview of 
the richness and diversity of new effects that nonequilibrium conditions induce on dynamic processes in complex systems.

The thesis has been divided into two different parts. The first part, which comprises chapters 2, 3, 4 and 5, is devoted to 
the study of the metastability problem and its associated dynamics in nonequilibrium systems with short-range interactions.
In particular, we have studied a nonequilibrium ferromagnet defined on a two-dimensional lattice. On the other
hand, in the second part, which comprises chapters 7, 8 and 9, we have studied respectively the kinetics of phase separation
in an anisotropic nonequilibrium lattice gas, an absorbing (dynamic) phase transition in a biological inspired model,
and the microscopic foundations of normal heat conduction and Fourier's law in a one-dimensional particle chain.
In what follows we summarize the results obtained in each chapter, and the possible follow-ups of this work.

In {\bf chapter \ref{capMotiv}} we present our motivation for studying metastability in magnetic systems, which is two-fold.
On one hand, this problem is very interesting from the technological point of view, since impure (i.e. nonequilibrium)
magnetic particles are at the basis of many modern and incipient technologies, as for instance dense recording magnetic 
materials, possible quantum computers, etc. On the other hand, the study of metastability in short-range nonequilibrium
magnets yields much information about the link between microscopic and macroscopic physics in nonequilibrium systems.
In this chapter we also present the model we study in the first part of the thesis, paying special attention to the
model properties and the way in which nonequilibrium conditions enter the model. In brief, the model is defined on a lattice 
with binary spins at the nodes interacting via the Ising Hamiltonian, and subject to a competing dynamics where two different
Glauber rates (one at finite temperature, $T$, and the other at ``infinite'' temperature) compete weighted by a nonequilibrium
parameter $p$. This impure dynamics, generically observed in real materials, drives the system towards a nonequilibrium 
steady state.

The first approach to the problem of metastability in this bidimensional nonequilibrium magnet 
is presented in {\bf chapter \ref{capMedio}}, where a first-order mean field approximation
is implemented. 
This approximation, which allows us to study both the static and dynamic properties of metastable states, is based on three
main hypothesis: absence of fluctuations, kinetic isolation of domains, and homogeneity. Using these hypothesis 
on the general master equation we are able to calculate the system phase diagram, $T_c(p)$, and both the stable and metastable 
state magnetizations.
On the other hand, studying the intrinsic coercive magnetic field, which is the field for which metastable states become unstable, we
find that under strong nonequilibrium conditions (in particular, $p>\pi_c \approx 0.0315$) a {\it non-linear cooperative 
phenomenon} between the thermal noise (parameterized by $T$) and the non-thermal noise (parameterized by $p$) emerges:
although both noise sources independently add disorder to the system, which implies the attenuation, or even destruction of 
existing metastable states, the combination of both noises parameterized in the microscopic dynamics does not always involves a 
larger disorder, giving rise to parameter space regions where there are no metastable states for low and high temperatures, but 
they appear for intermediate temperatures. All these predictions are confirmed via Monte Carlo simulations. 
In order to investigate the dynamics of the metastable-stable transition
using mean field approximation, we extend this theory to include fluctuations, since they are at the basis of the metastable state 
exit mechanism. In this way we build up a mean field stochastic dynamics based on the mean field predictions for the
stable phase growth and shrinkage rates. In spite of including fluctuations, the extended mean field theory fails to describe the
metastable-stable transition. This is so because this transition is fully inhomogeneous (it proceeds via droplet nucleation and 
growth), and hence cannot be described by the homogeneous mean field theory. In order to describe this inhomogeneous dynamic 
process the properties of the interface separating the stable and metastable phases must be studied.
Two possible continuations of the research presented in this chapter should be the following:
\begin{itemize}
\item It should be desirable to understand in a deeper way the mechanism which gives rise to the non-linear cooperative
phenomenon between the thermal and non-thermal noise sources for $p>\pi_c$.
\item It should be also worthy to study in depth the mean field stochastic dynamics proposed in this chapter as a natural way 
to include fluctuations in mean field theory. A more rigorous justification of this method, which seems natural, must 
be developed. Also a comparison of the predictions associated to fluctuations derived from this method with Monte Carlo results
should be welcome.
\end{itemize}

{\bf Chapter \ref{capSOS}} is devoted to the study of the properties of the interface separating the metastable and stable phases
in the nonequilibrium ferromagnet. In this chapter we develop a generalization of the Solid-On-Solid approximation \cite{BCF}
in order to understand the effects that nonequilibrium conditions induce on the microscopic and macroscopic properties of the 
interface. The generalization is based on the concept of effective temperature. It is found that different spin classes
suffer different effective temperatures for the nonequilibrium case, the more ordered spins suffering the higher effective 
temperatures. In this way interfacial spins suffer different effective temperatures for $p\neq 0$, depending on the spin class
to which they belong to. Using this observation, and neglecting the presence of overhangs in the interface and the interactions
between the interface and bulk fluctuations, we build up the generalization of the SOS theory. The microscopic structure
codified in the step probability function and predicted by the generalized SOS approximation matches almost perfectly Monte Carlo
results, finding that the larger $p$ is, the rougher the interface is. On the other hand, the macroscopic structure is 
captured by the surface tension. We find in this case that while the equilibrium surface tension monotonously increases as 
temperature decreases, the nonequilibrium surface tension exhibits a maximum at certain nonzero temperature, converging towards zero
in the low temperature limit. Such low temperature anomalous behavior is a consequence of the dominant character of the 
non-thermal (nonequilibrium) noise source in this limit, as can be deduced from the interface effective temperature, which becomes
independent of $T$ in the low temperature limit. The non-monotonous behavior of surface tension in the nonequilibrium case
will be fundamental when understanding the properties of the metastable-stable transition in the nonequilibrium ferromagnet,
since this anomalous behavior will be inherited by most of the relevant observables in this problem.
Using the explicit expression derived for the nonequilibrium surface tension via the generalized SOS theory, we also study in this
chapter the shape and form factor of a nonequilibrium spin droplet using Wulff construction. 
The non-zero interfacial effective temperature
induced by nonequilibrium conditions in the low temperature limit implies that the droplet shape is no more a square in this limit
for $p\neq 0$ (as opposed to what happens in equilibrium systems), but a differentiable curve in between a square and a circle.
In order to extend the investigations developed in this chapter, we propose the following lines:
\begin{itemize}
\item The generalized SOS approximation must be further tested against Monte Carlo simulations, both on its microscopic and
macroscopic aspects. In particular, one should study the interfacial roughness derived from this theory, since it is a 
relevant observable in many surface problems. Also the importance of overhangs and interactions between the interface and bulk 
fluctuations should be addressed.
\item The effects that the suppression of the surface tension at low temperature in the nonequilibrium case induce on 
many natural phenomena controlled by an interface must be investigated, since many real interfaces are subject to
nonequilibrium conditions as those captured by our competing dynamics.
\end{itemize} 

In {\bf chapter \ref{capNuc}} we present a nonequilibrium extension of nucleation theory.\cite{Langer1,Langer2,Rikvold}
We hypothesize the existence of a nonequilibrium potential associated to a spin droplet, equivalent to the equilibrium free 
energy, which controls the exit from the nonequilibrium metastable state. Moreover, we assume that such droplet nonequilibrium 
potential can be written as a competition between a surface term, proportional to the (nonequilibrium) surface tension and which
hinders the droplet growth, and a volume (bulk) term, which favours droplet growth and depends on the spontaneous magnetization.
Using the results obtained in previous chapters for both the nonequilibrium surface tension and the nonequilibrium spontaneous
magnetization, we build up our nonequilibrium generalization of nucleation theory. This extended approximation yields
correct predictions for the relevant observables in this problem, namely the critical droplet size, the droplet radial growth
velocity, the metastable state mean lifetime, etc. All these magnitudes inherit the anomalous, non-monotonous behavior of the
nonequilibrium surface tension. In particular, we observe that the critical droplet size and the metastable state mean lifetime 
exhibit a maximum as a function of temperature for any $p\neq 0$, decreasing for lower temperatures. On the other hand, the droplet 
radial growth velocity shows a minimum as a function of temperature. Thereby a main conclusion of our analysis is that the
properties of the interface separating the metastable and stable (nonequilibrium) phases determine in a fundamental way
the metastable state exit dynamics. All these results are verified by extensive Monte Carlo simulations. On the other hand, the 
morphology of the metastable-stable transition is also highly affected by nonequilibrium conditions, mainly at low temperatures. In
particular, it is found that finite nonequilibrium systems may demagnetize from the metastable state through the nucleation of 
multiple critical droplets both at high and low temperatures, while there is an intermediate temperature range where this process 
proceeds through the nucleation of a single critical droplet, as opposed to equilibrium systems, where the multidroplet
mechanism only emerges at high temperatures, being the single droplet nucleation process the relevant one at low temperatures.
This is in fact checked in Monte Carlo simulations. A principal conclusion derived from the results presented in this chapter
is that the hypothesis of existence of a nonequilibrium potential which controls the metastable-stable transition dynamics
accurately describes this nonequilibrium dynamic phenomenon. Hence, although we do not know how to construct such nonequilibrium
potential from the microscopic point of view, this result points out possible paths in order to build up such potential.
In addition to their theoretical interests, these results might also be relevant from the technological point of view,
since a main technological concern is to retain for as long as possible 
the actual (possibly metastable) states of impure magnetic particles
in storage magnetic materials. In order to continue investigating the lines developed in this chapter, we propose:
\begin{itemize}
\item Investigate in depth the validity of the hypothesis of existence of a nonequilibrium potential controlling
the exit from the metastable state. From a formal point of view, we do not know even whether this potential exists.
\item It should be worthwhile to compare our theoretical results with experiments in real magnetic materials. The chances are 
that the phenomenology here described can be in fact observed in actual magnets.
\item It should be also interesting to investigate metastability in one-dimensional nonequilibrium systems with ordered phase at 
low temperatures. Here the characterization of the interface is trivial, and the analysis of the metastability problem and
the effects of nonequilibrium conditions on it should be much simpler. 
\end{itemize}

In {\bf chapter \ref{capAval}} we study the effects that circular free borders induce on the properties of the metastable-stable 
transition studied in previous chapters. This transition proceeds now through the heterogeneous nucleation
of droplets near the border, due to obvious energetic effects. With this exception, all the nucleation properties found in 
previous chapters remain qualitatively valid when free boundaries are present. However, under the combined action of both open
borders and nonequilibrium conditions, the evolution of a stable phase nucleus inside the parent metastable phase
proceeds by avalanches. These burst-like events characterize the dynamics of many complex nonequilibrium systems. Once 
subtracted the extrinsic noise, the measured avalanche size and lifetime distributions show power law behavior, 
up to an exponential cutoff which depends algebraically on system size.
In addition, the size and lifetime of an avalanche are also power-law related. 
A detailed analysis of these scale free avalanches reveals that they are in fact the combined result of many avalanches 
of different well-defined \textit{typical} size and duration. That is, the simplicity and versatility of our 
model system allows us to identify many different types of avalanches, each type characterized by a 
probability distribution with well defined typical size and duration, associated with a particular 
\textit{curvature} of the domain wall. Due to free borders and the microscopic impurity the system 
visits a broad range of domain wall configurations, and thus the combination of these avalanches 
generally results in a distribution which exhibits several decades of power law behavior and an exponential 
cutoff. However, this apparent scale-free behavior does not mean that avalanches are critical, in the sense of a 
second order phase transition where diverging correlation lengths appear. 
The deep insight derived from this chapter comes when we compare our results with experiments on Barkhausen Noise in 
particular and $1/f$ Noise in general. In fact, our measured exponents are almost identical to those measured by Spasojevi\'c
et al\cite{Spaso} in Barkhausen experiments in quasi-bidimensional VITROVAC. Moreover, the exponents we measure 
exhibit finite size corrections similar to those observed in real avalanche systems,
and the algebraic dependence of cutoffs with system size is also a main feature of real systems. Avalanches in our model
show also some properties, as for instance reproducibility, observed in real Barkhausen materials. 
On the other hand, all actual theoretical approaches to Barkhausen Noise are based on the assumption of the existence of an 
underlying critical point (plain old one or SOC one), responsible of the observed scale invariance. However, all these
explanations imply that universality must hold in Barkhausen experiments, which is not observed in practice. 
Therefore, all the similarities found between avalanche properties in our model and Barkhausen experiments,  together with the 
fact that experimental observations do not support the existence of universality in Barkhausen Noise, led us to suspect that 
Barkhausen Noise might also come from the superposition of more elementary events with well-defined typical scales, which is 
the underlying mechanism in our model. The chances are that our observation that scale invariance originates in a combination 
of simple events, which we can prove in our model cases, is a general feature of similar phenomena in many complex systems. 
Several follow-ups of the work developed in this chapter can be proposed:
\begin{itemize}
\item It should be desirable to export the analysis method introduced in this chapter in order to identify the origin of
different avalanches and the superposition of different typical scales to many experimental situations,
simplifying in this way the investigation about the origin of Barkhausen emissions in particular, and $1/f$ Noise in general.
\item A general mathematical framework and a more complete theoretical approach to our observation that scale invariance 
originates in a combination of simple events is needed in order to generalize this idea.
\end{itemize}

In {\bf chapter \ref{capDLG}} we perform a theoretical and computational study of phase segregation under anisotropic 
nonequilibrium conditions. In particular, we study the driven lattice gas (DLG) in two dimensions, since it is a good microscopic 
metaphor of many real situations. The resulting picture holds for a class of highly anisotropic nonequilibrium
phenomena in Nature: ionic superconductors, fluids under shear flow, or subject to external electric fields or gravity, 
vibrated granular materials, etc. The methods developed in this chapter help in the analysis of such situations.  
Coarsening in DLG evolves as follows. After a quench from an initially fully disordered state, anisotropic grains develop,
which quickly give rise to strings. These strings further coarsen until well-defined narrow stripes form which percolate
in the field direction. While the grain and string coarsening stages last for very short as compared to the total evolution 
time, the stripe coarsening process involves most of the system evolution. In fact this is due to the hydrodynamic slowing 
down appearing as a consequence of particle conservation and local dynamics. Hence this last stage is the relevant (observable) 
one from the experimental point of view. The stripe coarsening proceeds through the effective diffusion and coalescence of
stripes. The effective diffusion of stripes can be understood in terms of two different single-particle processes: hole (particle)
diffusion within the stripe (HD), and surface evaporation/condensation (EC) of particles and holes. The EC mechanism is
dominant at the beginning of the stripe coarsening process, due to the large surface/volume ratio at these stages, yielding
a growth law for the stripe mean width of the form $\ell(t) \sim t^{1/4}$. However, as time goes on, the stripe surface/volume
ratio decreases, and eventually the HD mechanism becomes dominant, implying a $t^{1/3}$ growth law. The temporal crossover
between both growth trends appears for a time $\tau_{cross}(L_{\parallel})$. This temporal crossover implies the appearance
of a size crossover, since both $\tau_{cross}(L_{\parallel})$ and the time the system needs to reach the final steady state,
$\tau_{ss}(L_{\perp},L_{\parallel})$ depend on system size. It is found that, for small enough values of $L_{\perp}$, 
$\tau_{cross}(L_{\parallel}) > \tau_{ss}(L_{\perp},L_{\parallel})$, and therefore only $1/4$-behavior is expected at long times.
On the other hand, the $t^{1/3}$ growth law is the general one expected for macroscopic systems. All these theoretical results
are perfectly checked in Monte Carlo simulations. A relevant experimental observable is the structure factor.
We have studied such function in our system, due to its experimental importance. Since during the stripe coarsening stage there
is only one relevant scale, namely the mean stripe width, then dynamical scaling, i.e. time self-similarity during the 
segregation process, is expected for the structure factor. This is in fact confirmed in Monte Carlo simulations. Furthermore,
the shape of the scaling function shows the Guinier gaussian region, followed by the anisotropic Porod's region, $k_{\perp}^{-2}$, 
for large $k_{\perp}$ and a thermal tail $k_{\perp}^{-3}$ for very large $k_{\perp}$. The shape of the structure function,
and in particular the anisotropic extension of Porod's law and the thermal tail, are perfectly understood from the anisotropic, 
striped character of DLG clusters. Moreover, the anisotropy present in DLG is the key ingredient needed in order to understand
the whole coarsening process. Two possible extensions of the research carried out in this chapter are:
\begin{itemize}
\item The application of the theoretical analysis here developed to recent experiments on horizontally vibrated
granular materials\cite{Mullin} which show striped patterns very similar to those observed in our system should shed light
on the physical mechanisms behind such morphogenesis.
\item It should be also worthy to study phase separation from the field theoretical point of view, using a recently
proposed field equation which correctly describes the critical behavior of DLG for infinite field.
\end{itemize}

{\bf Chapter \ref{capLipo}} is devoted to the study of an absorbing (dynamic) phase transition in a biologically inspired 
lattice model, called Lipowski model. Phase transitions separating an active, fluctuating  phase from a frozen one are 
ubiquitously observed in Nature. Some examples are catalytic chemical reactions, disease and damage spreading, forest fires, 
pinning of surfaces, nonequilibrium wetting, sandpiles, etc. In particular, our motivation for studying Lipowski model is
to study the anomalous critical behavior reported for this system.\cite{Lip2} It has been claimed that this model shows a
sort of {\it dimensional reduction} or {\it superuniversality}, in such a way that both the one- and two-dimensional versions
of the model should belong to the one-dimensional directed percolation (DP) universality class. In this chapter we perform 
a finite size scaling analysis of the critical behavior of the bidimensional system together with spreading experiments in
order to address this question. Using these methods we calculate up to six different critical exponents, showing that
the two-dimensional Lipowski model does not belong to any known universality class. In particular, we do not find any
trace of dimensional reduction. Instead, a completely novel scaling is observed. We identify as the relevant ingredient
for the observed novel scaling the presence of {\it superabsorbing sites} (and cluster composed by them) in the system. A site
in the lattice that cannot be activated from some direction(s) by neighboring activity is called a superabsorbing site. A
site can be superabsorbing in one or more directions. The presence of superabsorbing sites in the system enormously slow down the
dynamics. The relevance of superabsorbing sites for the observed novel scaling behavior is strongly supported by two facts.
First, one-dimensional DP behavior is observed in the one-dimensional version of Lipowski model, where no superabsorbing sites are 
found at criticality. On the other hand, the two-dimensional Lipowski model defined on a honeycomb lattice shows 2d DP behavior,
while for this lattice coordination number the probability of finding superabsorbing sites at criticality is negligible. Hence
superabsorbing sites are at the basis of the novel scaling found. Depending on the system dimension, the lattice coordination 
number and other details, we identify different phenomenology. For instance, if complete frozen superabsorbing clusters exist 
above (below) certain threshold, and not below (above) such threshold, 
then a first order transition is expected just at the threshold point.
On the other hand, if {\it almost} frozen clusters of superabsorbing sites appear at criticality, we expect anomalous scaling
as the one reported in this chapter. Finally, if superabsorbing sites are not observed at the critical point, usual
DP scaling must be observed. There are still some open questions which might be addressed in future works:
\begin{itemize}
\item We might study in more realistic systems, as for instance catalytic dimer-dimer systems, whether effects similar to those 
uncovered in the present chapter play a relevant role.
\item It should be also interesting from the theoretical point of view to look for a field theoretical Langevin equation
describing from a coarse-grained point of view (in the spirit of the Reggeon Field Theory for the DP universality class) 
the novel scaling emerging due to the presence of superabsorbing sites.
\item Other interesting questions are related to the relevance of the new symmetry induced by the presence of superabsorbing sites
in the Thermodynamic Limit.
\end{itemize}

In {\bf chapter \ref{capFou}} we study another nonequilibrium dynamic phenomenon, related now with transport phenomena. In 
particular, we investigate the microscopic foundations of normal heat conduction and Fourier's law. In order
to do so we study a one-dimensional chain of point particles, interacting via elastic collisions, which are subject to
a temperature gradient induced by two deterministic heat baths at the extremes of the chain, working at different temperatures.
This problem is very important from the theoretical point of view, since an understanding of the microscopic mechanisms
governing heat conduction should shed much light on the connection between the microscopic fundamental physics 
and some macroscopic properties of many nonequilibrium systems, as for instance Local Thermodynamic Equilibrium, that are
far from being understood.
The general belief nowadays is that a one-dimensional system as the studied here, which conserves the total momentum and has a 
non-zero pressure, must exhibit an anomalous heat conductivity in the Thermodynamic Limit (TL). However, we show in this chapter
that, on the contrary to the popular belief, the system investigated here exhibits a finite
conductivity in the TL. In order to prove such result, we analyze the problem from several different, 
complementary points of view. First, we prove the existence of a non-trivial thermal profile in the TL, indicating that 
Local Thermodynamic Equilibrium (LTE) holds in the system, and thus it makes sense to ask about the validity of Fourier's law.
The thermal profile shows a linear central region, and gets curved near the heat reservoirs ({\it surface resistance}).
In order to verify whether Fourier's law holds or not for this system, we measure in a first step the heat flux $J$ through the 
system. If Fourier's law holds, $J$ should decrease as a function of the inverse number of particles in the system. However, 
finite size effects on $J$ are so strong that this analysis does not yield any definite conclusion about the system conductivity.
A different method, based on the energy current self-correlation function $C(t)$, from which thermal conductivity can be derived 
via the Green-Kubo formulae, is then used. The thermal conductivity $\kappa$ should be finite in the TL if $C(t)$ decays as
$t^{-1-\Delta}$, with $\Delta>0$. The long time analysis of $C(t)$ shows that it presents two different asymptotic behaviors, 
namely $t^{-1.3}$ and $t^{-0.85}$. However, we show that the $t^{-0.85}$ tail comes from system autocorrelations due to finite 
size effects, and hence the $t^{-1.3}$ is the true asymptotic one in the TL. Therefore, the thermal conductivity, as derived 
from the Green-Kubo formula, is finite in our system, so Fourier's law holds in this case.
The previous result depends critically on data analysis, and due to the slow decay of correlations and the strong finite size 
effects observed in this system, such analysis becomes very difficult. Hence further tests and a global consistent vision of the 
problem are needed in order to ensure about the validity of Fourier's law in this system. For this reason we also study in this 
chapter the dynamical aspects of Fourier's law. In particular we study the energy diffusion through the system, finding that
an initially localized energy pulse propagates in a diffusive manner through the system, thus confirming that Fourier's law
holds even in its dynamical aspects. Moreover, the energy diffusion study allows us ti identify the mechanism responsible 
of normal heat conduction in the system. We observe that light particles tend to dynamically absorb much more energy than 
heavy ones. In this way, the light degrees of freedom acquire energy easily, but release it at very long time scales, thus
giving rise to {\it hot-cold structures}. Furthermore, energy propagates in a diffusive way when many of those hot-cold 
structures are formed, thus indicating the cooperative character of the phenomenon. Cooperative phenomena as the one here 
described can do the job of local potentials, giving rise to normal heat conduction. In spite of our results, there are still
many open question associated to the conductivity problem. Some ideas are:
\begin{itemize}
\item The analysis of the energy current self-correlation function has provoked an exciting discussion among the
experts in the field. It should be desirable to obtain a clearer picture of the finite size effects which give rise to
the measured $t^{-0.85}$ spureous tail in $C(t)$.
\item Even simpler models must be proposed in order to study the problem of conductivity from a microscopic point of view.
In this sense, a model similar to the one studied here but with significant smaller finite size effects should be welcome. 
A good candidate is a one-dimensional ring with charged point particles subject to an electric field.
\item There are many more fundamental open question, as for instance: which are the necessary and sufficient conditions in
order to observe LTE in a system ?, which is the relation among LTE, energy equipartition and heat transport ?, etc.
\end{itemize}

As the reader will surely have realized, this thesis is highly heterogeneous. We have studied here many different systems,
and what is more significant, we have used many different theoretical methods 
and approximations in order to understand what is going on in each
problem. This fact points out one of the main problems of actual nonequilibrium physics: the lack of a general formalism,
equivalent in some sense to equilibrium Statistical Mechanics, in order to describe in a unified way nonequilibrium phenomena.
On the other hand, we have been able to understand all the observed phenomenology in all the nonequilibrium dynamic problems studied 
here using these incomplete theoretical approaches, many of them based on concepts derived from equilibrium statistical physics.
In order to do so, we have done many {\it reasonable} approximations, but which we are not able to prove. For these reasons
we think that theoretical physicists must put their effort nowadays on the rigorous connection between the microscopic physics 
of nonequilibrium systems and the meso-macroscopic assumptions which allow us to develop semi-phenomenologic theoretical
approaches to nonequilibrium phenomena, such as the Local Thermodynamic Equilibrium hypothesis, the existence of nonequilibrium
potentials controlling the system dynamic and static properties, etc. I think in this connection underlies the missing link of
nonequilibrium statistical physics.

\vskip 1cm

\begin{flushright}
Granada, November $6^{th}$, 2002
\end{flushright}

\appendix

\chapter{Monte Carlo with Absorbing Markov Chains Simulations and Rejection-Free Algorithms. Projective Dynamics and 
the Slow Forcing Approximation}
\label{apendMCAMC}
\markboth{Monte Carlo with Absorbing Markov Chains Algorithms}{}

In this appendix we present the foundations of Monte Carlo with Absorbing Markov Chains (MCAMC) algorithms, as well as the method of 
projective dynamics and the slow forcing approximation.\cite{reviewMCAMC,forcing,projective}

In general, Monte Carlo methods, first introduced by Metropolis, Rosenbluth, Rosenbluth, Teller and Teller\cite{Metrop} and
mainly characterized by the use of random numbers, are useful to study the static and dynamic properties of stochastic systems.
A Monte Carlo algorithm generates stochastic trajectories in the system's phase space, in such a way that the properties of the 
system are derived from averages over the different trajectories. If we want to study the static properties of a system,
there is a considerable freedom to choose the way in which we move through the phase space with the Monte Carlo algorithm.
However, if we want to study dynamic properties there is no such freedom, since the physical meaning of the dynamics is 
an essential part of the model. Since we want to understand here a dynamic process as metastability in ferromagnetic systems, 
the advanced simulation algorithms we will use must respect the system dynamics. This is the case for the Monte Carlo with Absorbing 
Markov Chains algorithms that we summarize in this appendix. 

Let's summarize the steps of an standard Monte Carlo algorithm for the Ising model with spin flip dynamics before going into the
functioning of MCAMC algorithms.\cite{Binderlibro} For a dynamics $\omega(\beta \Delta {\cal H})$, which yields the transition
probability per unit time between two configurations which differ in the state of a single spin, and which depends on the inverse
temperature $\beta$ and the energy increment between both configurations, $\Delta {\cal H}$, the steps the standard algorithm follows
are:
\begin{itemize}
\item Increase the time from $t$ to $t+1/N$.
\item Choose randomly a spin in the lattice.
\item Calculate a random number $\bar{r}$ with an homogeneous distribution in the interval $(0,1)$.
\item Calculate, or look up in a previously stored table, the energy $E_{old}$ of the current configuration, and the energy $E_{new}$
of the configuration of the system if we flip the selected spin. From these values, we calculate the energy increment 
$\Delta {\cal H}$ involved by this spin flip.
\item Accept the configuration change, i.e. flip the selected spin, if $\bar{r} \leq \omega(\beta \Delta {\cal H})$. Otherwise
keep the same configuration.
\end{itemize}
A Monte Carlo Step per Spin (MCSS) is defined as $N$ spin flip trials as the above described, where $N$ is the number of spins in
the lattice. As we explained in section \ref{capMotiv_apModel}, a MCSS corresponds to a physical time of order $10^{-13}$ seconds,
which is roughly the inverse frequency of the associated heat bath phonons. The fundamental problem of the previously described Monte 
Carlo scheme is that, with some probability, the algorithm rejects a spin flip, which involves a waste of computer time.  Furthermore,
for low enough temperatures the probability of accepting a spin flip against the local magnetic field a spin suffers is extremely 
small (the local magnetic field the spin $i$ suffers is defined as $h_i = -J \sum_j s_j -h$, where the sum runs over the nearest 
neighbors of spin $i$). Hence most of the trials are rejected in this case. This problem makes the standard Monte Carlo algorithm
inefficient in order to study the exit from a metastable state at low temperatures.

The advantage of MCAMC algorithms resides in that they are {\it rejection-free} algorithms: in this case, once we randomly select 
a spin in the lattice, it is flipped with unit probability, and time is incremented by the necessary amount for this spin flip
to take place. Hence, as opposed to classic Monte Carlo algorithms, in MCAMC algorithms the time increments are variable (and
stochastic, as we will see below).

MCAMC algorithms use the concept of absorbing Markov chains (as its name points out). Hence it is necessary to make a brief
introduction to the most relevant properties of absorbing Markov chains. Following ref. \cite{reviewMCAMC}, let's consider one
of these chains, with $s$ transient states and $r$ absorbing states. The generic system starts its evolution in one of the
transient states, and it remains in the transient state space up to it is trapped in one of the $r$ absorbing states. In order to 
completely define the absorbing Markov chain we only need to write the Markov matrix,
\begin{equation}
\mathbf{M}_{(r+s) \times (r+s)} = 
\left( \begin{array}{cc}
\mathbf{I}_{r \times r} & \mathbf{0}_{r \times s} \\
\mathbf{R}_{s \times r} & \mathbf{T}_{s \times s} \\
\end{array} \right)
\label{markovmatrix}
\end{equation}
The elements of this matrix yield the probability of evolving from the state $i$ to the state $j$ in each clock tic, 
$M_{i,j} \equiv M(i \rightarrow j)$. We must notice that here we are using the {\it mathematical} notation, where the state vector
is a row vector, on which the Markov matrix acts from the right. It is obvious that this matrix describes a Markov process, since 
the transition probabilities between states (the elements $M_{i,j}$) only depend on the initial state (given by $i$, the row index)
and the final state ($j$, the column index). Furthermore, the sum of the elements on each row is unity.

The size of the sub-matrices defining the Markov matrix is explicitely shown in the above expression. The matrix $\mathbf{I}$ is
the identity, $\mathbf{0}$ is a matrix with all its elements equal to zero, $\mathbf{T}$ is the transient matrix, which shows the
transition probabilities among the states belonging to the transient space, and $\mathbf{R}$ is the recurrent matrix, which shows
the transition probabilities among the $s$ transient states and the $r$ absorbing states. The matrix that governs the evolution
of the system after $m$ time steps is,
\begin{equation}
\mathbf{M}^m_{(r+s) \times (r+s)} = 
\left( \begin{array}{cc}
\mathbf{I}_{r \times r} & \mathbf{0}_{r \times s} \\
(\mathbf{I+T+ \ldots +T}^{m-1})_{s \times s} \mathbf{R}_{s \times r} & \mathbf{T}^m_{s \times s} \\
\end{array} \right)
\label{markovmatrix2}
\end{equation}
The system must initially lie in the transient space, so the initial state is represented by a vector $(\vec{0}^T,\vec{v}_I^T)$,
where the vector $\vec{v}_I^T$ has $s$ components and the super-index $^T$ denotes the vector transpose, i.e. a row vector.
Applying the matrix $\mathbf{M}^m$ to this initial state vector, we obtain the $(r+s)$-dimensional vector,
\begin{equation}
(\begin{array}{cc}
\vec{0}^T & \vec{v}_I^T
\end{array})
\mathbf{M}^m = 
(\begin{array}{cc}
\vec{v}_I^T(\mathbf{I+T+ \ldots +T}^{m-1}) \mathbf{R} & \vec{v}_I^T \mathbf{T}^m
\end{array})
\label{vectorestado}
\end{equation}
The components of this vector yield the probability of being in each one of the $(r+s)$ states of the system after $m$ time steps.
If we introduce the row vector $\vec{e}$, with dimension $s$, such the all its elements are equal to unity, the probability of
still being in the space of transient states after $m$ time step is,
\begin{equation}
p_{\textrm{transient}} = \vec{v}_I^T \mathbf{T}^m \vec{e}
\label{transit}
\end{equation}
The probability that the system has exited to each one of the $r$ possible absorbing states is determined by the components of the
following $r$-dimensional vector,
\begin{equation}
\vec{p}_{\textrm{abs. after m steps}} = \vec{v}_I^T (\mathbf{I+T+ \ldots +T}^{m-1}) \mathbf{R}
\label{vecabsorb}
\end{equation}
Here it is clearly observed that the probability that the system has exited to each one of the $r$ absorbing states after $m$ time
steps is equal to the probability that the exit takes place in the first step (determined by the term $\mathbf{IR}$), plus the
probability that the exit takes place in the second step ($\mathbf{TR}$), plus the probability that the exit takes place in the steps
$3, 4, \ldots, m$. Thus, the probability that the system exits to each one of the $r$ possible absorbing states, {\it given} that
the exit takes place in the $m$-th time step is,
\begin{equation}
\vec{p}_{\textrm{abs. in step m}} = \frac{\vec{v}_I^T \mathbf{T}^{m-1} \mathbf{R}}
{\vec{v}_I^T \mathbf{T}^{m-1} \mathbf{R} \vec{e}}
\label{vecabsorb2}
\end{equation}
Eqs. (\ref{transit}) and (\ref{vecabsorb2}) are the basic equations from which MCAMC algorithms are derived.
\begin{table}[t!]
\centerline{
\begin{tabular}{|c||c|c|c|}
\hline \hline
Class & Central spin & Number of up neighbors & $\Delta {\cal H}$ \\
\hline \hline
1 & +1 & 4 & 8J+2h \\
\hline
2 & +1 & 3 & 4J+2h \\
\hline
3 & +1 & 2 & 2h \\
\hline
4 & +1 & 1 & -4J+2h \\
\hline
5 & +1 & 0 & -8J+2h \\
\hline \hline
6 & -1 & 4 & -8J-2h \\
\hline
7 & -1 & 3 & -4J-2h \\
\hline
8 & -1 & 2 & -2h \\
\hline
9 & -1 & 1 & 4J-2h \\
\hline
10 & -1 & 0 & 8J-2h \\
\hline \hline
\end{tabular}
}
\caption[Spin classes for the two-dimensional Ising model.]
{\small Spin classes for the two-dimensional isotropic Ising model with periodic boundary conditions. The last column shows
the energy increment associated to each spin class.}
\label{tabclases}
\end{table}

In order to apply our knowledge about absorbing Markov chains to the bidimensional Ising model we must introduce the concept of 
{\it spin classes}. For a spin in the lattice, the spin class to which this spin belongs to is defined by the spin orientation 
($+1$ or $-1$) and the number of up nearest neighbors it shows. Hence, for the two-dimensional isotropic Ising model with periodic 
boundary conditions there are $10$ different spin classes, schematized in Table \ref{tabclases}. All spins belonging to the same
spin class involve the same energy increment when flipped (see Table \ref{tabclases}), so the transition rate for a spin depends
exclusively on the class $i \in [1,10]$ to which it belongs to, $\omega _i \equiv \omega (\beta \Delta {\cal H}_i)$, where
$\Delta {\cal H}_i$ is the energy increment associated to class $i$. On the other hand, if we have an up (down) spin in class
$i \in [1,5]$ ($i \in [6,10]$), when we flip this spin its class will change to $i+5$ ($i-5$). Equivalently its four nearest 
neighbor spins will change the class to which they belong to, increasing (decreasing) in one unit the class they were before
the spin flip.

Let's assume that $n_i$ is the number of spins in class $i$. Then $N=\sum_{i=1}^{10} n_i$. Thus the probability of flipping in a 
time step any of the spins in class $i$ will be  $n_i \omega _i /N$, since $n_i/N$ is the probability of selecting a
spin in class $i$, and $\omega _i$ is the probability per unit time of flipping the spin, once it has been selected. In order to
exit from a given state in the Ising system, one of the spins in the system, which belongs to some spin class, must be flipped. 
Consequently we can interpret this process as an absorbing Markov chain where the current state is the only transient state, and where 
there are $10$ different absorbing states, each one associated to a 
spin flip where the flipped spin belongs to one of the $10$ possible spin 
classes. Thus the absorbing state $3$ is the state to which the system exits if we flip in the current configuration a spin belonging 
to class $3$.

Therefore the system is governed by a Markov matrix $\mathbf{M}_{(10+1) \times (10+1)}$, with $s=1$ transient states and 
$r=10$ absorbing states. The recurrent matrix $\mathbf{R}$ is now a row vector with $10$ components,
\begin{equation}
\mathbf{R} = 
(\begin{array}{cccc}
\frac{n_1 \omega _1}{N} & \frac{n_2 \omega _2}{N} & \ldots & \frac{n_{10} \omega _{10}}{N}
\end{array})
\label{Isingrecurr}
\end{equation}
and the transient matrix $\mathbf{T}$ is now a scalar which yields the probability of remaining in the current configuration,
\begin{equation}
\mathbf{T} = 1- \frac{1}{N} \sum_{i=1}^{10} n_i \omega _i \equiv \lambda
\label{T}
\end{equation}
The next step consists in determining the number of time steps the system needs to exit the current configuration. The probability 
that the system is still in the same transient state (i.e. the current state) after $m$ time steps is $\lambda ^m$. This
probability does not depend on the absorbing state to which the system finally exits. The probability that the absorption occurs in 
the $m$-th step is $\lambda ^{m-1} (1-\lambda)$. Since we are speaking about probabilities, the number of time steps the 
system needs to exit the current state will be a stochastic variable determined through the inequality 
$\lambda^{m-1} \geq \tilde{r} > \lambda^m$, where $\tilde{r}$ is a random number homogeneously distributed in the interval 
$(0,1]$. From this inequality we obtain,
\begin{equation}
m = \lfloor \frac{\textrm{ln}(\tilde{r})}{\textrm{ln}\lambda} \rfloor + 1
\label{tiempo}
\end{equation}
where $\lfloor x \rfloor$ is the integer part of $x$. Once we determine the time the system needs to exit the current state, we only 
have to decide to which absorbing state the system exits to. In order to do so we select randomly, with 
probabilistic weights determined by the elements of the recurrent matrix $\mathbf{R}$, to which class the spin to flip belongs to.
We thus define $10$ partial sums $Q_i=\sum _{j=1}^{i} n_j \omega _j$, with $i \in [1,10]$. In order to decide the class
to which the spin to flip belongs to, we generate another random number $\bar{r}$ homogeneously distributed in the interval $[0,1)$
and we determine the index $k \in [1,10]$ that fulfills the following condition,
\begin{equation}
Q_{k-1} \leq \bar{r} Q_{10} < Q_k
\label{qs}
\end{equation}
The index $k$ yields the class where we must flip the spin. If there are more than one spin in class $k$, we must use another random 
number to decide, completely at random, the spin to flip in this class.

Summarizing, the steps to follow in a simulation of the Ising model using this rejection-free algorithm are,
\begin{itemize}
\item Calculate three random numbers $\tilde{r}$, $\bar{r}$ and $\hat{r}$.
\item Calculate the accumulated sums $Q_i=\sum _{j=1}^{i} n_j \omega _j$, with $i\in [1,10]$.
\item Using $\tilde{r}$, calculate from eq. (\ref{tiempo}) the number of time steps $m$ in order to exit the current configuration.
\item Using $\bar{r}$, calculate the class $k$ which satisfies the condition $Q_{k-1} \leq \bar{r}Q_{10} < Q_k$.
\item Using $\hat{r}$, randomly select one of the $n_k$ spins in class $k$, and flip this spin.
\item If the flipped spin was up (down) before the change, add one unit to the number of spins in class $k+5$ ($k-5$), 
$n_{k+5} \rightarrow n_{k+5}+1$ ($n_{k-5} \rightarrow n_{k-5}+1$). In the same way, subtract one unit from the number of spins
in class $k$, $n_k \rightarrow n_k - 1$.
\item If the flipped spin was up (down) before the change, each one of its four nearest neighbors changes from 
the class $l$ to which it belonged to the class $l+1$ ($l-1$). We increase the number of spins in class $l+1$ ($l-1$) in one unit,
and we decrease the number of spins in class $l$ in one unit.
\item Increase the time from $t$ to $t+m/N$  (we divide by $N$, the number of spins in the lattice, because we measure time in MCSS).
\end{itemize}
This algorithm is rejection-free, because it accepts with unit probability all proposed spin flips, and time increases in stochastic
intervals, depending on the system's current state. This algorithm, which is the one we have used when studying metastability,
is called in literature {\it s-1 Monte Carlo with Absorbing Markov Chains algorithm}. The name s-1 points out the fact that 
this algorithm uses a single transient state. Using eqs. (\ref{transit}) and (\ref{vecabsorb2}) it is possible to generalize 
this algorithm to transient subspaces with 2 (s-2 MCAMC) and 3 (s-3 MCAMC) states.\cite{reviewMCAMC} In general, these algorithms
are many order of magnitude faster than standard Monte Carlo algorithms when simulating systems with discrete state space at low 
temperatures. The efficiency of these algorithms increments as we enlarge the transient subspace.

There is an efficient way to computationally implement MCAMC algorithms. It is based on the creation of four different lists.
The vector $\texttt{NCLS(k)}$, with $k=1, \ldots, 10$, contains the number of spins $n_k$ in each spin class. The vectors 
$\texttt{ICLASS(i)}$ and $\texttt{ICPOS(i)}$ inform us about the class to which a spin situated in a lattice site $i$ belongs
to and the place that this spin occupies in the list of spins belonging to the same class, respectively. Finally, the matrix 
$\texttt{LOC(j,k)}$ yields the lattice position of a spin in class $k$ situated in the $j$-th position in the list of spins belonging
to this class. Hence we have that $\texttt{LOC(ICPOS(i),ICLASS(i))=i}$. Using these four vectors it is very simple to implement 
MCAMC algorithms, as well as to actualize the class populations after each spin flip.

When we want to study the problem of metastability in an Ising model at very low temperatures and weak magnetic fields, 
the local stability of the metastable state is so strong that simulations are not feasible even for MCAMC algorithms. Therefore it 
is necessary to go one step beyond MCAMC algorithms and design more advanced techniques able to simulate {\it rare events} as 
the exit from the metastable state in these extreme conditions. In our case the solution underlies in the so-called {\it slow forcing
approximation}.\cite{forcing} In this approximation the system is forced to evolve towards the stable state by a moving magnetization 
wall. That is, we define an upper bound for magnetization, which depends on time, $m_{lim}(t) = 1 - \phi t$, and such that
the system magnetization is forced to stay below this threshold at any time. This constraint imposed on magnetization clearly
modifies the system's original dynamics. Hence, given the dynamic character of the problem of metastability, such modification on the
dynamics would affect in principle the results of simulations performed with a nonzero forcing $\phi$. However, it has been shown
that for small enough values of the forcing $\phi$, a slow forcing limit exists\cite{forcing}, such that in this limit the system
observables are independent of the applied forcing, while the simulation is still significantly accelerated as compared to the
non-forced system. In some sense, this slow forcing limit is an {\it adiabatic limit}: although the system is forced, if $\phi$
is small enough the system has enough time to thermalize and select the same {\it typical} configurations that a system without forcing
should choose in its evolution from the metastable state to the stable one. In this way the phase space sampling in the slow forcing 
limit is almost indistinguishable from the sampling performed with non-forced algorithms, although in the first case the sampling 
can be done in a reasonable amount of time. 
It is important to realize that the forcing is only relevant in the neighborhood of the metastable state:
once the energy barrier has been overcome, the system rapidly evolves towards the stable state via nonequilibrium 
(non-thermalized) configurations, being the magnetization wall irrelevant in this process. The combination of the s-1 MCAMC algorithm
and the slow forcing approximation constitutes the basic computational scheme we have used to simulate our system.

If we now think, for instance, about the metastable state lifetime, it is obvious that the result obtained using the slow forcing
approximation will be much smaller than the real one. Thus we must develop a method that allows us to extract information, as for
instance the metastable state lifetime, from simulations performed using the slow forcing approximation. This method is based
on the so-called {\it projective dynamics method}.\cite{projective} The idea behind the projective dynamics method is that
one expects that a one-dimensional physical picture of the nonequilibrium potential which controls the process is valid even
for a complicated process as that of metastability in our nonequilibrium model. In this approximation the dynamics of the
(complex) system is projected on a one-dimensional system where the relevant variable is one of the {\it slow-evolving} observables
of the original system. In particular, all states with the same magnetization are projected on a single state, defined by the
value of magnetization. Therefore, instead of considering transitions among the $2^N$ possible states in the (complex) Ising system, 
which are captured by a Markov matrix with dimension $2^N \times 2^N$, we only consider transitions between projected states with well
defined magnetization, giving rise in this case to a projected Markov matrix with dimensions $(N+1) \times (N+1)$. It has been
mathematically proved that the original Markov matrix with $2^N$ states is not even weakly lumpable\cite{lump}, although it seems
to be lumpable with respect to the states in the escape route from the metastable state. Thus the original master equation which
governs the Ising dynamics, given by eq. (\ref{mastereq}) (see section \ref{capMotiv_apModel}) is projected on a master equation
for the $(N+1)$ magnetization states,
\begin{equation}
\frac{\textrm{d}P(n_{up};t)}{\textrm{d} t} = \sum_{n_{up}'} 
\Big[W(n_{up}' \rightarrow n_{up})P(n_{up}';t) - W(n_{up} \rightarrow n_{up}')P(n_{up};t)\Big]
\label{mastereqlumped}
\end{equation}
The variable $n_{up}$ is the number of up spins which identifies the projected state, and it is completely equivalent to 
magnetization $m$, $n_{up} = N(1+m)/2$. Here the projected transition rates between magnetization states, 
$W(n_{up}' \rightarrow n_{up})$, naturally appear. 
Only transitions between states which differ in one up spin are allowed, since our dynamics
is a single spin flip dynamics (see eq. \ref{rate}). Hence the transition rates $W(n_{up}' \rightarrow n_{up})$  and 
$W(n_{up} \rightarrow n_{up}')$ will be zero always that $n_{up}' -n_{up} \neq \pm 1,0$. This property implies that the associated 
projected Markov matrix is tridiagonal, and so very easy to treat analytically.

It is possible to measure these transition rates in a simple way in Monte Carlo simulations. We only have to notice that, for instance,
$W(n_{up} \rightarrow n_{up} -1)$ is just the probability per unit time that an up spin flips. An up spin must be in a class 
$k \in [1,5]$, and we know that the probability per unit time of changing a spin in class $k$ is given by $n_k \omega _k /N$,
where $n_k$ is the number of spins belonging to class $k$, and $\omega _k$ is the transition rate associated to this class.
Therefore we can write,
\begin{eqnarray}
W(n_{up} \rightarrow n_{up}-1) & \equiv & \frac{g(n_{up})}{N} = 
\frac{1}{N} \sum_{k=1}^5 n_k \omega _k  \nonumber \\
W(n_{up} \rightarrow n_{up}+1) & \equiv & \frac{s(n_{up})}{N} = 
\frac{1}{N} \sum_{k=6}^{10} n_k \omega _k \\
W(n_{up} \rightarrow n_{up})   & \equiv & 1-\frac{g(n_{up})}{N}-\frac{s(n_{up})}{N} \nonumber
\end{eqnarray}
These equations help us to define growth rate, $g(n_{up})$, and the shrinkage rate, $s(n_{up})$, of the stable phase in the state
with magnetization $n_{up}$. Using the s-1 MCAMC algorithm is very simple to measure both $g(n_{up})$ and $s(n_{up})$, since 
we know the classes populations at any time. We can obtain much information about the metastable-stable transition from these 
stable phase growth and shrinkage rates. For instance, if we perform $K$ demagnetization experiments with $h<0$ from the metastable 
state (with positive magnetization) to the stable one (with negative magnetization), we can write the following balance equation,
valid for all values of $n_{up}$,
\begin{equation}
N(n_{up} \rightarrow n_{up}-1) = K + N(n_{up}-1 \rightarrow n_{up})
\label{balance2}
\end{equation}
where $N(n_{up} \rightarrow n_{up} - 1)$ and $N(n_{up} - 1 \rightarrow n_{up})$ are, respectively, the number of times that,
in these $K$ experiments, we go from a state with $n_{up}$ up spins to other state with $n_{up} - 1$ up spins, and the number of
times we go from $n_{up}-1$ to $n_{up}$. If $\eta (n_{up})$ is the total time we spend in a state with magnetization $n_{up}$
in one experiment, we can write, $N(n_{up} \rightarrow n_{up} - 1) = K \eta (n_{up}) g(n_{up})$, and
$N(n_{up} - 1 \rightarrow n_{up}) = K \eta (n_{up}-1) s(n_{up})$. We can now write the following recurrence relation,
\begin{equation}
\eta (n_{up})= \frac{1 + \eta (n_{up}-1) s(n_{up}-1)}{g(n_{up})}
\label{hn}
\end{equation}
which relates the time we spend in states with $n_{up}$ up spins to the time we spend in states with $n_{up}-1$ up spins. On
the other hand, we must give a threshold in order to define the metastable state lifetime, such that the first passage time
through this threshold will yield the definition of the metastable lifetime. For lattice spin system as the one we study this 
threshold is usually defined by the zero magnetization condition. In this way we define the lifetime of the metastable state for 
the ferromagnetic spin system as the first passage time to $m=0$ (or, equivalently, $n_{up}=N/2$). We can calculate this 
mean lifetime using the above recurrence relation. In order to do so we must fix an initial condition for the recurrence. We can 
think in this process as a one-dimensional random walk, which starts at $n_{up}=N$ and finishes at $n_{up}=N/2$. Thus we have
that $\eta (N/2)=0$ so,
\begin{equation}
\tau = \sum _{n_{up}=\frac{N}{2}+1}^{N} \eta(n_{up}), \quad   \eta (n_{up}) = 
\frac{1 + \eta (n_{up}-1) s(n_{up}-1)}{g(n_{up})}, \quad    
\eta(\frac{N}{2})=0
\label{tauprojected}
\end{equation}
In this way we can obtain the metastable state mean lifetime from the stable phase growth and shrinkage rates, $g(n_{up})$ and 
$s(n_{up})$, respectively. An estimation of the error for $\tau$ can be obtained by a simple quadratic error propagation procedure,
from the errors associated to the times $\eta (n_{up})$, whose error derive from the statistical errors in the measure of
$g(n_{up})$ and $s(n_{up})$. The mean lifetime obtained from (\ref{tauprojected}) is exact up to statistical errors, as compared to 
the mean lifetime of the real system.\cite{reviewMCAMC} Finally, the projected growth and shrinkage rates, $g(n_{up})$ and 
$s(n_{up})$, allow us to study the shape of the nonequilibrium potential (or the fee energy in the equilibrium system) during
the escape from the metastable state. Thus the points $n_{up}^*$ for which $g(n_{up}^*)=s(n_{up}^*)$ identify the local extremes 
of this nonequilibrium potential. This fact allows us to measure, as explained in Chapter \ref{capMedio}, the magnetization of the
stable and metastable states with high precision, as well as the magnetization which defines the critical droplet.
We also are able to calculate from $g(n_{up})$ and $s(n_{up})$ the system's magnetic viscosity\cite{Vacas}.

\chapter{Calculation of the Probability $\Pi_2(T,p)$ of Finding an Interfacial Spin in Class 2} 
\label{apendPi2}
\markboth{Calculation of the Probability $\Pi_2(T,p)$}{}

In this appendix we present in detail the calculation of $\Pi_2(T,p)$, the probability of finding an interfacial spin belonging to
class 2, in our generalized Solid-On-Solid approximation for the nonequilibrium interface.

As explained in section \ref{capSOS_apGen}, the probability $\Pi_2(T,p)$ is defined by the average (see eq. (\ref{pii})),
\begin{equation}
\Pi_2(T,p)=\sum_{\delta,\epsilon=-\infty}^{+\infty} \pi_2(\delta,\epsilon) p(\delta,\epsilon)
\label{piibis}
\end{equation}
where $\pi_i(\delta,\epsilon)$ is the probability of finding a spin of class $i$ in an interfacial spin column characterized by
a left step with magnitude $\delta$ and a right step with size $\epsilon$, and where $p(\delta,\epsilon)$ is the probability of
finding an interfacial spin column in such configuration $(\delta,\epsilon)$. The two-body probability function $p(\delta,\epsilon)$
depends on the relative signs (including zero) of the steps $\delta$ and $\epsilon$. Table \ref{configs} shows the different functions
$p(\delta,\epsilon)$ for each one of the $9$ possible typical configurations for a interfacial spin column in the generalized SOS
approximation. On the other hand, as explained in section \ref{capSOS_apGen}, the probability function $\pi_2(\delta,\epsilon)$
can be written as,
\begin{equation}
\pi_2(\delta,\epsilon) = \frac{n_2(\delta,\epsilon)}{N(\delta,\epsilon)}
\label{probpiibis}
\end{equation}
where $n_2(\delta,\epsilon)$ is the number of spins in class 2 for an interfacial spin column characterized by $(\delta,\epsilon)$,
and where $N(\delta,\epsilon)$ is the total number of interfacial spins associated to this column. Table \ref{clasesconfig} shows
the different values of $n_2(\delta,\epsilon)$ and $N(\delta,\epsilon)$ for each typical column configuration. Attending to the 
entries of both tables, we can write for $\Pi_2(T,p)$,
\begin{eqnarray}
\Pi_2(T,p) & = & \frac{1}{\cal Q} \Big\{ \sum_{\delta=1}^{+\infty} \sum_{\epsilon=1}^{+\infty} (1-\frac{1}{\delta})
Y^{\delta+\epsilon}X_3^3X_2^{\delta+\epsilon-2} + \sum_{\delta=1}^{+\infty}(1-\frac{1}{\delta})Y^{\delta}
X_2^{\delta-1}X_3^2 \nonumber \\
 & + & \sum_{\delta=1}^{+\infty}\sum_{\epsilon=-\infty}^{-1} (1-\frac{\alpha}{\lambda})
Y^{\delta+\epsilon}X_4^3X_3^{2(\alpha-1)}X_2^{\lambda-\alpha} + \sum_{\epsilon=1}^{+\infty} Y^{\epsilon}
X_3^2X_2^{\epsilon-1} + X_2 \nonumber \\
 & + & \sum_{\epsilon=-\infty}^{-1} (1-\frac{1}{|\epsilon|})Y^{\epsilon}X_3^2
X_2^{|\epsilon|-1} + \sum_{\delta=-\infty}^{-1} \sum_{\epsilon=1}^{+\infty} Y^{\delta+\epsilon}X_4^3
X_3^{2(\alpha-1)}X_2^{\lambda-\alpha} \nonumber \\
 & + & \sum_{\delta=-\infty}^{-1} Y^{\delta} X_3^2X_2^{|\delta|-1} +
\sum_{\delta=-\infty}^{-1} \sum_{\epsilon=-\infty}^{-1} (1-\frac{1}{|\epsilon|}) Y^{\delta+\epsilon}
X_3^3X_2^{|\delta|+|\epsilon|-2} \Big\}
\label{Pi2sum}
\end{eqnarray}
where $X_i=\textrm{e}^{-2J\beta_{eff}^{(i)}}$ and $Y=\textrm{e}^{\gamma_p(\phi)}$. Writing the above equation we have used the
equalities $X_9=X_2$, $X_8=X_3$ and $X_7=X_4$, and we have defined $\alpha=\min(|\delta|,|\epsilon|)$ and 
$\lambda=\max(|\delta|,|\epsilon|)$. The factor ${\cal Q}$ is the normalization constant, already calculated in section
\ref{capSOS_apGen}, see eq. (\ref{Qnorma}).

We must use the classic results for the geometric sum and the geometric series, as well as some other results that can be 
easily derived from them, see eqs. (\ref{sum1})-(\ref{sum6}), in order to perform the sums involved in the calculation of
$\Pi_2(T,p)$. It is convenient to summarize again these results,
\begin{eqnarray}
\sum_{k=0}^{\infty} x^k & = & \frac{1}{1-x} \quad , \quad |x|<1 \label{sum1bis} \\
\sum_{k=0}^n x^k & = & \frac{1-x^{n+1}}{1-x} \quad , \quad x \neq 1 \label{sum2bis} \\
\sum_{k=1}^{\infty} \frac{x^k}{k} & = & \ln (\frac{1}{1-x}) \quad , \quad |x|<1 \label{sum3bis} \\
\sum_{k=1}^n \frac{x^k}{k} & = & \ln (\frac{1}{1-x}) - \int_0^x \textrm{d}y \frac{y^n}{1-y} \quad , 
\quad |x|<1 \label{sum4bis} \\
\sum_{k=1}^{\infty} k x^k & = & \frac{x}{(1-x)^2} \quad , \quad |x|<1 \label{sum5bis} \\
\sum_{k=1}^n k x^k & = & \frac{x(1-x^n)}{(1-x)^2} - \frac{nx^{n+1}}{1-x} \quad , \quad x \neq 1 \label{sum6bis}
\label{sumatorias1}
\end{eqnarray}
Using these expressions we can complete satisfactorily the calculation of $\Pi_2(T,p)$. To see in detail this calculation,
let's solve in a first step, as a simple case, the double sum appearing in the first term of the right hand side of eq. 
(\ref{Pi2sum}). This term is associated to an interfacial spin column characterized by $\delta,\epsilon > 0$ (type A in
Table \ref{configs}). If we denote this sum as $S_A$, then $S_A$ is written as,
\begin{eqnarray}
S_A & = & \sum_{\delta=1}^{+\infty} \sum_{\epsilon=1}^{+\infty} (1-\frac{1}{\delta})Y^{\delta+\epsilon}
X_3^3X_2^{\delta+\epsilon-2} \nonumber \\
 & = & \frac{X_3^3}{X_2^2} \sum_{\epsilon=1}^{+\infty} (X_2Y)^{\epsilon}
\sum_{\delta=1}^{+\infty} (1-\frac{1}{\delta})(X_2Y)^{\delta} \nonumber
\end{eqnarray}
Up to now we have only re-written the sum $S_A$. Applying eq. (\ref{sum1bis}) to the sum over $\epsilon$, decomposing the sum over 
$\delta$ in two different sums, and using for these sums the solutions expressed in eqs. (\ref{sum1bis}) and (\ref{sum3bis}),
we find the solution,
\begin{equation}
S_A=\frac{X_3^3Y^2}{(1-X_2Y)^2} - \frac{X_3^3X_2^{-1}Y}{1-X_2Y} \ln (\frac{1}{1-X_2Y})
\end{equation}
In the same way we have calculate this first sum, we can calculate the rest of terms on the right hand side of eq. (\ref{Pi2sum}),
being most of the calculations as simple as the one we have solved previously. The only sum which shows some degree of complexity 
is that associated to an interfacial spin column of type C (see Table \ref{configs}), i.e. $\delta >0$ and $\epsilon <0$.
This sum, called from now on $S_C$, corresponds to the third term on the right hand side of eq. (\ref{Pi2sum}),
\begin{equation}
S_C = \sum_{\delta=1}^{+\infty} \sum_{\epsilon=-\infty}^{-1} (1-\frac{\alpha}{\lambda})Y^{\delta+\epsilon}
X_4^3X_3^{2(\alpha-1)}X_2^{\lambda-\alpha} \nonumber
\end{equation}
where we must remember that $\alpha=\min(|\delta|,|\epsilon|)$ and $\lambda=\max(|\delta|,|\epsilon|)$. This sum can be written as,
\begin{eqnarray}
S_C & = & \frac{X_4^3}{X_3^2} \sum_{\delta=1}^{+\infty} \Big\{ (X_2Y)^{\delta} \sum_{\epsilon=1}^{\delta} 
(1-\frac{\epsilon}{\delta})(\frac{X_3^2}{X_2Y})^{\epsilon} \nonumber \\
 & & + (\frac{X_3^2Y}{X_2})^{\delta}
\sum_{\epsilon=\delta+1}^{+\infty} (1-\frac{\delta}{\epsilon})(\frac{X_2}{Y})^{\epsilon} \Big\}
\end{eqnarray}
Using eqs. (\ref{sum1bis})-(\ref{sum6bis}) we can perform the sum over $\epsilon$, yielding,
\begin{eqnarray}
S_C & = & \frac{X_4^3}{X_3^2} \sum_{\delta=1}^{+\infty} \Big\{ (X_2Y)^{\delta} 
\big[\frac{X_3^2X_2^{-1}Y^{-1} 
}{1-X_3^2X_2^{-1}Y^{-1}}
 - \frac{1}{\delta}\frac{X_3^2X_2^{-1}Y^{-1} - (X_3^2X_2^{-1}Y^{-1})^{\delta+1}}{(1-X_3^2X_2^{-1}Y^{-1})^2} 
\big] \nonumber \\
 & + & (\frac{X_3^2Y}{X_2})^{\delta}\big[\frac{(X_2Y^{-1})^{\delta+1}}{1-X_2Y^{-1}} - 
\delta \int_o^{X_2Y^{-1}}\textrm{d}y \frac{y^{\delta}}{1-y}\big]\Big\}
\end{eqnarray}
At this point we can perform easily the sums over $\delta$, taking into account that we can interchange the sum and the integral
(the Riemann integration is a linear application). Summing over $\delta$ we arrive to,
\begin{eqnarray}
S_C & = & \frac{X_4^3}{X_3^2} \Big\{ \frac{X_3^2}{(1-X_3^2X_2^{-1}Y^{-1})(1-X_2Y)} + 
\frac{X_3^2X_2^{-1}Y^{-1}}{(1-X_3^2X_2^{-1}Y^{-1})^2} \ln (\frac{1-X_2Y}{1-X_3^2}) \nonumber \\
 & + & \frac{X_3^2X_2Y^{-1}}{(1-X_2Y^{-1})(1-X_3^2)}
 - \int_0^{X_2Y^{-1}}\textrm{d}y \frac{X_3^2YX_2^{-1}y}{(1-y)(1-X_3^2YX_2^{-1}y)^2} \big\}
\label{Scint}
\end{eqnarray}
In order to finish the calculation of $S_C$ we only have to evaluate the integral. This integral is of the type,
\begin{equation}
zI \equiv z \int_0^x \textrm{d}y \frac{y}{(1-y)(1-zy)^2} \nonumber
\end{equation}
where in this case  $z=X_3^2X_2^{-1}Y$ and $x=X_2Y^{-1}$. Integrating $I$ by parts, choosing $u=y/(1-y)$ and 
$\textrm{d}v=\textrm{d}y/(1-zy)^2$, we obtain,
\begin{equation}
I=\frac{x}{z(1-x)(1-zx)} - \frac{1}{z} \int_0^{x} \textrm{d}y \frac{1}{(1-zy)(1-y)^2}
\label{Iint}
\end{equation}
We call $I_2$ to the integral appearing in the second term on the right hand side of the above equation. This last integral can be 
solved analytically, knowing that generically\cite{Gradshteyn},
\begin{eqnarray}
\int \textrm{d}y \frac{1}{(a+by)^n(\alpha+\beta y)^m} & = & - \frac{1}{(m-1)(a\beta-\alpha b)} 
\frac{1}{(a+by)^{n-1}(\alpha+\beta y)^{m-1}} \nonumber \\
 & - & \frac{(m+n-2)b}{(m-1)(a\beta-\alpha b)}
\int \textrm{d}y \frac{1}{(a+by)^n(\alpha+\beta y)^{m-1}} \nonumber
\end{eqnarray}
Applying this expression to $I_2$, taking into account that in this case $a=1$, $b=-z$, $\alpha=1$, $\beta=-1$, $m=2$ and $n=1$,
we arrive to,
\begin{equation}
I_2 = \frac{x}{(1-z)(1-x)} + \frac{z}{(1-z)^2} \ln (\frac{1-x}{1-zx}) \nonumber
\end{equation}
where we have already analytically solved the last remaining integral. Using this result in the expression for $I$, eq. (\ref{Iint}),
and applying the result of this substitution to the previously calculated expression for $S_C$, eq. (\ref{Scint}), and taking into
account that $z=X_3^2X_2^{-1}Y$ and $x=X_2Y^{-1}$, we finally obtain the sum $S_C$,
\begin{eqnarray}
S_C & = & \frac{X_4^3}{X_3^2} \Big\{ \frac{X_3^2}{(1-X_3^2X_2^{-1}Y^{-1})(1-X_2Y)}
 + \frac{X_3^2X_2^{-1}Y^{-1}}{(1-X_3^2X_2^{-1}Y^{-1})^2} \ln (\frac{1-X_2Y}{1-X_3^2}) \nonumber \\
 & + & \frac{X_3^2X_2Y^{-1}}{(1-X_2Y^{-1})(1-X_3^2)} + \frac{X_2Y^{-1}}{(1-X_2Y^{-1})(1-X_3^2X_2^{-1}Y)} \nonumber \\
 & + & \frac{X_3^2X_2^{-1}Y}{(1-X_3^2X_2^{-1}Y)^2} \ln (\frac{1-X_2Y^{-1}}{1-X_3^2}) 
 - \frac{X_2Y^{-1}}{(1-X_2Y^{-1})(1-X_3^2)} \Big\}
\end{eqnarray}
In a similar way we are able to calculate the rest of sums involved in the expression for $\Pi_2(T,p)$, finally obtaining the result
shown in eq. (\ref{pi2}). The calculations needed in order to evaluate $\Pi_4(T,p)$ are even simpler, so we do not include them
in this appendix.

\chapter{Avalanche Size Distribution for a Flat Domain Wall}
\label{apendAval}

In this appendix we calculate the avalanche size distribution associated to a flat
domain wall in our nonequilibrium ferromagnetic system (see Chapter \ref{capAval}).
These (small) avalanches characterize the so-called {\it extrinsic noise}\cite{Spaso} 
that appears in the avalanche size distributions associated to the ferromagnetic circular nanoparticle.
In order to perform the calculation we need again the concept of spin class. Remembering
from section \ref{capMedio_apEstat_subCoer}, if we have a spin $s$ in our lattice, the spin class 
to which this spin belongs to is defined once we  know the spin orientation, $s=+1$
or $s=-1$, and its number of up nearest neighbors, $n\in [0,4]$. Therefore, for the 
two-dimensional isotropic Ising model subject to periodic boundary conditions there are 
$10$ different spin classes (see Table \ref{tabclasesrates}).
The last column in Table \ref{tabclasesrates} shows the Glauber transition rate, see eq. 
(\ref{rate}), for each spin class once fixed the parameter values to $T=0.11T_{Ons}\approx 0.25$, 
$p=10^{-6}$ and $h=-0.1$. Remember also that $J=1$ and $\beta=1/T$.

\begin{table}[t!]
\centerline{
\begin{tabular}{|c||c|c|c|}
\hline \hline
Class & Central spin & Number of up neighbors & Transition rate\\
\hline \hline
1 & +1 & 4 & $\sim 10^{-6}$ \\
\hline
2 & +1 & 3 & $\sim 1.24\times 10^{-6}$ \\
\hline
3 & +1 & 2 & $\sim 0.69$ \\
\hline
4 & +1 & 1 & $\sim 1$ \\
\hline
5 & +1 & 0 & $\sim 1$ \\
\hline \hline
6 & -1 & 4 & $\sim 1$ \\
\hline
7 & -1 & 3 & $\sim 1$ \\
\hline
8 & -1 & 2 & $\sim 0.31$ \\
\hline
9 & -1 & 1 & $\sim 1.05\times 10^{-6}$ \\
\hline
10 & -1 & 0 & $\sim 10^{-6}$ \\
\hline \hline
\end{tabular}
}
\caption[Spins classes and Glauber rate for the 2d Ising model.]
{\small Spin classes for the two-dimensional isotropic Ising model 
with periodic boundary conditions. The last column shows the approximate value
of the Glauber transition rate, eq. (\ref{rate}), for $T=0.11T_{ons}$, $p=10^{-6}$ 
and $h=-0.1$.}
\label{tabclasesrates}
\end{table}

Let us assume now that we have our nonequilibrium magnetic system defined on a square lattice with size 
$L\times L$, subject to periodic boundary conditions along the horizontal direction and 
open boundary conditions in the vertical direction. The initial condition consists in a stripe
of down spins with height $L/2$ situated in the lower part of the system, and a complementary
up spins stripe situated in the upper part of the system. That is, the initial condition
comprises two bulk phases (up and down) and a perfectly flat interface between them. Under the action
of a negative magnetic field the interface moves upwards on average.
This system corresponds to the semi-infinite system introduced in sections \ref{capNuc_apExt_subVel} 
and \ref{capAval_apMC_subGlob} when studying the interface growth velocity and the extrinsic noise 
respectively. The probability per unit time of changing a spin in class $i$ is,
\begin{equation}
r_i(t)=\frac{n_i(t)}{N} \omega_i
\label{probapend}
\end{equation}
where $n_i(t)$ is the number of spins in class $i$ at time $t$, $\omega_i$ is the transition rate associated
to class $i$ (see section \ref{capMedio_apEstat_subCoer}) and $N=L^2$ is the total number of spins in the system.
For our initial condition the only occupied classes are class 1 (up bulk spins), class 2 (up interfacial spins),
class 9 (down interfacial spins) and class 10 (down bulk spins). For low temperatures and $-2J\leq h \leq 0$
the $\omega_i$ values for these four spin classes are very small (see Table \ref{tabclasesrates}). In fact,
we have that $\omega_{1,10}\sim 10^{-6}$, $\omega_2\sim 1.24\times 10^{-6}$ and $\omega_9\sim 1.05\times 10^{-6}$.
Since for the initial condition we have that $n_1, n_{10} \gg n_2, n_9$, there will be many bulk fluctuations 
before a fluctuation in the interface appears.

\begin{figure}[t]
\centerline{
\psfig{file=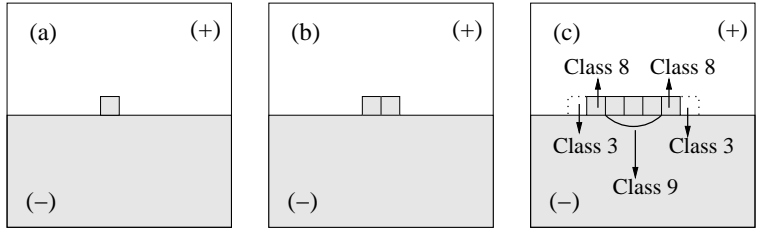, width=10cm}}
\caption[Origin and growth mechanism of a lateral interfacial avalanche.]
{\small Schematic plot of a flat interface and the origin and growth mechanism of a
lateral interfacial avalanche, as described in the main text. Notice in (c) that during the lateral avalanche
evolution the only relevant spins are the two spins in class 8 and the two spins in class 3.}
\label{avalsmall}
\end{figure}

However, when an interfacial fluctuation appears a lateral avalanche takes place. Let us assume that 
an interfacial fluctuation appears in the form depicted in Fig. \ref{avalsmall}.a, where one up interfacial
spin has been flipped. Now this spin belongs to class 7, 
and its two nearest neighbor spins in the direction of the 
interface belong to class 3. Since $\omega_{3} \approx 0.69$ and $\omega_{7} \approx 1$, in this case we have 
that $r_3 \approx 2\times 0.69/N$ and $r_7\approx 1/N$. These probabilities must be compared with 
$r_1 \approx N^{-1}[(L/2-1)L-1]\times 10^{-6}$, $r_2 \approx 1.24\times N^{-1}(L-2)\times 10^{-6}$, 
$r_9\approx 1.05\times N^{-1}(L-1)\times 10^{-6}$ and $r_{10} \approx N^{-1}[(L/2-1)L+1]\times 10^{-6}$. 
Hence, for feasible system sizes as those we use in our simulations, and for our fixed set of parameters, 
the probability of lateral growth of the initial fluctuation 
($2\times 0.69/N$) and the probability of destroying the 
fluctuation ($1/N$) are much larger than the probability of finding any other spin flip in the system.
This argument allows us to safely assume that once the interfacial perturbation has appeared, the system
dynamics can be reduced to the growth and shrinkage dynamics of the interfacial perturbation. Under this
assumption, the most probable process to be observed consists in the growth of the interfacial fluctuation
via the flipping of the lateral spins in class 3 which surround the fluctuation, until one of the two spins
in class 8 which delimit the lateral size of the interfacial fluctuation flips, stopping in this way
the fluctuation growth. We call this process and {\it avalanche}, and our aim now consists in calculating
the size distribution for these lateral interfacial avalanches.

Therefore we suppose that the system is in a state as that shown in Fig. \ref{avalsmall}.b, with two up 
interfacial spins flipped, and we want to know the probability of finding a lateral avalanche as
the one described above with size $\Delta_m<0$ 
(it is an avalanche {\it in the direction} of the magnetic field, $h<0$),
where the size of the avalanche is defined as the number of spins it involves. Following our assumptions, 
the only way an avalanche can grow is through the flip of any of the two spins in class 3 adjacent to the cluster
forming the avalanche (see Fig. \ref{avalsmall}.c). This avalanche will stop once any of the two spins
in class 8 at the lateral border of the avalanche cluster flips. All other spin flip processes in the system
have a negligible probability of being observed as compared to these two processes.

The restricted dynamics we propose here only involves four different spins (two of class 3 and two of class 8).
Hence the probabilities of flipping a spin in class 3 or 8 are now, respectively,
\begin{eqnarray}
r_3 & = & \frac{1}{2}\big[p+(1-p)\frac{\text{e}^{2\beta|h|}}{1+\text{e}^{2\beta|h|}}\big] \\
r_8 & = & \frac{1}{2}\big[p+(1-p)\frac{\text{e}^{-2\beta|h|}}{1+\text{e}^{-2\beta|h|}}\big]
\label{r3r8apend}
\end{eqnarray}
where we have used the general form of Glauber rate for arbitrary values of $T$, $p$ and $h$, 
once specified for classes 3 and 8. Since we have modified the system original dynamics, we must normalize 
again these probabilities. Applying the normalization condition, we obtain the avalanche growth and stop 
probabilities,
\begin{eqnarray}
p_{grow} & = & \frac{1}{1+p}\big[p+(1-p)\frac{\text{e}^{2\beta|h|}}{1+\text{e}^{2\beta|h|}}\big] \label{grow}\\
p_{stop} & = & \frac{1}{1+p}\big[p+(1-p)\frac{\text{e}^{-2\beta|h|}}{1+\text{e}^{-2\beta|h|}}\big]
\label{stop}
\end{eqnarray}
The probability of finding a lateral avalanche of size $\Delta_m <0$ is 
$P(\Delta_m)=p_{grow}^{|\Delta_m|}p_{stop}$, that is,
\begin{eqnarray}
P(\Delta_m<0) & = & \frac{\big[p+(1-p)\frac{\text{e}^{-2\beta|h|}}{1+\text{e}^{-2\beta|h|}}\big]}{1+p} \times
\Big\{\frac{1}{1+p}\big[p+(1-p)\frac{\text{e}^{2\beta|h|}}{1+\text{e}^{2\beta|h|}}\big] 
\Big\}^{|\Delta_m|} \nonumber \\
 & = & \frac{1}{1+p}\big[p+(1-p)\frac{\text{e}^{-2\beta|h|}}{1+\text{e}^{-2\beta|h|}}\big] \times 
\text{e}^{-|\Delta_m|/\bar{\Delta}_m^{(-)}}
\label{probavaldown}
\end{eqnarray}
where the typical size characterizing avalanches in the field direction can be written as,
\begin{equation}
\bar{\Delta}_m^{(-)} = \frac{1}{\displaystyle \ln \big[\frac{(1+p)(1+\text{e}^{2\beta|h|})}{p+
\text{e}^{2\beta|h|}}\big]}
\label{scaledown}
\end{equation}
\begin{figure}[t]
\centerline{
\psfig{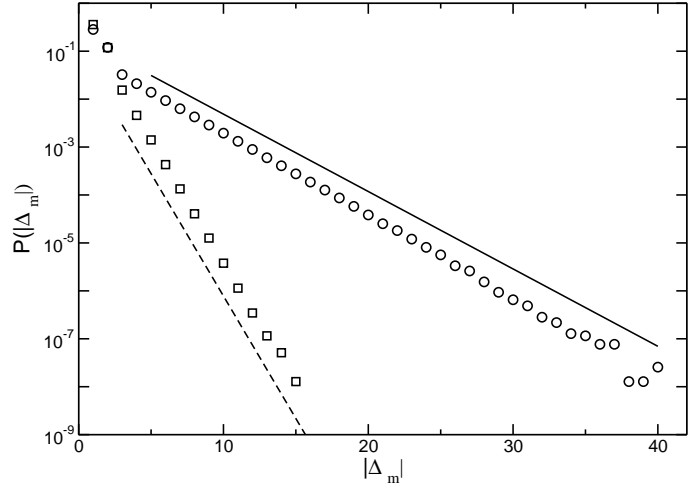}}
\caption[Small avalanche size distributions and theory.]
{\small Semilog plot of the avalanche size distributions $P(\Delta_m>0)$ (dashed line) 
and $P(\Delta_m<0)$ (continuous line) as obtained from our calculations for $T=0.11T_{ons}$, $p=10^{-6}$ and 
$h=-0.1$. The points are Monte Carlo results for the avalanche size distributions $P(\Delta_m>0)$ ($\square$) and 
$P(\Delta_m<0)$ ($\bigcirc$) for the semi-infinite system described in the main text with $L=53$. The agreement
between the theoretical prediction and Monte Carlo results is excellent.}
\label{distribupdown}
\end{figure}
In a similar way we can calculate the size distribution and the typical size of positive avalanches, 
$P(\Delta_m>0)$ and $\bar{\Delta}_m^{(+)}$ respectively. In this 
case the avalanche growth probability is given by eq. (\ref{stop}) and the avalanche stop probability is 
given by eq. (\ref{grow}). The result for the typical size is,
\begin{equation}
\bar{\Delta}_m^{(+)} = \frac{1}{\displaystyle \ln \big[\frac{(1+p)(1+\text{e}^{-2\beta|h|})}{p+
\text{e}^{-2\beta|h|}}\big]}
\label{scaleup}
\end{equation}
Fig. \ref{distribupdown} shows the avalanche size probability distributions $P(\Delta_m<0)$ and $P(\Delta_m>0)$
as obtained from the previous calculations, see eq. (\ref{probavaldown}) and its equivalent for $\Delta_m>0$,
for $T=0.11T_{ons}$, $p=10^{-6}$ and $h=-0.1$. This figure also shows Monte Carlo results for the avalanche size
distribution in the semi-infinite system described above with size $L=53$. As we observe in this figure, the 
agreement between the predictions and Monte Carlo results is excellent.

This agreement allows us to state the origin of the extrinsic noise in our nonequilibrium magnetic system.
As we have deduced previously in this appendix, the small avalanches which define the extrinsic noise are
just local random fluctuations of an advancing flat domain wall. The extrinsic noise is thus an intrinsic
property of the magnetic system, and it has nothing to do with the presence of free boundaries, as opposed
to large avalanches in the circular nanoparticle defined in Chapter \ref{capAval}, whose origin is intimately
related to the presence of free borders and its interplay with the domain wall.

\chapter{Variations of Lipowski Model}
\label{apendLipo}

In this appendix we briefly discuss some variations of Lipowski model not 
studied in Chapter \ref{capLipo}. In particular, we introduce some
results obtained for Lipowski model with parallel updating dynamics,
and we also comment on a multiplicative version of this model.

\section{Lipowski Model with Synchronous Updating}

 As an alternative attempt to speed up the dynamics, and examine further
some properties of the two-dimensional model, we have implemented
the microscopic dynamics replacing the original sequential updating by a
 synchronous or parallel one, i.e.
all active sites are ``deactivated'' simultaneously at each
Monte Carlo step, and all their associated bonds are replaced by
new random variables simultaneously. In this way, as random numbers do not
have to be extracted to sequentially select sites,
the dynamics is largely accelerated.
For this modified dynamics, we have examined some relatively
large system sizes, $L=256$, and concluded that the nature of the
transition is changed with respect to the sequential
updating case: in this case the transition is first order
and critical exponents cannot be defined. To show that this is the
case, in figure 9 we present the stationary activity curve.
The upper curve corresponds to simulations performed taking
an initial activity-density equal to unity. 
On the other hand, the lower curve is obtained by starting the system
with a natural absorbing configuration, and activating on the top 
of it a small percentage of sites (about a ten percent).   
\begin{figure}
\centerline{
\psfig{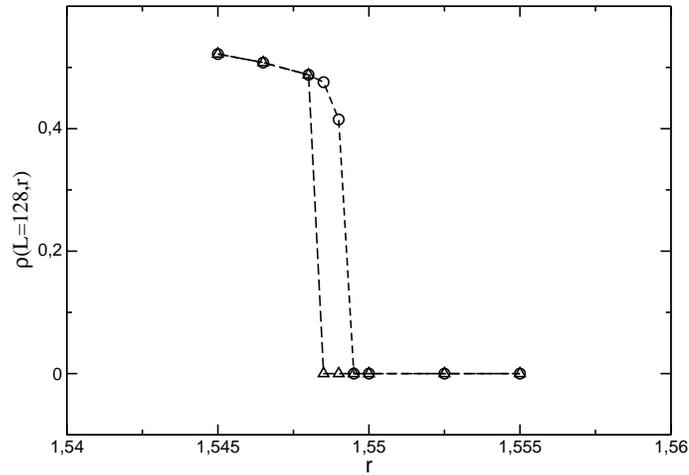}}
\caption[Discontinuous transition in parallel Lipowski model.]
{\small Order parameter as a function of $r$ in the case
of parallel updating. The transition appears to be
 discontinuous in this case, exhibiting also 
a hysteresis loop.
}
\label{hysteresis}
\end{figure}

For values of $r$ in the interval $[\approx 1.545,
 \approx 1.555]$ the system reaches
different states depending upon the initial condition.
The presence of a hysteresis loop is a trait of the
transition first-order nature.
First order absorbing state transitions have been observed in other
contexts \cite{firstorder}.  However, we caution the reader that,
as the transition is found to occur at a value of $r$ for which the
probability of creating superabsorbing
sites is very large  (much larger than in the sequential case),
 and the dynamics is therefore extremely
anomalous and slow, it could be the case that the first order
character of the transition is only apparent. Extracting clean, conclusive
results in the critical zone is a computationally very expensive task, that
we have not pursued.

\section{Multiplicative Version of Lipowski Model}

Very recently, Lipowski has introduced a multiplicative version of his model
on the square lattice in which sites are declared active
 if the product of the four adjacent bonds
is smaller than a certain value of the control parameter $r$  \cite{Lip3}.
  Bonds take uncorrelated values in the interval $[-0.5, 0.5]$
extracted from a homogeneous distribution.
For values of $r$ smaller than $r=0$ 
there is a finite (not small) probability to
generate superabsorbing sites.
 In this case, it is not difficult to see that
isolated superabsorbing sites remain frozen forever.
 In analogy with the discussion of the honeycomb-lattice model,
a first order transition is expected at
 $r_c=0$ (as discussed also in \cite{Lip3}).
However, in this case, as the probability to create superabsorbing sites
is not negligible, the first order transition is actually observable.
  Based on a numerical measurement of $\beta$
 Lipowski concludes that the model
shares first-order properties with second-order features. 
In particular, 
the transition is clearly shown to be discontinuous,
 there is no diverging correlation length,
but $\beta$ is claimed to be however in the two-dimensional DP class.
 Our guess
is  that this apparent puzzle is simply due to a numerical 
coincidence, and that in fact there is no trait of any second-order phase
transition feature (observe that the fit
for beta in \cite{Lip3} spans for less than half a decade 
in the abscise of the log-log plot).

\backmatter
\chapter{Published work}

\begin{itemize}
\item Pablo I. Hurtado, Miguel A. Mu\~noz, \emph{Systems with Superabsorbing States},
Phys. Rev. E {\bf 62}, 4633 (2000).

\item Pedro L. Garrido, Pablo I. Hurtado, Bjoern Nadrowski, \emph{Simple One-dimen-sional Model 
of Heat Conduction which Obeys Fourier's Law}, Phys. Rev. Lett. {\bf 86}, 5486 (2001).

\item Pedro L. Garrido, Pablo I. Hurtado, \emph{Reply to Comment on Simple One-dimensional 
Model of Heat Conduction which Obeys Fourier's Law by A. Dhar}, Phys. Rev. Lett. {\bf 88}, 249402 (2002).

\item Pedro L. Garrido, Pablo I. Hurtado,
\emph{Reply to Comment on Simple One-dimensional Model of Heat Conduction which Obeys 
Fourier's Law by H. Li et al},
Phys. Rev. Lett. {\bf 89}, 079402 (2002).

\item Joaqu\'{\i}n Marro, Jes\'us Cort\'es, Pablo I. Hurtado,
\emph{Modeling Nonequilibrium Phase Transitions and Critical Behavior in Complex Systems},
Comp. Phys. Comm. {\bf 147}, 115 (2002).

\item Pablo I. Hurtado, J. Marro, Ezequiel V. Albano,
\emph{Growth and Scaling in Anisotropic Spinodal Decomposition},
Europhys. Lett. {\bf 59}(1), 14 (2002).

\item Pablo I. Hurtado, J. Marro, Pedro L. Garrido, Ezequiel V. Albano,
\emph{Coarsening under Anisotropic Conditions in a Lattice Gas Model},
in {\it Modeling complex systems: Seventh Granada Lectures on Computational Physics}, 
Pedro L. Garrido, Joaqu\'{\i}n Marro (eds.), AIP Conference Proceedings (2002).

\item Pablo I. Hurtado, J. Marro, Pedro L. Garrido,
\emph{Metastability and Avalanches in a Nonequilibrium Ferromagnetic System},
in {\it Modeling complex systems: Seventh Granada Lectures on Computational Physics}, 
Pedro L. Garrido, Joaqu\'{\i}n Marro (eds.), AIP Conference Proceedings (2002).

\item Pablo I. Hurtado, J. Marro, Pedro L. Garrido, Ezequiel V. Albano,
\emph{Kinetics of Phase Separation in the Driven Lattice Gas: Self-Similar
Pattern Growth under Anisotropic Nonequilibrium Conditions},
accepted for publication in Phys. Rev. B (2002).

\item Pablo I. Hurtado, J. Marro, Pedro L. Garrido,
{\it Origin of the $1/f$ Noise in a Model of Disorder}, submitted to Phys. Rev. Lett. (2002).

\end{itemize}

\vskip 0.3cm

{IN PREPARATION}

\begin{itemize}

\item Pablo I. Hurtado, Joaqu\'{\i}n Marro, Pedro L. Garrido,
\emph{Metaestability in Ferromagnetic Systems under Nonequilibrium Conditions} (2003).

\item Pablo I. Hurtado, Joaqu\'{\i}n Marro, Pedro L. Garrido,
\emph{Generalization of the Solid-on-Solid Approximation to Nonequilibrium
Interfaces: Microscopic Structure, Surface Tension and Crystal Shape} (2003).

\item Pablo I. Hurtado, Joaqu\'{\i}n Marro, Pedro L. Garrido,
{\it Hysteresis and the Intrinsic Coercive Field in Nonequilibrium Ferromagnetic Systems} (2003).


\end{itemize}

\vskip 0.3cm

{ABSTRACTS}

\begin{itemize}

\item Pablo I. Hurtado, Miguel A. Mu\~noz,
\emph{Systems with Superabsorbing States},
Proc. of the Workshop on Surface Science and Porous Media, San Luis, Argentina (2000).

\item Pablo I. Hurtado, J. Marro, P.L. Garrido,
\emph{Impure Ferromagnetic Nanoparticles: on the 1/f Noise during Decay from Metastable
States}, in {\it SIMU Conference: Bridging the Time Scale gap}, Abstract Book (2001).

\item Pablo I. Hurtado, Miguel A. Mu\~noz,
\emph{Lipowski Model and Superabsorbing States},
in {\it Modeling complex systems: Sixth Granada Lectures on Computational Physics}, 
Pedro L. Garrido, Joaqu\'{\i}n Marro (eds.), AIP Conference Proceedings {\bf 574} (2001).

\item Pablo I. Hurtado, J. Marro, E.V. Albano,
\emph{Kinetics of Phase Separation in the Driven Lattice Gas: Pattern Gowth and
Scaling Laws under Anisotropic Nonequilibrium Conditions},
in {\it CMD19CMMP 2002-Condensed Matter and Materials Physics}, Europhysics Conference 
Abstracts {\bf 26A} (2002).

\end{itemize}

\end{document}